\documentclass[a4paper,11pt]{article}

\PassOptionsToPackage{authoryear}{natbib}
\usepackage{jheppub}
 
\makeatletter
\def\@fpheader{}
\makeatother
\setcitestyle{authoryear,round,semicolon,aysep={,},yysep={,}}
\makeatother

\makeatletter
\providecommand{\@fpheader}{} % prevent maketitle crash if style file lacks it
\makeatother

\usepackage[dvipsnames]{xcolor}
\definecolor{skyblue}{rgb}{0.229,0.708,0.922}
\hypersetup{
  colorlinks=true,
  linkcolor=black,
  citecolor=skyblue,
  urlcolor=skyblue
}

\usepackage{amsmath,amssymb,mathtools}
\usepackage{graphicx}
\usepackage{caption}
\usepackage{longtable}

\usepackage[capitalize,nameinlink]{cleveref}

\usepackage{enumitem}
\usepackage{multirow}
\usepackage{float}
\usepackage{dcolumn}
\usepackage{booktabs}
\usepackage{transparent}
\usepackage{extpfeil}
\usepackage{import}
\usepackage[most]{tcolorbox}
\tcbuselibrary{breakable,skins}
\usepackage{empheq}
\usepackage{tablefootnote}

\usepackage{braket}

\newcommand{\nint}[1]{\left\lfloor #1 \right\rceil}

\usepackage{tabularx}
\crefname{equation}{Eq.}{Eqs.}
\crefname{figure}{Fig.}{Figs.}
\crefname{section}{Sec.}{Secs.}
\crefname{table}{Table}{Tables}

\preprint{DESY-25-162}
\newcommand{\bq}{\boldsymbol{q}}
\newcommand{\mr}{\mathrm}
\newcommand{\mpl}{M_{\mr{Pl}}}

\newcommand{\be}{\begin{eqnarray}}
\newcommand{\bse}{\begin{subequations}}
\newcommand{\ese}{\end{subequations}}\newcommand{\bea}{\begin{eqnarray}}
\newcommand{\eea}{\end{eqnarray}}
\newcommand{\Beq}{\begin{eqnarray}}
\newcommand{\Eeq}{\end{eqnarray}}
\newcommand{\ba}{\begin{array}}
\newcommand{\ea}{\end{array}}
\newcommand{\ee}{\end{eqnarray}}

\def\p{\partial}
\def\bz{{\bar z}}

\newcommand{\Th}{\tilde{h}}

\newcommand{\mH}{\mathcal{H}}

\newcommand{\bk}{\boldsymbol{k}}
\newcommand{\bx}{\boldsymbol{x}}

\def\dre_g{\delta\rho_g}
\def\dpe_g{\delta P_g}
\def\dqe_g{\delta q_g}
\def\dre{\delta\rho}
\def\dpe{\delta P}
\def\dqe{\delta q}

\def\mH{\mathcal{H}}

\def\YM1{\frac{\dot\phi^2}{a^2}}
\def\YM2{\frac{g^2\phi^4}{a^4}}

\def\p{\partial}
\def\bz{{\bar z}}

\newcommand{\where}{\quad \textmd{where} \quad}
\newcommand{\an}{\quad \textmd{and} \quad}

\usepackage{amsmath,amssymb,mathtools}
\usepackage{graphicx}
\usepackage{caption}

\usepackage{longtable}

\usepackage[capitalize,nameinlink]{cleveref}

\usepackage{enumitem}
\usepackage{multirow}
\usepackage{float}
\usepackage{dcolumn}
\usepackage{booktabs}
\usepackage{transparent}
\usepackage{extpfeil}
\usepackage{import}
\usepackage[most]{tcolorbox}
\tcbuselibrary{breakable,skins}
\usepackage{empheq}
\usepackage{tablefootnote}

\usepackage{braket}

\usepackage{tabularx}
\crefname{equation}{Eq.}{Eqs.}
\crefname{figure}{Fig.}{Figs.}
\crefname{section}{Sect.}{Sects.}
\crefname{table}{Table}{Tables}

\def\p{\partial}
\def\bz{{\bar z}}

\def\dre_g{\delta\rho_g}
\def\dpe_g{\delta P_g}
\def\dqe_g{\delta q_g}
\def\dre{\delta\rho}
\def\dpe{\delta P}
\def\dqe{\delta q}

\def\mH{\mathcal{H}}

\def\YM1{\frac{\dot\phi^2}{a^2}}
\def\YM2{\frac{g^2\phi^4}{a^4}}

\def\p{\partial}
\def\bz{{\bar z}}

\title{When Geometry Radiates:\\
\Large Gravitational Waves in Theory, Cosmology, and Observation}

\author[a,b,c]{Azadeh Maleknejad}

\affiliation[a]{Centre for Quantum Fields and Gravity, Swansea University,
Swansea SA2 8PP, United Kingdom}

\affiliation[b]{Deutsches Elektronen-Synchrotron DESY,
Notkestra\ss e 85, 22607 Hamburg, Germany}

\affiliation[c]{Institute of Theoretical Physics, Universit\"at Hamburg,
22761 Hamburg, Germany}

\emailAdd{azadeh.maleknejad@swansea.ac.uk}
\abstract{Gravitational waves provide a unique window into gravity, cosmology, and high-energy physics, enabling the exploration of fundamental phenomena across a wide range of scales. This review presents a coherent and pedagogical framework that bridges foundational theory with observational frontiers. We begin by developing the theory of gravitational radiation within linearized general relativity, deriving gravitational waves as solutions to the linearized Einstein equations and clarifying their physical interpretation, polarization states, and key properties.
We then deepen the discussion through a more rigorous geometric perspective, tracing the connection between gravitational radiation and the algebraic structure of the Weyl tensor. This naturally leads to the Bondi–Sachs formalism, which provides a precise definition of gravitational waves in asymptotically flat spacetimes and establishes their role as genuine carriers of energy and angular momentum. Extending beyond flat backgrounds, we examine gravitational waves in an expanding universe, following their evolution across cosmological epochs and their generation during inflation. Within this setting, we explore adiabatic modes and consistency relations, which illuminate the infrared structure and universal behavior of long-wavelength perturbations, and derive the inflationary spectrum of vacuum gravitational waves together with their contribution to the integrated Sachs–Wolfe effect. The review further surveys the main observational strategies for detecting gravitational waves across a broad frequency range, including cosmic microwave background polarization, pulsar timing arrays, laser interferometers on Earth and in space, and resonant cavity–based detectors. We then discuss the astrophysical and cosmological mechanisms responsible for generating gravitational radiation. We conclude by summarizing the current status of the field and outlining promising directions for future theoretical and observational developments.}

\keywords{Gravitational radiation, Asymptotically flat space, Cosmology, Geometrical \& algebraic aspects of gravitational waves, Sources of gravitational waves, Detection of gravitational waves}

\begin{document}

\maketitle
%\tableofcontents

%%%%%%%%%%%%%%%%%%%%%%%%%%%%%%%%

% Introduction

\section{Introduction and motivation}
\label{sec:intro}

Gravitational waves (GWs) were first predicted by Einstein in 1916 as ripples in spacetime propagating at the speed of light, arising from the dynamical degrees of freedom of the metric in general relativity \citep{Einstein:1916cc}. They embody a fascinating interplay of physics, mathematics, and phenomenology, yet their physical reality was debated for decades. In the 1930s, Einstein himself briefly doubted the physical reality of gravitational waves due to a misinterpretation of early solutions.
He later revised his view, though questions about their true physicality remained. By the late 1950s and early 1960s, work by Bondi, Pirani, and others had clarified the physical nature of gravitational radiation.
Bondi’s analysis at null infinity demonstrated that gravitational waves carry energy and can have measurable physical consequences \citep{Bondi:1957dt}.
This marked the turning point in establishing gravitational waves as real, observable phenomena in general relativity that carry energy and angular momentum away from isolated systems.

With their physical interpretation rigorously established, attention turned toward their astrophysical sources and their detection. Indirect evidence arrived with the Hulse–Taylor binary pulsar in 1974 \citep{Hulse:1975}, whose orbital decay precisely matched the prediction for GW emission. Early detection efforts began with Weber’s resonant bars and gradually evolved into large-scale interferometers, propelled by the vision of Weiss, Drever, and Thorne. These decades of theoretical and experimental progress culminated in the landmark direct detection by  Advanced LIGO in 2015 \citep{Abbott:2016blz}, an achievement that inaugurated the era of GW astronomy.

Gravitational waves provide a powerful probe of high-energy phenomena beyond the Standard Model (BSM) and a unique window into the unexplored frontiers of gravity and the early Universe. As messengers from the earliest moments of cosmic history, they can access energy scales far beyond the reach of terrestrial accelerators, offering potential insights into new physics and even the quantum nature of spacetime. Stochastic gravitational wave backgrounds may originate from inflationary fluctuations \citep{Komatsu:2022nvu,LISACosmologyWorkingGroup:2022jok}, violent post-inflationary dynamics such as preheating \citep{Caprini:2018mtu}, or relics of high-energy phase transitions and topological defects \citep{Hindmarsh:2013xza,Dimitriou:2025bvq}. Intriguingly, in certain scenarios, GWs may do more than carry information about early-Universe processes; they may also influence the production of matter \citep{Alexander:2004us,Maleknejad:2014wsa,Garny:2015sjg,Maleknejad:2016dci,Caldwell:2017chz,Alexander:2018fjp,Bernal:2018qlk,Clery:2021bwz,Maleknejad:2024hoz,Maleknejad:2024ybn,Garani:2025qnm,Maleknejad:2024vvf}.

\begin{figure}[h]
\centering
\includegraphics[height=0.35\textwidth]{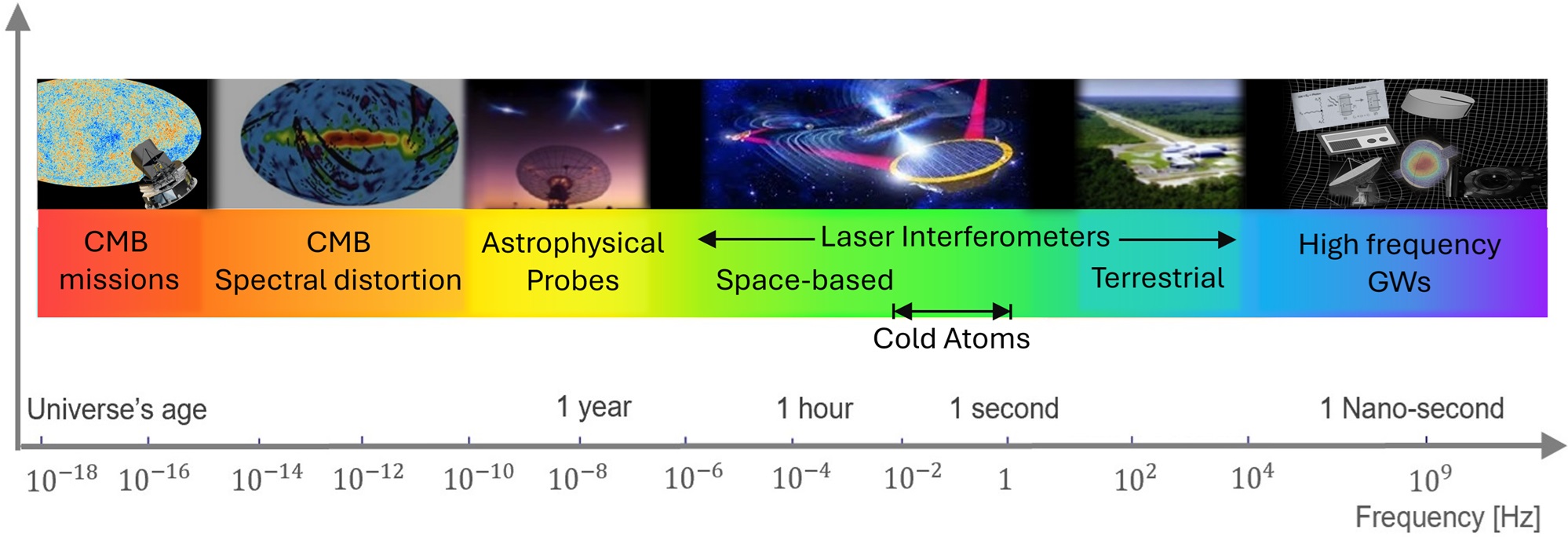}
\caption{Gravitational rainbow.  At the lowest frequencies, measurements of the cosmic microwave background (CMB) provide the most sensitive probe of gravitational waves. In the nanohertz range, pulsar timing arrays serve as powerful astrophysical detectors. Laser interferometers probe frequencies up to the kilohertz range, while atom-interferometer concepts target the mid-frequency band. Finally, we enter the realm of high-frequency gravitational waves above 10 kHz, where detection demands innovative techniques.}\label{fig:GW-rainbow}
\end{figure}

Like electromagnetic radiation, gravitational waves possess their own spectrum (see \cref{fig:GW-rainbow}). Although their frequency spectrum can, in principle, extend over an enormous range of frequencies, only a finite range is observationally accessible. At the low-frequency end, cosmological observations probe primordial tensor modes with wavelengths of order the present Hubble scale, corresponding to frequencies of order $10^{-18}$ Hz, while detector sensitivity currently limits the upper end to the 10~kHz range. Within this observational window, different frequency bands reveal complementary science: very-low-frequency GWs imprint the CMB polarization \citep{Campeti:2024JCAP,CMB-S4:2019WhitePaper}; nanohertz signals are tracked by pulsar timing arrays \citep{NANOGrav:2023gor,Antoniadis:2023rjw,Reardon:2023gzh} and astrometric observations \citep{Caliskan:2023cqm,Garcia-Bellido:2021zgu}; millihertz waves will be measured by space interferometers such as LISA \citep{Colpi:2024lisa}; and the kilohertz regime is already accessed by ground-based detectors such as LIGO \citep{Abbott:2016blz,KAGRA:2018plz,Saleem:2021iwi}. Future third-generation facilities, including the Einstein Telescope \citep{ET:2025xjr}, promise even greater sensitivity. Atom interferometry offers a complementary pathway to laser interferometers for GW detection \citep{Graham2013_PRL,Dimopoulos2008_PRD,Hogan2011_GRG}. Terrestrial and space-based cold-atom concepts under development \citep{MIGA,ZAIGA,MAGIS,ELGAR,AION,AEDGE,Alonso:2022oot,Abend:2023jxv,Abdalla:2024sst} aim to extend sensitivity into the mid-frequency band ($10^{-2}$--1 Hz), thereby bridging the gap between LISA and ground-based observatories. At the highest frequencies, innovative laboratory-scale concepts are being developed to extend gravitational wave sensitivity well beyond the MHz regime~\citep{Aggarwal:2025noe}, including a number of novel approaches based on quantum-sensing techniques~\citep{Tobar:2023ksi,Kharzeev:2025lyu}. This is not all: there are several further detectors and proposals for probing GWs across different frequency windows. For an overview of the global network of laser-interferometry detectors on Earth and in space, see \cref{fig:laser-}.  Together, these approaches build a multi-frequency gravitational wave astronomy, analogous to the multi-wavelength program of electromagnetic observations.

%{\textcolor{red}{\begin{itemize} 
%\item Intro
%    \item Lec II
%    \item III-1 Penrose diagram of de Sitter
%    \item III-3
%     \item further reading
%\end{itemize}}}

\begin{figure}[t]
  \centering
   \includegraphics[width=\linewidth]{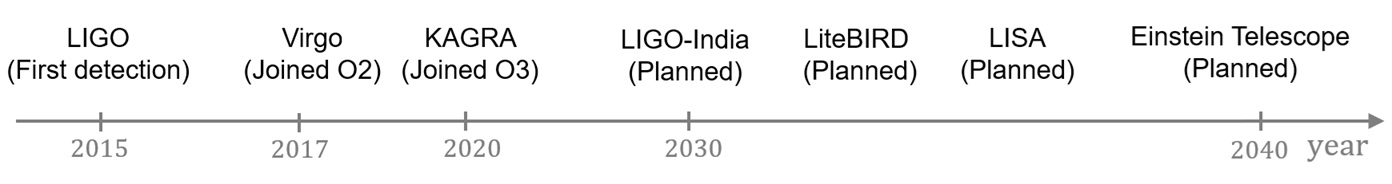}
  \caption{Timeline of GW observatories from the first detection (Advanced LIGO, 2015) to planned facilities through the 2040s. Virgo joined in 2017 and KAGRA in 2020, establishing the first global ground-based network. Forthcoming projects—LIGO–India (early 2030s) \citep{Saleem:2021iwi}, LiteBIRD (2032) \citep{LiteBIRD:2022cnt}, LISA (2035) \citep{Colpi:2024lisa}, and the Einstein Telescope (expected from $\sim$2035--2040) \citep{ET:2025xjr} —extend coverage across frequency bands and improve sky localization, charting the roadmap for GW astronomy into the next decades.}
  \label{fig:gw_timeline}
\end{figure}

\begin{figure}[!ht]
\centering
\includegraphics[width=\textwidth]{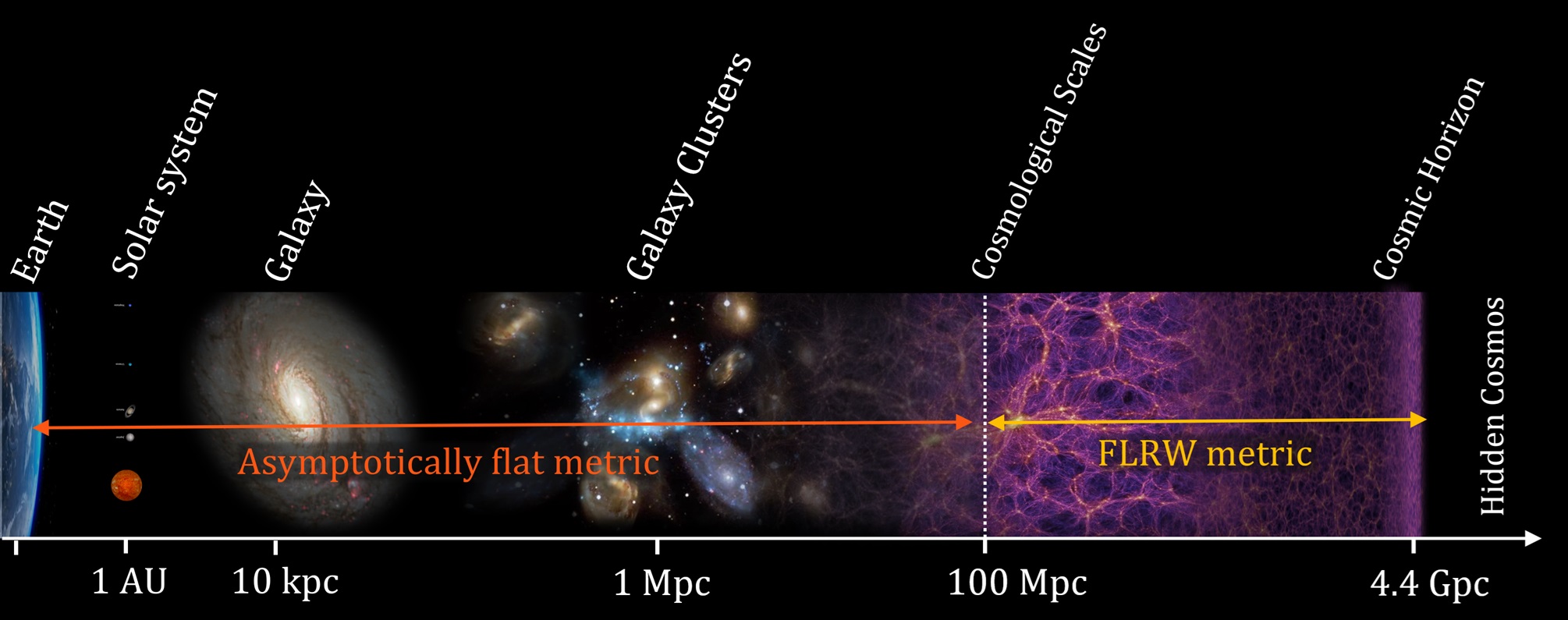}
\caption{Hierarchy of astrophysical and cosmological scales. 
Observations reveal that our Universe is expanding and becomes statistically isotropic and homogeneous 
on scales larger than about $100~\mathrm{Mpc}$, defining the cosmological regime. 
Gravitational systems up to these scales are well described by asymptotically flat geometries, 
whereas on larger scales the spacetime must be modeled by the FLRW metric 
to capture the expansion of the Universe. The observable Universe (Hubble radius) marks the limit of observation, not the limit of existence.}\label{fig:hierarchy}
\end{figure}

The Hubble radius is a characteristic scale in an expanding Universe; it is distinct from the particle horizon, which helps define the observable region. The particle horizon is the maximum comoving distance from which light could have reached an observer by time \(t\), given by
\[
\chi_{\mathrm{p}}(t) =  \int_{t_{\mathrm{i}}}^{t} \frac{\mathrm{d}t'}{a(t')},
\]
where \(a(t)\) is the scale factor and \(t_{\mathrm{i}}\) is the beginning of the relevant cosmological history.

\begin{figure}[t]
    \centering
    \includegraphics[width=0.98\textwidth]{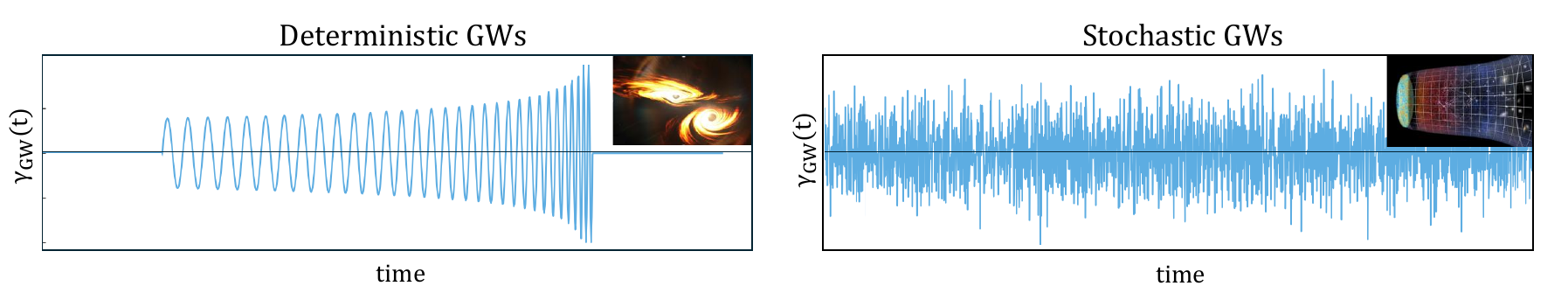}
    \caption{Comparison between deterministic and stochastic classical gravitational waves.
Left: A smooth, deterministic waveform from an astrophysical binary merger. Right: A zero-mean Gaussian stochastic background.
    }
    \label{fig:gw_comparison}
\end{figure}

Since gravitational waves are generated in and propagate through the gravitational backgrounds of the Universe, 
it is essential to describe them within the appropriate spacetime geometry relevant to the physical scale under consideration. 
The behavior of gravity manifests differently across scales. 
On relatively small scales, up to roughly $100\,\mathrm{Mpc}$, the expansion of the Universe is negligible, 
and gravitational dynamics can be accurately described within asymptotically flat geometries. 
On larger, cosmological scales, however, the expansion of space itself becomes dynamically relevant, 
and spacetime must instead be modeled by an FLRW background (see \cref{fig:hierarchy}). In addition to this geometric distinction, there is also a conceptual difference between the sources of gravitational waves across these regimes. 
Individual astrophysical sources, such as binary mergers or supernovae, are typically deterministic, producing coherent signals with non-vanishing mean amplitude, $\langle \gamma_{ij} \rangle \neq 0$. 
By contrast, unresolved superpositions of sources, whether astrophysical or cosmological, are naturally described statistically as stochastic GW backgrounds (GWB) characterized by vanishing mean but nonzero two-point correlations, $\langle \gamma_{ij} \gamma_{kl} \rangle \neq 0$ (see Fig.~\ref{fig:gw_comparison}). 
This distinction underlines why the statistical treatment of gravitational waves becomes crucial in cosmological contexts. This fundamental separation, both geometric and statistical, motivates the structure of this review. It is therefore divided into two parts: the first devoted to gravitational physics in asymptotically flat spacetimes, 
and the second to that in expanding cosmological geometries.

%%%%%%%%%%%%%%%%%%%%%%%%%%%%%%%%%%%%%%%%%%%%%%%

\medskip

\noindent
This review is structured as follows. \cref{sec:1} develops the framework of linearized general relativity, deriving GWs as solutions to the linearized Einstein equations and examining their physical meaning and properties. \cref{sec:2} deepens the discussion with a more rigorous treatment, tracing how gravitational radiation is encoded in the algebraic structure of the Weyl tensor. This naturally leads to the Bondi–Sachs formalism, which clarifies the physical nature of GWs as real carriers of energy and angular momentum. \cref{sec:3} examines GWs in an expanding Universe, following their evolution across different cosmological epochs and their generation during inflation. We then explore Weinberg’s adiabatic modes and Maldacena’s consistency relations, which shed light on the infrared structure and the universal behavior of long-wavelength perturbations. The section concludes with a derivation of the inflationary spectrum of vacuum gravitational waves and a discussion of how they contribute to the integrated Sachs–Wolfe effect.
 In \cref{sec:detect} we survey the principal observational techniques used to detect gravitational waves, including CMB polarization measurements, pulsar timing arrays, ground- and space-based laser interferometry, and proposed photon-regeneration cavity techniques for high-frequency GW detection. \cref{sec:5} is devoted to the physical mechanisms that generate gravitational radiation, beginning with astrophysical sources and followed by cosmological sources. Finally, in \cref{sec:summary} we summarize the main results and outline possible future directions. 

\paragraph*{Notation and Conventions: \qquad} Throughout this review we set $c = \hbar= 1$ while keeping $G$ explicit, and we work
within four-dimensional General Relativity unless stated otherwise. 

\renewcommand{\arraystretch}{1.25}
\begin{longtable}{ll}
\hline
\textbf{Quantity} & \textbf{Definition / Convention} \\
\hline
\endfirsthead

\hline
\textbf{Quantity} & \textbf{Definition / Convention} \\
\hline
\endhead

\hline
\endfoot

\hline
\endlastfoot

Reduced Planck mass 
& $M_{\mathrm{Pl}}^{2} \equiv (8\pi G)^{-1}$ \\

Metric signature 
& $\eta_{\mu\nu} = \mathrm{diag}(-,+,+,+)$ \\

Spacetime indices 
& Greek letters $\mu,\nu,\rho,\sigma = 0,\ldots,3$ \\

Spatial indices 
& Latin letters $i,j,k = 1,2,3$ \\

Partial derivative 
& $\partial_\mu \equiv \partial/\partial x^\mu$ \\

Lie derivative 
& $\mathcal{L}_{\xi} X^{\mu}{}_{\nu}
= \xi^{\rho}\nabla_{\rho} X^{\mu}{}_{\nu}
- (\nabla_{\rho}\xi^{\mu}) X^{\rho}{}_{\nu}
+ (\nabla_{\nu}\xi^{\rho}) X^{\mu}{}_{\rho}$ \\

Riemann tensor 
& $R^{\rho}{}_{\sigma\mu\nu}
= \partial_\mu \Gamma^{\rho}_{\nu\sigma}
- \partial_\nu \Gamma^{\rho}_{\mu\sigma}
+ \Gamma^{\rho}_{\mu\lambda}\Gamma^{\lambda}_{\nu\sigma}
- \Gamma^{\rho}_{\nu\lambda}\Gamma^{\lambda}_{\mu\sigma}$ \\

Ricci tensor 
& $R_{\mu\nu} = R^{\rho}{}_{\mu\rho\nu}$ \\

Ricci scalar 
& $R = g^{\mu\nu} R_{\mu\nu}$ \\

Gravitational waves in TT gauge
& $\gamma_{ij}$ \\

Cosmic scale factor
& $a(t)$ \\

Hubble parameter 
& $H \equiv \dot a / a$ \\

Cosmic time 
& $t$ \\

Bondi time 
& $u$ \quad ($u = t - r$) \\

Conformal time 
& $\tau$ \quad ($d\tau \equiv dt/a(t)$) \\
\end{longtable}

Throughout the text, shaded boxes are used to introduce definitions essential for understanding the material in the current or subsequent sections. Finally, we will frequently use the following acronyms.

\noindent\begin{tabular}{@{}llll}
\textbf{BBN} & Big Bang Nucleosynthesis 
& \textbf{LSS} & Large-Scale Structure \\
\textbf{BH} & Black Hole 
& \textbf{PND} & Principal Null Direction \\
\textbf{BMS} & Bondi--Metzner--Sachs group 
& \textbf{PTA} & Pulsar Timing Array \\
\textbf{BSM} & Beyond the Standard Model 
& \textbf{GWB} &  Gravitational Wave Background \\
\textbf{CMB} & Cosmic Microwave Background 
& \textbf{SLS} & Surface of Last Scattering \\
\textbf{GR} & General Relativity 
&  \textbf{SM} & Standard Model \\
\textbf{GW} & Gravitational Wave 
& \textbf{SNR} & Signal-to-Noise Ratio  \\
\textbf{HFGW} & High-Frequency Gravitational Waves 
&  \textbf{TT} & Transverse-Traceless  \\
\end{tabular}

%%%%%%%%%%%%%%%%%%%%%%%%%%%%%%%%%

%\addcontentsline{toc}{section}{\large{Part I \,  Asymptotically Flat Specetimes }}

\section{Linearized general relativity and GWs}
\label{sec:1}

Shortly after formulating the theory of general relativity, Albert Einstein linearized his field equations and discovered that the theory admits wave-like solutions \citep{Einstein:1916cc}, small perturbations of the Minkowski spacetime, that propagate at the speed of light. These solutions, known today as gravitational waves, appear in the weak-field approximation, where the spacetime metric is only slightly perturbed from flat space. Although this linearized approach may initially appear as a mere simplification, it has proven to be remarkably powerful in both theoretical and observational contexts within astrophysics and cosmology. It enables the analysis of gravitational radiation from compact astrophysical sources such as binary black holes and neutron stars, and forms the theoretical foundation for modern gravitational wave detectors. We begin this section by considering the weak-field limit of general relativity and deriving the general solution to the linearized Einstein equations. We next investigate the properties of gravitational waves generated by localized sources, and analyze their polarization states and observable effects on freely falling test particles.

\subsection{The weak-field approximation and wave solutions} \label{sec:1-1}

A weak gravitational field corresponds to a region of spacetime that is only mildly curved. In such regions, there exist coordinate systems in which the spacetime metric can be expressed as
\begin{equation}
g_{\mu\nu} = \eta_{\mu\nu}+ \Th_{\mu\nu},
\end{equation}
where $\eta_{\mu\nu}$ is the Minkowski metric
\begin{equation}
\eta_{\mu\nu} = \text{diag}(-1,1,1,1),
\end{equation}
and $\Th_{\mu\nu}$ represents a small perturbation such that both it and its partial derivatives remain small
\begin{equation}
\Th_{\mu\nu}\ll 1 \an \p_{\lambda}^n \Th_{\mu\nu}\ll 1 \quad (\text{for all } \, n>1).
\label{eq:weak-field}
\end{equation}
This intuitive characterization will be given a mathematically rigorous formulation in \cref{sec:2}. 
The formulation in \cref{eq:weak-field} sets the stage for linearizing Einstein’s equations 
\begin{equation}
R_{\mu\nu} - \frac{1}{2} R \, g_{\mu\nu} = 8\pi G \, T_{\mu\nu}, \label{eq:Einstein}
\end{equation}
and studying the dynamics of gravitational waves as perturbations propagating in flat spacetime.
Note that we can well consider small perturbations about some other background metric, such that $g_{\mu\nu} = \bar{g}_{\mu\nu}+ \Th_{\mu\nu}$. In particular, later in \cref{sec:3}, we will use a similar weak field approach for cosmological perturbations around the FLRW metric. 

In the weak-field limit of general relativity, we expand the field equations in powers of $\tilde{h}_{\mu\nu}$ and keep the linear terms. In fact, from the Einstein equation in \cref{eq:Einstein}, we find the linearized gravitational field equation as
\be
\p_{\alpha}\p^{\alpha} \Th_{\mu\nu} + \p_{\mu}\p_{\nu} \Th -\p_{\nu}\p_{\lambda} \Th^{\lambda}_{\mu} -\p_{\mu}\p_{\lambda} \Th^{\lambda}_{\nu} - \eta_{\mu\nu} (\p_{\alpha}\p^{\alpha}  \Th - \p_{\sigma}\p_{\lambda} \Th^{\lambda\sigma} ) = -16\pi G T_{\mu\nu},
\ee
where $\Th$ is the trace of the field, $\Th=\Th^{\mu}_{\mu}$. Although linearized, the equation above does not yet resemble a wave equation. However, it can be significantly simplified through an appropriate field redefinition 
\be\label{trace-inv-h}
h_{\mu\nu} \equiv \Th_{\mu\nu} - \frac12 \eta_{\mu\nu} \Th,
\ee
which is the trace-reverse cousin of $\Th_{\mu\nu}$, \textit{i.e.} $h=- \Th$. The trace-reversed perturbation $h_{\mu\nu}$ possesses a gauge freedom that originates from the diffeomorphism invariance of general relativity.

Under an infinitesimal gauge transformation
\begin{equation}
x^\mu \rightarrow x^\mu + \xi^\mu(x),
\end{equation}
where $\xi^\mu$ is a smooth vector field, we keep the background metric fixed,
\begin{equation}
\eta_{\mu\nu}\rightarrow\eta_{\mu\nu},
\end{equation}
and absorb the transformation into the metric perturbation,
\begin{equation}
\Th_{\mu\nu}\rightarrow\Th_{\mu\nu}+\mathcal L_\xi\eta_{\mu\nu}
=\Th_{\mu\nu}+\partial_\mu\xi_\nu+\partial_\nu\xi_\mu.
\end{equation}
Consequently,
\begin{equation}
\Th\rightarrow\Th+2\partial_\rho\xi^\rho,
\end{equation}
and the trace-reversed perturbation transforms as
\begin{equation}
h_{\mu\nu}\rightarrow h_{\mu\nu}+\partial_\mu\xi_\nu+\partial_\nu\xi_\mu-\eta_{\mu\nu}\partial_\rho\xi^\rho.
\end{equation}
This transformation expresses the gauge freedom: the physical content of $h_{\mu\nu}$ remains unchanged under such a shift. To partially fix this gauge ambiguity, it is conventional to impose the Lorenz gauge condition
\begin{equation}
\partial^\mu h_{\mu\nu} = 0 \quad (\text{Lorenz gauge}).
\end{equation}
Residual gauge transformations satisfying $\Box\xi^\mu=0$ remain. In this gauge, the linearized Einstein field equations take the particularly simple and covariant form
\begin{equation} 
\Box h_{\mu\nu} = -16\pi G \, T_{\mu\nu},
\label{eq:linear-E}
\end{equation}
where $\Box = \partial^\alpha \partial_\alpha$ is the d'Alembertian operator in flat spacetime. 

We now turn to the vacuum case of the linearized Einstein field equations, obtained by setting $T_{\mu\nu} = 0$ in \cref{eq:linear-E}. This equation admits plane wave solutions of the form
\begin{equation}
h_{\mu\nu}(x) = e_{\mu\nu}(k) \, e^{ik \cdot x},
\end{equation}
where $e_{\mu\nu}(k)$ is a constant, symmetric tensor representing the amplitude and polarization of the wave, and $k_\mu$ is the wave 4-vector. Substituting into the wave equation shows that the wave vector must satisfy the null condition $k^\mu k_\mu = 0$, while the Lorenz condition further implies $k^\mu e_{\mu\nu}=0$,
which confirms that gravitational waves propagate at the speed of light.

\vskip 0.5cm

%\subsubsection*{General Solution of the Linearized Field Equations:}

In the presence of matter or energy, the stress-energy tensor $T_{\mu\nu}$ acts as a source for the gravitational field, and the linearized Einstein equation becomes inhomogeneous. To solve it, we employ the Green's function method for the wave operator
\begin{equation} \label{lEq-}
\Box G(x^\mu - y^\mu) = \delta^{(4)}(x^\mu - y^\mu),
\end{equation}
where \( G(x^\mu - y^\mu) \) is the Green's function of the flat-space d'Alembertian. The physically relevant solution is the retarded Green's function, which ensures causal propagation and takes the form
\begin{equation}
G(x^\mu - y^\mu) = -\frac{1}{4\pi \lvert \boldsymbol{x} - \boldsymbol{y} \rvert} \, \delta\left(x^0 - y^0 - \lvert \boldsymbol{x} - \boldsymbol{y} \rvert\right), \quad \text{for } x^0 > y^0.
\end{equation}

Using this, the solution to the sourced linearized gravitational field equations is given by
\begin{equation} \label{h-Tmunu}
h_{\mu\nu}(\boldsymbol{x}, t) = 4G \int d^3y \, \frac{T_{\mu\nu}(\boldsymbol{y}, t - \lvert \boldsymbol{x} - \boldsymbol{y} \rvert)}{\lvert \boldsymbol{x} - \boldsymbol{y} \rvert}.
\end{equation}

\begin{figure}[h!]
\centering
\includegraphics[width=0.6\textwidth]{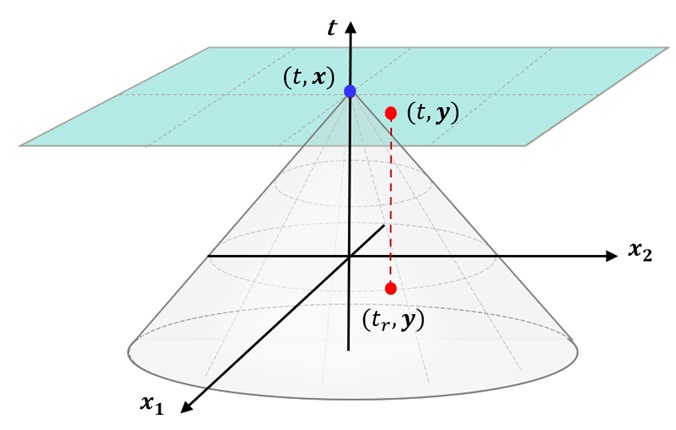}
\caption{The variation of the gravitational field at a spacetime point $(t,\boldsymbol{x})$ arises from the cumulative influence of the source energy–momentum tensor $T_{\mu\nu}$ evaluated at points $(t_r, \boldsymbol{y})$ lying on its past light cone, where the retarded time is given by $t_r = t - \lvert \boldsymbol{x} - \boldsymbol{y} \rvert$.}\label{fig:causal}
\end{figure}

This expression reveals that the gravitational field at the spacetime point $(t, \boldsymbol{x})$ is determined entirely by the values of the source $T_{\mu\nu}$ along the past light cone of that point. In other words, gravitational effects propagate causally at the speed of light, just as expected for massless fields (see \cref{fig:causal}).

\subsubsection{Gravitational radiation of compact objects}

 In many astrophysical scenarios, the gravitational source is spatially compact compared to the distance from the source to the observation point. In such cases, one can expand the integrand of the Green’s function solution \eqref{h-Tmunu} in powers of $|\boldsymbol y|/r$ about the source origin, retaining only the leading-order contribution
\begin{equation}
\frac{1}{\lvert \boldsymbol{x} - \boldsymbol{y} \rvert} = \frac{1}{r} +  \frac{y^i x_i}{r^3} + \frac{1}{2} y^i y^j \left( \frac{3x_i x_j - r^2 \delta_{ij}}{r^5} \right) + \dots,
\end{equation}
where $r = \lvert \boldsymbol{x} \rvert$ is the distance from the origin to the observation point. This is the multipole expansion of the Newtonian potential. In writing the retarded time as $t-r$ below, we also assume that the source is small compared with the characteristic wavelength, so that retardation across the source can be neglected at leading order. Applying this expansion to the linearized solution \eqref{h-Tmunu} and keeping only the leading term yields the far-field approximation
\begin{equation} \label{h-Tmunu-compact}
h_{\mu\nu}(\boldsymbol{x}, t) = \frac{4G}{r} \int d^3y \, T_{\mu\nu}(\boldsymbol{y}, t - r) + \mathcal{O}(r^{-2}),
\end{equation}
which gives the leading far-field contribution; its time-dependent transverse-traceless spatial part describes gravitational radiation. Each component of the integral above has a direct physical interpretation: \\
$\bullet$ $\displaystyle \int T^{00}\, d^3y = M c^2$ represents the total energy (mass) of the source, \\
$\bullet$  $\displaystyle \int T^{0i}\, d^3y = P^i c$ represents the total momentum of the source along the $x^i$ direction, \\
$\bullet$  $\displaystyle \int T^{ij}\, d^3y = \Pi^{ij}$ represents the spatially integrated internal stresses of the source, encompassing pressure and shear.

For an isolated system, conservation of the energy-momentum tensor implies $\nabla_\mu T^{\mu\nu} = 0$. In the flat-background linearized derivation considered here, $\nabla_\mu T^{\mu\nu}=\partial_\mu T^{\mu\nu}=0$.
From this, it follows that $M$ and $P^i$ are constants of motion in the linear theory. Moreover, for compact sources, the total momentum $P^i$ can be expressed as
\begin{equation}
P^i = \left. \frac{\partial}{\partial y^0} \left[ \int T^{00}(y^0, \boldsymbol{y}) \, y^i \, d^3y \right] \right|_{y^0 = t - r}.
\end{equation}
The total four-momentum $P^\mu=(E,P^i)$ transforms as a Lorentz vector under changes of inertial frame. By performing a Lorentz boost with velocity $ \boldsymbol{v} = \boldsymbol{P}/E $, we arrive in the center-of-momentum frame where
\begin{equation}
P^i = 0 \quad (\text{center-of-momentum frame}).
\label{eq:center-momentum}
\end{equation}
While the stress-energy tensor $ T^{\mu\nu}(x)$ transforms covariantly under coordinate changes, one cannot, in general, find a global inertial frame in which $T^{0i}(x) = 0$ at all points; only the integrated momentum $P^i$ can be set to zero by a suitable choice of frame.

In the center-of-momentum frame, the gravitational field reveals its most transparent form
\begin{align}
h_{00} &= \frac{4GM}{r}, \\
h_{0i} &= 0, \\
h_{ij} &= \frac{4G}{r} \int T_{ij}(t - r, \boldsymbol{y}) \, d^3y.
\end{align}
We can simplify this expression for compact sources. Using integration by parts and assuming that the stress-energy tensor vanishes outside the source, we write\footnote{Since $T^{\mu\nu}$ has compact support, surface terms such as $\int \partial_k (T^{ik} y^j)\,d^3y$ vanish at spatial infinity.}
\begin{equation}
\int T^{ij} \, d^3y = -\frac{1}{2} \int \left[ \partial_k (T^{ik}) y^j + \partial_k (T^{jk}) y^i \right] \, d^3y.
\end{equation}
Now, applying energy-momentum conservation, $\partial_\mu T^{\mu\nu} = 0$, we obtain
\begin{equation}
\int \partial_k (T^{ik}) y^j \, d^3y = -\frac{d}{dy^0} \left[ \int T^{i0}(y^0, \boldsymbol{y}) y^j \, d^3y \right].
\end{equation}
Using conservation again and integrating by parts, we find
\begin{equation}
\int \left( T^{i0} y^j + T^{j0} y^i \right) d^3y = \frac{d}{dy^0} \left[ \int T^{00}(y^0, \boldsymbol{y}) y^i y^j \, d^3y \right].
\end{equation}
This motivates introducing the (mass) quadrupole moment tensor of the source
\begin{equation}
I_{ij}(y^0) = \int T_{00}(y^0, \boldsymbol{y}) \, y^i y^j \, d^3y.
\end{equation}
Thus, in the far-field limit, the spatial components of the trace-reversed perturbation become
\begin{equation}
h_{ij}(t, \boldsymbol{x}) = \frac{2G}{r}  \ddot{I}_{ij}(y^0),
\end{equation}
where the dot denotes a derivative with respect to $y^{0} = t - r$.
It is often convenient to decompose this expression into its trace and traceless parts
\begin{equation}
h_{ij}(t, \boldsymbol{x}) = h(t, r) \, \delta_{ij} + \gamma_{ij}(t, \boldsymbol{x}),
\end{equation}
where the traceless component is given by
\begin{equation}
\gamma_{ij}(t, \boldsymbol{x}) = \frac{2G}{ r} \ddot{Q}_{ij}(y^0) ,
\label{eq:gamma-Q}
\end{equation}
and $Q_{ij}$ is the reduced quadrupole moment
\begin{equation}
Q_{ij}(y^0) = I_{ij}(y^0) - \frac{1}{3} \delta_{ij} I^k_k(y^0).
\end{equation}
The perturbation $\gamma_{ij}$ is traceless, but it is not in general transverse. The physical radiative field is obtained by taking its transverse-traceless part,
\begin{equation}
h^{\text{TT}}_{ij}(t,\boldsymbol{x})=\gamma^{\text{TT}}_{ij}(t,\boldsymbol{x})=\frac{2G}{r}\ddot Q^{\text{TT}}_{ij}(y^0),
\end{equation}
where the TT projection is given explicitly in \cref{sec:1-polarization}. In the vacuum wave zone, the residual gauge freedom within the Lorenz gauge can be used to impose the TT gauge, characterized by
\begin{equation}
h^{\text{TT}}_{0\mu}=0,\qquad h^{\text{TT}}=0,\qquad \partial^i h^{\text{TT}}_{ij}=0.
\end{equation}
Consequently, the gravitational wave is fully described by the spatial components of a symmetric, transverse, and traceless tensor, leaving exactly two dynamical degrees of freedom. We return to this discussion in \cref{sec:1-polarization}, where a more rigorous treatment is provided.

\begin{tcolorbox}[colback=gray!10, colframe=gray!10, boxrule=0pt,
                  enhanced, breakable, halign=justify]
                  
\subsubsection*{Radiation in Electromagnetism vs Gravity}
Just as Maxwell's equations predict the existence of electromagnetic waves, the linearized Einstein equations reveal that gravity, too, supports wave-like solutions. However, unlike electromagnetic radiation, which begins at the dipole level, GWs first arise at the quadrupole order, a distinction that may seem counterintuitive at first glance. Let us pause for a moment and explore why. This fundamental difference stems from the nature of the sources and the symmetries of the underlying field equations. In electromagnetism, dipole radiation is ubiquitous because charges, positive and negative, can accelerate independently, even within an isolated system. For a localized charge distribution $\rho(t,\mathbf y)$, the radiative electric field in the far zone can be expressed in terms of the electric dipole moment $p^i(t)\equiv \int d^3y\,\rho(t,\mathbf y)\,y^i $ and takes the form
\be
\mathbf E^{\mathrm{rad}}(t,\mathbf x)
=
\frac{1}{4\pi\varepsilon_0\,c^2}\,\frac{1}{r}\,
\hat{\mathbf n}\times\!\Big(\hat{\mathbf n}\times \ddot{\mathbf p}(t-r)\Big),
\ee
where $\hat{\mathbf n}=\mathbf x/r$ denotes the direction of propagation. Gravity, however, couples to energy-momentum. For an isolated system, momentum conservation implies that the mass dipole has vanishing second time derivative; in the center-of-momentum frame it is constant. Together with conservation of the total mass-energy, this forbids monopole and dipole gravitational radiation, so the leading radiative multipole is the quadrupole.  

%% floating tables and figures in boxes are tricky...
\bigskip
\captionsetup{hypcap=false}
%%\begin{table}[h]
\captionof{table}{Comparison between electromagnetic and gravitational radiation in the linearized (weak-field) regime.}
\label{tab:rad-comparison}
\centering
\renewcommand{\arraystretch}{1.6}
\setlength{\tabcolsep}{6pt} % slightly tighter columns

\begin{tabularx}{\linewidth}{>{\centering\arraybackslash}m{3.5cm}
                             >{\centering\arraybackslash}m{4cm}
                             >{\centering\arraybackslash}X}
\toprule
\textbf{Feature} & \textbf{EM} & \textbf{Linearized Gravity} \\
\midrule
Source & $j^\mu$ & $T^{\mu\nu}$ \\
%%\hline
Wave equation & $\Box A^\mu = j^\mu$ & $\Box h_{\mu\nu} = -16\pi G\,T_{\mu\nu}$ \\
%%\hline
Conservation law & $\nabla_\mu j^\mu = 0$ & $\nabla_\mu T^{\mu\nu} = 0$ \\
%%\hline
Leading radiative  & Dipole & Quadrupole \\
%%\hline
Monopole & $\times$ (charge conservation) & $\times$ (energy conservation) \\
%%\hline
Dipole & $\checkmark$ (leading order) & $\times$ (momentum conservation) \\
%%\hline
Quadrupole & $\checkmark$ & $\checkmark$ (leading order) \\
%%\hline
Radiated field & $E_i^{\mathrm{rad}} \propto \ddot p_i$ & $h_{ij}^{\mathrm{TT}} \propto \ddot Q_{ij}^{\mathrm{TT}}$ \\
%%\hline
Spin of carrier & 1 (photon) & 2 (graviton) \\
\bottomrule
%Physical degrees of freedom & 2 (transverse) & 2 (transverse--traceless) \\
%\hline
\end{tabularx}
%%\end{table}

\end{tcolorbox}

 %%%%%%%%%%%%%%%%%%%%%%%%

 %%%%%%%%%%%%%%%%%%%%%%%%%%%

\begin{figure}[h]
\includegraphics[width=0.88\textwidth]{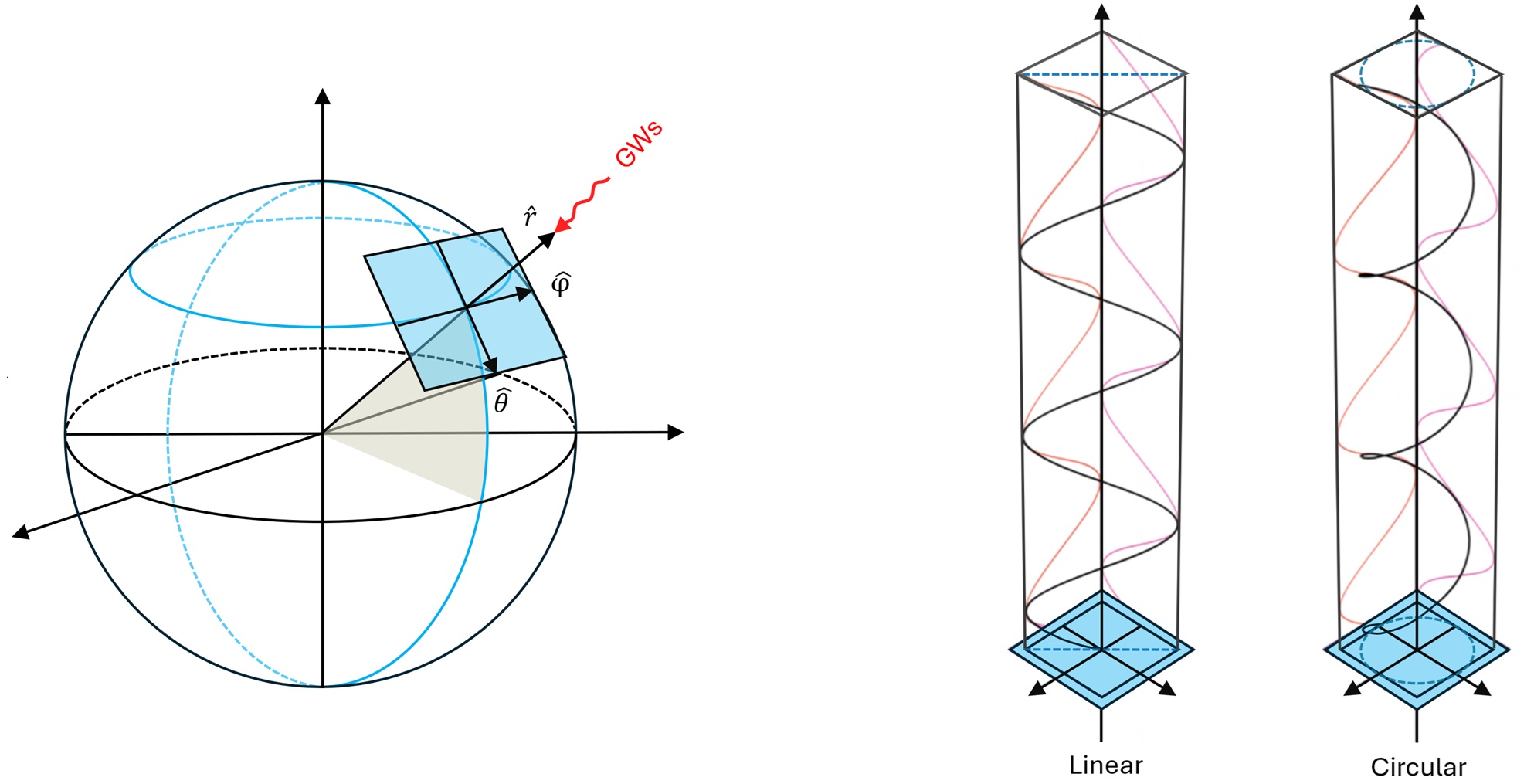} 
\caption{Left panel: the spherical coordinate frame along with its associated unit normal vectors. Right panel: linearly and circularly polarized waves: in linear polarization, the deformation pattern in the transverse plane oscillates along fixed axes, while in circular polarization, the deformation rotates uniformly within the transverse plane.}\label{fig:linear-circular}
\end{figure}

\subsection{Polarization states of GWs and free particles} \label{sec:1-polarization}

For a gravitational wave propagating in an arbitrary direction $\bk = k \hat{n}$, the wave’s spatial polarization tensor in the TT gauge can be extracted using the projection operator
\begin{equation}
e_{ij}^{\text{TT}}(\hat{n}) = \left(P_i^{\ k} P_j^{\ l} - \frac{1}{2} P^{kl} P_{ij}\right) e_{kl}(k) \quad \quad (\text{polarization tensor}),
\end{equation}
where the spatial projection tensor $P_{ij}$ is defined as $P_{ij} = \delta_{ij} - \hat{n}_i \hat{n}_j$,
and serves to project any tensor onto the 2d plane orthogonal to the wave propagation direction $\hat{n}$. Much like in electromagnetism, the two independent dynamical modes of the gravitational wave can be decomposed into distinct polarization states. Expressing $\hat{n}=-\hat{r}$ in terms of polar and azimuthal angles $(\theta, \phi)$ as $\hat{r} = (\sin\theta \cos\phi,\, \sin\theta \sin\phi,\, \cos\theta)$,
we define the complex polarization basis vectors
\begin{equation}
e^{\pm}_{i}(\hat{n}) = \frac{1}{\sqrt{2}} \left( \hat{\theta}_i \pm i \hat{\phi}_i \right) \quad \quad (\text{polarization vector}),
\end{equation}
where $\hat{\theta}$ and $\hat{\phi}$ are orthonormal unit vectors spanning the transverse plane, aligned respectively with increasing $\theta$ and $\phi$. Since $\hat{\mathbf r}$ points opposite to the direction of photon propagation, the helicity (defined as the projection of the spin angular momentum along the direction of motion $h=\boldsymbol{S}.\hat{n}$) is the negative of the projection along $\hat{\mathbf r}$. For this reason, the labels $e_i^{\pm}$ correspond to the circular polarization basis vectors of helicity $\mp 1$.
These two helicity states correspond to circular polarizations as 
\begin{equation}
e_i^{\pm}(\hat{n}) = e_i^{L,R}(\hat{n}).
\end{equation}
From these, one can construct two equivalent bases for the polarization tensors (see \cref{fig:linear-circular}): 
\begin{itemize}[label=$\bullet$]
\item Linear polarization tensors, i.e. the familiar plus and cross modes
\begin{equation}
e_{ij}^{+}(\hat{n}) = \frac{1}{\sqrt{2}} \left( \hat{\theta}_i \hat{\theta}_j - \hat{\phi}_i \hat{\phi}_j \right), \quad
e_{ij}^{\times}(\hat{n}) = \frac{1}{\sqrt{2}} \left( \hat{\theta}_i \hat{\phi}_j + \hat{\phi}_i \hat{\theta}_j \right),
\end{equation}
\item Circular polarization tensors, i.e. the right- and left-handed helicity states
\begin{equation}
e_{ij}^{R}(\hat{n}) = e^{R}_i \otimes e^{R}_j, \quad \text{and} \quad e_{ij}^{L}(\hat{n}) = e^{L}_i \otimes e^{L}_j .
\end{equation}
\end{itemize}
These two bases are related through the transformation:
\begin{equation}
e^{R,L}_{ij}(\hat{n}) = \frac{1}{\sqrt{2}} \left( e^{+}_{ij} \mp i \, e^{\times}_{ij} \right),
\end{equation}
showing that circular polarizations are simply complex combinations of the linear ones.

In the special case where the wave propagates along the $z$-axis, i.e., $\hat{n} = \hat{z}$, the linear polarization tensors simplify to the following non-zero components:
\begin{equation}
e_{11}^{+} = -e_{22}^{+} = \frac{1}{\sqrt{2}}, \quad
e_{12}^{\times} = e_{21}^{\times} = \frac{1}{\sqrt{2}},
\end{equation}
with all other components vanishing. For circular polarization, the non-zero components are:
\begin{equation}
e_{11}^{R,L} = -e_{22}^{R,L} = \frac{1}{2}, \quad
e_{12}^{R,L} = e_{21}^{R,L} = \mp \frac{i}{2},
\end{equation}
with vanishing longitudinal components.

\vskip 0.5cm

\begin{tcolorbox}[colback=gray!10, colframe=gray!10, boxrule=0pt,
                  enhanced, breakable, halign=justify]
                  
\subsubsection*{GW Polarizations in $d$ Dimensions and Extended Frameworks }

In $d$-dimensional General Relativity (GR), a massless spin-2 graviton carries
\begin{equation}
    N_{\text{pol}} = \frac{d(d-3)}{2},
\end{equation}
independent polarization (radiative) degrees of freedom.  
This immediately implies that GR admits propagating gravitational wave modes only for $ d \ge 4$. In particular, for $d = 3$ one obtains $N_{\text{pol}} = 0$, meaning that
three-dimensional GR has no local gravitational radiation and no graviton
polarization states. In contrast, many modified theories of gravity introduce additional dynamical degrees of freedom, which generically lead to extra gravitational wave polarization modes beyond the two transverse $(+,\times)$ modes of four-dimensional General Relativity (GR). In particular, in four dimensions the metric begins with ten degrees of freedom; fixing the coordinate system removes four, leaving six physical degrees of freedom corresponding to the six possible gravitational wave polarizations.
 Thus, up to four supplementary modes (two scalar and two vector polarizations) may arise, although a given theory of gravity need not include all of them simultaneously. For a thorough review of how additional degrees of freedom arise and propagate
in modified gravity frameworks, see
\citet{deRham:2014zqa}. More general discussions of gravitational-wave
polarizations in alternative theories can be found in
\citet{Will:2014kxa,Ezquiaga:2017ekz}. 

\end{tcolorbox}

\subsubsection{Geodesic distortion by passing GWs}\label{sec:geodesic-distortion}

To gain a clearer understanding of the physical manifestation of each GW polarization state, it is instructive to examine their effect on the geodesic deviation between freely falling test particles.
Consider two nearby test particles that interact only through gravity, initially at rest and separated by the spacelike vector  
\begin{equation}
    X^{\mu}(0) = (0,\,\xi^i_0).
\end{equation}
The passage of a gravitational wave perturbs their geodesic motion, inducing a nonzero contribution to the geodesic--deviation equation. 
Recalling that the evolution of the separation four--vector \(X^{\mu}\) between neighboring geodesics with tangent four--vector \(u^{\mu}\) is governed by  
\begin{equation}
    \frac{D^2 X^{\mu}}{D\tau^2} = R^{\mu}{}_{\nu\lambda\sigma}\,u^{\nu}u^{\lambda}X^{\sigma},
\end{equation}
where  
\begin{equation}
    \frac{D}{D\tau} \equiv u^{\mu}\nabla_{\mu}
\end{equation}
is the covariant derivative along the particle’s worldline. 
In the local proper (detector) frame associated with the reference geodesic, the motion is static to zeroth order and receives corrections only at $\mathcal{O}(h_{\mu\nu})$, so that 
\begin{equation}
    u^{\mu} = (1,0,0,0) + \mathcal{O}(h_{\mu\nu}) .
\end{equation}
Therefore, the geodesic deviation of the nearby particles is 
\be
\frac{D^2 X^i}{D\tau^2} =- R^{i}_{~0j0} X^j.
\ee

\begin{figure}[h]
\centering
\includegraphics[width=\textwidth]{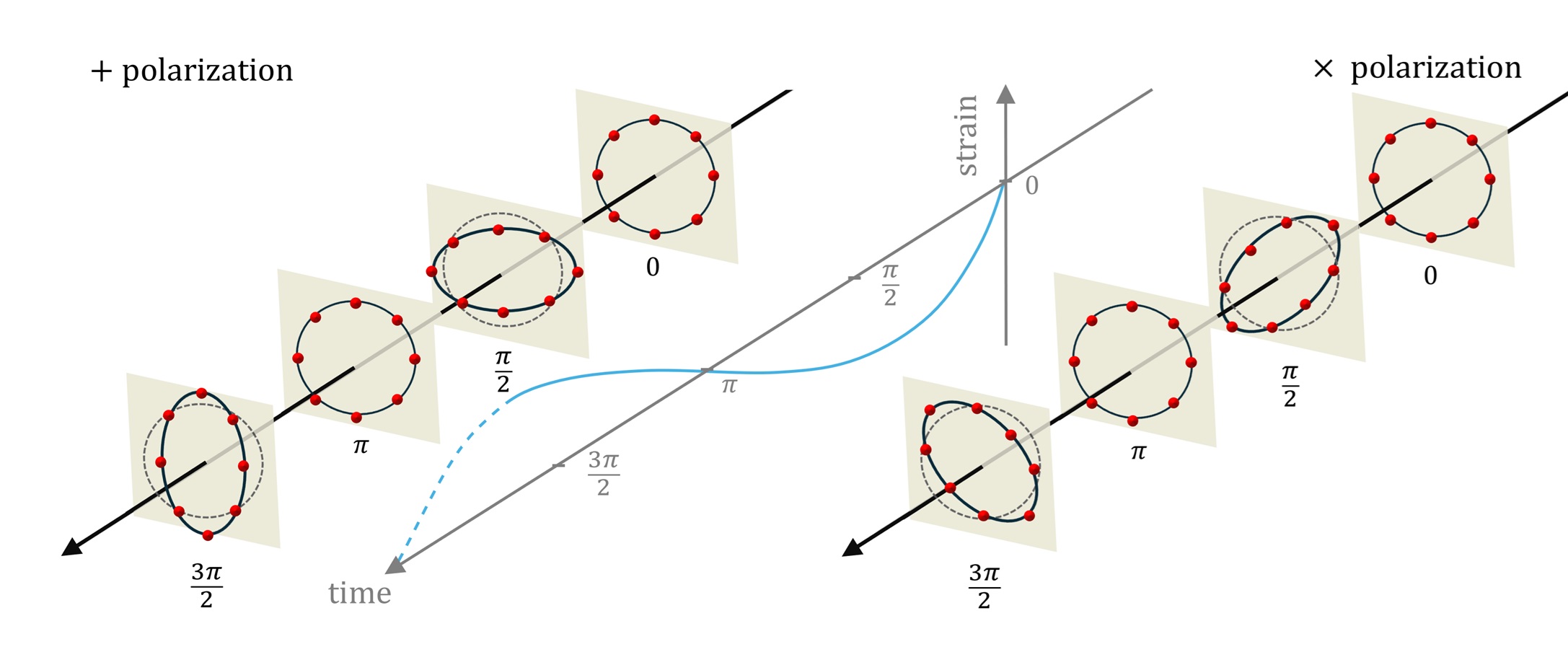} \\
\includegraphics[width=\textwidth]{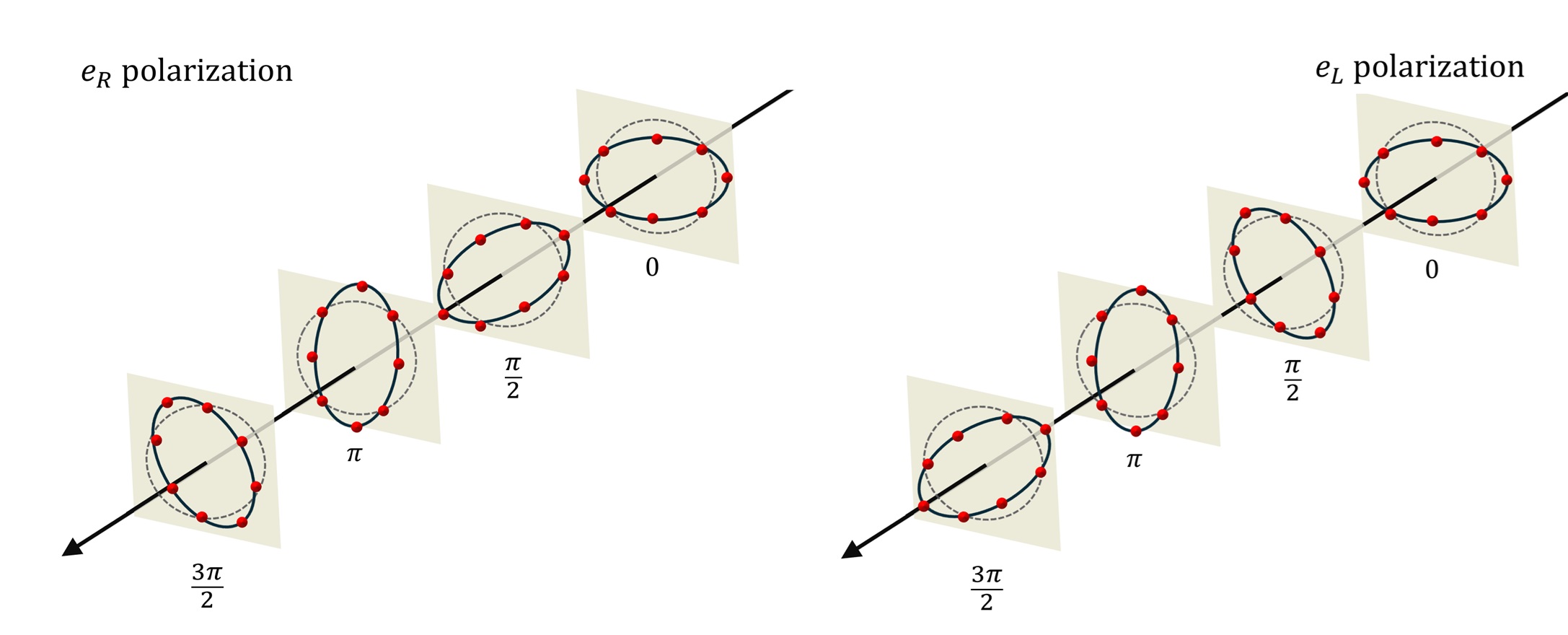} 
\caption{Effect of GW polarization states on a ring of freely falling particles. Solid lines and red dots show particle positions over time; dashed lines mark their unperturbed locations. Top: plus and cross polarizations. Bottom: right- and left-handed circular polarizations. Image credit: Adapted and modified from \citet{Bishop:2016lgv}. }\label{fig:pol-GWs}
\end{figure}

In the rest frame of the particle, we have $dx^i = 0$, and the line element reduces to 
\begin{equation}
ds^2 = -d\tau^2 = -\big(1 + \Th_{00}\big)\, dt^2.
\end{equation}
In the TT gauge, $\Th_{00}^{\mathrm{TT}} = 0$ at leading order, so the coordinate time and the particle’s proper time coincide up to second order in the perturbation,
\begin{equation}
\tau = t + \mathcal{O}\!\left(h_{ij}^{\mathrm{TT}\, 2}\right).
\end{equation}
Using this, the geodesic deviation simplifies to
\be
\frac{d^2 X^i(t)}{dt^2} =\frac12 \p_t^2 h_{ij}^{TT} \xi^j_0.
\ee
For particles initially at rest relative to one another, choosing the initial time before the wave arrives so that $h_{ij}^{TT}=0$, this integrates to the displacement of the separation vector
\begin{equation}
X^i(t) = \xi^{j}_0 \big( \delta^{i}{}_{j} + \frac12 h^{TT\,i}{}_{j}(t,\bx)\big).
\label{eq:dis-geodesic-sep}
\end{equation}
That shows a passing gravitational wave induces a deformation of the separation vector between freely falling test particles.
\cref{fig:pol-GWs} illustrates this tidal effect for the two linear and the corresponding two circular polarization basis states, showing how the ring of freely falling particles is periodically stretched and compressed as the wave propagates through spacetime.

%\vskip 0.2cm

\begin{tcolorbox}[colback=gray!10, colframe=gray!10, boxrule=0pt,
                  enhanced, breakable, halign=justify]
                  
\subsubsection*{Gravitational Memory Effect\footnotemark} Before proceeding to the next part, let us pause to discuss a fascinating phenomenon: the remarkable way in which spacetime itself remembers the passage of GWs, leaving behind a permanent trace of the universe’s most violent events. 
When a burst of gravitational radiation carries a non-oscillatory component, i.e., the waveform exhibits a net offset 
$\Delta h^{\mathrm{TT}}_{ij} \neq 0$, the spacetime curvature leaves a lasting imprint on freely falling test particles.  
After the wave has passed, the relative separation discussed in \cref{eq:dis-geodesic-sep} does not return to its initial value but retains a permanent displacement,  
\begin{equation}
\Delta X^{i} \;\simeq\; \tfrac{1}{2}\, \Delta h^{\mathrm{TT}\,i}{}_{j}\, \xi^{j}(0) \;\neq\; 0.
\end{equation}
This enduring change in the relative separation of particles, known as the gravitational memory effect
 \citep{ZeldovichPolnarev1974,Braginsky:1985vlg,Christodoulou:1991cr}, reveals that spacetime itself retains memory of the passing gravitational wave.  
Although one may formally remove this step in the metric by performing a large diffeomorphism, such as a late-time spatial shear,
\begin{equation}
x^{i} \;\longrightarrow\; x^{i} + \tfrac{1}{2}\, \sigma^{i}{}_{j}\, x^{j},
\qquad
\sigma^{i}{}_{i} = 0,
\qquad
\sigma^{i}{}_{j} = -\, \Delta h^{\mathrm{TT}\,i}{}_{j},
\end{equation}
this transformation is not a mere gauge redundancy.  
Because it does not vanish at null infinity, it maps the geometry between inequivalent asymptotic vacua \citep{Bondi:1962px, Sachs:1962wk,Strominger:2014pwa}.  Consequently, the memory effect stands as a genuine observable: a permanent, measurable shift in relative distance or interferometric phase, encoding the asymptotic difference between the early and late gravitational wave configurations. The gravitational-wave memory effect was first explored by \citep{ZeldovichPolnarev1974} and further studied by \citep{Braginsky:1985vlg}. \citep{Braginsky:1987kwo} later discussed bursts with memory and their experimental prospects, while \citep{Christodoulou:1991cr} developed the nonlinear memory effect. Observational prospects for detecting the memory effect have been proposed using ground-based interferometers such as LIGO \citep{Lasky:2016knh}, space-based detectors including LISA \citep{Inchauspe:2024ibs}, as well as through observations with Pulsar Timing Array experiments \citep{Wang:2014zls}. We will return to this connection when we examine the infrared structure of 
asymptotically flat spacetimes in \cref{sec:why-BMS}, and again in the context 
of cosmology and Weinberg’s adiabatic modes in \cref{sec:IR}.

\end{tcolorbox}

% Part 2 -------------------------------------------

\section{Theoretical aspects of gravitational waves}
\label{sec:2}

%\section{Theoretical Aspects}
%\section{Physicality of Gravitational radiation?!}

In the previous section, we employed the weak-field approximation to derive gravitational waves. This approach dates back to Einstein himself, who was the first to identify gravitational waves as solutions within the linearized GR theory. Yet, due to the conceptual and mathematical ambiguities of the time, he later questioned their physical reality within the full nonlinear theory. This uncertainty slowed progress for decades, until the pioneering works of Bondi, Pirani, Robinson, and Trautman finally resolved it, showing, through a fully consistent physical and mathematical treatment of asymptotically flat spacetimes, that gravitational waves are real, radiative solutions that carry energy and angular momentum. In this section, we will first discuss the theoretical issues and ambiguities of linearized gravity. Next, we will examine the algebraic structure of the Weyl tensor and its geometric interpretation as the carrier of tidal forces. Finally, we will conclude by exploring how spacetime geometry can radiate, leading to a precise definition of GWs.
%\subsection{Gravitation vs Electromagnetic radiation}

\subsection{Theoretical issues of linear gravity}

Einstein’s field equations are intrinsically nonlinear
\be
G_{\mu\nu}[g] = R_{\mu\nu} - \tfrac{1}{2} g_{\mu\nu} R = 8\pi G\, T_{\mu\nu}.
\ee
The Einstein tensor $G_{\mu\nu}[g]$ is a nonlinear functional of the metric $g_{\mu\nu}$ and its derivatives.
Hence, spacetime curvature couples to itself through the very structure of the equations. Let us now take a more rigorous look at the weak-field approximation introduced in \cref{sec:1}. We consider small perturbations $h_{\mu\nu}$ around flat spacetime
\be
g_{\mu\nu} = \eta_{\mu\nu} + h_{\mu\nu}, 
\qquad |h_{\mu\nu}| \ll 1.
\ee
Expanding the Einstein tensor in powers of $h_{\mu\nu}$ gives
\be
G_{\mu\nu}[g] 
= G^{(1)}_{\mu\nu}[h] 
+ G^{(2)}_{\mu\nu}[h] 
+ \mathcal{O}(h^3),
\ee
where $G^{(1)}_{\mu\nu}$ is linear and $G^{(2)}_{\mu\nu}$ is quadratic in $h_{\mu\nu}$.
Substituting this into Einstein’s equations yields
\be
G^{(1)}_{\mu\nu}[h] 
= 8\pi G\, T_{\mu\nu} - G^{(2)}_{\mu\nu}[h] + \mathcal{O}(h^3).
\ee
The quadratic term $G^{(2)}_{\mu\nu}$ acts as an additional source, 
and, to quadratic order, one may define the gauge-dependent effective stress-energy tensor of the gravitational field as
\be
t_{\mu\nu}^{(\mathrm{grav})} 
= -\frac{1}{8\pi G}\, G^{(2)}_{\mu\nu}[h].
\ee
The field equations can then be written in the form
\be
G^{(1)}_{\mu\nu}[h]
= 8\pi G \left(T_{\mu\nu} + t_{\mu\nu}^{(\mathrm{grav})}\right) + \mathcal{O}(h^3).
\ee
This relation makes manifest that the gravitational perturbation itself contributes to the total source of curvature, encapsulating the idea that gravity gravitates. Owing to its universal coupling to all forms of energy, including its own, the gravitational field renders general relativity inherently nonlinear. A further subtlety concerns the condition $|h_{\mu\nu}|\ll 1$ (and similarly for its derivatives). 
The metric, however, is not a directly observable quantity; its components depend on the chosen coordinates. 
By an appropriate coordinate transformation, the same physical spacetime can appear with metric components that are either large or small. 
Thus, the weak-field limit must be defined in terms of coordinate-invariant physical quantities.

These considerations lead to several key conceptual questions:
\begin{enumerate}[label=\roman*)]
    \item Do the full, nonlinear Einstein equations admit propagating solutions 
          that can meaningfully be interpreted as gravitational waves?
    \item In the weak-field limit, do these solutions reduce to the familiar linearized waves?
    \item How can one consistently define gravitational energy and momentum 
          in a generally covariant theory where energy cannot be localized?
    \item Given that the metric is not itself an observable, how should one formulate 
          a physically meaningful and coordinate-invariant definition of the weak-field limit?
\end{enumerate}

For decades, the physical reality of gravitational waves remained clouded due to the above conceptual ambiguities. A decisive turning point came with insightful analysis by \cite{Pirani:1956wr}, which linked the Weyl tensor to measurable tidal effects and inspired a new understanding of radiation in general relativity. Building on these ideas, \cite{Bondi:1957dt} introduced the first mathematically precise definition of gravitational waves. His pioneering work, together with subsequent contributions by Robinson and Trautman, established that the full, nonlinear Einstein equations admit genuine radiative solutions that carry energy and angular momentum to infinity. In what follows, we briefly review these seminal developments.

\subsection{The Weyl tensor: geometry of tidal forces}

\begin{figure}[h!]
\centering
\includegraphics[width=\textwidth]{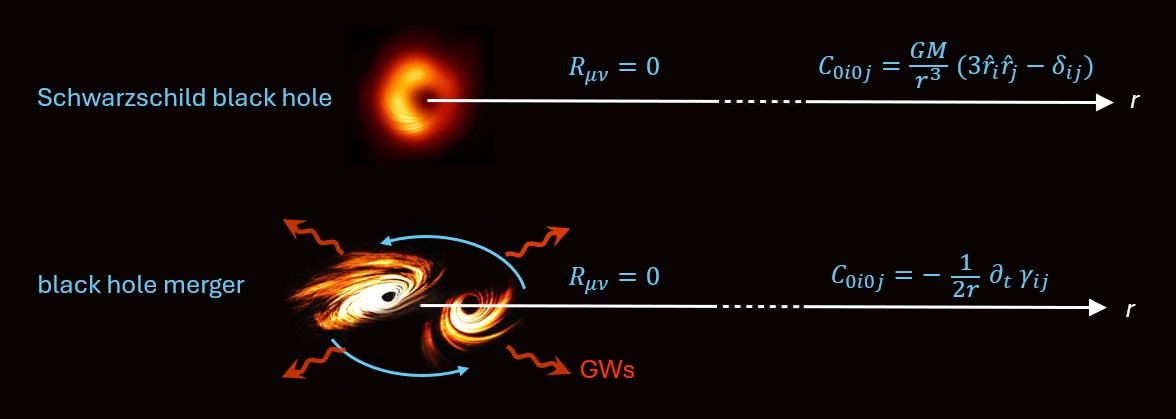}
\caption{The Weyl tensor encodes the free gravitational field, tidal shear and distortion, independent of matter. To highlight the behavior of the Weyl tensor in different regimes, we consider two illustrative examples: the Schwarzschild solution, which is non-radiative and exhibits a leading-order Weyl tensor decay of $\mathcal{O}(1/r^3)$, and a radiative binary black hole merger, where the leading-order Weyl tensor component falls off as $\mathcal{O}(1/r)$ in the far-field (null infinity) limit.
}\label{fig:Weyl}
\end{figure}

Even before Bondi’s seminal Nature paper \citep{Bondi:1957dt}, Felix Pirani made a brilliant early attempt to define gravitational waves in purely geometric terms~\citep{Pirani:1956wr}. His insight was to analyze the algebraic structure of the Weyl tensor through the Petrov classification as a means to identify radiative regions of spacetime. In doing so, Pirani proposed that the physicality of gravitational waves should be anchored not in coordinate-dependent metrics, but in the measurable tidal effects encoded in the curvature tensor, i.e., in the relative acceleration of nearby geodesics governed by the Weyl tensor in vacuum. This idea elegantly shifted the focus from perturbative metrics to intrinsic geometry, offering a criterion that applies even in exact solutions. The approach is well motivated: far from the source, the Ricci tensor vanishes, and the full Riemann tensor reduces to its traceless part, which is the Weyl tensor (see \cref{fig:Weyl}). In what follows, we briefly outline Pirani’s idea and the resulting geometric characterization of radiative spacetimes.

In 4d, the Riemann curvature tensor possesses 20 independent components. Of these, 10 correspond to the Ricci tensor, while the remaining 10 are captured by the traceless Weyl tensor, which encodes the conformal structure of spacetime.
\be
C_{\mu\nu\lambda\sigma}=R_{\mu\nu\lambda\sigma}-(g_{\mu[\lambda}R_{\sigma]\nu}-g_{\nu[\lambda}R_{\sigma]\mu})+ \frac13 R~g_{\mu[\lambda}g_{\sigma]\nu}.
\ee
 More precisely, it is unchanged under a conformal transformation of the metric
\be
g_{\mu\nu} & \mapsto & g'_{\mu\nu} =\Omega^2 g_{\mu\nu},\\
C_{\mu\nu\lambda}^{~~~\sigma} & \mapsto & C_{\mu\nu\lambda}^{'~~\sigma} = C_{\mu\nu\lambda}^{~~~\sigma}.
\ee
Note that $C'_{\mu\nu\lambda\sigma}= \Omega^2 C_{\mu\nu\lambda\sigma}$. More intuitively, the Weyl tensor expresses the tidal forces that a free-falling body feels along a geodesic. However, unlike the Ricci tensor, it does not have information about the change of the volume, but only how the shape of the body is distorted by the tidal forces.

%\subsubsection*{Petrov classification for Weyl tensor}

In four dimensions, the Weyl tensor can be viewed as a linear operator acting on the 
six-dimensional space of antisymmetric rank-two tensors (bivectors),
\begin{equation}
    (C\!\cdot\!B)_{\mu\nu} = \tfrac{1}{2}\, C_{\mu\nu}{}^{\alpha\beta} B_{\alpha\beta},
    \qquad B_{\mu\nu} = -B_{\nu\mu}.
    \label{eq:weyl--B}
\end{equation}
A bivector represents an oriented two-plane in the tangent space and can be expressed as
\begin{equation}
    B_{\alpha\beta} = v_{[\alpha} w_{\beta]} = \frac{1}{2}\left(v_\alpha w_\beta - v_\beta w_\alpha\right).
\end{equation}
Formally, one considers \cref{eq:weyl--B} and seeks its eigenbivectors and corresponding eigenvalues,
\begin{equation}
    \tfrac{1}{2} C^{\mu\nu}{}_{\alpha\beta} \, X_{\mu\nu} = \lambda\, X_{\alpha\beta}.
    \label{eq:eigenbivector}
\end{equation}
Among them, eigenbivectors constructed from null vectors play a privileged role: they are invariant under rescaling 
of $l^\mu$ and encode the geometry of lightlike directions. Equivalently, in four dimensions a nonvanishing Weyl tensor possesses four principal null directions (PNDs), counted with multiplicity, and their degeneracies determine the Petrov type. Physically, their importance stems from the fact that the causal structure of a Lorentzian geometry is determined entirely by its null cones: null vectors specify the propagation of signals, gravitational radiation, and the horizons of black‐hole spacetimes.
 In particular, by examining how the Weyl tensor maps null bivectors into themselves, one probes directly how the free gravitational field influences the focusing, twisting, and shearing of null congruences.

We now show how a single null direction $l^\mu$ can be completed into 
a full null tetrad.  First choose a second
null vector $n^\mu$ such that $l_\mu n^\mu = -1$. The 2-dimensional subspace orthogonal to both $l^\mu$ and $n^\mu$ is
spacelike and admits an orthonormal basis $\{e^\mu_{1},
e^\mu_{2}\}$. From these one constructs two complex null vectors
\be
m^\mu = \frac{1}{\sqrt{2}} \left( e^\mu_{1} + i\, e^\mu_{2} \right),
\qquad
\bar{m}^\mu = \frac{1}{\sqrt{2}} \left( e^\mu_{1} - i\, e^\mu_{2} \right),
\ee
satisfying $m^\mu m_\mu = \bar{m}^\mu \bar{m}_\mu = 0$, $m^\mu \bar{m}_\mu = 1$, and 
$l_\mu m^\mu = n_\mu m^\mu = 0$. The four null vectors $\{ l^\mu,\; n^\mu,\; m^\mu,\; \bar{m}^\mu \}$ form a null tetrad. \footnote{The null tetrad formalism is the central geometric structure of the Newman–Penrose approach \citep{NewmanPenrose1962} and underlies its spin-coefficient representation of curvature and gravitational radiation.} Once the tetrad is constructed, we can introduce bivectors built from $l^\mu$ as 
\be
X_{\mu\nu} = l_{[\mu} m_{\nu]},
\label{eq:Xmunu}
\ee
and its square satisfies $X_{\mu\nu} X^{\mu\nu} = 0$. Similarly, its complex conjugate
\be
\bar{X}_{\mu\nu} = l_{[\mu} \bar{m}_{\nu]},
\ee
is also a simple null bivector.

\begin{figure}[h!]
\centering
\includegraphics[width=0.7\textwidth]{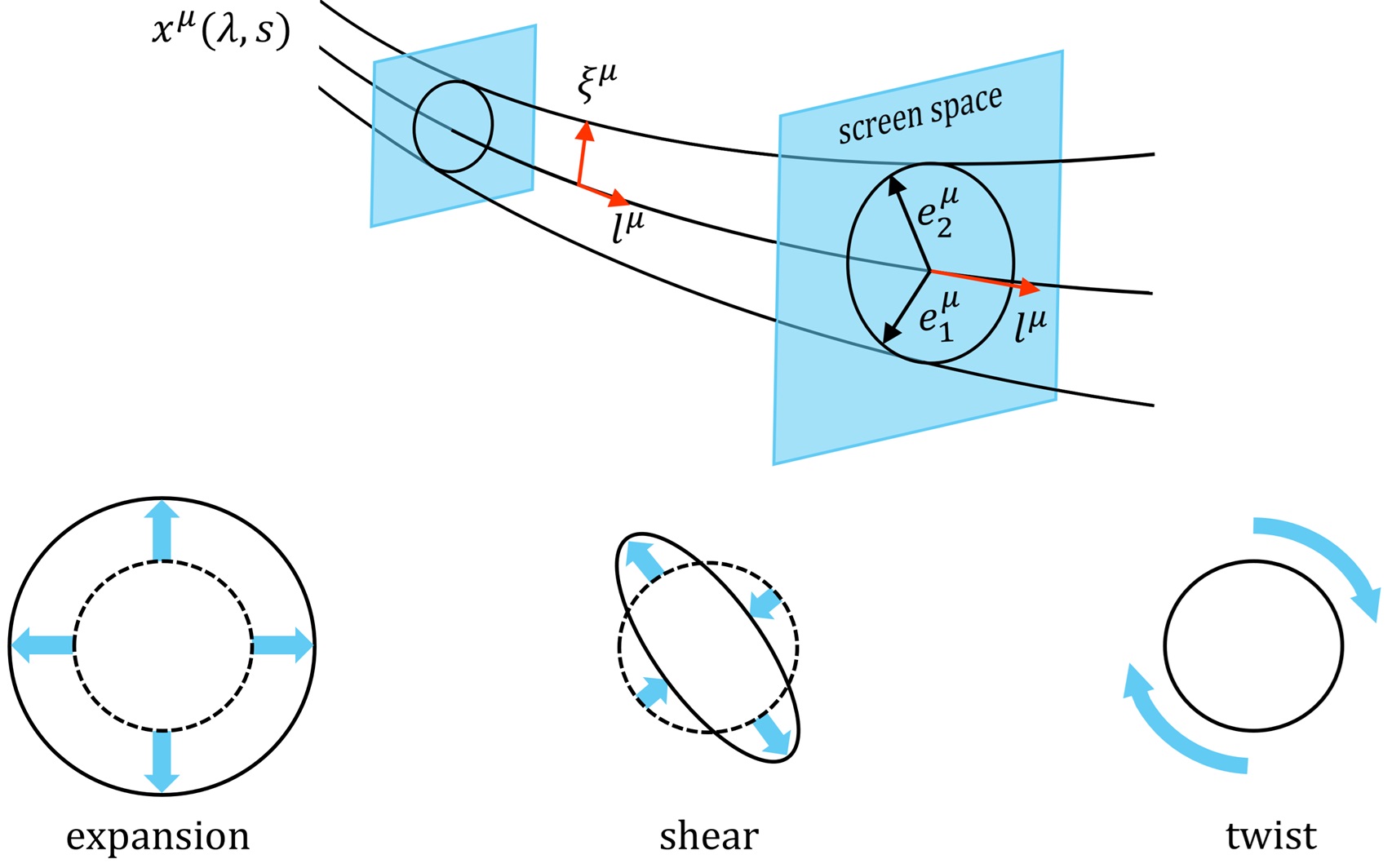}
\caption{
A congruence of null geodesics $x^\mu(\lambda, s)$ labelled by the affine
parameter $\lambda$ and the transverse parameter $s$.
The tangent vector along each geodesic is 
$l^\mu$.
The separation between neighbouring geodesics is described by the
geodesic--deviation vector 
$\xi^\mu$.
The screen space is the two-dimensional subspace orthogonal to both
$l^\mu$ and the auxiliary null vector $n^\mu$, which at each fixed value of 
$\lambda$ is spanned by an orthonormal basis $\{e^\mu_1,e^\mu_2\}$.
Within this screen the cross-section of the light bundle is defined,
along with the expansion, shear, and twist.
}
\label{fig:sacks}
\end{figure}

The principal null directions and their degeneracies give rise to the 
Petrov classification, thereby linking the algebraic structure of curvature 
to the causal geometry of spacetime. Originally introduced by \cite{Petrov:1954} and later 
reinterpreted by \cite{Pirani:1956wr}, the Petrov classification characterizes the 
possible algebraic symmetries of the Weyl tensor at each spacetime point. 
Now you may ask why do these special null directions matter physically? The answer lies in how the curvature acts on the relative motion of neighboring geodesics—particularly
null geodesics that represent bundles of light rays or gravitational waves. Consider a smooth one-parameter family of null geodesics $x^\mu(\lambda,s)$,
where $\lambda$ is an affine parameter along each geodesic and $s$ labels the
distinct geodesics in the congruence. The tangent vector is obtained by differentiating with respect to $\lambda$ 
at fixed $s$, while the geodesic--deviation vector arises from differentiating 
with respect to $s$ at fixed $\lambda$, i.e.
\be
l^\mu \equiv \frac{\partial x^\mu}{\partial\lambda}\Big|_{s},
\qquad
\xi^\mu \equiv \frac{\partial x^\mu}{\partial s}\Big|_\lambda.
\ee
These vectors satisfy the commutation relation
\be
[l,\xi]^\mu = l^\nu \nabla_\nu \xi^\mu
              - \xi^\nu \nabla_\nu l^\mu = 0,
\ee
which holds for any smooth congruence. 
The screen space (or optical plane) is the 2-dimensional subspace 
orthogonal to both $l^\mu$ and $n^\mu$. It is spanned by a pair of orthonormal vectors $\{e^\mu_1, e^\mu_2\}$, which serve as a basis for projecting tensors and defining the transverse degrees of freedom. 
The screen projector onto this 2D subspace is defined by
\be
q_{\mu\nu} = g_{\mu\nu} + l_\mu n_\nu + n_\mu l_\nu,
\ee
which projects any vector onto the space orthogonal to both $l^\mu$ and $n^\mu$ (see \cref{fig:sacks}). Next we define the optical tensor as the doubly projected derivative
\be
\Theta_{\mu\nu} \equiv q_\mu{}^{\alpha} q_\nu{}^{\beta} \nabla_{\beta} l_{\alpha}.
\ee
This is the object that decomposes into
\be
\Theta_{\mu\nu} = \tfrac12 \theta\, q_{\mu\nu} + \sigma_{\mu\nu} + \omega_{\mu\nu},
\ee
giving expansion, shear, and twist of the null congruence as \footnote{Frobenius' theorem states that a vector field $v^\mu$ is hypersurface orthogonal 
iff it satisfies the integrability condition $v_{[\mu}\nabla_{\nu}v_{\rho]}=0$. 
Consequently, hypersurface–orthogonal geodesic congruences possess vanishing 
vorticity, $\omega_{\mu\nu}=0$ \citep{Wald:1984rg}.}
\be
\theta \equiv q^{\mu\nu} \nabla_\nu l_\mu, \quad  \sigma_{\mu\nu} \equiv 
\Theta_{(\mu\nu)} - \tfrac12 \theta\, q_{\mu\nu}, \quad
\omega_{\mu\nu} \equiv \Theta_{[\mu\nu]}.
\ee
In vacuum, the Riemann tensor reduces to the Weyl tensor, so the relative acceleration between two nearby
null geodesics is governed by
\begin{equation}
    \frac{D^2\xi^\mu}{d\lambda^2} = -\, C^\mu{}_{\nu\rho\sigma}\, l^\nu \xi^\rho l^\sigma.
\label{eq:geodesic-devi}
\end{equation}
The right-hand side encodes how spacetime curvature focuses, shears, or twists the light bundle.
When projected onto the two-dimensional screen space orthogonal to $l^\mu$ (the subspace spanned by the complex null vectors $m^\mu$ and $\bar{m}^\mu$),
this equation splits into an isotropic focusing (the expansion) and an anisotropic distortion (the shear). 
A generic PND does not by itself define a shear-free null congruence. In a non-conformally-flat vacuum spacetime, the Goldberg--Sachs theorem states that a null congruence is geodesic and shear-free if and only if its tangent is a repeated PND. Thus this property characterizes the algebraically special Petrov types II, D, III, and N \citep{GoldbergSachs1962, StephaniExactSolutions2003}.

\begin{figure}[h]
\centering
\includegraphics[width=0.95\textwidth]{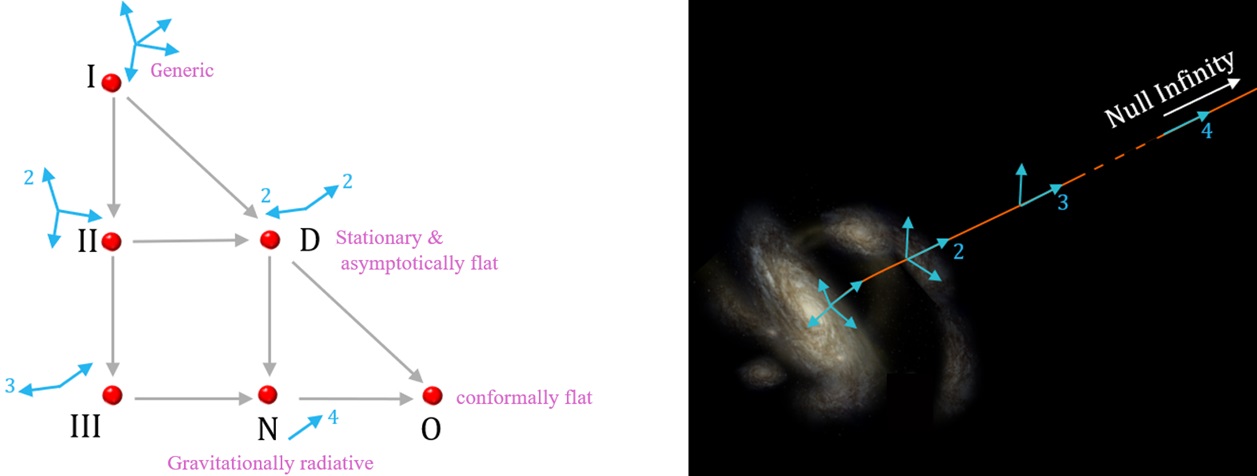}
\caption{Left panel: The schematic form of the Petrov classification of the Weyl tensor. The blue arrows correspond to each of the principal null directions and the numbers on the arrows represent the number of coinciding PNDs. Right panel: The Weyl tensor peels along outgoing null directions: the leading asymptotic component is type N at future null infinity, with type III, II, and I contributions appearing at successively higher inverse powers of $r$. The blue arrow indicates the approach toward null infinity in an asymptotically flat spacetime. }\label{fig:Petrov}
\end{figure}

%% floating tables and figures in boxes are tricky...
\bigskip
\begin{tcolorbox}[colback=gray!10, colframe=gray!10, boxrule=0pt,
                  enhanced, breakable, halign=justify]

\subsubsection*{Petrov Classification} 
It organizes the possible algebraic symmetries of the Weyl tensor into precisely six distinct categories, known as the Petrov types presented in \cref{tab:petrov} and summarized in the left panel of \cref{fig:Petrov}. 

\medskip

%%\begin{table}[h]
\captionsetup{hypcap=false}
\captionof{table}{Petrov Classification of Spacetime, ordered from the most general to the most symmetric.}
\label{tab:petrov}
%%\centering
\renewcommand{\arraystretch}{1.4}
{\small
\begin{tabular}{cc>{\centering\arraybackslash}m{5.3cm}}
\hline
\textbf{Petrov Type} & \textbf{Conditions on PNDs} & \textbf{Typical example / interpretation} \\
\hline
I & 4 simple & Generic Weyl field \\
\hline
II & 2 simple + 2 coincide &  \\
\hline
D & 2 pairs coincide & Schwarzschild/Kerr  \\
\hline
III & 1 simple + 3 coincide & \\
\hline
N & 4 coincide & Pure radiation / radiative leading field \\
\hline
O & Weyl tensor vanishes & Conformally flat \\
\hline
\end{tabular}
}
%%\end{table}

\medskip

  Near localized massive sources, the spacetime is typically of Petrov type I, 
    reflecting its most general algebraic structure. 
    As $r\to\infty$, successive terms in the asymptotic Weyl tensor peel off: the leading $1/r$ contribution is of type N, followed by type III, II, and I terms at higher inverse powers of $r$. Thus the leading asymptotic Weyl curvature is type N, where only the radiative component of the gravitational field remains. Type D includes important isolated vacuum solutions such as Schwarzschild and Kerr. The two repeated PNDs represent radially ingoing and outgoing null congruences in these examples. Type O regions are conformally flat places with zero Weyl tensor, e.g., exact Minkowski and FRW. In this case, any gravitational effects must be due to the immediate presence of matter or the field energy of some non-gravitational field. Type N regions are those regions with transverse gravitational radiation. A spacetime region is type N, if and only if there exists a null vector, $k_{\mu}$, such that
\be
C_{\mu\nu\lambda\sigma}k^{\sigma}=0.
\ee
The four coinciding PNDs are given by this null vector which is the wave vector of the propagating gravitational wave.

\subsubsection*{Peeling Theorem} The Petrov type of a spacetime need not be uniform: it may vary from region to region. The right panel of \cref{fig:Petrov} illustrates the hierarchy of the asymptotic Weyl components along outgoing null directions. This behavior is described by the peeling theorem in general relativity \citep{Sachs1961}, which describes the asymptotic behavior of the Weyl tensor as one goes to null infinity. Let $\gamma$  be a null geodesic from a point $p$ to null infinity, with affine parameter $r$. The peeling theorem states that, as $r$  approaches infinity
\begin{equation}
C_{abcd}=\frac{C_{abcd}^{(1)}}{r}+\frac{C_{abcd}^{(2)}}{r^{2}}+ \frac{C_{abcd}^{(3)}}{r^{3}} + \frac{C_{abcd}^{(4)}}{r^{4}} + ...,
\label{eq:peeling}
\end{equation}
where $C_{abcd}^{(1)}$ is type N, $C_{abcd}^{(2)}$ is type III, $C_{abcd}^{(3)}$ is type II and $C_{abcd}^{(4)}$ is type I \citep{Wald:1984rg}. To summarize, the leading Weyl curvature of a radiative spacetime is of type N at future null infinity.

\end{tcolorbox}

\subsection{When geometry radiates: defining gravitational waves}\label{sec:Bondi}

In his Nature paper, \cite{Bondi:1957dt}, followed by a subsequent paper by \cite*{Bondi:1958aj}, he provided the first mathematically precise definition of gravitational waves in the full Einstein equation. Moreover, he proved that gravitational radiation carries energy. In what follows, we review these results in the context of asymptotically flat spacetimes.  Throughout this discussion in \cref{sec:Bondi,sec:BMS}, we work in Bondi retarded coordinates $(u,r,z,\bar z)$, which are well suited for describing radiation propagating toward null infinity. These coordinates are related to the standard spherical coordinates $(t,r,\theta,\phi)$ by
\be
u = t - r, \qquad 
z = \cot\frac{\theta}{2}\, e^{i\phi}, \qquad 
\bar z = \cot\frac{\theta}{2}\, e^{-i\phi}.
\ee
The coordinate $u$ denotes the Bondi (retarded) time and labels outgoing null hypersurfaces. For comparison, the advanced time in Minkowski spacetime is given by
\be
v = t + r.
\ee
The angular coordinates on the sphere are collectively denoted by $\Theta^A = (z,\bar z)$ and provide a complex parametrization of the unit two-sphere. In these coordinates, the metric on the sphere takes the form
\be
\eta_{z\bar z} = \frac{2}{(1+z\bar z)^2}.
\ee
In this coordinate system, the Minkowski spacetime metric is
\begin{equation}
ds^2 = - du^2 - 2 du dr + 2r^2 \eta_{z\bz} dz d\bz.
\end{equation}

\begin{figure}[h]
\centering
\includegraphics[width=0.75\textwidth]{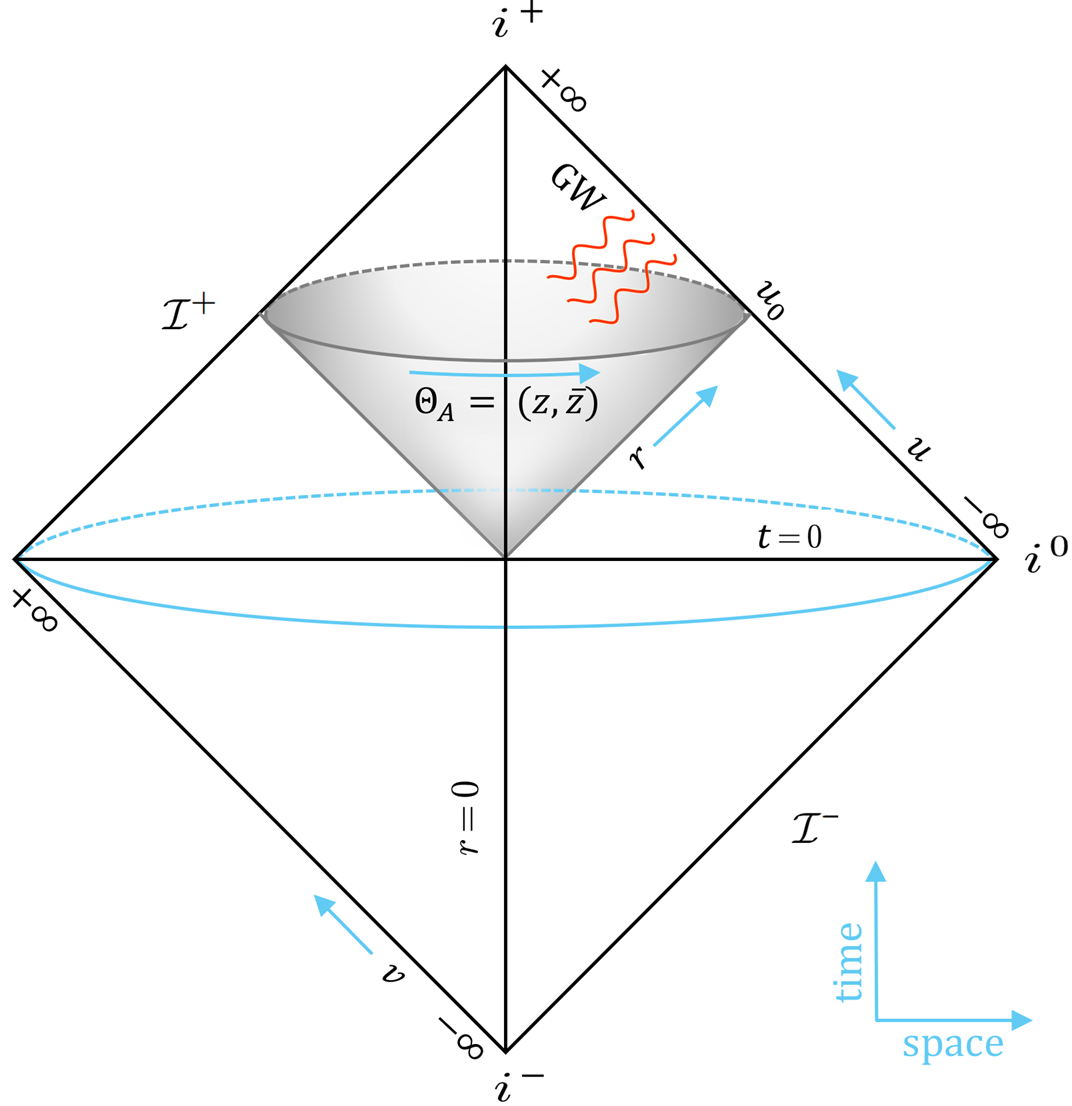}
\caption{Penrose diagram of an asymptotically flat spacetime. 
    The future and past null infinities, $\mathcal{I}^{\pm}$, are parametrized by the retarded and advanced Bondi times $u$ and $v$. 
    The blue ring, $i_0$, denotes spatial infinity. 
    The gray cone in the upper half represents a null surface at $u = u_0$.     The orange wavy lines indicate outgoing gravitational waves (GWs) propagating toward future null infinity.
     }\label{fig:Penrose}
\end{figure}

Before proceeding further, we need to introduce three essential geometric notions that underlie the analysis of gravitational radiation in general relativity.
\begin{itemize}[label=$\bullet$]
\item Asymptotic flatness: An asymptotically flat spacetime is a Lorentzian manifold in which the curvature vanishes at large distances from some region so that the geometry becomes indistinguishable from Minkowski. Outside the source, the Ricci tensor is zero, and therefore, the asymptotic flatness imposes asymptotic falloff conditions on the Weyl tensor as shown in \cref{eq:peeling}.
\item Future null infinity: $\mathcal{I}^+$ is defined as the endpoints of all future-directed null geodesics along which $r\rightarrow \infty$. This null surface is the product of  $S^2$ with a null line $u$ taking values in $\mathbb{R}$. Each null hypersurface, $\sigma_{u_0}$, intersects $\mathcal{I}^+$ in a 2-sphere with $u=u_0$. (See  \cref{fig:Penrose})
\item The topology of  $\mathcal{I}^+$, is that of a null 
hypersurface with the structure $\mathbb{R} \times S^2$, where the 
$\mathbb{R}$ factor corresponds to the retarded time and the $S^2$ factor 
represents angular directions. At each fixed retarded time, the cross-sections 
of $\mathcal{I}^+$ are 2-spheres, forming the so-called celestial sphere 
$\mathrm{CS}^2$.

\end{itemize}

Now, we want to study gravitational theories in which the metric is asymptotic to, but not exactly equal to, the flat metric, and we abbreviate $\Theta^A=(z,\bz)$. A natural starting point is to impose the Bondi gauge,
\begin{align}\label{eq:gauge-bondi}
g_{rr} = 0, \qquad g_{rA} = 0, \qquad 
\partial_r \det\!\left( \frac{g_{AB}}{r^2} \right) = 0,
\end{align}
which fixes the four local diffeomorphism degrees of freedom associated with 
$(r,u,\Theta^A)$. Under these conditions, the most general four-dimensional 
asymptotically flat metric takes the form
\begin{equation}
\begin{split}\label{eq:bondimetric}
ds^2 = - U du^2 - 2 e^{2\beta} du dr + g_{AB} \left( d\Theta^A + \frac{1}{2} U^A du  \right)\left( d\Theta^B + \frac{1}{2} U^B du  \right).
\end{split}
\end{equation}
where $r$ is the luminosity distance and the metric coefficients 
appear in the six functions. Here $U^A$ contributes two angular components and $g_{AB}$, constrained by 
$\det(g_{AB}/r^2)=\det \eta_{AB}$, carries two independent degrees of freedom. 
Thus the Bondi gauge fixes four diffeomorphisms and leaves precisely six 
independent metric functions, the expected number for a general spacetime metric 
modulo coordinate freedom.
Any geometry can be described locally by the metric \cref{eq:bondimetric}. In Bondi gauge, the hypersurface Einstein equation $R_{rr}=0$ (equivalently $G_{rr}=0$) fixes the function $\beta$ rather than leaving it as free data. Defining the normalized angular metric $\hat g_{AB}\equiv g_{AB}/r^2$, the determinant condition is $\det \hat g_{AB}=\det \eta_{AB}$, and the asymptotic expansion is
\be
\hat g_{AB}=\eta_{AB}+\frac{\gamma_{AB}}{r}+O(r^{-2}),
\ee
The constraint equation $R_{rr}=0$ implies the standard hypersurface relation
\be
\partial_r\beta=\frac{r}{16}\,\hat g^{AC}\hat g^{BD}(\partial_r \hat g_{AB})(\partial_r \hat g_{CD}) ,
\label{eq:beta-hypersurface}
\ee
as derived in \cite{Madler:2016xju,Sachs:1962wk, Winicour:2008vpn} 
which integrates to
\be
\beta(u,r,x^A)
= \beta_\infty(u,x^A)
- \frac{1}{32}\frac{\gamma_{AB}\gamma^{AB}}{r^{2}}
+ O(r^{-3}).
\label{eq:beta-integrated}
\ee
Asymptotic flatness fixes the integration function $\beta_\infty(u,x^A)$: since $g_{ur}= -1$ as $r\to\infty$, hence $\beta_\infty=0$. Thus $\beta$ is not an independent degree of freedom but is completely determined by the shear $\gamma_{AB}$, vanishing at null infinity and beginning at order $1/r^{2}$. The remaining hypersurface equations similarly determine the non-radiative metric functions, while the two independent components of the tracefree shear $\gamma_{AB}$ contain the radiative data.

Imposing the asymptotic flatness condition at large $r$ with fixed $(u,z,\bz)$ leads to falloff conditions on the metric components. For the natural choice made by Bondi, van der Burg, Metzner, and Sachs (BMS) \citep{Bondi:1962px}, the large-$r$ structure of the metric is constrained to be \footnote{Compared with the notation adopted in Strominger's book~\citep{Strominger:2017zoo}, we make a slight adjustment for consistency with the conventions used throughout this review. In particular, what is written as $C_{zz}$ in \citet{Strominger:2017zoo} is denoted here by $\gamma_{zz}$ for gravitational waves, and their $\gamma_{z\bar z}$ is written here as $\eta_{z\bar z}$. The conventions of \citet{Strominger:2017zoo} follow those standard in the infrared-gravity and BMS literature, whereas in this work we adopt the notation commonly used in cosmology, expressed in terms of complex spherical–angle coordinates.}

\begin{equation}
\begin{split}
ds^2 &= - du^2 - 2 du dr + 2 r^2 \eta_{z\bz} dz d\bz \\
&+ \frac{2m_B}{r}du^2 + r \gamma_{zz} dz^2 + r \gamma_{\bz\bz}d\bz^2 +D^z\gamma_{zz}d u d z+D^\bz \gamma_{\bz\bz} d u d\bz \\
&  + \frac{1}{r}\left(\frac{4}{3}(N_z+u\p_z m_B)-\frac{1}{4}\p_z(\gamma_{zz}\gamma^{zz})\right)d u d z+c.c.+\ldots, 
\end{split}
\label{eq:Bondi-metric}
\end{equation}
where  $D_z$ is the covariant derivative with respect to  $\eta_{z\bz}$ while $\gamma_{zz}$, $m_B$ and $N_z$ are $r$ independent and functions of $(u,z,\bz)$.  The first three terms in \cref{eq:Bondi-metric} are simply the flat Minkowski metric, and the remaining terms are the leading corrections:
\begin{itemize}[label=$\bullet$]
\item{The first quantity $m_B(u,z,\bz)$ is the Bondi mass aspect (for Kerr BH, $m_B=GM$),}
\item{The next one is $N_z(u,z,\bz)$, which is the Bondi angular momentum aspect (for a Kerr BH, $N_z=2GmL$).}
\item{The last term is $\gamma_{zz}$ which describes the gravitational waves. This quantity is transverse to $\mathcal{I}^{+}$ and $r^{-1}$-suppressed comparing to the dominant orders. The Bondi news tensor is defined as
\be
N_{zz} = \p_{u} \gamma_{zz},
\ee
which is the gravitational analogue of the field strength in gauge field theories, i.e. $F_{uz}=\p_u A_z$.}
\end{itemize}
For later convenience, let us recall a simple but useful identity.  
On each cut of future null infinity $S^{2}_{u}$, endowed with the standard
measure $d^{2}z\, \eta_{z\bar z}$, any covariant total derivative on the sphere
integrates to zero.  
Explicitly, for any smooth vector field $V^{A}$ on $S^{2}$,
\begin{equation}
    \int_{S^{2}} d^{2}z\, \eta_{z\bar z}\, D_{A} V^{A} = 0 \,.
\end{equation}

\bigskip

\begin{tcolorbox}[colback=gray!10, colframe=gray!10, boxrule=0pt,
                  enhanced, breakable, halign=justify]
\subsubsection*{Post-Newtonian Weak-Field, Slow-Rotation Geometry (Lense--Thirring Metric)}
To build intuitive insight into the structure of the Bondi metric, it is useful to begin with the line element describing the weak-field, slow-rotation geometry of an isolated gravitating system \citep{Lense:1918zz} In this post-Newtonian regime, the metric takes the form
\begin{equation}
ds^{2}
= -\bigl(1 - \frac{2GM}{r}\bigr)\, dt^{2}
+ \frac{4G}{r^{2}}\, (\hat{r} \times \boldsymbol{J}) \cdot d\vec{r}\, dt
+ \bigl[(1 + \frac{2GM}{r})\delta_{ij}\bigr]\, dx^{i} dx^{j}.
\label{eq:post-newton}
\end{equation}
The above metric contains the following terms
\begin{itemize}[label=$\bullet$]
\item $M$: the total ADM/post-Newtonian mass (monopole),
\item $\boldsymbol{J}$: the total ADM angular momentum (dipole).
\end{itemize}
Unlike the Bondi metric introduced in \cref{eq:Bondi-metric}, the spacetime above is stationary: neither $M$ nor $\boldsymbol{J}$ varies in time. In this sense, $M$ represents the static, isotropic limit of the Bondi mass aspect $m_B(u,z,\bar z)$. The gravitomagnetic interaction responsible for frame dragging
\be
g_{0i} = \frac{2G}{r^{2}} \big( \hat{r} \times \boldsymbol{J} \big)_{i},
\ee
encodes the characteristic dipolar $1/r^{2}$ falloff produced by the angular momentum $\boldsymbol{J}$. In the Bondi description, the corresponding contribution appears in
\be
g_{uz} \supset \frac{2N_{z}}{3r} + \cdots,
\ee
where $N_{z}$ is the Bondi angular-momentum aspect, carrying precisely the dipole moment associated with $\boldsymbol{J}$.

\end{tcolorbox}

\subsubsection{GWs carry energy}  
The Bondi mass at a Bondi time, $u_1$, is defined as the integral over $S^2_{u_1}$ (the sphere with $u_1$ at $\mathcal{I}^{+}$), as
\begin{equation}\label{MB}
M_B(u_1) = \frac{1}{4\pi G} \int_{S^2} \,  d^2z \, \eta_{z\bz} \, m_B(u_1,z,\bz),
\end{equation}
which is positive and time-dependent, such that it is always non-increasing with time. Moreover, in the limit $u\rightarrow -\infty$, $S^2_{u}$ asymptotically approaches the spatial infinity, $i^0$, and the Bondi mass is equal to the (conserved) ADM mass \citep{Wald:1983ky, Bieri:2016fjs}
\begin{equation}
M_\text{ADM} = \lim_{u \rightarrow -\infty} M_B(u).
\end{equation}
The time evolution of $m_B$ is given by the Einstein equation component $G_{uu}$ at $\mathcal{I}^{+}$ as 
\begin{equation}
\begin{split}\label{mbconstraint}
\p_u m_B = \frac{1}{4} \left[ D_z^2 N^{zz} + D_\bz^2 N^{\bz\bz} \right] - \frac{1}{4} N_{zz} N^{zz} - 4\pi G \lim_{r\to\infty} \left[ r^2 T^M_{uu} \right] , 
\end{split}
\end{equation}
where $T^M_{uu}$ is the matter field's energy-momentum tensor.
Using the Einstein equation \cref{mbconstraint} in \cref{MB} and considering a compact source with $r^2 T^M_{uu}\sim \mathcal{O}(r^{-1})$ at future null infinity, we have
\begin{equation}
M_{B}(u_2)-M_{B}(u_1) = - \frac{1}{16\pi G} \int_{u_1}^{u_2} du \int d^2z \, \eta_{z\bz} \,  N_{zz} N^{zz},   \quad (u_2>u_1),
\label{eq:mass--radiation}
\end{equation}
in which the $D_z^2 N^{zz}$ terms vanish under the $S^2$ integral.  This is the famous Bondi mass-loss formula, which measures the mass lost by gravitational radiation through $\mathcal{I}^{+}$ (see the Penrose diagram in \cref{fig:Penrose}). This vanishes when the Bondi news vanishes. Otherwise, the Bondi mass decreases with time as energy is carried away by gravitational radiation.\footnote{The Bondi mass and the mass loss formula were generalized for spacetimes with non-zero cosmological constant in \cite{Saw:2016isu}.}

\subsubsection{GWs carry angular momentum} 
The Bondi angular momentum associated with the $i$-th rotation generator 
$Y^A_{(i)}$ at a Bondi time $u_1$ is defined by
\be
J_i(u_1)
= \frac{1}{8\pi G} \int_{S^2} d^2z\, \eta_{z\bar{z}} \,  
Y^{A}_{(i)}(z,\bar{z})\, N_A(u_1,z,\bar{z}), 
\ee
The three rotation Killing vectors $Y^{A}_{(i)}(z,\bar{z})$ of the round sphere can be written in the
geometric form
\be
Y^{A}_{(i)}(z,\bar{z})
= \varepsilon_{ijk}\, \hat r^{\,j}\, 
D^{A}\hat r^{\,k},
\ee
where $\hat r^{\,i}$ is the unit radial vector on $S^2$. They generate the $\mathrm{SO}(3)$ isometries of $S^{2}$ and each $Y^{A}_{(i)}$ corresponds to a rotation about one of the 
three Cartesian axes in $\mathbb{R}^{3}$. The evolution of the angular-momentum aspect is governed by the Bondi
constraint equation
\begin{align}
\p_u N_z &= \frac14 \p_z (D^2_z \gamma^{zz} - D_{\bz}^2 \gamma^{\bz\bz}) - u \p_{u}\p_{z}m_B + \frac14 \p_z (\gamma_{zz}N^{zz}) + \frac12 \gamma_{zz} D_z N^{zz} \nonumber\\
& - 8\pi G \lim_{r\rightarrow \infty} [r^2T^M_{uz}].
\end{align}
Integrating the above, similar to the mass aspect, one can find the amount of angular momentum carried by the GWs as
\begin{equation}
\p_u J_i(u)
= -\frac{1}{16\pi G} \int_{S^2} d^2z\, \eta_{z\bar{z}} \,  
 \gamma_{zz}(u,z,\bar{z}) \, Y^{A}_{(i)}(z,\bar{z})\, D_A N^{zz}(u,z,\bar{z}) + \mathrm{c.c.}
 \label{eq:angular-momentum-rad}
\end{equation}
Integrating this relation over retarded time yields the total change in
Bondi angular momentum between two cuts $u_1$ and $u_2$ of future null
infinity
\begin{align}
J_i(u_2) - J_i(u_1)
&= -\frac{1}{16\pi G}
\int_{u_1}^{u_2}\! du
\int_{S^2} d^2z\, \eta_{z\bar z}\,
Y^{A}_{(i)}(z,\bar z)\,
\gamma_{zz}(u,z,\bar z)\, D_A N^{zz}(u,z,\bar z) \nonumber\\
&+ \mathrm{c.c.}    
\end{align}
The negative of this change represents the net angular momentum carried away by GWs
through future null infinity during the interval $[u_1,u_2]$.  In
particular, a nonzero change implies that the outgoing radiation transports
angular momentum from the source to $\mathcal{I}^+$, thereby reducing the
Bondi angular momentum of the system.  Thus, gravitational waves, in
addition to carrying energy and momentum, also carry angular momentum to
null infinity.

\begin{tcolorbox}[colback=gray!10, colframe=gray!10, boxrule=0pt,
                  enhanced, breakable, halign=justify]

\subsubsection*{Energy--Momentum Tensor and Flux of GWs}

To get a better understanding of the results we found above, let us define the effective energy--momentum tensor of gravitational waves and the associated energy and angular-momentum fluxes. 
In the short-wavelength (Isaacson) approximation, gravitational waves behave like an effective radiative field whose energy--momentum tensor is
\be
T^{_\text{GW}}_{\mu\nu}
= \frac{1}{32\pi G}
\bar{\nabla}_\mu \gamma_{\alpha\beta}\,
\bar{\nabla}_\nu \gamma_{\alpha\beta}.
\ee
For an observer with four-velocity $u^\mu$, the energy density of gravitational waves is defined as
\be
\rho_{_\text{GW}} = T^{_\text{GW}}_{\mu\nu} u^\mu u^\nu .
\ee
In nearly flat spacetime, choosing $u^\mu = (1,0,0,0)$ gives the standard expression
\be
\rho_{_\text{GW}}
= \frac{1}{32\pi G} \, 
\dot \gamma_{ij}\, \dot \gamma_{ij}.
\ee
That is related to the right-hand side of \cref{eq:mass--radiation}.
The angular momentum carried by gravitational waves is characterized through the associated flux.  
For rotations about the $i$-axis, the angular-momentum flux is
\be
\frac{dJ_i}{dt}
= -\frac{1}{32\pi G}\,
\epsilon_{ijk}
\, \int  d\Omega \, r^2
\dot \gamma_{j\ell}\,
\gamma_{k\ell}.
\ee
The above flux is related to the right-hand side of \cref{eq:angular-momentum-rad}. 
These relations encode the essential energetic and dynamical content of gravitational radiation in the wave zone and provide the foundation for computing the energy and angular momentum carried away from compact astrophysical systems by the gravitational field.

\end{tcolorbox}

%%%%%%%%%%%%%%%%%%%%%%%%%%%%%%
\subsection{At the edge of spacetime: from Poincar\'e to BMS}\label{sec:BMS}

In their pioneering analysis of gravitational waves, \cite*{Bondi:1962px}, together with \cite{Sachs:1962zza}, set out to characterize 
the asymptotic symmetry group of spacetimes approaching Minkowski space at future 
null infinity~$\mathcal{I}^+$. Rather than recovering the expected Poincar\'e symmetry, 
they uncovered an infinite-dimensional symmetry, the BMS group. This surprising 
result revealed that the gravitational field possesses a far richer asymptotic 
structure than previously anticipated, with far-reaching implications for radiation, 
gravitational memory, and the infrared behavior of gravity \citep{Strominger:2014pwa}. 
Crucially, this richness is intimately connected to the presence of gravitational 
waves and their imprints at null infinity—a connection we will return to and 
explore in more detail below. In what follows, we present a brief summary of this 
historical development. For a more in-depth exposition, see 
\citet{Strominger:2017zoo,Madler:2016xju}.

\subsubsection{Asymptotic symmetries of gravitational fields}

The asymptotic symmetry group of gravitational fields in asymptotically flat spacetimes is defined as the set of vector fields whose Lie derivatives preserve both the Bondi gauge \cref{eq:gauge-bondi} and asymptotic flatness conditions. More concretely, for a vector field $\xi^\mu$ of the form
\be
\xi = \xi^u \partial_u + \xi^r \partial_r + \xi^A \partial_A ,
\ee
the Bondi gauge and fall-off preserving conditions require the following asymptotic expansions
\begin{align}
\xi^u &= f(z,\bz) + \frac{u}{2} D_A Y^A(z,\bz) + O(r^{-1}),\\
\xi^A &= Y^A(z,\bz) - \frac{1}{r} D^A f(z,\bz) + O(r^{-2}),\\
\xi^r &= -\frac{r}{2} D_A Y^A(z,\bz) + \frac{1}{2} D^2 f(z,\bz) + O(r^{-1}),
\end{align}
where $f(z,\bz)$ is an arbitrary function on $CS^2$ and the vector field $Y^A(z,\bar z)$ is a conformal Killing vector on the
2-sphere, obeying
\begin{equation}
 D_A Y_B + D_B Y_A = \eta_{AB}\, D_C Y^C .
\label{eq:killing-conf}
  \end{equation}
This equation implies that the Lie derivative of the metric takes the conformal form
\be
\mathcal{L}_Y \eta_{AB} = \Omega\, \eta_{AB},
\qquad 
\Omega \equiv D_C Y^C ,
\ee
so that the trace gives the conformal factor explicitly. On the round sphere $S^2$, \cref{eq:killing-conf} admits exactly six independent
solutions. This follows from the fact that
\be
\mathrm{Conf}^{+}(S^2) \cong PSL(2,\mathbb{C}) \simeq SO^{+}(3,1),
\ee
whose Lie algebra is the six-dimensional Lorentz algebra $\mathfrak{so}(3,1)$, consisting of three rotations and three boosts. Its double cover is $SL(2,\mathbb{C})$. The three rotational generators appear as true Killing vectors on the celestial sphere:
they preserve the round metric and therefore satisfy
\be
\mathcal{L}_Y \eta_{AB} = 0 \quad \textmd{(rotations)}.
\ee
The remaining three generators correspond to Lorentz boosts. Boosts do not preserve 
the spatial metric, since they mix time and space, but they induce M\"obius 
transformations on $S^2$ and therefore are conformal Killing vectors. Hence the six generators of the Lorentz group, forming $\mathfrak{so}(3,1)$, act on 
the celestial sphere as the six global conformal symmetries of $CS^2$.

\begin{figure}[ht]
\centering
\includegraphics[width=0.5\textwidth]{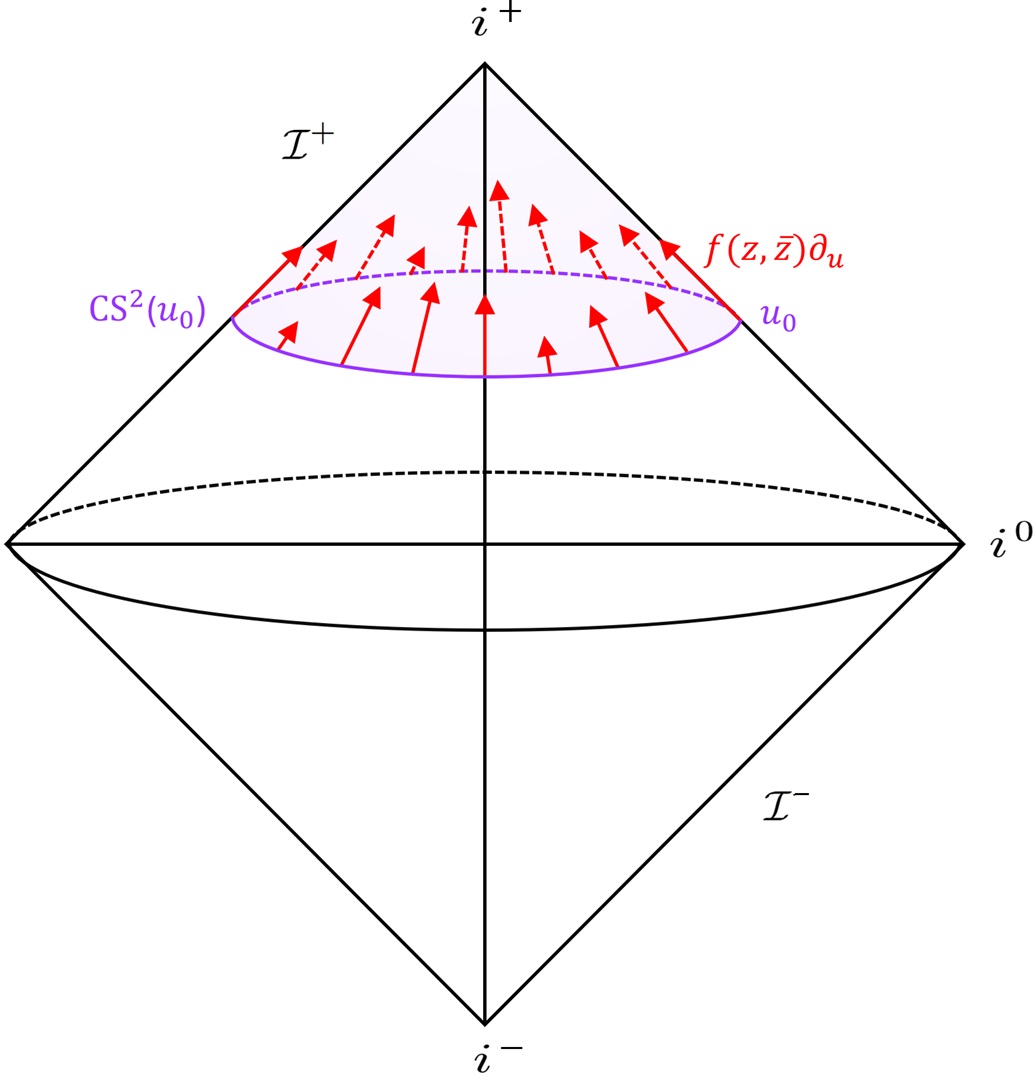}
\caption{Illustration of the supertranslation action $f(z,\bar z)\,\partial_u$ on the 
celestial sphere at fixed Bondi time $u_0$. Each direction on $\mathcal{I}^+$ 
undergoes an angle-dependent shift along the null generators, highlighting the 
geometric meaning of $\mathrm{BMS}_4$ supertranslations.
}\label{fig:BMS4}
\end{figure} 

Having identified the Lorentz 
algebra $\mathfrak{so}(3,1)$ on $CS^2$, one may naturally ask: 
where are the translations of 4d Minkowski space? Remarkably, 
translations arise from the 
scalar function $f(z,\bar z)$ that parametrizes supertranslations. Unlike the 
Lorentz transformations, which are fixed by the conformal geometry of the 
sphere, the function $f(z,\bar z)$ is completely arbitrary. It assigns to each 
null direction on the celestial sphere an independent shift along the null 
generators of $\mathcal{I}^+$. In this sense, translations correspond to the 
lowest spherical-harmonic modes of an infinite-dimensional space of functions 
on $S^2$. Since $f(z,\bar z)$ is an arbitrary function of angle on the celestial sphere, 
it can be expanded in spherical harmonics,
\begin{equation}
f(z,\bar z) = \sum_{\ell=0}^\infty \sum_{m=-\ell}^{\ell} f_{\ell m}\, Y_{\ell m}(z,\bar z).
\label{eq:f-decom}
\end{equation}
The ordinary spacetime translations arise from the four lowest modes with 
$\ell = 0$ and $\ell = 1$. All higher multipoles $\ell \ge 2$ generate genuine 
supertranslations, which deform the structure of $\mathcal{I}^+$ in an 
angle-dependent manner (see \cref{fig:BMS4}). More precisely, a supertranslation is defined by a smooth function $f(\hat{n})$ on the celestial sphere that shifts retarded time according to
\be
u \mapsto u + f(\hat{n}),
\ee
linking the geometry of $CS^2$ to the infinite-dimensional BMS group.

\bigskip

\begin{tcolorbox}[colback=gray!10, colframe=gray!10, boxrule=0pt,
                  enhanced, breakable, halign=justify]

\subsubsection*{Structure of the $\mathrm{BMS}_4$ Algebra}
The four-dimensional BMS algebra is a semidirect product between 
the Lorentz algebra and the infinite-dimensional abelian algebra of 
supertranslations. Let $T_f$ denote a supertranslation with parameter 
$f(\theta,\phi)$ and let $Y$ denote a Lorentz generator. Their commutation 
relations are
\be
[T_f, T_{f'}] = 0, 
\qquad 
[Y, T_f] = T_{Y\!\cdot f},
\ee
where
$Y\!\cdot f \equiv Y^A D_A f - \frac12 f D_A Y^A$.
This shows that Lorentz transformations act nontrivially on the space of 
supertranslations by mixing the spherical-harmonic components of $f$. 
This yields the semidirect product\footnotemark
\be
\mathrm{BMS}_4 = SL(2,\mathbb{C}) \ltimes \mathcal{T},
\ee
where $\mathcal{T}$ is the group of smooth functions on the celestial sphere 
$CS^2$ under addition. Thus 
$\mathrm{BMS}_4$ extends the Poincar\'e group by an infinite-dimensional set of 
transformations acting nontrivially under Lorentz rotations and boosts.

\end{tcolorbox}

\footnotetext{Definition (Semidirect product): 
Let $N$ and $H$ be groups, and let $\rho: H \to \mathrm{Aut}(N)$ be a homomorphism describing an action of $H$ on $N$ by automorphisms. The semidirect product of $N$ by $H$, denoted $N \rtimes_{\rho} H$, is defined as the set $N \times H$ equipped with the group law
\be
(n_1, h_1)(n_2, h_2)
= \big( n_1 \,\rho(h_1)(n_2),\, h_1 h_2 \big).
\ee
With this multiplication, $N$ is a normal subgroup of $N \rtimes_{\rho} H$, while $H$ embeds as a (not necessarily normal) subgroup. The product is semidirect because the group law in $N$ is twisted by the action of $H$, encoding the nontrivial way in which elements of $H$ act on elements of $N$.
}

\subsubsection{Why supertranslations matter in radiative spacetimes}\label{sec:why-BMS}

Up to now we showed that the asymptotic symmetry group of asymptotically 
flat spacetimes in Bondi gauge is the BMS group. Yet a natural question remains: why does its infinite-dimensional supertranslation sector become physically relevant in radiative spacetimes? The BMS group follows from the asymptotic boundary conditions themselves. Gravitational radiation makes its supertranslation sector physically nontrivial through the Bondi shear, memory, and transitions between asymptotic vacua. In exact Minkowski spacetime there is no radiation, whereas an asymptotically flat spacetime can carry gravitational waves whose imprints remain at null infinity.

In exact Minkowski spacetime the asymptotic metric at $\mathcal{I}^+$ is globally 
Lorentz invariant, and the Bondi shear vanishes,
\be
\gamma_{AB}^{\rm Mink} = 0.
\ee
Under a supertranslation with parameter $f(z,\bz)$, the shear transforms as
\begin{equation}
\delta_f \gamma_{AB} = f\,\partial_u \gamma_{AB} -2 D_A D_B f + \eta_{AB} D^2 f.
\label{eq:deltaf--}
\end{equation}
 Demanding that Minkowski spacetime remain invariant forces 
this variation to vanish. Using \cref{eq:f-decom}, the only solutions are the $\ell =0,1$ spherical 
harmonics—corresponding to the usual time and space translations. Thus the 
subgroup that leaves this Minkowski vacuum invariant is precisely the Poincaré group. The picture changes dramatically once gravitational radiation is present. If gravitational radiation passes through future null infinity, the shear changes 
between early and late retarded times. Integrating the news gives the net change
\be
\Delta \gamma_{AB} 
= \gamma_{AB}(u = +\infty) - \gamma_{AB}(u = -\infty) 
= \int_{-\infty}^{+\infty} N_{AB}\, du .
\ee
This nonzero shift is the gravitational memory effect: freely falling observers 
experience a permanent relative displacement even after the wave has passed. 
What is striking is that the mathematical form of this shift is highly 
constrained. On a news-free vacuum cut of the two–sphere at null infinity, the inhomogeneous part of \cref{eq:deltaf--} is the symmetric trace-free tensor constructed from a scalar function $f(z,\bar z)$. Thus, if the memory-induced shift $\Delta \gamma_{AB}$ 
matches this expression for some $f$, then the initial and final vacuum 
configurations are related by a supertranslation. Radiative processes therefore 
move the system between inequivalent vacua connected by supertranslations. These 
transformations cease to act trivially: instead, they map an infinite continuum 
of physically distinct gravitational vacua into one another. The physical relevance of the supertranslation sector is, in this sense, a direct imprint of gravitational waves—the memory they 
leave, the resulting infinite vacuum degeneracy, and the refined geometric 
structure of spacetime at future null infinity~$\mathcal{I}^+$. 

%%%%%%%%%%%%%%%%%%5

\bigskip

\begin{tcolorbox}[colback=gray!10, colframe=gray!10, boxrule=0pt,
                 enhanced, breakable, halign=justify]

\subsubsection*{Infrared Triangle}
{\centering
\includegraphics[width=0.45\textwidth]{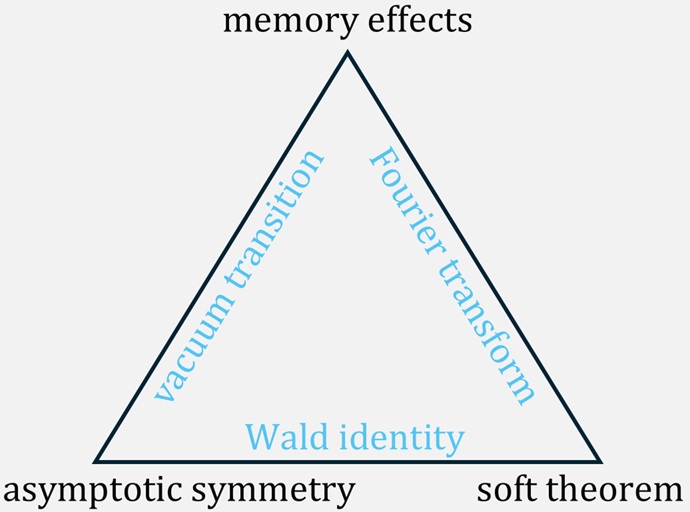}
\par}\vspace{0pt}
The infrared triangle expresses the equivalence between asymptotic symmetries, soft theorems, and memory effects in gauge theory and gravity. The first corner is formed by the infinite-dimensional Bondi--Metzner--Sachs (BMS) asymptotic symmetry group, discovered by \cite*{Bondi:1962px} and independently by \cite{Sachs:1962wk}.  The second corner contains the universal soft theorems, beginning with Weinberg's 1965 soft photon and graviton theorems~\citep{weinberg1965}. It was later revealed by Strominger that gravitational memory is equivalent to the soft graviton theorem \citep{Strominger:2013jfa}. The third corner comprises the family of gravitational wave memory effects, encompassing both linear and nonlinear memory, as originally described by \cite{ZeldovichPolnarev1974}, \cite{Christodoulou:1991cr}, and further developed \cite{Braginsky:1985vlg,Thorne:1992sdb}. \cite{Strominger:2014pwa} showed that gravitational memory corresponds to a transition between BMS vacua and that the Fourier transform of the memory formula yields Weinberg's soft theorem. The structure was later explicitly named the infrared triangle in \citet{Strominger:2017zoo}.

\vskip 0.2cm

Analogous IR effects also arise in QED, where an infrared triangle relates large gauge symmetries, soft photon theorems, and electromagnetic memory effects. Large gauge transformations at null infinity give rise to conserved charges whose Ward identities reproduce the universal soft photon theorem \citep{Bieri:2013hqa,Strominger:2013lka,He:2014cra,Strominger:2014pwa,Kapec:2015ena}. These symmetries have a direct physical imprint in the form of a permanent electromagnetic memory effect. In the usual electromagnetic memory effect, a pulse of radiation imparts a net
velocity kick to a distant test charge \citep{Bieri:2013hqa}. This kick is the classical
counterpart of the leading soft-photon theorem \citep{Pasterski:2015zua}. A complementary observable,
photon chiral memory, is sourced by the photon's helicity (topological) current
at future null infinity and so tracks the net helicity radiated in each direction
\citep{Maleknejad:2023nyh}.  Memory-type observables have also
been related to angle-dependent electric--magnetic duality rotations on the
celestial sphere \citep{Henneaux:2020nxi, Oblak:2023axy, Liu:2023qtr}.

\end{tcolorbox}

%\cite{Seraj:2021rxd,Faye:2024utu}

%%%%%%%%%%%%%%%%%%%%%%%%%%%%%

\section{Expanding Universe and gravitational waves}
\label{sec:3}

In \cref{sec:1,sec:2}, we explored gravitational radiation in Minkowski and asymptotically flat spacetimes. In this part, we shift our attention to the generation and evolution of gravitational waves in a cosmological background. 
On scales much smaller than the Hubble radius, cosmic expansion can often be neglected, and spacetime may be treated approximately as asymptotically flat.
On larger, cosmological scales, the expansion becomes dynamically relevant, and the spacetime must instead be modeled by the FLRW metric (see \cref{fig:hierarchy}). We begin this section with the expansion history of the Universe and the sequence of cosmological eras, then examine gravitational waves produced during inflation and their behavior in (quasi)–de Sitter geometry. Finally, we turn to the infrared structure of inflation, discuss Weinberg’s adiabatic theorem, and conclude with Maldacena’s consistency relation for cosmological correlators.

\subsection{Expansion history and cosmological eras}
\label{sec:cosmological-background}

Modern cosmology is based on two key observations: i) the universe is expanding, and ii) on large scales ($> 100$ Mpc) the matter distribution is homogeneous and isotropic. The average spacetime is then described by the Friedmann--Lema\^{i}tre--Robertson--Walker (FLRW) metric 
\be
ds^2_\mathrm{{FLRW}} = -dt^2+ a^2(t) (\frac{dr^2}{1-K r^2} + r^2 d\Omega^2),
\ee
where $t$ is the cosmic time, $a(t)$ is the scale factor, and $K=0,+1,-1$ describe flat, positively curved and negatively curved spacelike 3-hypersurfaces, respectively. This spacetime is conformally flat, with a vanishing Weyl tensor (Petrov type O). 
From now on, we restrict our discussion to the case of the flat universe with $K=0$, which is favored by present observations. \footnote{The curvature density parameter, 
$\Omega_K \equiv -\dfrac{K}{a^2H^2}$ quantifies the fractional contribution of spatial curvature to the total energy density. 
Current observations constrain it to $|\Omega_K| < 0.002$ $98\%~\mathrm{C.L.}
$ \citep{Planck:2018vyg}.} Now, we define the conformal time $\tau$ in terms of the physical time $t$ as 
\be
d\tau \equiv \frac{1}{a(t)} dt,
\ee
We also define the Hubble parameter, $H$, which measures the expansion rate of the Universe
\begin{equation}
H \equiv \frac{\dot{a}}{a}.
\end{equation}
Using this relation, we find that the flat FLRW geometry is conformally equivalent to Minkowski spacetime
\begin{equation}
g_{\mu\nu}(\tau,\boldsymbol{x}) = a^2(\tau)\,\eta_{\mu\nu}.
\end{equation}
The factor $a^2$ uniformly rescales the Minkowski geometry, stretching both space and time by the scale factor $a(\tau)$. Light rays (null geodesics) remain unaffected by this conformal rescaling, since for $ds^2 = 0$ the scale factor cancels out. Consequently, the causal structure of the Universe in conformal coordinates is identical to that of flat spacetime, making the conformal time $\tau$ a natural variable for describing wave propagation, such as GWs, CMB photons, and cosmological perturbations, in an expanding universe.

Let us now consider a perfect fluid
\be
T_{\mu\nu} = (\rho + P)\, u_\mu u_\nu + P\, g_{\mu\nu},
\ee
where 
$\rho$ is the energy density measured in the rest frame of the fluid, 
$P$ is the isotropic pressure, and
$u^\mu$ is the 4–velocity of the fluid (normalized as $u_\mu u^\mu = -1$). For an equation of state 
\be
P=w\rho,
\ee
the Einstein field equations reduce to the Friedmann equations, i.e.
\be\label{Friedmann}
3\mpl^2(\frac{\dot a}{a})^2 = \rho \an \mpl^2 \frac{\ddot a}{a}= - \frac16 (1+3w)\rho,
\ee
where $\mpl^2 \equiv (8\pi G)^{-1}$ is the reduced Planck mass.
Solving these equations for a constant equation-of-state parameter $w\neq -1$, we have 
\be
a(t) = \bigg(\frac{t}{t_I}\bigg)^{\frac{2}{(1+3w)+2}} \an a(\tau) = \bigg(\frac{\tau}{\tau_I}\bigg)^{\frac{2}{(1+3w)}},
\ee
where $t_I$ and $\tau_I$ are some positive constants. The behavior of the scale factor during different cosmological eras is summarized in \cref{Table:T0}.

\begin{table}[h]
\caption{Equation of state, scale factor, and Hubble parameter for each cosmological era. The last column shows the validity of the Strong Energy Condition (SEC), i.e., whether $\rho + 3P > 0$. Here, $\mH \equiv aH$. }
\label{Table:T0}
\centering
\renewcommand{\arraystretch}{2.5}
\begin{tabular}{ m{3cm} m{2cm} m{2.5cm} m{2cm} m{1cm} } 
 \toprule
 \textbf{Cosmological era} & \textbf{Eq. of state} & \textbf{Scale factor} & \textbf{Hubble} & \textbf{SEC} \\ 
 \midrule
 \multirow{2}{*}{Cosmic inflation} 
 & \multirow{2}{*}{$w=-1$ } 
 & $a(t)=\exp(Ht)$ 
 & $H(t)=H_{\text{inf}}$ 
 & \multirow{2}{*}{No} \\
 &  & $a(\tau)=-\frac{1}{H\tau}$ 
    & $\mH(\tau)=-\frac{1}{\tau}$ 
    & \\
 \hline

 \multirow{2}{*}{Stiff era \tablefootnote{Stiff era is a hypothetical phase between inflation and the radiation era, appearing in models where a scalar field becomes kinetic-energy dominated \citep{Spokoiny:1993,Joyce:1996,Ferreira:1998}. Although unconfirmed, it is consistent with current bounds if it ends before BBN and is phenomenologically notable for enhancing high-frequency primordial gravitational waves \citep{Gouttenoire:2021jhk}.}
} 
 & \multirow{2}{*}{$w = 1$} 
 & $a(t)=\left(\frac{t}{t_S}\right)^{\frac{1}{3}}$ 
 & $H(t)=\frac{1}{3t}$ 
 & \multirow{2}{*}{Yes} \\
 & & $a(\tau)=\left(\frac{\tau}{\tau_S}\right)^{\frac{1}{2}}$ 
   & $\mH(\tau)=\frac{1}{2\tau}$ 
   & \\
 \midrule

 \multirow{2}{*}{Radiation era} 
 & \multirow{2}{*}{$w=\frac{1}{3}$} 
 & $a(t)=\left(\frac{t}{t_I}\right)^{\frac{1}{2}}$ 
 & $H(t)=\frac{1}{2t}$ 
 & \multirow{2}{*}{Yes} \\
 & & $a(\tau)= \frac{\tau}{\tau_I}$ 
   & $ \mH(\tau)=\frac{1}{\tau}$ 
   & \\
 \midrule
 \multirow{2}{*}{Matter era} 
 & \multirow{2}{*}{$w=0$} 
 & $a(t)=\left(\frac{t}{t_J}\right)^{\frac{2}{3}}$ 
 & $H(t)=\frac{2}{3t}$ 
 & \multirow{2}{*}{Yes} \\
 & & $a(\tau)= \left(\frac{\tau}{\tau_J}\right)^{2}$ 
   & $ \mH(\tau)=\frac{2}{\tau}$ 
   & \\
 \bottomrule
\end{tabular}
\end{table}

%For more details about the stiff era see the footnote. 

To gain insight into the causal structure of cosmological spacetime, 
let us consider the propagation of light, characterized by the null condition $ds^2 = 0$. 
For a radial null geodesic, this condition leads to
\begin{equation}
r(\tau) - r(\tau_I) 
   = \int_{\tau_I}^{\tau} d\tau' 
   = \int_{a_I}^{a} \frac{d\ln a'}{a' H(a')},
\label{eq:horizon-int}
\end{equation}
where $\tau$ is the conformal time. 
This integral defines the comoving distance that light can travel between $\tau_I$ and $\tau$, 
and hence determines the size of the observable region of the Universe at a given epoch. 
In terms of the comoving Hubble radius, $(aH)^{-1}$, we can write
\begin{equation}\label{eq:aH}
(aH)^{-1} = \frac{(1 + 3w)}{2}\, \tau_I\, a^{(1 + 3w)/2},
\end{equation}
which depends on the scale factor $a$ and the equation-of-state parameter $w$. 
The characteristic physical distance scale of the expanding Universe is set by the Hubble radius,
\be
R_H = \frac{1}{H},
\ee
which sets the characteristic scale separating sub-Hubble and super-Hubble evolution. 
A perturbation mode with comoving wavenumber $k$ has a corresponding physical length scale $
\lambda_{\mathrm{phys}}(t) = \frac{a(t)}{k} $, which grows with the cosmic expansion and can be compared with $R_H$ to determine whether the mode lies inside or outside the Hubble horizon. Modes satisfying $\lambda_{\mathrm{phys}} \le R_H$ (or equivalently, $k \ge aH$) are said to be sub-Hubble, 
while those with $\lambda_{\mathrm{phys}} > R_H$ are super-Hubble; the relevant adiabatic modes are approximately frozen until re-entry \citep{Weinberg:2003sw}. 
Today, the Hubble radius is approximately $R_H^{(0)} \simeq 4.4~\mathrm{Gpc}$, corresponding to the present Hubble parameter $H_0 \simeq 67.7~\mathrm{km\,s^{-1}\,Mpc^{-1}}$ \citep{Planck:2018vyg}. Hence, the present Hubble scale corresponds to
\begin{equation}
k_0 = H_0\simeq 2.25 \times 10^{-4}~\mathrm{Mpc}^{-1}, 
\qquad 
\lambda_{0} \equiv k_0^{-1} \simeq 4.4~\mathrm{Gpc}.
\label{eq:k0-lambda0}
\end{equation}
The observed statistical homogeneity and isotropy of the cosmic microwave background (CMB) imply that perturbations on scales as large as $\lambda_0$ must have originated from a causally connected region.

%%%%%%%%%%%%%%%%%%%%%%%%%%%%%%%%%
%However, within the standard Big Bang framework, regions separated by such large distances could never have communicated before recombination, as their past light cones do not overlap. This apparent contradiction constitutes the horizon problem, whose natural resolution arises through a period of accelerated expansion, cosmic inflation, which stretches causally connected quantum fluctuations to cosmological scales observed today.

\begin{figure}[h]
\centering
\includegraphics[width=0.4\textwidth]{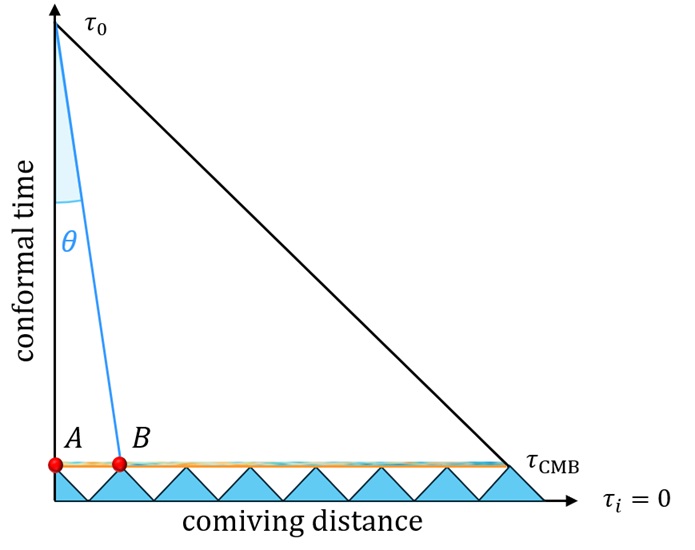}
\includegraphics[width=0.58\textwidth]{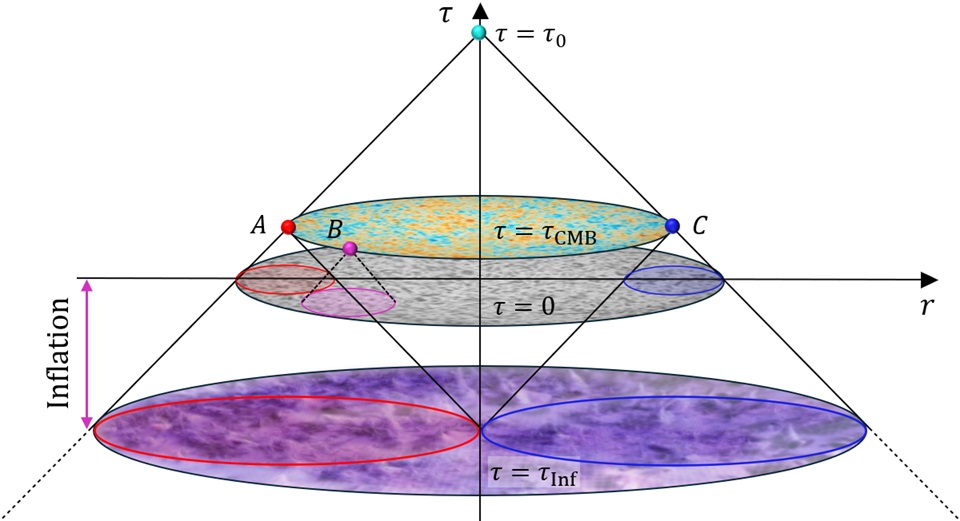}
\caption{Left panel: The (comoving) causal past of an observer today at $\tau_0$ (redshift $z_0=0$), in FLRW spacetime made of ONLY ordinary matter, i.e. $1+3w>0$. The orange line shows the last scattering surface, $\tau_\text{CMB}$ (at redshift $z_{rec}\simeq 1090$). $A$ and $B$ are two causally disconnected points at the last scattering surface. In fact, the angle spanned by the shaded area at the surface of last scattering is the comoving horizon at recombination. Right panel: The causal past of an observer today, including the epoch of cosmic inflation. 
Inflation extends conformal time to $\tau_\mathrm{inf} < 0$, enlarging the causal domain of the Universe. 
Consequently, regions that are outside each other's horizons at $\tau = 0$ were in causal contact during inflation, 
thereby resolving the horizon problem and explaining the observed uniformity of the CMB.
 }\label{fig:r-tau}
\end{figure} 

\bigskip

\begin{tcolorbox}[colback=gray!10, colframe=gray!10, boxrule=0pt,
                  enhanced, breakable, halign=justify]

\subsubsection*{Horizon Problem\footnotemark} Ordinary forms of matter, with positive pressure, satisfy the strong energy condition (SEC), i.e. $\rho+3P > 0$ ($1+3w>0$). Thus, from \cref{eq:aH}, the comoving Hubble radius increases as the universe expands. Now, let us compute the angle spanned by the comoving horizon at recombination, $\theta$. As we see in  \cref{fig:r-tau}, $\theta$ is given as
\begin{equation}
\label{eq:thetaaa}
\sin\theta = \frac{2(\tau_\text{CMB}-\tau_i)}{\tau_0-\tau_\text{CMB}}.
 \end{equation}
On the other hand, we can read the $\tau$ integral in \cref{eq:horizon-int} as
\begin{equation}\label{eq:tau--}
\tau - \tau_I = - \int_{z_1}^{z_2} \frac{dz}{H(z)},
\end{equation}
where $z$ is the redshift parameter $1+z = \frac{1}{a(z)}$, in which we set the scale factor today to unity, $a_0=1$. Moreover, the Hubble parameter is
\begin{equation}\label{eq:Hz}
H(z) = H_0\sqrt{\Omega_m (1+z)^3 + \Omega_{r}(1+z)^4+\Omega_{\Lambda}},
\end{equation}
where $\Omega_{m}=0.3$, $\Omega_{r}=\frac{\Omega_m}{1+z_{eq}}$, and $\Omega_{\Lambda}=1-\Omega_{m}-\Omega_r$ are the matter, radiation, and (late time) dark energy fractions today. Using \cref{eq:tau--,eq:Hz} in \cref{eq:thetaaa} and solving the integral, we find
\be
\theta_c \simeq 2.3^{\circ}.
\ee
At recombination, the causal wavenumber is 
\be
k_{\mathrm{causal}} \simeq a_* H_* \approx 4\times10^{-3}~\mathrm{Mpc^{-1}},
\ee
corresponding to a comoving length scale $\lambda_{\mathrm{causal}}\simeq 250~\mathrm{Mpc}$. Only perturbation modes with $k \gtrsim k_{\mathrm{causal}}$ could have been in causal contact and reached thermal equilibrium by the time of recombination (\cref{fig:r-tau}, left panel). In contrast, the mode corresponding to our present Hubble scale, with $\lambda_0 \simeq 4.4~\mathrm{Gpc}$ given by \cref{eq:k0-lambda0}, gives
\be
\frac{k_{\mathrm{causal}}}{k_0} \simeq 17.
\ee
That implies the region encompassing our present Hubble volume 
would have consisted of roughly $(17)^3 \sim 10^4$ causally disconnected domains at recombination. 
On the other hand, the observed homogeneity of the CMB implies that even scales as large as $\lambda_0$ 
were once in causal contact, an apparent contradiction known as the horizon problem.  As we see in the following, cosmic inflation, an early period of accelerated expansion in which the SEC is violated, solves the horizon problem dynamically. 

\end{tcolorbox}
\footnotetext{The covariant form of the strong energy condition (SEC) is 
$\left(T_{\mu\nu} - \tfrac{1}{2} g_{\mu\nu}T\right)u^\mu u^\nu \ge 0$ 
for all timelike vectors $u^\mu$.
}

\subsubsection{Cosmic inflation}  
\begin{figure}[h!]
\centering
\includegraphics[width=\textwidth]{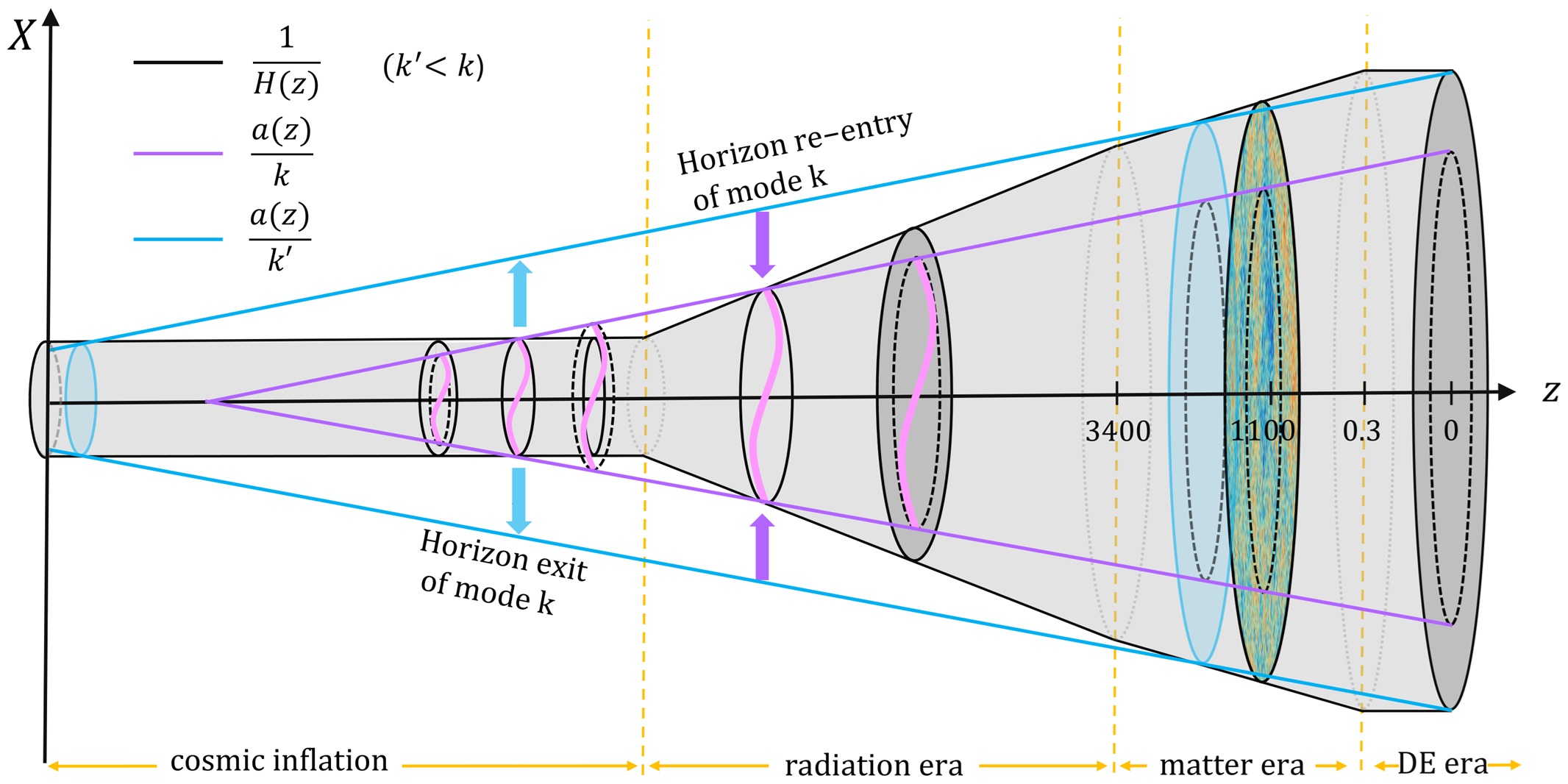}
\includegraphics[width=\textwidth]{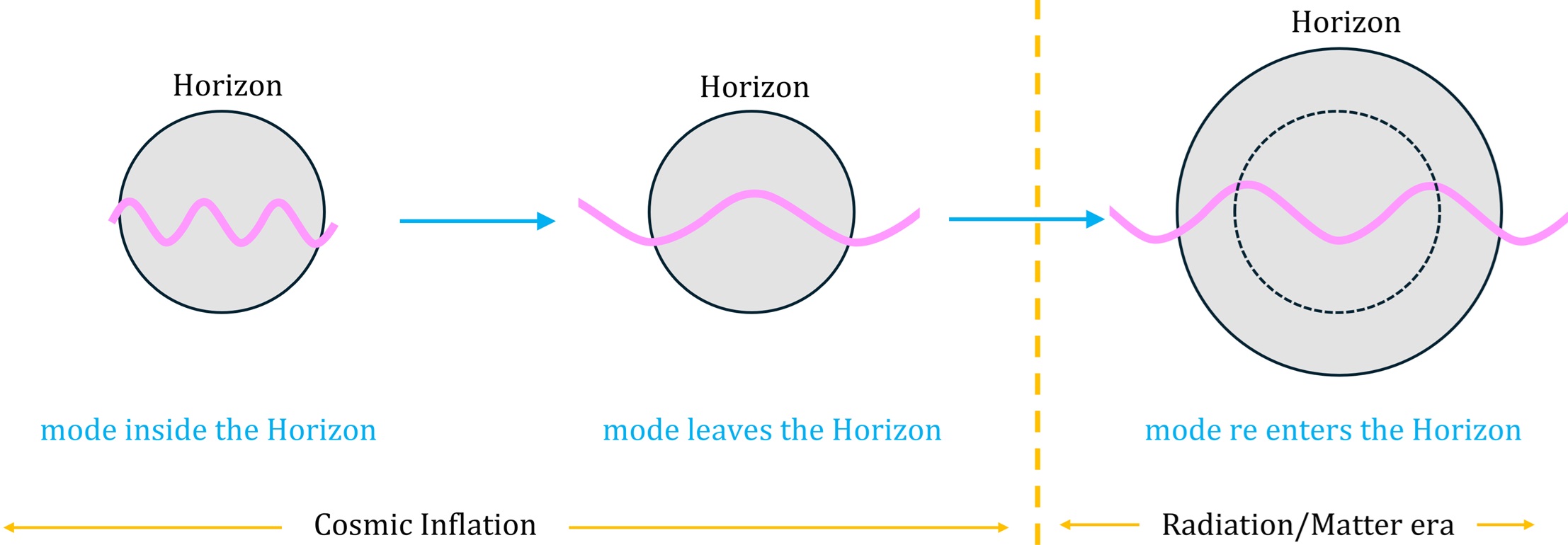}
\caption{Top panel: The Hubble radius, $\frac{1}{H}$, and the physical length scale, $\frac{a}{k}$, as functions of the scale factor. Notice that $X$ is the physical coordinate, $X=ax$. The black line shows $1/H$, and the red and blue lines denote the physical length scales of two modes that re-entered the Hubble radius during radiation domination (RD) and matter domination (MD) eras, respectively. The shaded gray region shows the sub-Hubble region. Bottom panel: close-up view of the mode relative to the Hubble radius.}\label{fig:scale-Hubble-I}
\end{figure}

Our discussion of the horizon problem was based on the validity of the SEC ($1+3w>0$) and, therefore, the growing Hubble sphere of the standard Big Bang cosmology. A simple solution, therefore, is a phase of decreasing comoving Hubble radius in the early
history of the universe,
\be
\frac{d(aH)^{-1}}{dt} <0 \where 1+3w<0.
\ee
If this lasts long enough, the horizon problem can be solved (see the right panel of \cref{fig:r-tau} and \cref{fig:scale-Hubble-I}).

To quantify this idea, let us compute the total amount of expansion that took place between the moment a comoving mode $k$ crossed the Hubble radius during inflation and the end of inflation, a quantity known as the number of e-folds, $N_k$. It is defined as $N_k \equiv \ln\!\left(\frac{a_{\mathrm{end}}}{a_k}\right)$, where $a_k$ and $a_{\mathrm{end}}$ denote the scale factor at horizon exit of mode $k$ and at the end of inflation, respectively. For the present Hubble scale, $k_0 = 2.25\times10^{-4}~\mathrm{Mpc}^{-1}$, the required number of e-folds is approximately 
\be
N \simeq 61.4 + \frac{1}{2}\ln\!\left(\frac{H_{\mathrm{inf}}}{10^{14}~\mathrm{GeV}}\right),
\ee
which sets the minimal duration of inflation required to solve the horizon problem and to ensure that our present Hubble volume originated from a single, causally connected patch before inflation began. For typical inflationary energy scales $H_{\mathrm{inf}}\!\sim\!10^{13}\!-\!10^{14}~\mathrm{GeV}$, the Universe must have expanded by roughly $60$ e-folds during inflation.

The inflation paradigm postulates a brief period (within $t_\text{inf} \approx \frac{N_0}{H_\text{inf}}$) of quasi-exponential accelerated expansion during which the scale factor increased by about 60 e-folds \citep{Starobinsky:1980, Sato:1981, Guth:1981, Linde:1982}. This considerable expansion is sourced by a negative pressure component in the energy-momentum of the matter contents and drives the universe towards almost perfect homogeneity, isotropy, and flatness that we have observed.
During inflation, the spacetime is very close to a de Sitter space $w\simeq -1$, and the Hubble parameter, $H$, is almost constant. For a recent pedagogical discussion of de Sitter geometry, causal structure, and quantum fields, see \cite{Maleknejad:2026dS}. The scale factor grows exponentially, $a(t) \propto e^{Ht}$, and the deviation from a perfect de Sitter space is quantified in terms of two slow-roll parameters  \footnote{
Besides the definitions introduced in \Cref{eq:def-slow-roll}, two other slow–roll conventions are widely used in the literature.  
The potential slow–roll parameters are  
$\epsilon_V \equiv \tfrac{M_{\rm Pl}^2}{2}\!\left(\tfrac{V'}{V}\right)^{\!2}$  
and  
$\eta_V \equiv M_{\rm Pl}^2 \tfrac{V''}{V}$;  
the Hubble slow–roll parameters are  
$\epsilon_H \equiv -\tfrac{\dot H}{H^{2}}$  
and  
$\eta_H \equiv \tfrac{\dot\epsilon_H}{\epsilon_H H}$.  }
\begin{equation}
\epsilon \equiv -\frac{\dot H}{H^2} \an \eta = -\frac{\ddot{H}}{2\dot{H}H},
\label{eq:def-slow-roll}
\end{equation}
which should be small during the slow-roll inflation. This causes the comoving Hubble radius $(aH)^{-1}$ to shrink rapidly, 
so physical scales that were once much smaller than the Hubble radius ($k = aH$) 
are stretched to super-horizon sizes. 
Modes that were initially causally connected (inside the horizon) 
are driven far apart, and after inflation these same comoving scales 
re-enter the horizon during the radiation and matter eras (see \cref{fig:scale-Hubble-I}). 

\bigskip

\begin{tcolorbox}[colback=gray!10, colframe=gray!10, boxrule=0pt,
                  enhanced, breakable, halign=justify]

\subsubsection*{Beyond Cosmic No-Hair Theorem} 
At cosmological scales, the Universe exhibits an extraordinary degree of homogeneity and isotropy. Since cosmic evolution may begin from generic initial conditions beyond our control, this striking simplicity naturally invites a dynamical explanation—one in which a homogeneous and isotropic state arises as an attractor of the cosmic evolution. The cosmic no-hair conjecture---that expanding cosmologies with a positive cosmological constant approach de Sitter spacetime at late times---was motivated by \cite{Gibbons:1977mu} and subsequently discussed by \cite{BoucherGibbons:1982}. Wald then established the result for a broad class of initially expanding Bianchi cosmologies (except type IX), with a positive cosmological constant and matter satisfying the strong and dominant energy conditions~\citep{Wald:1983ky}.
\vskip 0.2cm

Inflation was originally proposed to explain the observed flatness and large-scale homogeneity of the Universe, offering a dynamical mechanism through which horizon-scale correlations are generated. This motivation led to a reassessment of Wald’s theorem in inflationary contexts, culminating in a generalized cosmic no-hair theorem for inflation presented in \citet{Maleknejad:2012as}. This result demonstrates that, although anisotropies may grow during inflation in the presence of spinning fields, i.e., evading the traditional cosmic no-hair conjecture, their amplitude remains parametrically small and is subject to a universal upper bound. An example of inflationary models with anisotropic hair is the dilaton–Maxwell theory \citep{Watanabe:2009ct}. East and collaborators~\citep{East:2015ggf} extended the discussion of the cosmic no-hair conjecture to include initial anisotropies and demonstrated, through fully nonlinear numerical simulations, that inflation can emerge from highly inhomogeneous, gradient-dominated initial conditions, provided the inflaton field remains within a sufficiently flat region of its potential. Their analysis, however, is restricted to scalar-field models, and further investigations are required to assess the impact of spinning fields, e.g., gauge fields interacting with the inflaton, on the persistence or decay of cosmic inhomogeneities during inflation.

\end{tcolorbox}

\subsubsection{From the inflationary paradigm to precision cosmology} 
Early accelerated-expansion scenarios were explored by \cite*{Brout:1978,Brout:1979}, \cite{Starobinsky:1980}, \cite{Sato:1981}, and \cite{Guth:1981}, followed by the new-inflation scenarios of \cite{Linde:1982} and \cite{Albrecht:1982wi}. These developments established accelerated expansion as a mechanism for addressing the horizon, flatness, homogeneity, and monopole problems.
A profound realization soon followed: if inflation stretches all preexisting irregularities, it must also amplify the inevitable quantum fluctuations of matter and spacetime. Bardeen established the systematic gauge-invariant formulation of linear cosmological perturbations \citep{Bardeen:1980kt,KodamaSasaki1984}. \cite{Mukhanov:1981xt} gave the first explicit derivation of the quantum-generated primordial scalar spectrum in an early quasi-de Sitter stage, followed by further developments in the inflationary setting by \cite{Hawking:1982cz}, \cite{Starobinsky:1982ee}, \cite{Guth:1982ec}, and \cite*{Bardeen:1983qw}. The canonical gauge-invariant description of scalar perturbations was subsequently developed by \cite{Mukhanov:1985,Sasaki:1986}, and its comprehensive treatment of classical and quantum cosmological perturbation theory is given by  \cite{Mukhanov:1988jd,Mukhanov:1990me}.

Together, these developments established the nearly scale-invariant, approximately Gaussian and adiabatic primordial scalar spectrum as a central prediction of inflation. As their wavelengths exceed the Hubble radius, these quantum fluctuations lose their quantum coherence and become effectively classical, but inherently stochastic fields. These perturbations serve as the seeds of the large-scale structure of the Universe, eventually growing under gravity into the cosmic web of galaxies and clusters we observe today. 
In addition to tensor modes, inflation also predicts a stochastic background of scalar (curvature) perturbations. 
These fluctuations seed the temperature anisotropies of the cosmic microwave background and eventually grow into the large-scale structure of the Universe. 
Their statistical properties are described by the scalar power spectrum $\Delta_\zeta^2(k)$, which quantifies the variance of curvature perturbations per logarithmic interval in wavenumber. For slow-roll inflation, the scalar power spectrum takes the form
\be
\Delta_\zeta^2(k) \;=\; \frac{1}{8\pi^2 \epsilon_*}\,\frac{H_*^2}{ M^2_{\rm Pl}}
\left(\frac{k}{k_*}\right)^{n_s-1},
\ee
where $H_*$ and $\epsilon_*$ are evaluated at the pivot scale $k_*$, and $n_s=1-4\epsilon+2\eta$ is the scalar spectral index.

More than a decade later, the \textsc{COBE} satellite detected intrinsic CMB temperature anisotropies for the first time~\citep{COBE:1992syq}, providing direct observational evidence for primordial fluctuations consistent with inflationary predictions. 
Subsequent missions refined this picture with ever-increasing precision. 
The \textsc{WMAP} satellite \citep{WMAP:2012nax} mapped the CMB sky with degree-scale resolution, 
measured the spectral index $n_s < 1$, and confirmed the predicted slight red tilt of the primordial spectrum. 
The \textsc{Planck} mission \citep{Planck:2018vyg} further improved the precision, finding that the fluctuations are consistent with being nearly Gaussian and predominantly adiabatic. The observations precisely determine the amplitude of the scalar power spectrum at the pivot scale as
\be
\Delta_\zeta^2(k_*) \simeq 2.1\times10^{-9},
\ee
providing one of the most important observational anchors for inflationary cosmology. 
The \textsc{Planck} 2018 results \citep{Planck:2018vyg} give
\be
n_s = 0.9649 \pm 0.0042,
\ee
confirming, with high statistical significance, the deviation from exact scale invariance predicted by slow-roll inflation. This remarkable progression represents one of the central achievements of modern cosmology, establishing a direct link between the primordial fluctuations of the early Universe—most plausibly arising from quantum fluctuations during inflation—and the large-scale structure we observe today. Continued advances from \textsc{ACT}, \textsc{BICEP/Keck}, and the \textsc{BICEP Array} have substantially improved CMB constraints, especially on $B$-mode polarization~\citep{ACT:2025DR6,BICEP:2021xfz,Hui:2018cvg}. Observations from the \textsc{Simons Observatory}~\citep{SimonsObservatory:2018koc}, together with future space missions such as \textsc{LiteBIRD}~\citep{Campeti:2024JCAP} and the proposed \textsc{CMB-S4} experiment~\citep{CMB-S4:2019WhitePaper}, aim to detect the subtle imprint of primordial gravitational waves---an observation that would provide the next decisive test of inflation. Taken together, these measurements have transformed inflation from a bold theoretical proposal into the leading framework for early-universe cosmology, offering compelling empirical support for the view that the cosmic structures we see today originated from quantum fluctuations stretched to macroscopic scales during an epoch of accelerated expansion.

\subsection{Inflation and GWs}
\label{sec:Inf-GWs}

%Cosmic inflation generates primordial scalar perturbations that seed all structure formation in the observable universe. More precisely, most of the inflationary models consistent with the data predict an adiabatic and almost non-Gaussian scalar perturbation, the comoving curvature perturbation $\zeta$  with a nearly scale-invariant power spectrum as
%\be
%\Delta^2_{\zeta} = A_s(k_{*}) \big(\frac{k}{k_*}\big)^{n_s-1},
%\ee
%which specifies in terms of two numbers $A_s$ and the spectral tilt
%\be
%n_s \equiv 1+ \frac{d\ln P_{\zeta}}{d\ln k},
%\ee 
%which is of the order of slow-roll parameters.

\begin{figure}[ht]
\centering
\includegraphics[width=0.75\textwidth]{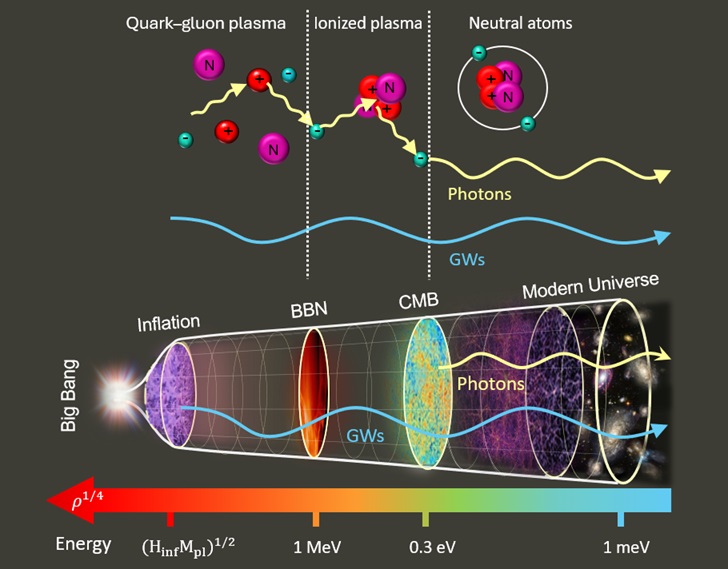}
\caption{
Schematic illustration of the thermal and structural history of the Universe, from cosmic inflation and Big Bang nucleosynthesis (BBN) to the formation of the cosmic microwave background (CMB), 
the emergence of large-scale structure, and the modern Universe. 
After inflation ended, the Universe remained for a long period filled with a hot, charged plasma. 
In this phase, photons were repeatedly scattered by free charged particles and could not propagate freely. Only when protons and electrons combined to form neutral atoms did the Universe become transparent, 
allowing CMB photons to decouple and free-stream across space, carrying a snapshot of the Universe at that epoch. In contrast, gravitational waves, though universally coupled to all forms of energy, interact so weakly that they were never in thermal equilibrium. 
The cosmic medium has always been transparent to them, meaning that once detected, these primordial gravitational waves could reveal direct information 
about the Universe’s highest energy scales and earliest moments, far beyond the reach of any electromagnetic probe. The horizontal axis in this illustration represents the energy scale of the Universe, defined as $\rho^{1/4}$, which decreases as the Universe expands and cools.
}
\label{fig:cosmic-history}
\end{figure}

 A key prediction of the inflationary paradigm is the existence of a stochastic background of primordial gravitational waves (PGWs) generated by tensor fluctuations of the spacetime metric in the very early Universe. Quantum amplification of gravitational waves in an expanding Universe was studied early by \cite{Grishchuk:1974}, while \cite{Starobinsky:1979ty,Rubakov1982} calculated the relic tensor spectrum for an early de Sitter-like state. For reviews of inflationary gravitational waves and their observational prospects, see \cite{Guzzetti:2016mkm,Bartolo:2016ami}.
Just like the temperature anisotropies of the CMB, this relic GW background forms a random field with no distinct features in either the time or frequency domain. 
However, GWs carry a crucial advantage over photons: while the CMB decoupled roughly $4\times10^{5}$ years after the Big Bang, primordial gravitational waves have been free-streaming since inflation, potentially tracing back to energy scales near the Planck regime (see \cref{fig:cosmic-history}). For detailed discussions of the thermal history of the Universe, including the evolution from inflation and reheating through Big Bang nucleosynthesis and recombination, see \citet{Starobinsky:1979ty,KolbTurner:1990,DodelsonSchmidt:2020,Brandenberger:2010bpq,Cline:2018fuq,Mukhanov:2005sc,Cosmology-W}.

We are interested in the tensor perturbations of the metric that correspond to gravitational waves observed today. 
At linear order, the perturbed FLRW metric can be written as 
\be
ds^2 = a^2(\tau)\!\left[-d\tau^2 + \big(\delta_{ij} + \gamma_{ij}(\boldsymbol{x},\tau)\big)dx^i dx^j \right],
\ee
where the symmetric, traceless, and transverse tensor $\gamma_{ij}$ represents the two physical polarization states of gravitational waves. The quadratic action for 
$\gamma_{ij}$ is
\begin{equation}
S^{(2)}_\gamma = \frac{M_{\rm Pl}^2}{8} \int d\tau\, d^3x\, a^2(\tau)
\left[ (\gamma'_{ij})^2 - (\partial_\ell \gamma_{ij})^2 \right],
\label{eq:2ndaction}
\end{equation}
where primes denote derivatives with respect to conformal time $\tau$. The evolution of GW modes is governed by the linearized Einstein equations, which in conformal time for the FLRW background reduce to
\be\label{eq:GW}
\gamma''_{ij} + 2\frac{a'}{a}\gamma'_{ij} - \nabla^2 \gamma_{ij}
= 16\pi G\,a^2\,\pi^{T}_{ij}.
\ee
Here $\pi^{T}_{ij}$ is the transverse and traceless part of the anisotropic stress tensor.\footnote{
The anisotropic stress tensor can be decomposed as
\be
\pi_{ij} \equiv \delta T_{ij} - \frac13 \delta_{ij}\,\delta T^k_{~k}
= \partial_{ij}^2 \pi^S - \frac13 \partial^2 \pi^S \delta_{ij} 
+ 2\partial_{(i}\pi^V_{j)} + \pi^{T}_{ij},
\ee
with $\partial_i\pi^V_i = \partial_i \pi^T_{ij}=0$. 
Here, $\pi^S$, $\pi^V_i$, and $\pi^T_{ij}$ denote the scalar, vector, and tensor components, respectively, with $\pi^T_{ij}$ capturing the transverse and traceless part of the anisotropic stress \citep{Cosmology-W}.} In what follows, we study the evolution of the initial quantum fluctuations of these tensor modes in the expanding background during inflation. 
Since inflation is dominated by a nearly homogeneous scalar field, any anisotropic stress from matter fields can be neglected, so we set $\pi^T_{ij} = 0$.

The field operator can be expanded in Fourier modes as
\be
\gamma_{ij}(\tau, \boldsymbol{x}) = \sum_{\lambda}
\int \frac{d^3k}{(2\pi)^3}
\left[ \gamma_\lambda(\tau,\bk)\, \hat{b}_\lambda(\boldsymbol{k})\, e^{i\boldsymbol{k}\cdot\boldsymbol{x}}
\, \epsilon^{\lambda}_{ij}(\hat{\bk}) + \gamma_\lambda^*(\tau,\bk)\, \hat{b}_\lambda^\dagger(\boldsymbol{k})\, e^{-i\boldsymbol{k}\cdot\boldsymbol{x}} \, \epsilon^{\lambda *}_{ij}(\hat{\bk})\right],\,
\ee
where $\epsilon^{\lambda}_{ij}(\hat{\boldsymbol{k}})$ denotes the polarization tensor, and 
$\hat{b}_\lambda(\boldsymbol{k})$ and $\hat{b}_\lambda^\dagger(\boldsymbol{k})$ are the
annihilation and creation operators, respectively, satisfying
\be
[\hat{b}_\lambda(\boldsymbol{k}), \hat{b}_{\lambda'}^\dagger(\boldsymbol{k}')]
= (2\pi)^3 \delta_{\lambda\lambda'}\, \delta^{(3)}(\boldsymbol{k}-\boldsymbol{k}').
\ee
The tensor perturbations can be treated as a stochastic field, whose statistical properties are captured by the two-point correlation function of the corresponding quantum operators,
\be
\langle\, \hat{\gamma}_{\lambda}(\boldsymbol{k})\, 
\hat{\gamma}_{\lambda'}^{\dagger}(\boldsymbol{k}')\,\rangle
= (2\pi)^3 \delta^{(3)}(\boldsymbol{k}-\boldsymbol{k}')\, 
\delta_{\lambda\lambda'}\, P_{\gamma}(k),
\ee
where  $P_{\gamma}(k)$ denotes the tensor power spectrum per polarization. For convenience, one often defines the dimensionless tensor power spectrum summed over both polarizations as
\be
\Delta_T^2(k) \equiv 2\,\frac{k^3}{2\pi^2}\, P_{\gamma}(k),
\ee
which directly characterizes the amplitude of the stochastic gravitational wave background generated during inflation.
In the scale-invariant limit, $\Delta_T^2(k)$ becomes nearly constant, reflecting the uniform excitation of tensor modes across different wavelengths.

As is evident from \cref{eq:2ndaction}, the field $\gamma_{ij}$ is not yet canonically normalized. 
To introduce a canonically normalized variable, we define
\be
v_{ij} \equiv \frac{a M_{\rm Pl}}{2}\, \gamma_{ij}.
\ee
In terms of $v_{ij}$, the action becomes
\be
S^{(2)} = \frac{1}{2} \int d\tau\, d^3x 
\left[ (v'_{ij})^2 - (\partial_\ell v_{ij})^2 + \frac{a''}{a}\, v_{ij}^2 \right],
\ee
and leads to the field equation
\be
v''_\lambda + \left(k^2 - \frac{a''}{a}\right) v_\lambda = 0.
\ee
This equation describes the propagation of quantum tensor modes in a time-dependent background, where the term $\frac{a''}{a} \simeq \frac{2+3\epsilon}{\tau^2}$ acts as an effective mass induced by cosmic expansion. Its general solution can be expressed in terms of Hankel functions as
\be
v_\lambda(\tau, k) = \frac{\sqrt{\pi\vert\tau\vert}}{2}e^{i(1+2\nu_{T})\pi/4} \,\left[ c_1 H^{(1)}_{\nu_T}(-k\tau)
+ c_2\, H^{(2)}_{\nu_T}(-k\tau) \right],
\ee
where $c_1$ and $c_2$ are constants determined by the choice of initial conditions and $\nu_{T}$ is \footnote{
The asymptotic behavior of the Hankel function is as \citep{NIST:DLMF}
\begin{align}
    \lim_{z\rightarrow \infty} H^{(1)}_{\nu}(z) = \sqrt{\frac{2}{\pi \, z}}\, e^{-i(2\nu+1)\pi/4} \, e^{i z}, \quad \lim_{z\rightarrow 0} H^{(1)}_{\nu}(z) = -\frac{i}{\pi} \, (\frac{2}{z})^{\nu} \, \Gamma[\nu].
\end{align}
}
\be
\nu_{T}\simeq  
\frac32+\epsilon.
\ee
 Assuming seeds of perturbations to be from the quantum fluctuations, the so-called Bunch--Davies vacuum~\citep{Bunch:1978yq} imposes the initial value in the asymptotic past as
\be
\lim_{\tau \rightarrow -\infty}  v_{\lambda}(\tau,\bk) = \frac{1}{\sqrt{2k}} e^{-i k\tau}.
\ee
Consequently, the corresponding mode function $\gamma_{\lambda}$ 
 (up to an overall phase) takes the form 
\be
\gamma_\lambda(\tau, k) = \sqrt{\pi \, |\tau|^3} \, \frac{H}{\mpl} \, H^{(1)}_{\nu_T}(-k\tau).
\ee

\begin{figure}[t]
\centering
\includegraphics[width=0.6\textwidth]{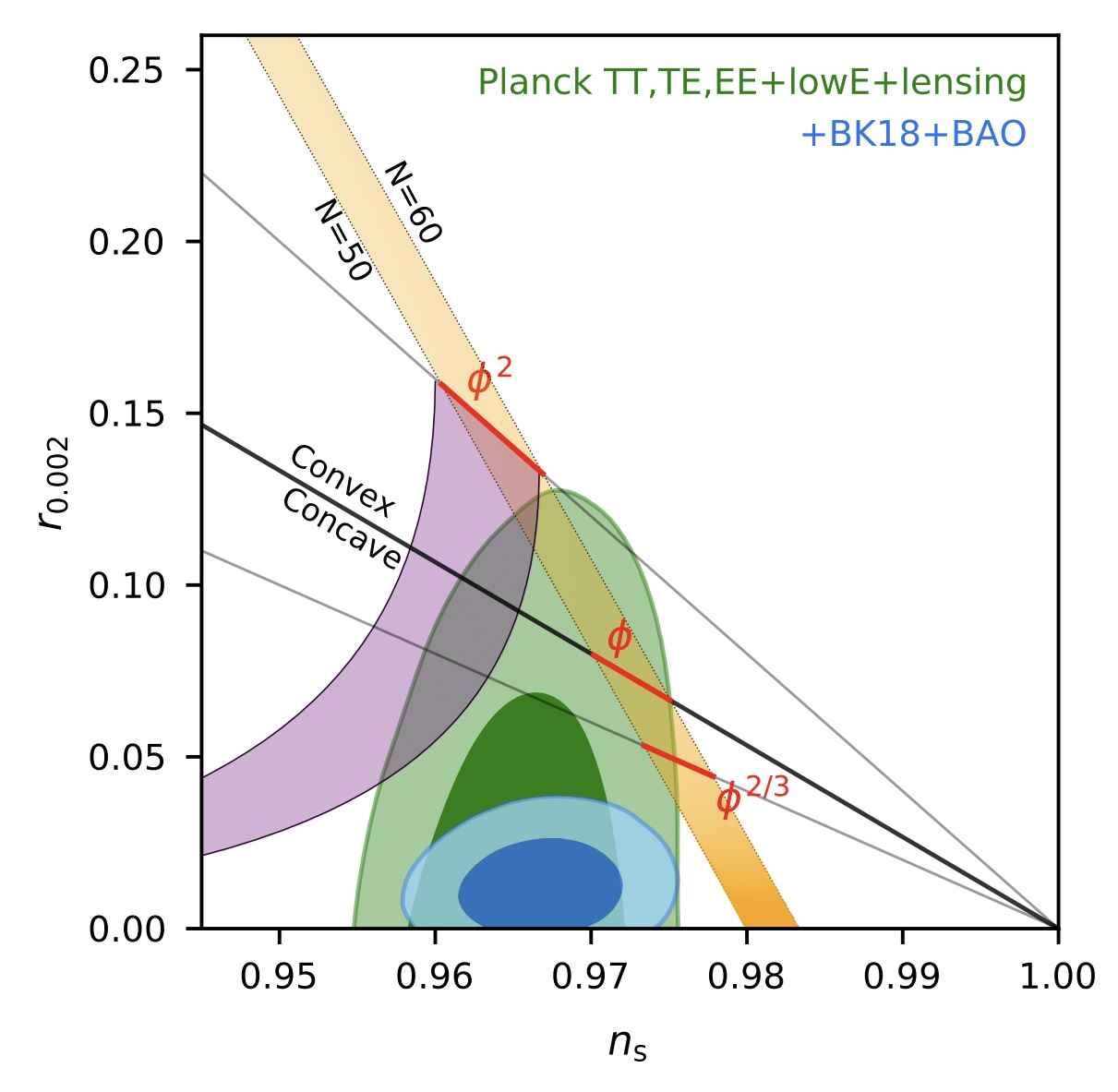} 
\caption{Constraints on the scalar spectral index and the tensor-to-scalar ratio $(n_s,\,r)$ at $k = 0.002~\mathrm{Mpc}^{-1}$ from \textsc{Planck},  with BK18 \& BAO data, 
compared with predictions of representative inflationary models. Image credit: BICEP/Keck Collaborations
 \citep{BICEP:2021xfz}.
}\label{ns-r}
\end{figure} 

Now, let us compute the tensor power spectrum for modes after horizon crossing. 
During inflation, the quantum fluctuations of the metric are stretched to super-horizon scales 
($k \ll aH$), where they effectively freeze and behave as classical perturbations. 
The amplitude of each mode at horizon exit ($k = aH$) determines the primordial tensor spectrum. 
The dimensionless tensor power spectrum is given by
\begin{equation} \label{power-tensor-single-scalar}
\Delta^2_{T}(k) 
= \frac{2}{\pi^2}\,\frac{H_*^2}{M_{\rm Pl}^2}\,
\left(\frac{k}{k_*}\right)^{n_T},
\end{equation}
where $H_*$ is the Hubble parameter evaluated when the pivot mode $k_*$ exits the horizon. Each mode is characterized by the value of $H$ when it left the horizon. The quantity $n_T$ in \cref{power-tensor-single-scalar} is the tensor spectral index, defined as
\begin{equation}\label{tilt-tensor-single-scalar}
n_T \equiv \frac{d\ln \Delta^2_{T}}{d\ln k} = -2\epsilon.
\end{equation}
A strictly scale-invariant spectrum corresponds to $n_T = 0$, in which case the power 
$\Delta_T^2$ is independent of the wavenumber $k$.  
Inflation predicts a slight red tilt ($n_T < 0$), meaning that longer-wavelength modes 
have slightly larger amplitudes. The overall amplitude of the tensor spectrum is directly related to the energy scale of inflation, while its nearly scale-invariant shape reflects the quasi-de Sitter nature of the inflationary expansion.

A key observable quantity in inflationary cosmology is the tensor-to-scalar ratio
\be
r_* \;=\; \frac{\Delta^2_T(k_*)}{\Delta^2_\zeta(k_*)},
\ee
which measures the relative amplitude of primordial GW perturbations to scalar (curvature) perturbations. For single-field slow-roll (single-clock) inflation, the tensor and scalar amplitudes are directly linked through the slow-roll parameter, $r_* = 16\,\epsilon_*$,
providing a direct probe of the energy scale and dynamics of inflation. The most recent joint analysis by the \textsc{Planck} and BICEP/Keck collaborations 
places a stringent upper bound on the tensor-to-scalar ratio \citep{BICEP:2021xfz}
\be
r_{0.05} < 0.036 \quad \text{(95\% C.L.)}.
\ee

%% floating tables and figures in boxes are tricky...
\bigskip

% in your preamble (once):
% \usepackage{caption}

\begin{tcolorbox}[colback=gray!10, colframe=gray!10, boxrule=0pt,
                  enhanced, breakable, halign=justify]

\subsubsection*{The Energy Scale of Inflation}
Combining the upper limit on $r$ with the measured scalar amplitude yields an upper bound on the Hubble scale during inflation,
\begin{equation}
H_\text{inf} \;=\; \pi M_{\rm Pl}
\sqrt{\frac{r\,\Delta_\zeta^2(k_*)}{2}}
\;\lesssim\; 4.7 \times 10^{13}~{\rm GeV},
\label{eq:Hmax}
\end{equation}
corresponding to an inflationary energy scale of 
$V_*^{1/4} \lesssim 10^{-2} \, \mpl$ . 
Thus, even a small value of $r$ may probe physics near the grand-unified scale and 
provide a direct window into the energy scale of inflation. On the other hand, Big Bang Nucleosynthesis (BBN) requires the Universe to be thermalized before light-element formation, 
implying a minimum reheating temperature $T_\mathrm{reh} \gtrsim 5~{\rm MeV}$. 
Assuming instantaneous reheating, this sets a lower bound on the inflationary Hubble scale
\begin{equation}
H_\mathrm{inf} = \sigma_\text{reh} \, \sqrt{\frac{\pi^2 g_*}{90}}\,\frac{T_\mathrm{reh}^2}{M_\mathrm{pl}} \gtrsim 1.1\times10^{-23}~{\rm GeV},
\label{eq:Hmin}
\end{equation}
where $g_*$ is the effective number of relativistic degrees of freedom, 
which at $T_\mathrm{reh} \simeq 5~{\rm MeV}$ takes the value $g_* = 10.75$. The parameter $\sigma_\text{reh}$ is a phenomenological factor that encodes the efficiency of the reheating process and equals unity in the case of instantaneous reheating (see \cref{fig:inflation_scale}).

%%\begin{figure}[h]
%%    \centering
\centering
  \includegraphics[width=0.7\textwidth]{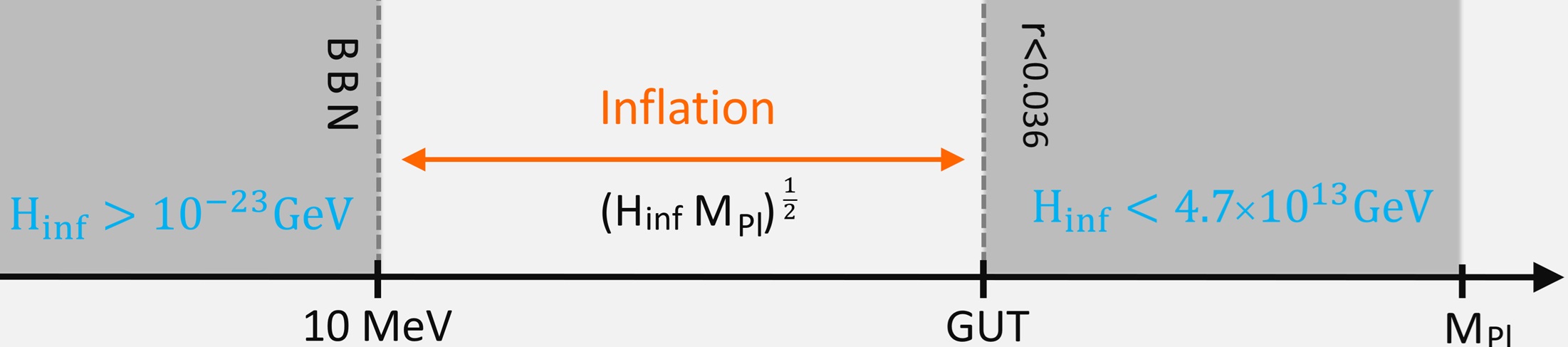}
  \captionsetup{type=table}
  \captionof{figure}{%
  %%\caption{
    Current allowed window for the inflationary energy scale,
    $\sqrt{M_\mathrm{pl} H_\mathrm{inf}}$.
    The upper bound is determined by the observational limits on the
    tensor-to-scalar ratio, as given in \cref{eq:Hmax}, while the lower bound is set by the
    requirement that reheating occurs before Big Bang Nucleosynthesis (BBN), as shown in \cref{eq:Hmin}. This corresponds to an enormous uncertainty of about $18$ orders of magnitude,
    spanning from the Grand Unified Theory (GUT) scale at
    $10^{16}~\mathrm{GeV}$ down to around $10~\mathrm{MeV}$.%
  }
  \label{fig:inflation_scale}
%%\end{figure}

\end{tcolorbox}

\subsection{IR structure of cosmological perturbations}\label{sec:IR}

Here, we briefly review Weinberg's adiabatic modes theorem and Maldacena's inflationary consistency relations. The adiabatic mode theorem plays a central role in connecting the primordial perturbations generated in the very early Universe to the cosmological perturbations observed today, providing a bridge between inflationary initial conditions and late–time cosmological observables. The inflationary consistency relation, which fundamentally relies on the presence of these adiabatic modes, encapsulates the symmetry constraints that govern correlations between long-wavelength (soft) and short-wavelength perturbations in single-field inflation.

\subsubsection{Adiabatic modes in cosmology} 
To extract information about inflation from observations, we must connect the primordial fluctuations produced during inflation to the structures observed in the later universe (see \cref{fig:cosmic-history}). This task is complicated by our limited understanding of the post-inflationary epoch—when the energy stored in scalar fields was converted into matter and radiation—and of any subsequent, poorly constrained eras. Although these stages occurred while the relevant fluctuations were stretched far beyond the horizon (see \cref{fig:scale-Hubble-I}), they may still have affected their amplitude or scale dependence. Therefore, to relate the inflationary perturbations to those observed in the CMB or large-scale structure, we need a conservation law that remains valid on super-horizon scales and is independent of the detailed dynamics of the universe’s evolution. The gauge-invariant formulation of cosmological perturbations was established by \cite{Bardeen:1980kt}. The conservation of adiabatic curvature perturbations on large scales had been studied before Weinberg's theorem; in particular, \cite{Wands:2000dp} gave a simple proof using only local energy-momentum conservation. \cite{Weinberg:2003sw} later proved the general existence of adiabatic modes under broad conditions.

We start with the construction of Weinberg's adiabatic modes in the Newtonian gauge, which is essential for the derivation of the consistency relations. Fixing the gauge uniquely specifies all the modes with finite momentum, $k\neq0$. However, there are residual gauge symmetries for the very long-wavelength modes, which remain a symmetry of the gauge-fixed action. Starting with the flat FLRW metric as the unperturbed metric
\be
ds^2=-dt^2+a^2d\textbf{x}^2.
\ee
Here we focus on the implications of the diffeomorphism invariance on the nature of the very long-wavelength modes with $k/a\ll H$. Therefore, only the homogeneous diffeomorphisms are relevant. 
The general spatially homogeneous perturbed metric in the Newtonian gauge is
\begin{equation}
ds^2=-(1+2\Phi(t))dt^2+a^2\bigg((1-2\Psi(t))\delta_{ij}+\gamma_{ij}(t)\bigg)dx^idx^j,
\label{eq:metric-pert}
\end{equation}
where $\Phi$ and $\Psi$ are the Bardeen potentials and $\gamma_{ij}$ is the traceless tensor perturbation describing gravitational waves. \footnote{Here we neglect the vector perturbations due to their damping nature in most of the inflationary models, including our axion-gauge field model.} One can decompose the general spatially homogeneous perturbed energy-momentum in the Newtonian gauge as
\bse
\begin{align}
T_{00}&=-\bar{\rho}g_{00}+\delta\rho \quad \textmd{and} \quad T_{0i}=-(\bar{\rho}+\bar{P})\partial_i\delta u,\\\label{Tij}
T_{ij}&=\bar{P}(t)g_{ij}+a^2\bigg(\delta_{ij} \delta P+\partial_{ij}^2\pi^S+\pi^T_{ij}\bigg),
\end{align}
\ese
where a bar denotes an unperturbed quantity, $\delta \rho$, $\delta P$ and $\delta u$ are the perturbed density, pressure and velocity potential, respectively. Moreover, $\pi^S$ and $\pi^T_{ij}$ are the scalar and tensor anisotropic stresses which characterize departures of $T_{\mu\nu}$ from the perfect fluid form. \footnote{It is interesting to note that in the decomposition \eqref{Tij}, the effects of bulk viscosity are included in $\delta p$ \citep{Weinberg:2003sw}. }
In the scalar sector, the perturbed energy conservation equation,  
$\nabla_\mu T^{\mu 0}=0$, is
\be
\delta\dot\rho + 3H(\delta\rho + \delta p)  - 3(\bar{\rho} + \bar{P})\dot\Psi  +\nabla^2\left( \frac{1}{a^2}(\bar{\rho}+\bar{P})\delta u + H \pi^S\right) = 0.
\ee
In terms of the curvature perturbation on uniform-density slices as formulated by \cite{Bardeen:1988},
\be
\zeta = -\Psi + \frac{\delta\rho}{3(\rho + p)}.
\ee
and the non-adiabatic pressure perturbation
\begin{equation}
\delta P_\mathrm{nad}
 \;\equiv\;
\delta P - \frac{\dot{\bar p}}{\dot{\bar\rho}}\,\delta\rho,
\label{eq:nonadiabatic_pressure}
\end{equation}
we can write the continuity equation as
\be
\dot{\zeta}
=
-\,\frac{H}{\bar{\rho} + \bar{P}}\,\delta P_{\mathrm{nad}}
+\frac{k^{2}}{3}\left(
\frac{\delta u}{a^2} + \frac{H\,\pi^{S}}{\bar{\rho}+\bar{P}}
\right).
\ee
In the super-horizon limit, $k\ll aH$, the gradient term is suppressed as $a^{-2}$, and for adiabatic perturbations with $\delta P_{\mathrm{nad}}=0$, the curvature perturbation is conserved \citep{Wands:2000dp}
\be
\lim_{\frac{k}{aH} \rightarrow 0}\dot{\zeta} = 0.
\ee
Consequently, the long-wavelength dynamics of $\zeta$ admits two independent 
adiabatic modes:
\begin{equation}
\lim_{\frac{k}{aH}\rightarrow 0}\zeta_{1}(t)=\mathrm{const},
\qquad
\lim_{\frac{k}{aH}\rightarrow 0}\zeta_{2}(t)=0.
\end{equation}
In the absence of anisotropic stress, $\pi^{S}=0$, one has 
$\Phi = \Psi$, and these two adiabatic modes correspond to the following 
super-horizon behaviors of the Bardeen potential
\begin{equation}
\lim_{\frac{k}{aH}\rightarrow 0} \Psi_{1}(t)
= \zeta_{1}\!\left[
 -1 + \frac{H(t)}{a(t)}
   \int_{t_{0}}^{t} a(t')\, dt'
 \right],
\qquad
\lim_{\frac{k}{aH}\rightarrow 0} \Psi_{2}(t)
= C\,\frac{H(t)}{a(t)},
\end{equation}
where $C$ is an integration constant.

For tensor perturbations, in the regime where the physical wavelength 
is much larger than the Hubble radius, $k/a \ll H$, we have 
\be
\ddot\gamma +3H \dot \gamma= \mathcal{O}\!\left(\frac{k^2}{a^2}\gamma\right),
\ee
and the gravitational wave 
equation admits two independent super-horizon homogeneous solutions
\begin{equation}
\gamma_{1}(t)=\text{const},
\qquad
\gamma_{2}(t)= C \,\int_{t_0}^{t}
\frac{dt'}{a^{3}(t')} .
\end{equation}
The second solution is a decaying mode in the standard inflationary, radiation-, and matter-dominated eras.  
Since tensor perturbations satisfy a second-order differential equation, 
these two modes exhaust the space of possible long-wavelength solutions.  
Consequently, for generic initial conditions, the tensor perturbations at 
late times are dominated by the constant mode $\gamma_{1}$.  
These are precisely the tensor-sector adiabatic modes guaranteed by 
Weinberg’s adiabatic mode theorem \citep{Cosmology-W}. Note that in the presence of a non-vanishing $\pi^{T}_{ij}$, gravitational waves acquire inhomogeneous solutions sourced by this anisotropic stress, in addition to the adiabatic modes. For explicit examples, see \citet{Maleknejad:2012fw, Adshead:2013qp, Dimastrogiovanni:2012ew, Maleknejad:2016qjz, Maleknejad:2018nxz, Komatsu:2022nvu}.

Having completed the derivation of adiabatic modes in cosmology, we now comment on the role of gauge transformations. In our analysis, we worked in the Newtonian gauge and solved the Einstein equations in the infrared limit,
$\frac{k}{aH} \ll 1$, where Weinberg's theorem guarantees the existence of two adiabatic solutions (a constant mode and a decaying one) for both the scalar perturbations and GWs. However, an important subtlety remains: even after fixing the local gauge completely to the Newtonian gauge, there exist residual diffeomorphisms that preserve the Newtonian-gauge form of the metric. These transformations generate solutions that are physically adiabatic modes but arise purely from large coordinate transformations. Thus, adiabatic modes can be understood as originating from these residual large gauge symmetries, rather than from independent dynamical degrees of freedom. Under the action of diffeomorphism transformations 
\be
x^\mu\mapsto \tilde{x}^{\mu}=x^\mu+\epsilon^\mu(t,\textbf{x}),
\ee
there is an $\epsilon^\mu$ which generates spatially homogeneous transformations on the metric and preserves the Newtonian gauge  \citep{Cosmology-W}
\bse
\begin{align}
&\epsilon^0(t,\textbf{x})=-f(t)-\chi(\textbf{x}),\\
&\epsilon^i(t,\textbf{x})=(\theta\delta_{ij}+\sigma_{ij})x^j-\partial_i\chi(\textbf{x})\int \frac{dt}{a^2(t)}.
\end{align}
\label{eq:large-diff}
\ese
Here $\theta$ is a constant scalar and $\sigma_{ij}$ is a constant, traceless and symmetric matrix\footnote{In general, $\epsilon^i(t,\textbf{x})$ can have a constant term $\epsilon^i_0$ as well as a term like $\omega_{ij}x^j$ where $\omega_{ij}=-\omega_{ji}$. Here, however, we ignored them because (due to the spatial translational and rotational symmetry of the background metric) they do not have any effects on the linear perturbed metric. }, $\text{tr}\sigma_{ij}=0$. Although infinitesimal locally, these transformations become large as one approaches infinity and are therefore called large gauge transformations. The scalar quantities $f(t)$, $\chi(\textbf{x})$, and $\theta$ act only on the scalar perturbations 
\be
\Phi(t)\mapsto \Phi(t)+\dot{f}(t) \quad \textmd{and} \quad \Psi(t)\mapsto \Psi(t)+\theta-Hf(t),
\ee
 while leaving the tensor perturbations unchanged. On the other hand, $\sigma_{ij}$ acts only on the gravitational waves as
\be
\gamma_{ij}(t)\mapsto \gamma_{ij}(t)-2\sigma_{ij}.
\ee

If $\Phi(t)$, $\Psi(t)$, and $\gamma_{ij}(t)$ solve the spatially homogeneous 
Einstein equations, then any large gauge transformation acting on these 
quantities produces new, spatially homogeneous solutions as well.  
In particular, consider a redefinition of the background metric
\begin{equation}
\bar{g}_{\mu\nu}(t)\;\mapsto\;
\bar{g}_{\mu\nu}(t)+\delta g_{\mu\nu}^{\rm A}(t),
\end{equation}
where the perturbation is generated by a diffeomorphism. 
For a scalar-type residual diffeomorphism, the induced metric perturbations in Newtonian gauge are
\begin{equation}\label{phi-g}
\Phi_{\rm A}(t) = \dot{f}(t),
\qquad
\Psi_{\rm A}(t) = \theta -H f(t).
\end{equation}
The corresponding transformations of the background energy--momentum tensor are
\begin{equation}
\delta\rho_{\rm A}(t) = \dot{\bar{\rho}}\,f(t),
\qquad
\delta P_{\rm A}(t)   = \dot{\bar{P}}\,f(t),
\qquad
\delta u_{\rm A}(t)   =  - f(t),
\end{equation}
with a vanishing scalar anisotropic stress, $
\pi^{S}_{\rm A}(t) = 0$.
This leads to a constant curvature perturbation,
\begin{equation}\label{R-g}
\zeta_{\rm A}(t) = -\theta .
\end{equation}
There is likewise a spatially homogeneous solution in the tensor sector,
\begin{equation}\label{gamma-g}
\gamma^{\rm A}_{ij}(t) = -2\sigma_{ij},
\end{equation}
with vanishing tensor anisotropic stress, $\pi^{T{\rm A}}_{ij}(t) = 0$.  
At this stage, it appears that the adiabatic modes we derived earlier are nothing more than gauge artifacts associated with the $k=0$ limit.

This raises the question: how does Weinberg's theorem elevate these seemingly pure-gauge solutions to physically meaningful modes?  The essential step is the following. When both the anisotropic stress, $\pi_{ij}(t,\mathbf{k})$, and the non-adiabatic pressure perturbation, $\delta P_\mathrm{nad}$, vanish on super-horizon scales, the spatially homogeneous ($k=0$) solutions can be continuously extended to solutions with $k\neq 0$.  Since modes with nonzero wavenumber possess no residual gauge freedom in Newtonian gauge, these extended solutions represent genuine physical perturbations. This seemingly simple observation has a profound physical consequence.  
During inflation, many Fourier modes exit the Hubble horizon and re-enter during the radiation or matter eras.  
Weinberg's argument shows that the homogeneous adiabatic solution can be continuously extended to finite $k$; the constant adiabatic mode is conserved on large scales.  
This remarkable robustness allows us to relate the very early Universe to cosmological observables in the late Universe.  
Therefore, in the absence of super-horizon entropy and anisotropic-stress perturbations, inflation predicts that the primordial fluctuations are adiabatic. Note that these scalar modes satisfy 
$\delta \rho_\alpha / \dot{\bar\rho}_\alpha$ equally for all individual 
constituents $\alpha$ of the universe, regardless of whether energy is 
separately conserved for each component. Perturbations with this property 
are referred to as adiabatic. Any solutions that do not satisfy this 
condition are correspondingly called entropic.  Present cosmological data agree strikingly well with the adiabatic initial conditions predicted by simple single-field inflation. CMB temperature and polarization anisotropies constrain the fractional 
entropy contribution to the primordial power spectrum,
\be
\frac{P_\mathrm{iso}(k_\ast)}{P_\mathrm{ad}(k_\ast)+P_\mathrm{iso}(k_\ast)}  \lesssim 0.03,
\ee
with $k_\ast = 0.05\ {\rm Mpc}^{-1}$ \citep{Akrami:2018odb}.

\subsubsection{Maldacena's consistency relation \& primordial non-Gaussianity} 

Now, we turn to the consistency relations, which are a powerful probe of the early universe physics and hold under very general conditions, i.e., in the presence of a long-wavelength adiabatic mode. Maldacena's consistency relation is, at its core, a statement about the structure 
of cosmological $n$-point functions and the pattern of primordial 
non-Gaussianity generated during inflation.  At the level of linear approximation, vacuum inflationary fluctuations have a Gaussian probability.

\begin{tcolorbox}[colback=gray!10, colframe=gray!10, boxrule=0pt,
                  enhanced, breakable, halign=justify]  

\subsubsection*{Primordial non-Gaussianity}
Primordial non-Gaussianities are most generally characterized by the connected
$n$-point correlation functions of the primordial scalar perturbation
$\varphi$,
\be
\langle \varphi_{\bk_1} \cdots \varphi_{\bk_n} \rangle_c
= (2\pi)^3 \delta^3\!\left(\sum_{i=1}^n \bk_i \right)
\, F_n(\bk_1,\ldots,\bk_n),
\ee
where statistical homogeneity enforces momentum conservation and the functions
$F_n$ encode deviations from Gaussian statistics. The statistical properties of an isotropic Gaussian field are completely determined by the 2-point function, while any odd $N$-point functions with $N\ge 3$ are exactly zero, and any even $N$-point functions with $N\ge 4$ are given in terms of the 2-point functions (called disconnected parts of $N$-point functions).

The leading contribution to primordial non-Gaussianity arises from the
three-point function, or bispectrum. The deviation from Gaussianity can be
formulated in terms of the bispectrum as
\be\label{bispec}
\langle \varphi_{\bk_1} \varphi_{\bk_2} \varphi_{\bk_3} \rangle
= (2\pi)^3 \delta^3(\bk_1+\bk_2+\bk_3)\,
B_{\varphi}(k_1,k_2,k_3).
\ee

If the power spectrum is scale invariant, dimensional analysis implies that the
bispectrum can be expressed in terms of two independent ratios,
\be
B_{\varphi}(k_1,k_2,k_3)
= k_1^{-6} B_{\varphi}(1,x_2,x_3),
\ee
where $x_2 = k_2/k_1$ and $x_3 = k_3/k_1$. In \cref{fig:shape}, we present the possible momentum configurations of the
bispectrum. 
Different inflationary scenarios produce bispectra that peak in different configurations  \citep{Komatsu:2009kd}. Inflationary three-point functions and second-order non-Gaussianity had been studied earlier by \cite{Gangui:1993} and \cite{Acquaviva:2002ud}. \cite{Komatsu:2001rj} developed the CMB bispectrum as a practical framework for constraining primordial non-Gaussianity~\citep{Komatsu:2010hc}. 

\begin{figure}[H]
\centering
\includegraphics[width=0.7\textwidth]{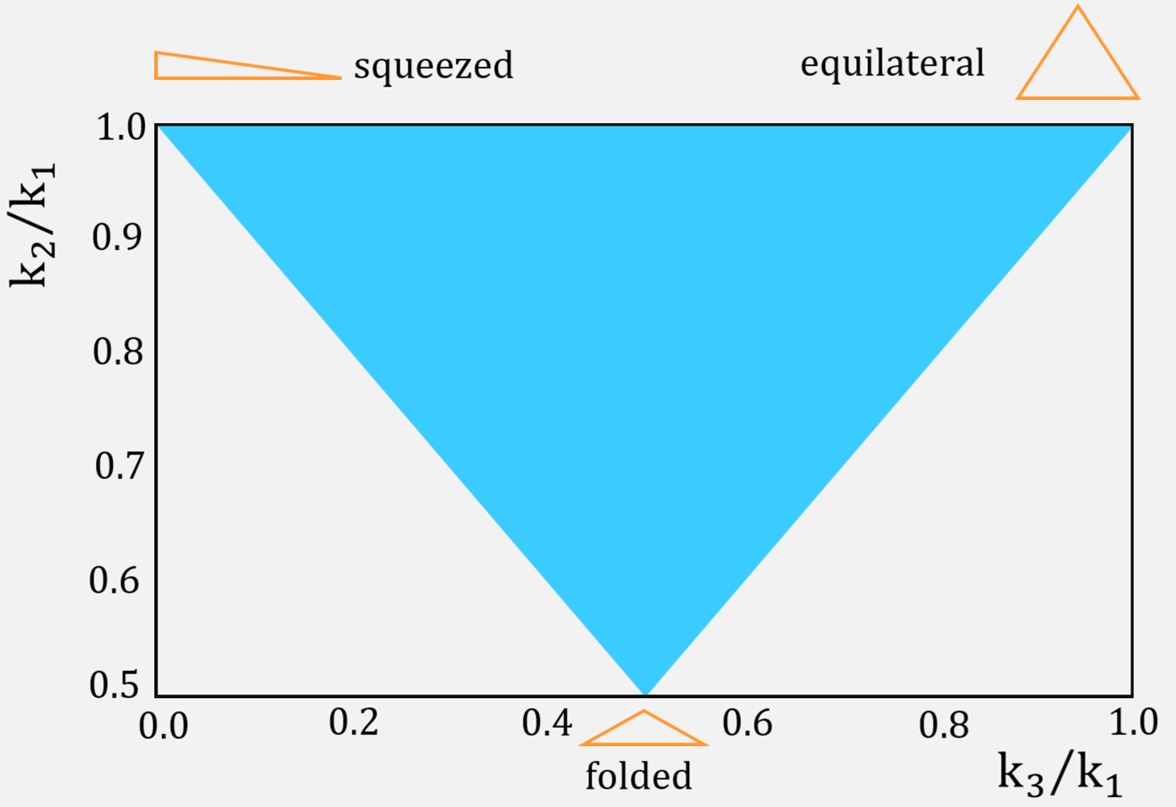} 
\caption{The shaded area shows the possible momentum configurations of the bispectrum. The relevant
limiting configurations are:
(i) the squeezed limit $k_1 \ll k_2 \simeq k_3$,
(ii) the equilateral limit $k_1 = k_2 = k_3$, and
(iii) the folded (or flattened) limit $k_1 \simeq k_2+k_3$ (up to permutations).}\label{fig:shape}
\end{figure} 

\end{tcolorbox}

%\subsubsection*{Cosmological n-point functions and Primordial Non-Gaussianities}

%Therefore, assuming Banch-Davis' initial condition, the scalar consistency relation only holds for single-clock inflationary models in which the entropy perturbation is zero. However, gravitational waves' consistency relations hold for more general inflationary models. More precisely, assuming Banch-Davis initial condition, only models in which $\pi^{T}_{ij}\neq0$ at super-horizon scales can violate tensor consistency relations, e.g., solid inflation \citep{Endlich:2013jia} and anisotropic inflation \citep{Bartolo:2012sd}. As shown in \citet{Maleknejad:2012as}, such inflationary models violate the cosmic no-hair conjecture.

%Computing the gravitons 3-point function as well as the combination of 2 scalars and one graviton Bispecturms in the squeezed limit for the vacuum GWs in general relativity, one realizes that they are both slow-roll suppressed. The reason for that is the fact that the self-interaction terms are subleading by slow-roll parameters. In the squeezed limit, an adiabatic long (classical) wave gravitational wave acts as a coordinate transformation for the other short wavelength modes, either gravitons or scalars. That then leads to Maldacena's powerful consistency relation, which we review in the next section.

While the non-Gaussianity must be computed separately for each specific inflationary model, 
an important universal result emerges: for single-clock attractor inflationary scenarios, the 
bispectrum in the squeezed limit satisfies Maldacena's consistency relation.

Since the tensor modes are the main focus of this work, here we present the consistency relation for the gravitational waves. The scalar consistency relation follows analogously from the scalar adiabatic dilation generated by $\zeta_{\rm A}$. The key physical point behind the consistency relations is the observation that the adiabatic long-wavelength modes, $\delta g_{\mu\nu}^{\rm A}(t)$, can be removed by large coordinate transformations, i.e.
\begin{equation}
\bar{g}_{\mu\nu}(t)+\delta g_{\mu\nu}^{\rm A}(t)\;\mapsto\;
\tilde{g}_{\mu\nu}(\tilde t) =\bar{g}_{\mu\nu}(t).
\end{equation}
Hence, adiabatic long-wavelength modes act as a classical background for the short-wavelength modes, which freeze out much later than the long-wavelength modes. In particular, an n-point correlation function of the short modes can be written as
\be\label{n-point-cor}
\langle  \zeta(x_1)\zeta(x_2)\cdots\zeta(x_n) \rangle _{\gamma_{ij}^{\rm A}(x)}=\langle \zeta(\tilde{x}_1)\zeta(\tilde{x}_2)\cdots\zeta(\tilde{x}_n) \rangle,
\ee
where $x^i$ and $\tilde x^i$ are related as
\be
\tilde x^i=x^i+\frac12\gamma^{\rm A}_{ij}x^j.
\ee
Now, Taylor expanding the RHS around $x^i$, we find the change of the short-distance n-point correlation function as  
%\be
%\langle \zeta(\tilde{x}_1)\zeta(\tilde{x}_2)...\zeta(\tilde{x}_n) \rangle=\langle \zeta(x_1)\zeta(x_2)...\zeta(x_n) \rangle+\sum_{\rm I=1}^{n}\delta \boldsymbol{x}_{\rm I}.\vec{\nabla}_{\rm I}\langle \zeta(x_1)\zeta(x_2)\cdots\zeta(x_n) \rangle +\cdots,
%\ee
\be
\delta\langle \zeta(\tilde{x}_1)\zeta(\tilde{x}_2)\cdots\zeta(\tilde{x}_n) \rangle=\sum_{\rm I=1}^{n}\delta \boldsymbol{x}_{\rm I}.\vec{\nabla}_{\rm I}\langle \zeta(x_1)\zeta(x_2)\cdots\zeta(x_n) \rangle +\cdots.
\ee
As a result, the (n+1)-point correlation function including the long-wavelength mode is given as
\be
\langle \gamma_{ij}^{\rm A}(x) \zeta(x_1)\zeta(x_2)\cdots\zeta(x_n) \rangle\simeq\frac12\bigg\langle \gamma_{ij}^{\rm A}(x)\gamma_{kl}^{\rm A}(x)\sum_{\rm I=1}^{n}x^k_{\rm I}\partial_{l}\langle \zeta(x_1)\zeta(x_2)\cdots\zeta(x_n) \rangle \bigg\rangle,
\ee
in which we only keep the dominant term that has the relevant contribution. The above equality is the consistency relation in real space. Going to Fourier space, we arrive at Maldacena's consistency relation
\be
\lim_{q\rightarrow 0}\,\langle \gamma_{\lambda}(\textbf{q}) \zeta_{\textbf{k}_1}\zeta_{\textbf{k}_2}\cdots\zeta_{\textbf{k}_n} \rangle'\simeq-\frac12P^{\rm vac}_{\gamma}(q)\sum_{\rm I=1}^n \textbf{e}^{\lambda}_{ij}(\hat{q}) k_{{\rm I}i}\partial_{k_{{\rm I}j}}\langle \zeta_{\textbf{k}_1}\zeta_{\textbf{k}_2}\cdots\zeta_{\textbf{k}_n} \rangle',
\ee
where the prime in $\langle \cdots \rangle$ indicates that we extracted the prefactor  $(2\pi)^{3}\delta^{(3)}(\textbf{q}+\sum_{\rm I}^n \textbf{k}_{\rm I})$ associated to momentum conservation. Note that the above result follows directly from the fact that adiabatic long-wavelength gravitational waves are equivalent to a change of coordinate for the short-wavelength mode, regardless of the super-horizon behavior of the short modes. Therefore, as far as our inflationary system generates adiabatic tensor perturbations, the above consistency relation holds. It is important to stress that Maldacena’s squeezed-limit non-Gaussianity is not a directly observable quantity.  As shown rigorously in \citet{Pajer:2013ana}, the primordial squeezed bispectrum predicted by single-field inflation does not imprint an observable modulation of small-scale power in late-time cosmological data. Maldacena’s original relation was elevated to a conformal consistency relation in \citet{Creminelli:2012ed} and to an infinite hierarchy of Ward identities in \citet{Hinterbichler:2013dpa}. %The consistency conditions for models of multi-field scalar inflation has been studied in \citet{Dimastrogiovanni:2015pla}.

\bigskip

\begin{tcolorbox}[colback=gray!10, colframe=gray!10, boxrule=0pt,
                  enhanced, breakable, halign=justify]  

\subsubsection*{Cosmological Collider Physics}
In inflationary settings, heavy particles can be spontaneously produced by the expansion of the Universe \citep{Schrodinger, Parker:1968mv,Kolb:2023ydq}. Their interactions with curvature perturbations generate characteristic signatures which encode the masses, spins, and interaction structures of particles. Cosmological collider physics views the inflationary universe as a natural
high-energy probe in which massive fields present during inflation imprint
characteristic oscillatory and angular structures in primordial
non-Gaussianity. Building on early studies of inflationary correlators and quasi-single-field models by \cite{Chen:2009zp,Chen:2009we}, as well as subsequent developments by \cite{Noumi:2012vr}, the idea was formulated explicitly by \cite{Arkani-Hamed:2015bza}. In this picture, heavy particles act as virtual states, encoding their
masses, spins, and chemical potential in specific shapes of correlation functions accessible to
observation, thereby probing physics far beyond the
reach of terrestrial accelerators \citep{Chen:2015lza, Lee:2016vti,Arkani-Hamed:2018kmz, Jazayeri:2022kjy,Bodas:2020yho,Jazayeri:2023kji,deRham:2025mjh}, among many others in a rapidly developing literature. Motivated by these theoretical developments, there have been important observational searches for cosmological collider signals in both CMB \citep{Sohn:2024xzd} and Large Scale Structure (LSS) \citep{Cabass:2022wjy}. This program is naturally linked to the cosmological bootstrap, which seeks
to determine primordial correlation functions from symmetry, analyticity,
and consistency conditions alone \citep{Baumann:2022jpr}. Bootstrap methods classify all allowed
inflationary signatures, while the cosmological collider provides their
physical interpretation in terms of massive spinning particles in the early
universe. 

\end{tcolorbox}

\subsection{Inflationary gravitational waves at late times} \label{sec:4}

%We now turn to one of the most profound intersections of early-Universe physics and observational cosmology: the imprint of gravitational waves on the cosmic microwave background (CMB). These primordial ripples in spacetime leave subtle yet distinctive fingerprints on the CMB anisotropies, offering a rare observational window into the physics of the very early Universe. Although primordial gravitational waves can also be probed through a variety of other observables, as highlighted in \cref{fig:GW-rainbow}, our discussion here will focus solely on their CMB signatures for the sake of time and clarity. 

The primordial gravitational waves produced during inflation leave observable imprints only after going through the universe’s subsequent expansion history. Following our derivation of the inflationary tensor spectrum and its infrared structure, we now translate these primordial fluctuations into their present-day form. %This provides a direct link between inflationary physics and observable signatures, including the contribution of gravitational waves to the integrated Sachs–Wolfe effect in the cosmic microwave background.
We begin this part by deriving the spectrum of primordial gravitational waves sourced by vacuum fluctuations during inflation. This spectrum spans an enormous range of frequencies—from as low as $10^{-17}$ Hz, corresponding to wavelengths comparable to the size of the observable Universe, to as high as $10^8$ Hz, associated with microscopic scales near the end of inflation and the reheating epoch.  We then explore the contribution of gravitational waves to the integrated Sachs–Wolfe effect in the cosmic microwave background.

%We then explore how these waves shape the CMB and how their signatures can be uncovered through precise measurements of temperature and polarization anisotropies.

\subsubsection{The present-day spectrum of gravitational waves from inflation}

\begin{figure}[h!]
\centering
\includegraphics[width=0.6\textwidth]{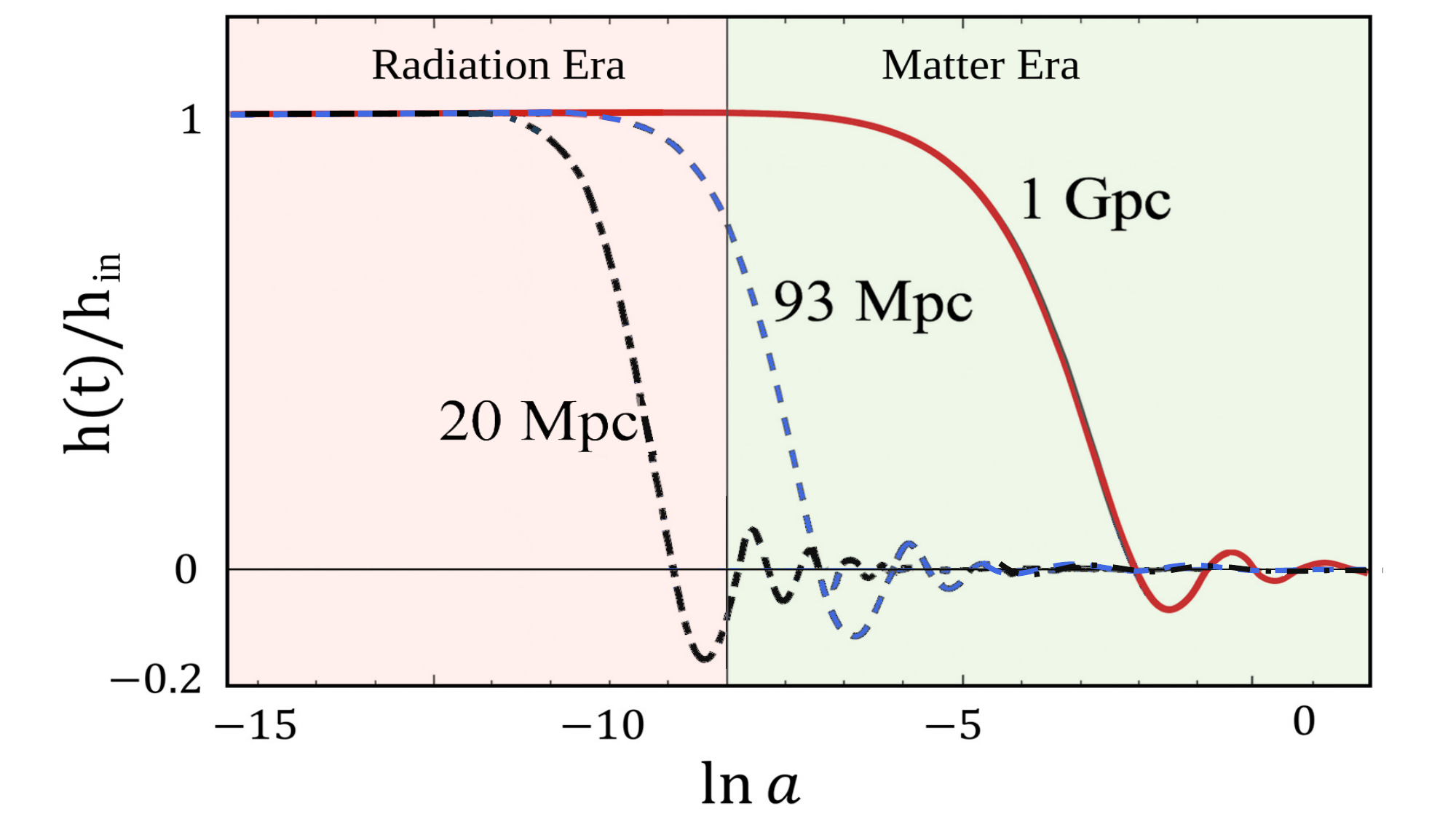} 
\caption{The time evolution of $\gamma_{+,\times}(\tau,k)$, normalized to its initial value as a function of $\ln a$, for three different values of the wavelengths.  Image credit: Adapted from \cite{Maggiore:2018sht}.}\label{fig:GW-ev}
\end{figure}

Earlier in  \cref{sec:cosmological-background}, we introduced the concept of horizon crossing and re-entry, and outlined the major cosmological epochs. In \cref{sec:Inf-GWs}, we discussed how gravitational waves generated during inflation arise from the amplification of quantum fluctuations in spacetime. These tensor perturbations are nearly scale-invariant at horizon exit, with a power spectrum 
\begin{equation}
\mathcal{P}_h(k) \simeq \frac{2}{\pi^2} \frac{H_{\text{inf}}^2}{M_{\text{Pl}}^2},
\end{equation}
where $ H_{\text{inf}} $ is the Hubble parameter during inflation, and $M_{\text{Pl}}$ is the reduced Planck mass. We are now ready to explore how the primordial GWs generated during inflation evolve across these eras, and how they shape the gravitational waves spectrum today. The field equation of vacuum $\gamma_{\lambda}(\tau,\bk)$ in the expanding background (regardless of the cosmological era) is
\be\label{GW-eq}
\gamma_{\lambda}''(\tau,\bk) + 2\mH \gamma_{\lambda}'(\tau,\bk) + k^2 \gamma_{\lambda}(\tau,\bk)=0,
\ee
where again $\lambda=+,\times$, a prime denotes a derivative with respect to the conformal time and $\mH=aH$. 
We can extract some general features of the solution analytically by using the field redefinition
\be
h(\tau,\bk) \equiv a\gamma_{\lambda}(\tau,\bk).
\ee
In the vacuum case, the two polarization states evolve identically, and the redefinition above applies uniformly to both polarization modes.
The field equation takes the simple form
\be\label{g-eq}
h''(\tau,\bk)+(k^2-\frac{a''}{a})h(\tau,\bk)=0,
\ee
where $'$ denotes the derivative with respect to the conformal time and the effective mass term $\frac{a''}{a}=\frac12(aH)^2(1-3w)$ is 
\be
\frac{a''}{a}= \begin{cases}
 (2-\epsilon)\mH^2 \simeq \frac{2+3\epsilon}{\tau^2} & \textmd{inflation} \\
 0  & \textmd{radiation-era}\\
 \frac12\mH^2  \simeq \frac{2}{\tau^2} & \textmd{matter-era}
\end{cases} .
\ee
When the mode functions are outside the Hubble radius ($\frac{k}{a}\ll H$), the gravitational wave is constant, while when it is well inside the Hubble radius ($\frac{k}{a}\gg H$), it is a simple oscillating function that scales like $1/a$
\be
\gamma(\tau,\bk) \propto \begin{cases}
 \gamma_0  \qquad & (\frac{k}{a}\ll H) \\
  \frac{1}{a}\sin(k\tau+\alpha) \qquad & (\frac{k}{a} \gg H)
\end{cases},
\ee
where $\gamma_0$ and $\alpha$ (a phase) are both given by the initial conditions; see \cref{fig:GW-ev}.
For freely propagating sub-Hubble gravitational waves, the oscillation-averaged energy density is
\be
\rho_{_\text{GW}} = \frac{1}{32\pi G}\sum_{\lambda=+,\times} \langle \dot\gamma_{\lambda}^2 \rangle,
\ee
where the expectation value denotes the statistical average. For gravitational waves inside the horizon, we have $\dot\gamma\propto k\cos(k\tau+\alpha)/a^2$ which gives 
$$\rho_{_\text{GW}}\propto a^{-4},$$
as is expected from any form of radiation.

%Within this accessible window, each frequency band unveils distinct observational prospects: very-low-frequency GWs leave imprints on the cosmic microwave background polarization \citep{LiteBIRD:2023zmo, CMB-S4:2019WhitePaper}, nanohertz waves are monitored through pulsar timing arrays \citep{NANOGrav:2023gor,Antoniadis:2023rjw,Reardon:2023gzh}, millihertz signals will be accessible to space-based interferometers such as LISA \citep{Colpi:2024lisa}, and the kilohertz regime is already probed by ground-based detectors like LIGO, Virgo, and KAGRA \citep{Abbott:2016blz,KAGRA:2018plz}, with LIGO-India expected to join the network in the coming years \citep{Saleem:2021iwi} (see \cref{fig:gw_timeline}). Looking ahead, third-generation observatories such as the Einstein Telescope \citep{ET:2025xjr} will dramatically extend the sensitivity reach. 

After inflation, each mode re-enters the horizon at different times, and the energy density in GWs evolves as radiation, redshifting as \( a^{-4} \). The present-day gravitational wave energy density per logarithmic interval in wavenumber is given by
\begin{equation}
\Omega_{_{\text{GW}}}(k) = \frac{1}{\rho_c} \frac{d\rho_{_{\text{GW}}}}{d\ln k} \propto k^2 \mathcal{P}_h(k) T^2(k),
\end{equation}
where \( T(k) \) is the transfer function encoding the evolution of each mode through the radiation- and matter-dominated eras.
The late-time gravitational-wave spectrum depends on the horizon re-entry history of inflationary tensor modes as follows.
\begin{itemize}[label=$\bullet$]
  \item Modes re-entering during the radiation era: $k^2T^2(k)$ is nearly constant, so the spectrum remains approximately scale-invariant
  $\Omega_{_{\text{GW}}}(k) \propto \text{const.}$  
  \item Modes re-entering during the matter era: the energy density is less suppressed, scaling as
  $\Omega_{_{\text{GW}}}(k) \propto k^{-2}$.
 \item Modes never left the horizon: At very high frequencies, the gravitational wave spectrum exhibits an exponential cutoff. These modes never exited the horizon during inflation and therefore were not stretched or amplified. This sets a natural upper bound on the frequency of primordial GWs,
 $k_\text{max} = a_\text{end} H_\text{end}$, where $a_\text{end}$ and $ H_\text{end}$ denote these quantities at the end of inflation. Assuming instantaneous reheating, the highest frequency of primordial gravitational waves is given by
  \begin{equation}
     f_\text{max}=\frac{k_\text{max}}{2\pi} \sim 10^{8} \, \left(\frac{r}{0.01}\right)^{\frac14} \, \text{Hz}.
  \end{equation}

\end{itemize}

\begin{figure}[h!]
\centering
\includegraphics[width=0.75\textwidth]{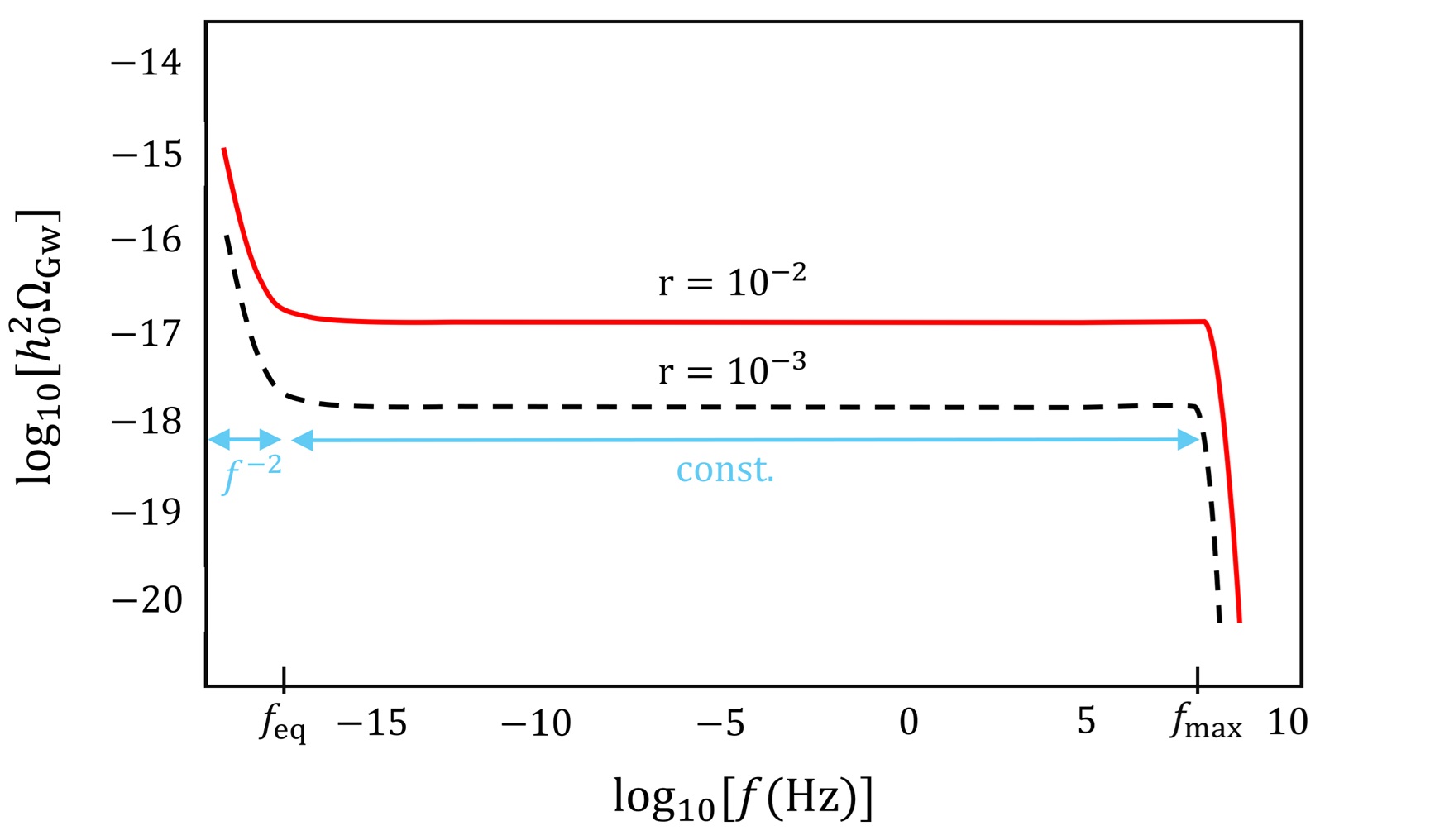} 
\caption{The primordial GW spectrum predicted by inflation (from quantum fluctuations) for $r=10^{-2}$ (solid red line) and $r=10^{-3}$ (dashed blue line).}\label{fig:GW-spectrum}
\end{figure} 

Modes re-entering during radiation domination start oscillating (and redshifting) earlier, leading to greater suppression in $\Omega_{_{\text{GW}}}$. Modes that re-enter during matter domination experience less redshift and thus appear less suppressed today.
The spectrum of gravitational waves today is typically expressed in terms of frequency (in Hz), which in natural units is related to the comoving wavenumber by $k=2\pi f$.
Hence, the overall shape of the primordial GW spectrum today in terms of $r$ and $f$ is a broken power law:
\begin{equation}
\Omega_{_{\text{GW}}}(f) \propto
\begin{cases}
r \, f^{-2}, & \text{for } f < f_{\text{eq}}, \\
r, & \text{for } f_{\text{eq}}< f < f_{\text{max}},\\
\text{exponential cut-off } & \text{for } f >f_{\text{max}},
\end{cases}
\end{equation}
where \( f_{\text{eq}}\approx 3 \times 10^{-17} \) Hz corresponds to the mode that re-entered the horizon at the time of matter-radiation equality. This spectral break encodes important information about the thermal history of the Universe and is shown in \cref{fig:GW-spectrum}. Note that the primordial gravitational wave spectrum from inflation can exhibit a richer structure than discussed here once thermalization effects in the early universe are taken into account. In particular, interactions with the thermal plasma and free-streaming SM particles can induce additional damping features and smooth transitions in the spectrum at high frequencies, as shown in \cite{Watanabe:2006qe,Saikawa:2018rcs}. More recently, scalar inhomogeneities during radiation domination have also been shown to enable quantum production of gravitons from vacuum fluctuations, with a characteristic GHz-scale signal~\citep{Mierna:2025}.

\bigskip

\begin{tcolorbox}[colback=gray!10, colframe=gray!10, boxrule=0pt,
                  enhanced, breakable, halign=justify]

\subsubsection*{BBN Bound on the Cosmological GW Background}
Big Bang Nucleosynthesis provides a robust upper bound on the total energy density stored in a stochastic background of cosmological gravitational waves. An excess radiation component increases the Hubble expansion rate during nucleosynthesis, thereby altering the neutron--proton freeze-out ratio and the primordial abundances of light elements. This requirement can be expressed as a constraint on the present-day gravitational wave energy density fraction \citep{Maggiore:2018sht}
\be
\int_{f_\mathrm{BBN}}^{\infty} d\ln f \, \Omega_{_\mathrm{GW},0}(f)
\;\lesssim\; 3.5 \times 10^{-6},
\ee
where $\Omega_{_\mathrm{GW},0}(f)$ denotes the spectral GW energy density today and 
$f_\mathrm{BBN} \sim 10^{-10}\,{\rm Hz}$ is the frequency corresponding approximately to the comoving Hubble scale at BBN. This bound limits the total gravitational wave radiation present prior to BBN and constrains a wide range of cosmological GW sources, including high-frequency backgrounds inaccessible to current detectors.

\end{tcolorbox}

\subsubsection{Integrated Sachs–Wolfe effect and primordial GWs}\label{sec:SW}

\begin{figure}[h!]
\centering
\includegraphics[width=0.75\textwidth]{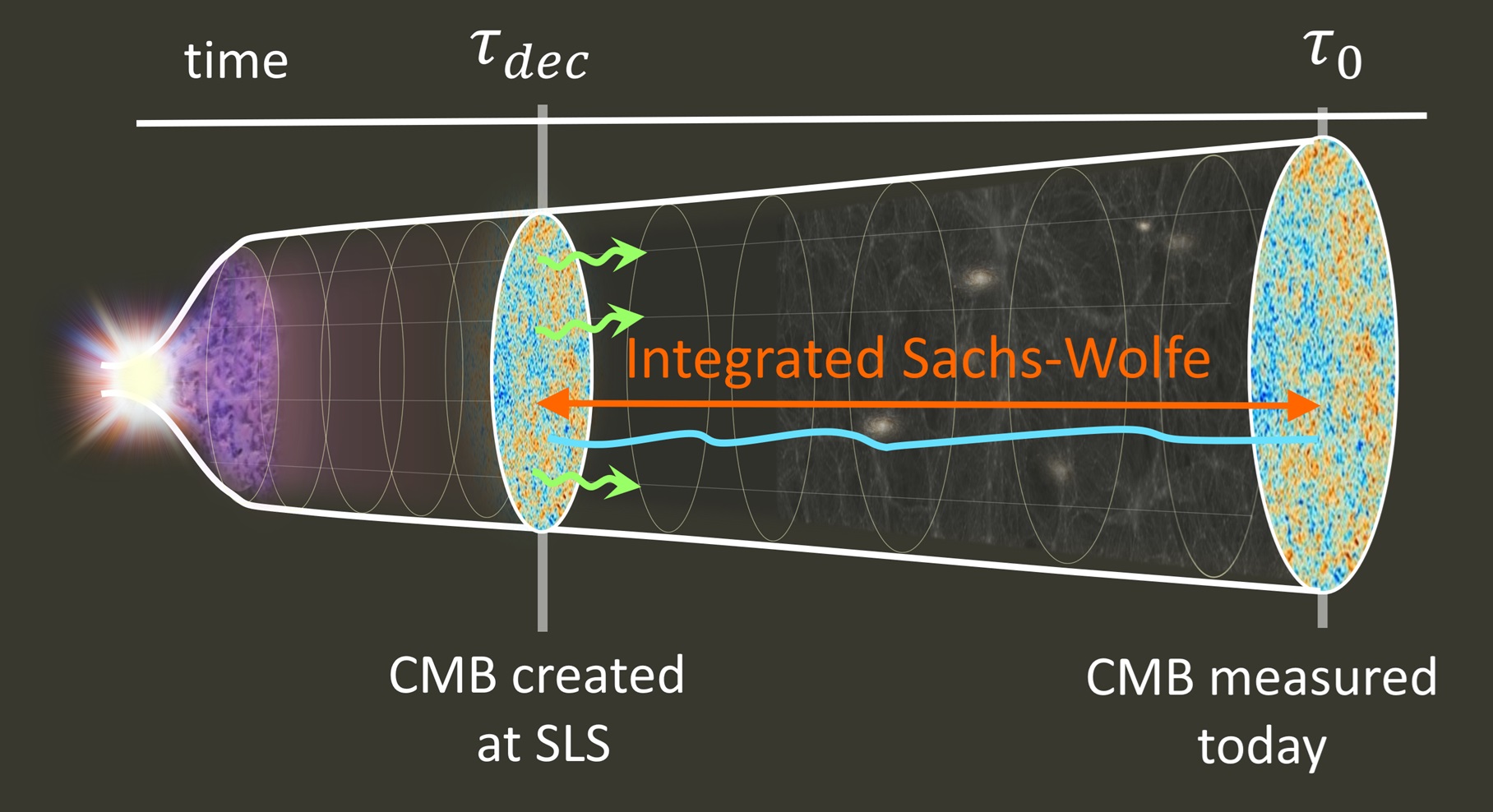} 
\caption{The photon geodesics are perturbed by gravitational fields, e.g., cosmic perturbations. This affects both the direction and energy of photons, contributing to the Integrated Sachs--Wolfe effect. The CMB anisotropies imprinted at the surface of last scattering (SLS) are subsequently altered by cosmic perturbations and large-scale structures encountered along their path from the decoupling time \( \tau_{\text{dec}} \) to the present epoch \( \tau_0 \), giving rise to the anisotropies we observe today.
}\label{fig:ISW}
\end{figure}

After decoupling, the photon mean free path becomes comparable to the size of the observable Universe, allowing photons to reach our detectors without further scattering (\cref{fig:cosmic-history}). Along their journey, photon geodesics are subtly perturbed by spacetime inhomogeneities (see \cref{fig:ISW}). These perturbations give rise to temperature fluctuations, an effect known as the Sachs–Wolfe effect, which arises from scalar and tensor modes of the underlying geometry.  In general, the total Sachs-Wolfe effect consists of two components: the original Sachs-Wolfe effect,  which arises at the surface of last scattering $\tau_{\text{dec}} $ by scalar perturbations, and the integrated Sachs-Wolfe (ISW) effect, which accumulates along the photon's trajectory, $\tau \in(\tau_{\text{dec}},\tau_0)$, due to the time variation of metric perturbations (both scalar perturbations and GWs). \footnote{Note that 
$\tau_{\mathrm{rec}}$ denotes the epoch of recombination, when protons and electrons form neutral hydrogen 
($z \sim 1100$--$1400$, $T_\gamma \sim 3000\,\mathrm{K}$). On the other hand, 
$\tau_{\mathrm{dec}}$ refers to the epoch of photon decoupling, when the Thomson scattering rate falls 
below the Hubble expansion rate and CMB photons begin to free–stream; this occurs slightly later, at 
$z \simeq 1050$ and $T_\gamma \simeq 2800\,\mathrm{K}$, and defines the true last–scattering surface relevant for 
primary anisotropies.  
For the Integrated Sachs--Wolfe (ISW) effect, the line–of–sight integral extends from $\tau_{\mathrm{dec}}$ to today, 
since the ISW contribution arises from the evolution of gravitational potentials after decoupling.
}
In the following, we only focus on GWs and neglect the contribution of scalar perturbations in the above discussion. 

The observed anisotropies in the CMB are related to how photons are seen from the cosmic fluid rest frame, i.e., the baryon–photon fluid. The photon's energy measured with respect to the rest frame of the baryon-photon fluid is
\be
E(x^{\nu}) \equiv -P^{\mu}(x^{\nu}) u_{\mu}(x^{\nu}),
\ee
where $P^{\mu}=\frac{dx^{\mu}}{d\lambda}$ is the photon's 4-momentum and $u^{\mu}$ is the baryon-photon fluid bulk velocity. The photon's four-momentum, $P^\mu P_\mu=0$,  can be expressed as  
\begin{align}
P^\mu= E(1, \frac1a\hat{n}),
\label{eq:photon-P}
\end{align}
where $\hat{n}$ denotes the unit vector along its direction of propagation. Note that $\hat{n}=-\hat{r}$ where $\hat{r}$ is a radial direction in the celestial sphere (line-of-sight direction), as illustrated in the left panel of \cref{fig:linear-circular}. 
Suppose that a photon emitted with energy $E_{E}$ at a point, $x^{\mu}_E$, and is observed at $x^{\mu}_O$ with $E_{O}$. Then, the temperature at $x^{\mu}_E$ and at $x^{\mu}_{O}$ in a rest frame of baryon-photon fluid are given as
\be
\frac{E_O}{E_E} = \frac{T_O}{T_E}.
\ee
To understand how evolving gravitational potentials leave imprints on the CMB temperature through the Integrated Sachs–Wolfe (ISW) effect, we must examine how these perturbations influence the propagation of photons. For that purpose, we need to compute the effect of the perturbations on the photon trajectories 
\be\label{geo}
\frac{dP^{\mu}}{d\lambda} + \Gamma^{\mu}_{\nu\sigma}(x^{\alpha}(\lambda)) P^{\nu} P^{\sigma} =0,
\ee
where $\Gamma^{\mu}_{\nu\sigma}(x^{\alpha}(\lambda))$ are the Christoffel symbols up to the first order of perturbations. Since our focus is on temperature perturbations, we are particularly interested in the projection of the above equation along the four-velocity $u^\mu$. For the perturbed metric \cref{eq:metric-pert}, we find the first order perturbed $u^\mu$ as $u^\mu=(1-\Phi,0,0,0)$. 

Here, we only consider the contribution of the GWs to the perturbed geodesic trajectory. The perturbed geodesic is
\be\label{geo-0}
\frac{dP^{0}}{dt} \frac{dt}{d\lambda} + \Gamma^{0}_{ij}(x^{\alpha}(\lambda)) P^{i} P^{j} =0,
\ee
in which $\frac{dt}{d\lambda}=P^0$ and $\Gamma^{0}_{ij}= a^2\left[H(\delta_{ij}+\gamma_{ij})+ \frac12 \dot{\gamma}_{ij}\right]$. At leading order, we find  $a(t)\bar{T}(t)= a_0T_0$ where $\bar{T}(t)$  is the mean CMB temperature at time $t$. At first order in perturbation, we arrive at \footnote{The change in photon energy along its trajectory, including the effects of scalar perturbations, is given by
\begin{align}
    \frac{1}{P^0}\frac{dP^0}{dt} = - \frac{\dot a}{a} - \frac{d\Phi}{dt} + \dot\Psi +  \dot\Phi   - \frac12 \dot\gamma_{ij} \hat{n}_i \hat{n}_j,
\end{align}
in which we used $\frac1a \hat{n}. \nabla \Phi = \frac{d\Phi}{dt} -\dot\Phi$ \citep{Sachs:1967er}. The full Sachs-Wolfe effect assuming adiabatic initial conditions is
\begin{align}
    \frac{\delta T}{\bar{T}} = \frac13 \Psi(\tau_\text{dec}) + \int^{\tau_0}_{\tau_\text{dec}} d\tau (\Phi' +\Psi' - \frac12 \gamma'_{ij}\hat{n}^i \hat{n}^j),
\end{align}
where the first term is the Sachs-Wolfe effect and the last three terms are the ISW effect. }
\begin{align}
 \frac{d}{d\tau}\left(\frac{\delta T(\tau, \hat{n})}{\bar{T}(\tau)}\right) = - \frac12 \gamma'_{ij}(\tau,\boldsymbol{x}) \hat{n}^i \hat{n}^j,
\end{align}
where $\delta T(\tau, \hat{n})$ denotes the perturbed CMB temperature at conformal time $\tau$. Integrating the above equation, we find the temperature perturbation today in terms of the perturbations at the time of the photon decoupling, $\tau_\text{dec}$, as (see \cref{fig:ISW})
\be
\frac{\Delta T(\hat{n})}{T} = - \frac12  \, \bigg[\int^{\tau_0}_{\tau_\text{dec}} \,  d\tau \,  n^i n^j \, \gamma_{ij}'(\tau,(\tau_0-\tau)\hat{n})\bigg],
\ee
where $r_L=\tau_0-\tau_\text{dec}$. For an intuitive and detailed explanation of the gravitational effects on temperature anisotropy, see \href{https://wwwmpa.mpa-garching.mpg.de/~komatsu/presentation/imprs2020-3-delivered.pdf}{here}. This discussion is taken up again in \cref{sec:CMB}, where we analyze the role of inflationary gravitational waves in shaping the cosmic microwave background and their manifestation in temperature and polarization anisotropies.

%DETECTION

\section{Detection techniques and observational frontiers} \label{sec:detect}

Gravitational waves span an enormous range of frequencies, the gravitational rainbow illustrated in \cref{fig:GW-rainbow}, from wavelengths comparable to the size of the observable Universe to frequencies of order $10$ kHz and beyond, where both compact astrophysical systems and early-Universe cosmology can act as sources. Each portion of this spectrum demands a distinct observational strategy, and taken together these complementary approaches enable a coherent, multi-band view of the gravitational-wave Universe. In this section, we begin by examining how cosmic microwave background polarization, particularly the search for primordial $B$-modes, provides access to gravitational waves on the largest observable scales, opening a unique window onto the physics of the earliest moments of cosmic history. We then turn to pulsar timing arrays, which exploit the remarkable rotational stability of millisecond pulsars to probe gravitational waves in the nanohertz regime. We summarize both the current observational status and the prospects for next-generation PTA experiments. Next, we outline the physical principles underlying laser interferometric detection of spacetime strain and introduce the current global network of ground-based and space-based interferometers. Finally, we discuss resonant and cavity-based detectors as potential probes of gravitational waves at much higher frequencies.

%For completeness, we note that we do not discuss high-frequency gravitational waves (above the interferometer band), whose detection remains technologically challenging but is an active and rapidly evolving area of research. Numerous novel concepts  \citep{Tobar:2023ksi,Kharzeev:2025lyu} are being explored to probe this regime. For a detailed and recent review of high-frequency GW detection proposals, see \citet{Aggarwal:2025noe} and references therein. 

%\subsection{Generation of gravitational waves}

%%%%%%%%%%%%%%%%%%%%%%%%%%%%%%%%%%%1-CMB

\subsection{CMB polarization: imprints of primordial gravitational waves}\label{sec:CMB}

%Up to now, we considered the temperature fluctuations in CMB which its two-point temperature correlation function provides an important characterization of CMB anisotropies. 
In addition to the temperature anisotropies, there is more information to be gained by measuring the CMB. More precisely, CMB photons are expected to be polarized due to Thomson scattering by free electrons before decoupling. The polarization of CMB photons provides one of the most promising methods to detect primordial gravitational waves. Before turning to CMB polarization, let us briefly pause to introduce the systematic description of light polarization in optics. %In this context, we will present the Stokes parameters.

%\vskip 0.5cm

%\subsubsection{A Framework for Describing Light Polarization}

\begin{figure}[h!]
\centering
\includegraphics[width=0.8\textwidth]{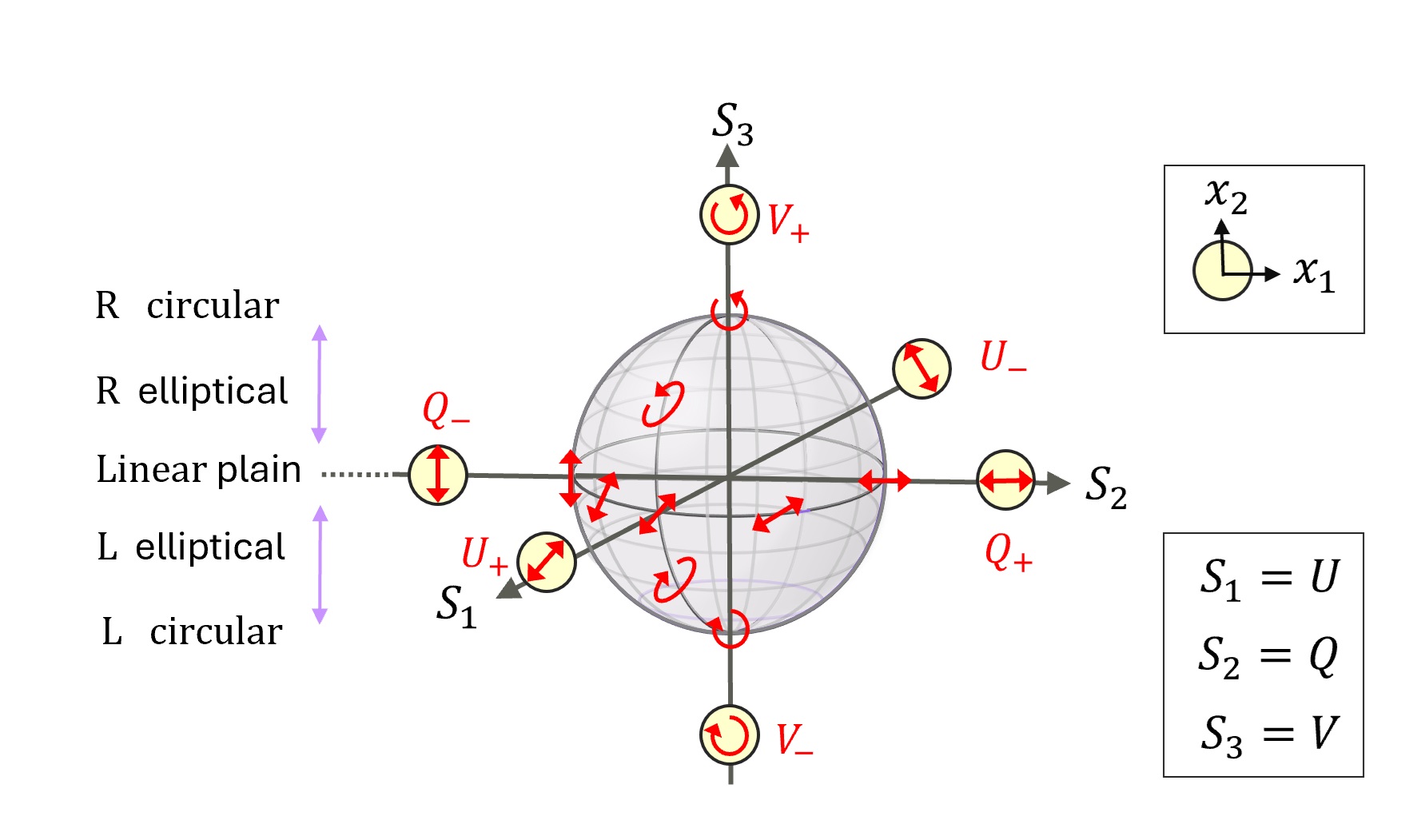} 
\caption{
The Poincaré sphere representation of the polarization state of light. Each point on the sphere corresponds to a unique polarization state. The three axes represent the Stokes parameters $Q$, $U$, and $V$, describing the two linear-polarization components and circular polarization, respectively. The radial vector from the center to a point on the sphere represents the full polarization state, and its length gives $I_\text{pol}$.
}
\label{fig:Poincare}
\end{figure}

The polarization of an electromagnetic wave is described in terms of the Stokes parameters. Consider a monochromatic electromagnetic plane wave that propagates along the direction $\hat{x}_3$. Its electric field can be written in terms of its two transverse components as
\be
\boldsymbol{E} = e^{-i(\omega t-\bk\cdot\boldsymbol{x})}
\begin{pmatrix}
\mathcal E_1\\
\mathcal E_2\\
0
\end{pmatrix},
\qquad
\mathcal E_1\equiv E_1 e^{i\theta_1},\quad
\mathcal E_2\equiv E_2 e^{i\theta_2}.
\ee
The coherency matrix is defined as the $2\times 2$ Hermitian matrix of the transverse field components,
\be
C = \begin{pmatrix}
\langle |\mathcal E_1|^2\rangle & \langle \mathcal E_1\mathcal E_2^*\rangle\\
\langle \mathcal E_1^*\mathcal E_2\rangle & \langle |\mathcal E_2|^2\rangle
\end{pmatrix},
\ee
where $\langle ...\rangle$ denotes the ensemble average. The coherency matrix  can be expanded in the basis of Pauli matrices \footnote{The Pauli matrices $\sigma_\mu$ are
$\sigma_0 = 
\begin{pmatrix}
1 & 0 \\
0 & 1
\end{pmatrix}$, $
\sigma_1 = 
\begin{pmatrix}
0 & 1 \\
1 & 0
\end{pmatrix}$, $
\sigma_2 = 
\begin{pmatrix}
0 & -i \\
i & 0
\end{pmatrix}$, and $
\sigma_3 = 
\begin{pmatrix}
1 & 0 \\
0 & -1
\end{pmatrix}$.
}
as
\begin{align}
C = \frac{1}{2} \left( I \sigma_0 + Q \sigma_3 + U \sigma_1 + V \sigma_2 \right) = \frac{1}{2}
\begin{pmatrix}
I + Q & U - iV \\
U + iV & I - Q
\end{pmatrix}.
\end{align}
Here, $I, Q, U$, and $V$ are the Stokes parameters, i.e.,
\be
I &\equiv& \langle E_1^2\rangle + \langle E_2^2\rangle,\\
Q &\equiv& \langle E_1^2\rangle - \langle E_2^2\rangle,\\
U &\equiv& 2\langle E_1 E_2 \cos(\theta_1-\theta_2)\rangle ,\\
V &\equiv& -2\langle E_1 E_2 \sin(\theta_1-\theta_2)\rangle.
\ee
See \cref{fig:Poincare} for the polarization corresponding to each of the Stokes parameters.

\begin{itemize}[label=$\bullet$]
\item{The quantity $I$ is the total intensity of the wave, here the CMB intensity.}
\item{$Q$ describes the difference between the linear polarization in $\hat{e}_1$ and $\hat{e}_2$ directions.}
\item{ The $U$ and $V$ parameters give information on the phases. In particular, expanding the wave in the $\pm 1$ helicity polarization states, $\hat{e}^{R,L}_i = 1/\sqrt{2} \, (\hat{e}_1 \pm i \hat{e}_2)_i$, we have
\be
V = \langle E_{R}^2 \rangle  - \langle E_{L}^2 \rangle,
\ee
which represents the difference between the positive and negative helicity intensities.}
%\item{Note that the mechanisms that generate CMB polarization produce only linear polarizations, and no circular one. Therefore, in the absence of any parity violation interaction during inflation, we can set $V=0$.}
\item{Here \(Q\) and \(U\) describe the linear polarization, while \(V\) quantifies the circular polarization.}
\item{The square of the total polarized intensity of light is
\begin{align}
   I^2_{\text{pol}} = Q^2 + U^2 + V^2, 
\end{align}
which quantifies how much of the light is polarized, regardless of whether it is linearly or circularly polarized. From that, we can define the degree of polarization as
\begin{align}
    P = \frac{I_{\text{pol}}}{I}, 
\end{align}
which is a quantity equal to or less than one. This leads to the constraint
\begin{align}
 I^2 \geq Q^2 + U^2 + V^2 ,
\end{align}
with equality satisfied in the case of fully polarized light.}
\item{For fully polarized light, the most general polarization state is elliptical, which is described by two parameters; 
polarization angle $\theta$
\be
\theta = \frac12 \tan^{-1}(\frac{U}{Q}),
\ee
and the  ellipticity angle 
$\chi$
\be
\sin(2\chi) =  \frac{V}{I_{\text{pol}}}.
\ee
These two parameters fully describe the shape of the ellipse.}
\end{itemize}

\bigskip

\begin{tcolorbox}[colback=gray!10, colframe=gray!10, boxrule=0pt,
                  enhanced, breakable, halign=justify]

\subsubsection*{Spin-Weight of the Stokes parameters}  
A spin-$s$ function on the sphere under a rotation of angle $\phi$ around $\hat{n}$ transforms as
\begin{align}
{}_sf(\hat{n}) \rightarrow e^{-is\phi} {}_sf(\hat{n}).
\end{align}
Analogous to ordinary spherical harmonics $Y_{\ell m}(\theta,\phi)$ (basis for \(s=0\)), 
there exist spin-weighted spherical harmonics ${}_sY_{\ell m}(\theta,\phi)$ which form a complete orthonormal basis for spin-$s$ functions
\be
{}_s f(\hat{n}) = \sum_{\ell = |s|}^\infty \sum_{m=-\ell}^{\ell} 
a_{\ell m}^{(s)} \, \, {}_sY_{\ell m}(\theta,\phi).
\ee
For \(s \neq 0\), they are obtained by acting with the 
spin-raising $\eth^+$ and spin-lowering $\eth^-$ operators on the ordinary $Y_{\ell m}$. The operators $\eth^{\pm}$ have the explicit action
\be
\eth^{\pm} \, {}_s f
= -(\sin\theta)^{\pm s}
\left( \frac{\partial}{\partial \theta}
\pm \frac{i}{\sin\theta}\frac{\partial}{\partial \phi} \right)
\big[ (\sin\theta)^{\mp s} \, {}_s f \big],
\ee
and shift the spin weight as
\be
\eth^{\pm} \, {}_s f(\hat{n}) \propto {}_{s\pm1} f(\hat{n}).
\ee
The spin-weighted spherical harmonics are related to the scalar (spin-0) spherical harmonics as
\be\label{d+-1}
{}_sY_{lm}(\hat{n}) &\equiv \sqrt{\frac{(l-s)!}{(l+s)!}} (\eth^{+})^s Y_{lm}(\hat{n}) \qquad \quad & \textmd{for} \quad ~0 \leq s \leq l\\ \label{d+-2}
{}_sY_{lm}(\hat{n}) &\equiv \sqrt{\frac{(l+s)!}{(l-s)!}} (-1)^{s}(\eth^{-})^{-s} Y_{lm}(\hat{n}) & \textmd{for} \quad -l \leq s \leq 0.
\ee
This implies that only the spherical harmonics with $l \geq |s|$ can give rise to ${}_sY_{lm}$. 
Under the action of parity, $\hat{n} \rightarrow -\hat{n}$, the spin-weighted harmonics transform as
\be
{}_sY_{lm}(-\hat{n}) = (-1)^l {}_{-s}Y_{lm}(\hat{n}).
\ee

\end{tcolorbox}

\begin{figure}[h!]
\centering
\includegraphics[width=0.8\textwidth]{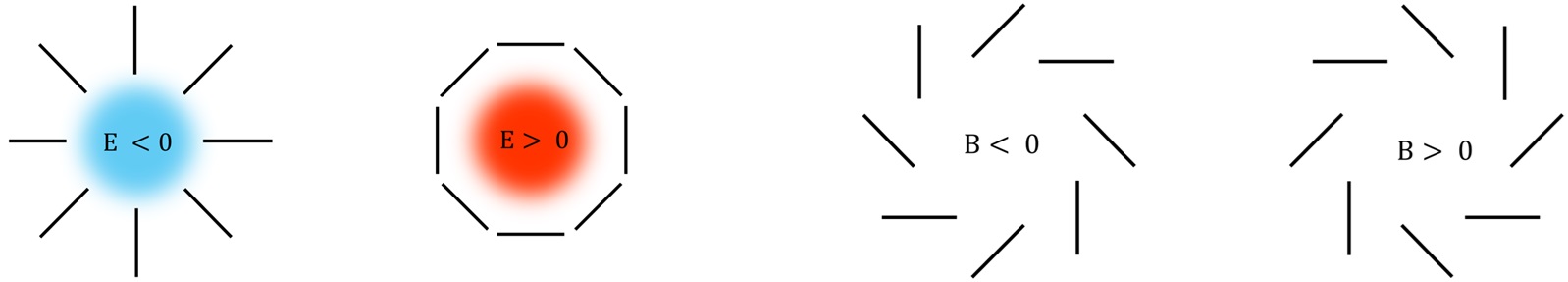} 
\caption{E-mode and B-mode patterns of polarization. The cold and hot spots are shown in blue and red, respectively. }\label{fig:E-B-}
\end{figure} 

Let us now take a second look at the Stokes parameters. Under a rotation of the polarization basis in the transverse plane by angle $\phi$, the Pauli matrices transform as
\begin{align}
R(\phi)\,\sigma_0\,R^{-1}(\phi) = \sigma_0, \quad
R(\phi)\,\sigma_2\,R^{-1}(\phi) = \sigma_2,
\end{align}
and
\begin{equation}
R(\phi)\,(\sigma_3 \pm i\sigma_1)\,R^{-1}(\phi)
= e^{\mp 2 i \phi}\,(\sigma_3 \pm i\sigma_1).
\end{equation}
The coherency matrix can be written as
\begin{align}
C &= \frac{1}{2} \left( I \sigma_0 + V \sigma_2 + \frac12(Q +i U) (\sigma_3 - i \sigma_1) + \frac12(Q -i U) (\sigma_3 + i \sigma_1)\right).
\end{align}
As a result, the intensity, $I$, and the helical (circular) polarization, $V$, remain invariant under rotation, i.e. spin-0. However, the linear polarizations $Q$ and $U$ transform like spin-2 fields $\big(Q \pm i U\big) \rightarrow e^{\mp 2i \varphi} \big(Q \pm i U\big)$, which implies 
\begin{align}
(Q\pm iU)(\hat{n}) = \sum_{\ell m}  \, \, a^{\pm 2}_{lm}\, {}_{\pm2}Y_{\ell m}(\hat{n}) .
\label{eq:QU}
\end{align}
Following the original polarization decompositions~\citep{Seljak:1996gy,Kamionkowski:1996ks}, we define a scalar quantity (E) and a pseudo-scalar quantity (B) (see \cref{fig:E-B-}): 
\begin{align}
(Q\pm iU)(\hat{n}) &= -\sum_{\ell m} \big(a_{\ell m}^E \pm i\,a_{\ell m}^B\big)\; {}_{\pm2}Y_{\ell m}(\hat{n}).
\end{align}
%See the E-mode and B-mode patterns of polarization in \cref{fig:E-B-}.

\begin{figure}[H]
\centering
\includegraphics[width=0.6\textwidth]{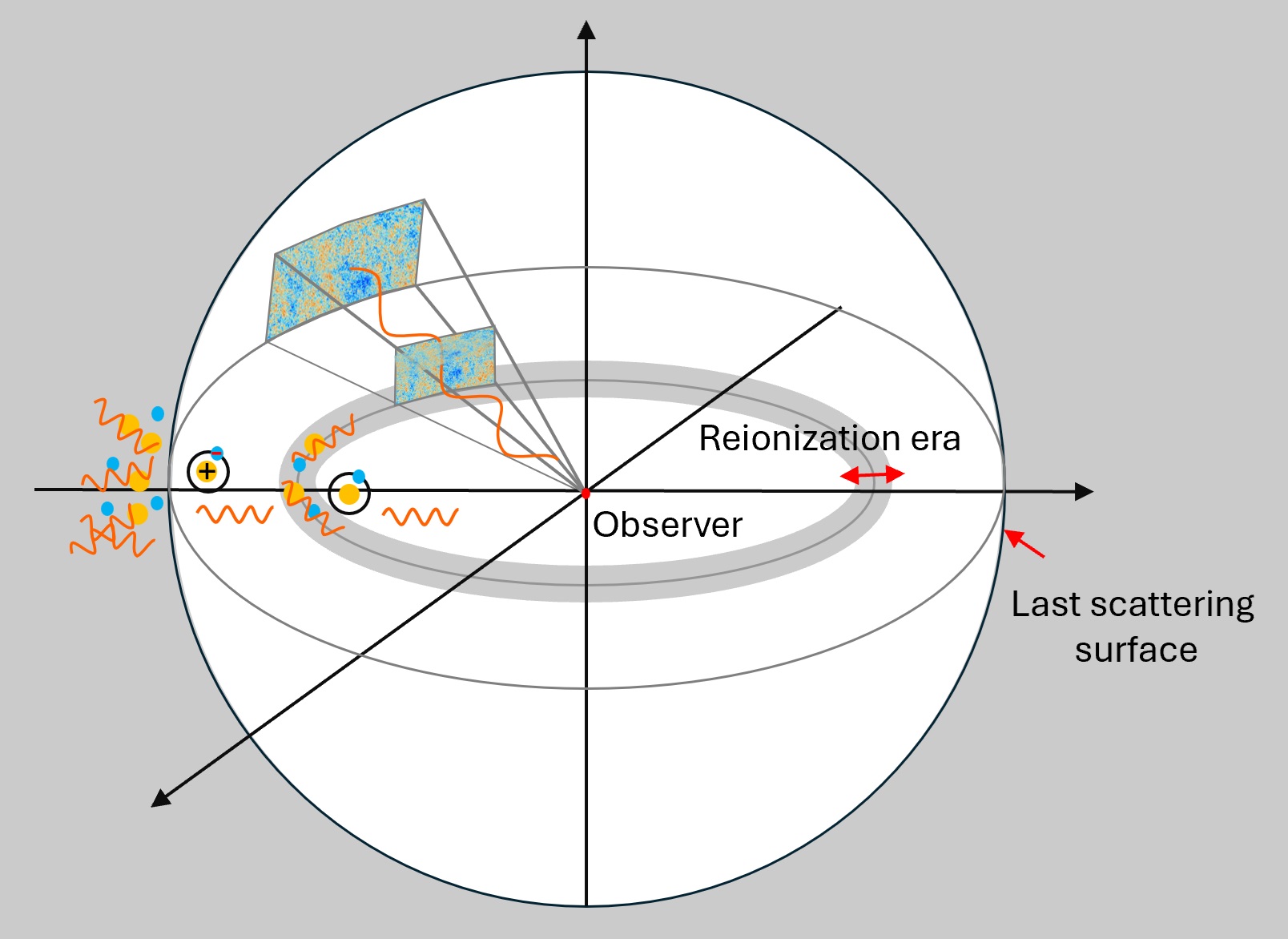} \includegraphics[width=0.32\textwidth]{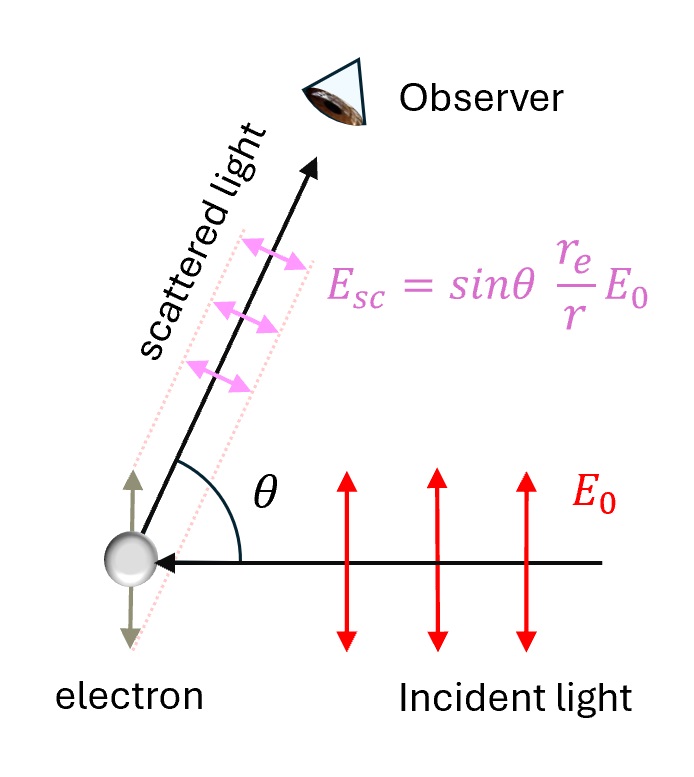} 
\caption{Left panel: Schematic of the celestial sphere showing the observable Universe, 
where the radial coordinate $r$ corresponds to cosmological redshift, 
ranging from the surface of last scattering ($z \approx 1100$) to the observer today ($z = 0$), 
with the reionization era marked in the interval $6 \lesssim z \lesssim 10$. The shaded regions in gray denote the redshifts at which the Universe was ionized. Right panel: Thomson scattering of an electromagnetic wave by a free electron. The incident photon with wavevector $\boldsymbol{k}$ is scattered into direction $\boldsymbol{k}'$ at polar angle $\theta$. %The classical electron radius is $r_e \equiv \frac{e^2}{4\pi\varepsilon_0 m_e c^2}$ (Gaussian units: $r_e=\frac{e^2}{m_e c^2}$). In the Thomson limit ($h\nu \ll m_e c^2$), the differential cross section is $\mathrm{d}\sigma/\mathrm{d}\Omega=\tfrac{1}{2}r_e^2(1+\cos^2\theta)$, with total $\sigma_T=\tfrac{8\pi}{3}r_e^2$. https://www.roe.ac.uk/ifa/postgrad/pedagogy/2007_memari.pdf
}
\label{shape-}
\end{figure}

%\bigskip

\begin{tcolorbox}[colback=gray!10, colframe=gray!10, boxrule=0pt,
                  enhanced, breakable, halign=justify]
\subsubsection*{Thomson Scattering in an Anisotropic Background}

After decoupling at the surface of last scattering ($z \simeq 1100$), 
CMB photons free-stream toward us. Along the way, mainly during 
reionization, they can rescatter off free charges via Thomson scattering. 
In the Thomson limit ($h\nu \ll mc^2$), the differential cross section is
\begin{equation}
\frac{d\sigma_T}{d\Omega}
= \frac{r_q^2}{2}\,(1+\cos^2\theta),
\qquad 
r_q \equiv \frac{q^2}{4\pi\varepsilon_0 m c^2}.
\end{equation}
Here $\theta$ is the scattering angle, given by 
$\cos\theta=\hat{\boldsymbol{k}}\cdot\hat{\boldsymbol{k}}'$, 
and $q$, $m$, and $r_q$ are the particle’s charge, mass, and classical radius. 
Given that $(m_e/m_p)^2\approx 10^{-7}$ and $\sigma_T\propto 1/m^2$, 
electrons are far more efficient scatterers than protons.

A single Thomson scattering converts a local quadrupole anisotropy of 
the incident radiation into linear polarization of the CMB. In the non-relativistic limit, the 
electron’s equation of motion is
\[
m_e\ddot{\mathbf{x}}(t) = -e\,\mathbf{E}_\mathrm{inc}(t),
\qquad
\mathbf{a}(t)=-(e/m_e)\,\mathbf{E}_\mathrm{inc}(t),
\]
so the acceleration is parallel to $\hat{\boldsymbol{\epsilon}}$. 
The far-zone (radiation) electric field from an accelerating charge is
\begin{equation}
\mathbf{E}_\mathrm{sc}(\mathbf{r},t)
=\frac{e}{4\pi\varepsilon_0 c^2}\frac{1}{r}\,
\hat{\mathbf{n}}\times\big(\hat{\mathbf{n}}\times\mathbf{a}(t_r)\big)
= \frac{r_e}{r}\, E_0\, \sin\vartheta,
\end{equation}
where $\hat{\mathbf{n}}$ points from the electron to the observer and 
$t_r$ is the retarded time (see \cref{shape-}).

%some ref about CMB polarization https://astro.uni-bonn.de/~kbasu/CMB/Online_Slides/CMB.07%20-%20Polarization-I.pdf
%https://www.scribd.com/document/84929591/Polarization-by-Scattering
%https://background.uchicago.edu/~whu/polar/webversion/node9.html

%%\begin{figure}[h]
%%    \centering
\begin{center}
\includegraphics[width=0.8\textwidth]{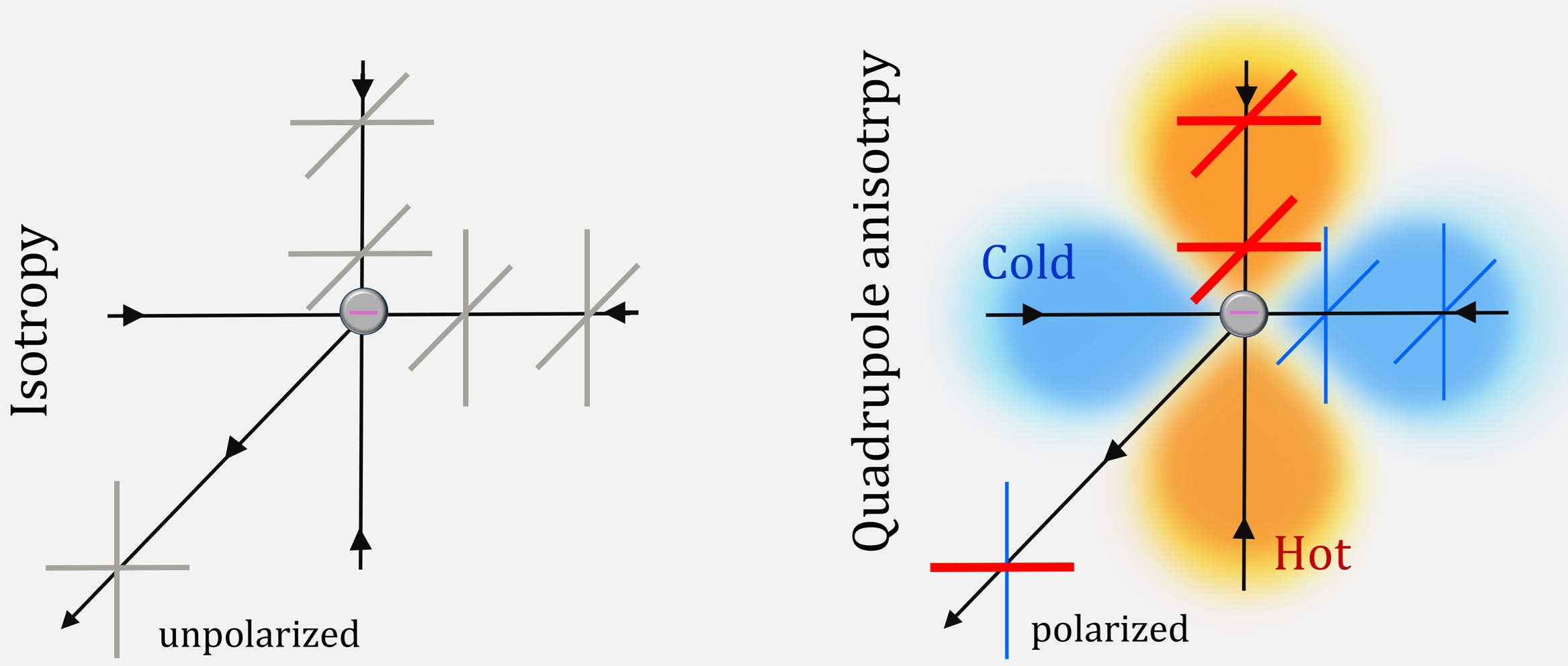}
\captionsetup{type=table}
  \captionof{figure}{%
 %%\caption{
   Illustration of Thomson scattering of initially unpolarized CMB photons. For an isotropic radiation field or one with only dipole anisotropy, the scattered radiation remains unpolarized. Polarization arises only when a quadrupole anisotropy is present, making the quadrupole the lowest-order multipole that generates CMB polarization. 
  }\label{thomson-CMB}
\end{center}
%%\end{figure}

In the last equality, we used $\mathbf{a}\!\parallel\!\hat{\boldsymbol{\epsilon}}$, while $\vartheta$ is the angle between $\hat{\boldsymbol{\epsilon}}$ (the acceleration direction) and $\hat{\mathbf{n}}$,
and $r_e$ is the classical electron radius. For fixed incident polarization, the corresponding intensity  is
$\,I_\mathrm{sc}(\vartheta)\propto |E_\mathrm{sc}|^2\propto \sin^2\vartheta$,
the characteristic dipole radiation pattern. 
Now, let $\hat{\mathbf{n}}$ be the observed line of sight and $(\vartheta,\phi)$ the angles of the
incident direction $\hat{\mathbf{n}}'$ in the scattering frame.
For a generic unpolarized incident intensity $I(\hat{\mathbf{n}}')$, the linear polarization generated per unit path length is
\begin{equation}
\frac{d(Q\pm iU)(\hat{\mathbf{n}},\tau)}{d\ell}=\frac{3\sigma_T}{16\pi}\,n_e(\tau)
\!\int\! d\Omega'\;\sin^2\vartheta\,e^{\pm 2i\phi}\; I(\hat{\mathbf{n}}',\tau) ,
\label{eq:QpmiU-kernel}
\end{equation}
where $e^{\pm 2i\phi}$ is the spin-$\pm2$ phase from the scattering geometry, $\sigma_T$ is the Thomson cross section, and $n_e(\tau)$ is the electron number density at time $\tau$. Expanding the incident field in spherical harmonics, $I(\hat{\mathbf{n}}')=\sum_{\ell m} a_{\ell m}Y_{\ell m}(\hat{\mathbf{n}}')$, the angular kernel in \cref{eq:QpmiU-kernel} picks out the $\ell=2$ quadrupole. More precisely, the linear polarization, 
$Q\pm iU$, transforms as a spin-$\pm2$ field on $S^2$, so it admits an expansion in
${}_{\pm2}Y_{\ell m}$ and therefore vanishes unless $\ell\ge |s|=2$. 
Thus, neither the monopole $(l=0)$ nor the dipole $(l=1)$ components of the incident radiation generate linear polarization. The leading contribution arises from a nonzero quadrupole anisotropy, as illustrated in \cref{thomson-CMB}. 
In other words, Thomson scattering converts a quadrupole intensity anisotropy of the incident radiation
into linear polarization on the sky. 

\end{tcolorbox}

\begin{figure}[h!]
\centering
%%\includegraphics[width=0.9\textwidth]{EE-BB.jpg} 
%%FS: use original vector fig from arXiV
\includegraphics[width=\textwidth]{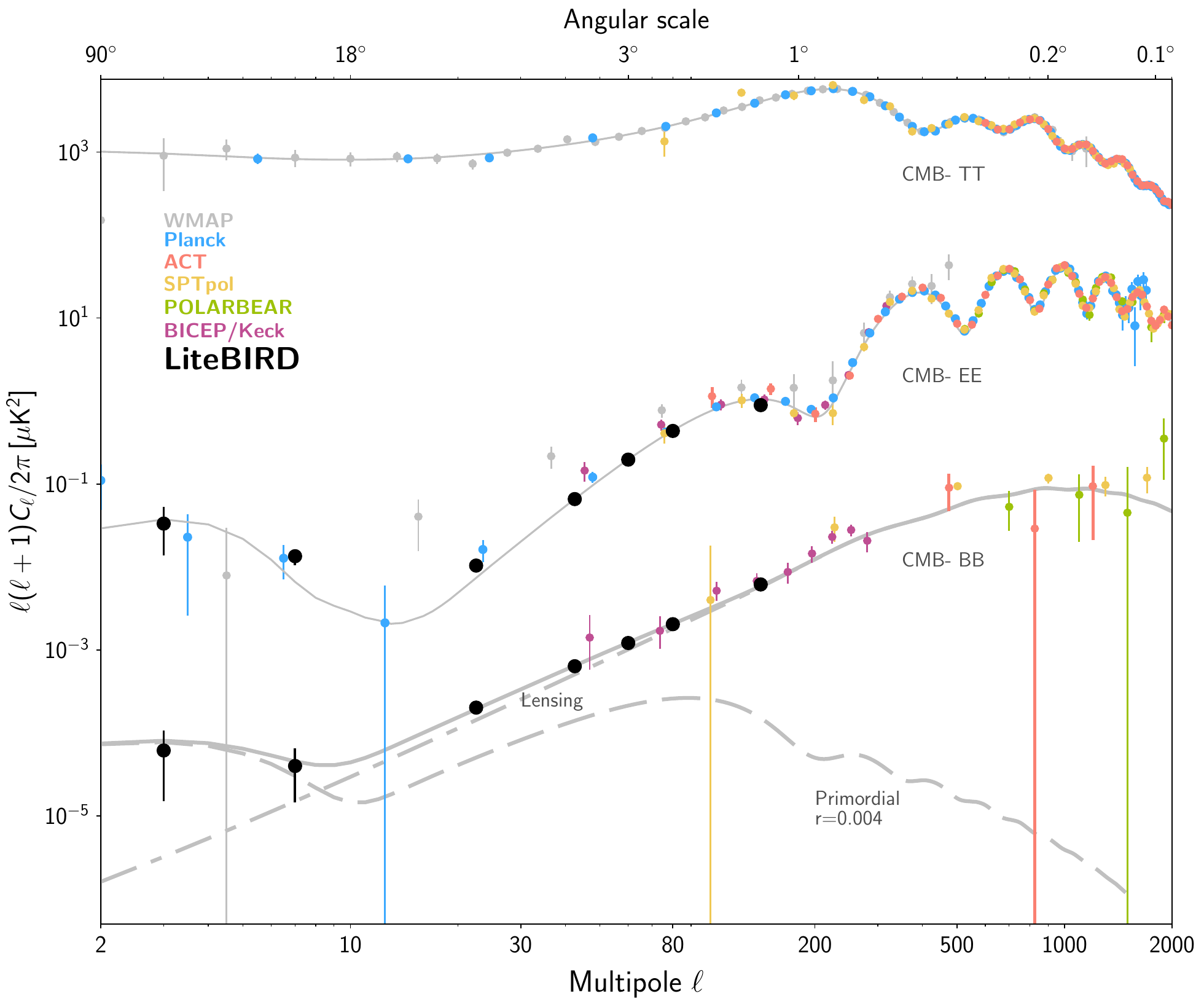} 
\caption{CMB power spectra for temperature anisotropy (CMB-TT), E-mode polarization (CMB-EE), and B-mode polarization (CMB-BB) are shown. The solid curves represent the predictions of the best-fit $\Lambda CDM$ model, including a scale-invariant tensor perturbation with a tensor-to-scalar ratio $r=
0.004$. For comparison, the thin dashed line indicates the contribution of tensor modes to the B-mode spectrum alone. Also displayed are current measurements of the CMB power spectra (colored points) and the projected polarization sensitivity of LiteBIRD (black points). Image credit: LiteBIRD mission \citep{LiteBIRD:2022cnt}.}\label{fig:E-B}
\end{figure}

%\subsubsection{CMB Polarization and Primordial GWs}
Along the line of sight from last scattering to today, polarization is sourced wherever free
electrons see a local temperature quadrupole. In essence, the observed polarization anisotropies can be expressed as an integral over conformal time
\begin{equation}
(Q\pm iU)(\hat{\mathbf{n}})
=\int_{\tau_{\text{SLS}}}^{\tau_0}\! d\tau\; g(\tau)\,\mathcal{S}_{\pm2}(\tau,\hat{\mathbf{n}}),
\end{equation}
where $ g(\tau) = a(\tau) n_e(\tau) \sigma_T e^{-\tau_e(\tau)} $ is the visibility function and $\mathcal{S}_{\pm2}(\tau,\hat{\mathbf{n}})$ is the polarization source term. Given that the polarization is related to the electron number density in the Universe, there are two key epochs:
(i) recombination, when the mean free path grows rapidly, and a finite quadrupole develops,
creating the acoustic $E$-mode pattern; and
(ii) late reionization (see left panel of \cref{shape-}). To compute $C_\ell^{EE}$ and $C_\ell^{BB}$, cosmologists first solve the linear cosmic perturbation equations to obtain the polarization source from primordial anisotropies.  
This source feeds into the line-of-sight integral, which is then projected onto spin-2 spherical harmonics to extract E- and B-mode components.  
Tools like CAMB and CLASS perform this numerically \citep{Lesgourgues:2011rg}. A detailed computation is beyond the scope of the current review, and the final result of these impressive calculations is shown in \cref{fig:E-B}, where the $TT$, $EE$, and $BB$ spectra (both lensing and primordial) are presented.  
In the following, we summarize the main features of the CMB polarization spectra.

\begin{itemize}[label=$\bullet$]
\item{The scalar quantities $E$ and $B$, completely determine the linear polarization fields. }
\item{The $E$-mode is curl-free and even under the action of parity. Its polarization vectors are radial around cold spots (under-dense) and tangential around hot spots (over dense). }
\item{The $B$-mode is the divergence-free field and odd under parity. Its polarization vectors have vorticity around the under- and over-dense areas. }
\item{The angular power spectra are defined as 
\be
C^{XY}_{l} \equiv \frac{1}{2l+1} \sum_{m} \langle a_{lm}^{X}a_{lm}^{Y*}\rangle  \where X,Y=T,E,B.
\ee
The autocorrelations of $E$- and $B$-modes, denoted by $EE$ and $BB$, are presented in  \cref{fig:E-B}. }
\item{At linear order, scalar (density) perturbations source only E-modes, while tensor perturbations (gravitational waves) generate both E- and B-modes.  }
\item{At nonlinear order, gravitational lensing by large-scale structure distorts $E$-modes into $B$-modes, producing a lensing-induced $B$-mode signal at small angular scales.
}
\end{itemize}

Primordial gravitational waves and primordial B-modes have not yet been detected. At linear order, scalar perturbations do not generate B-modes, while tensor (gravitational wave) perturbations do, a detection of primordial B-modes would be a distinctive fingerprint of primordial gravitational waves and, if consistent with inflationary predictions, would provide powerful evidence for inflation. This is the ambitious goal of the upcoming LiteBIRD satellite mission: to measure the CMB polarization with unprecedented sensitivity and probe tensor-to-scalar ratios as low as $r=10^{-3}$, pushing the frontier of our understanding of the early Universe. For additional information on LiteBIRD, see \cite{LiteBIRD:2022cnt}.  CMB B-modes also offer a unique window into a broad range of 
new physics, including parity violation and relic particle species. We refer the interested reader to \cite{Komatsu:2022nvu} for a recent comprehensive overview. Finally, CMB spectral distortions act as an intermediate probe of primordial gravitational waves, bridging $B$-mode polarization measurements and pulsar timing array observations \citep{Kite:2020uix}.

%%%%%%%%%%%%%%%%%%%%%%%%%%%%%%%%%%%%%%%%%2-PTA

\subsection{Pulsar timing arrays: precision cosmic clocks for GW detection} \label{sec:PTA}

A population of exceptionally stable millisecond pulsars, timed with precisions approaching $100\,\mathrm{ns}$ over many years, acts as an array of cosmic clocks. Their remarkably regular pulses are subtly and coherently disturbed by gravitational waves propagating along the Earth--pulsar line of sight: the passing waves perturb the null geodesics along the photon’s path to Earth. A global collaboration now monitors dozens of these pulsars with ever-increasing precision, effectively transforming the Milky Way into a galaxy-scale detector for nanohertz gravitational waves through pulsar timing arrays (PTAs).

By tracking the arrival times of pulses from a network of pulsars, PTAs are sensitive to GWs in the nanohertz band. Crucially, PTAs are sensitive not only to individual deterministic sources, but also to stochastic gravitational wave backgrounds (SGWBs). Indeed, the strongest existing bounds on the primordial nanohertz background arise from PTA measurements, providing leading constraints on the gravitational wave energy density in this frequency band.

\begin{figure}[h]
\centering
\includegraphics[width=\linewidth]{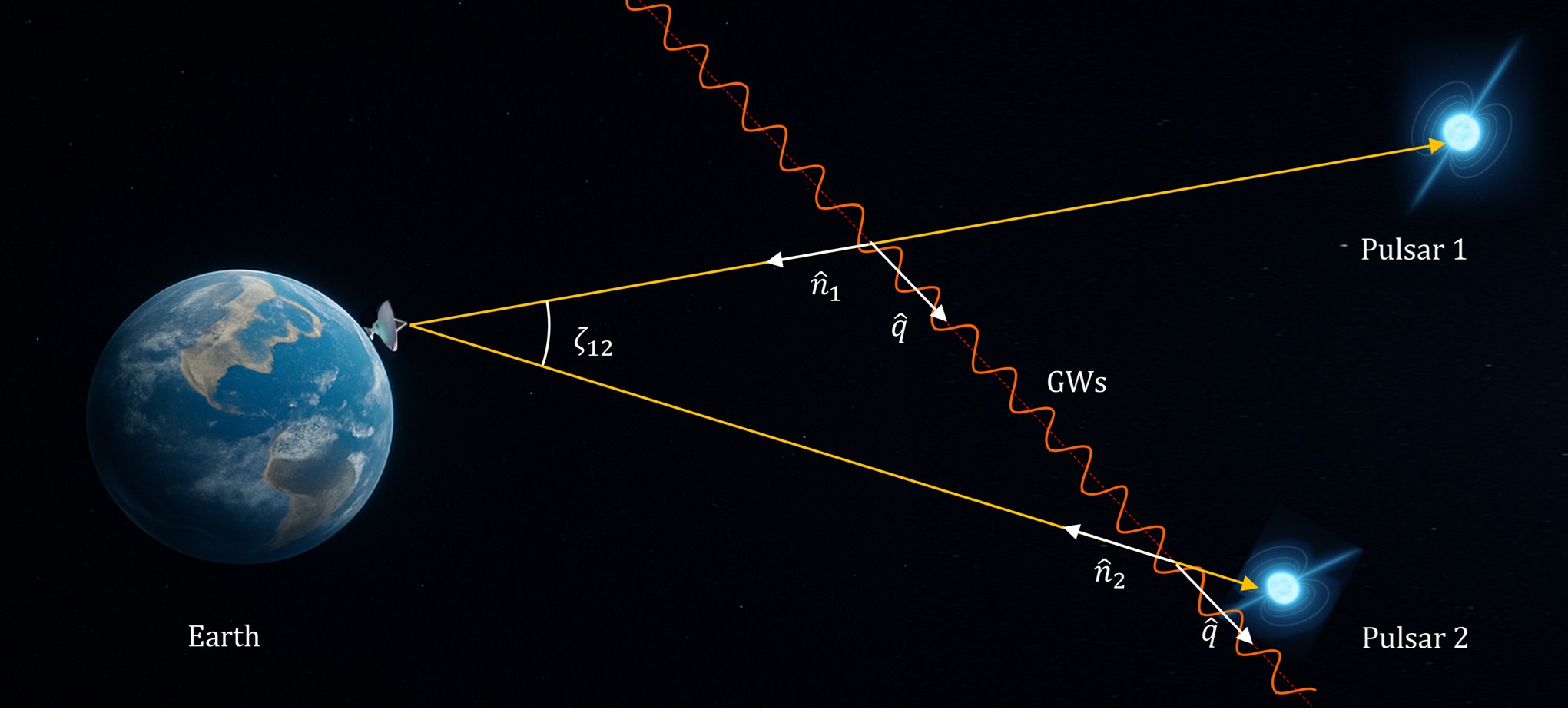}
%%\captionof{figure}{
\caption{
Pulsar timing arrays detect nanohertz gravitational waves by measuring
timing residuals across widely separated millisecond pulsars.}
\label{fig:PTA}
\end{figure}

Pulsar timing arrays operate by monitoring the times of arrival (TOAs) of radio pulses from their ensemble of millisecond pulsars, whose exceptional rotational stability makes them superb astrophysical clocks \citep{FosterBacker1990,Manchester2013}. A passing GW propagating in direction $\hat{q}$ perturbs the spacetime metric along the photon trajectory, inducing a small but coherent shift in the observed pulse frequency. The measured fractional frequency shift at Earth can be expressed as follows~\citep{Detweiler:1979wn}:
\begin{equation}
    \frac{\delta\nu(t)}{\nu}
    = \frac{1}{2}\, \frac{\hat{n}^i \hat{n}^j}{1-\hat{q}\!\cdot\!\hat{n}}
    \big[ \gamma_{ij}(t_{\mathrm{e}},\bx_{\mathrm{e}}) - \gamma_{ij}(t_{\mathrm{p}},\bx_{\mathrm{p}}) \big],
\end{equation}
where $\hat{n}$ points from the pulsar to Earth, $t_{\mathrm{e}}$ is the reception time at Earth, and 
$t_{\mathrm{p}} = t_{\mathrm{e}} - L/c$ is the photon emission time at the pulsar, with $L$ the pulsar distance and $\bx_{\mathrm p}=\bx_{\mathrm e}-L\hat n$. Here $\gamma_{ij}(t_{\mathrm{e}},\bx_{\mathrm{e}})$ denotes the gravitational wave metric perturbation evaluated at Earth when the pulse is received, the Earth term, while $\gamma_{ij}(t_{\mathrm{p}},\bx_{\mathrm{p}})$ denotes the perturbation at the pulsar at the time of emission, constituting the pulsar term. The corresponding timing residual is obtained by integrating the frequency shift as
\begin{equation}
    R(t) = \int^{t} dt' \,\frac{\delta\nu(t')}{\nu}.
\end{equation}
Gravitational waves in the nanohertz band ($f\sim 10^{-9}$--$10^{-7}$ Hz) induce correlated timing residuals across the pulsars in a PTA. The essential observable for a stochastic background is the two-point function
\begin{equation}
    C_{ab} \equiv \langle R_a(t)\, R_b(t) \rangle,
\end{equation}
which captures the GW-induced covariance between pulsars $a$ and $b$. This geometry is illustrated schematically in \cref{fig:PTA}. The pulsar--pair angular separation $\zeta_{ab}$ is defined as the angle 
between the sky directions of pulsars $a$ and $b$
\begin{equation}
\cos \zeta_{ab} = \hat{\boldsymbol{n}}_{a} \cdot \hat{\boldsymbol{n}}_{b}.
\end{equation}

\begin{figure}
\centering
\includegraphics[width=0.99\linewidth]{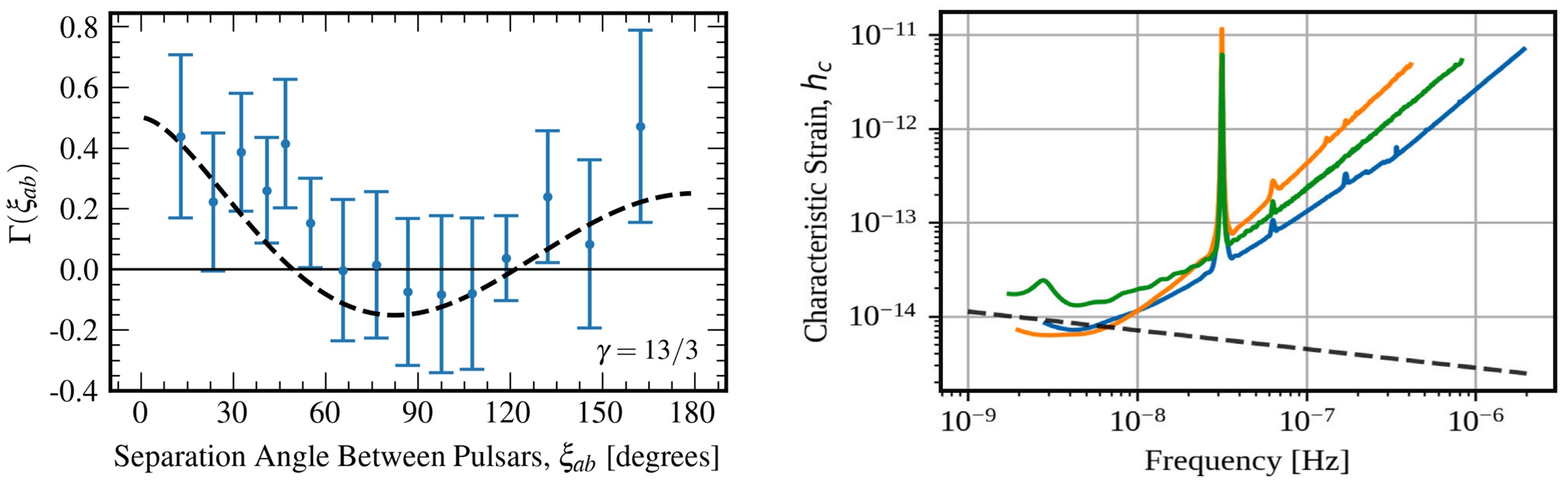}
\caption{Left panel: Angular correlation of pulsar timing residuals measured by 
NANOGrav (blue points with uncertainties) \citep{NANOGrav:2023gor}, shown as a function of the 
separation angle between pulsar pairs. The dashed black curve is the 
Hellings--Downs prediction for an isotropic GWB.
The data exhibit the characteristic 
quadrupolar trend expected from gravitational waves, whereas the absence 
of a signal would yield correlations consistent with zero across all 
angles. %The observed angular dependence therefore provides evidence for the spatial correlations induced by a stochastic gravitational wave background.
Image credit: The NANOGrav Collaboration \citep{NANOGrav:2023gor}. Right panel: Characteristic strain sensitivity curves for current pulsar timing array (PTA) collaborations. The blue, orange, and green curves show the sensitivities of EPTA+InPTA, NANOGrav, and PPTA, respectively. The dashed black line indicates the joint-median sensitivity obtained by combining information from all PTA datasets. %These curves illustrate the common sensitivity band of PTAs to nanohertz gravitational waves and highlight differences in timing precision and pulsar populations across the arrays 
Image credit: \citet{Verbiest:2024nid}.}
\label{fig:NanoGrav}
\end{figure}

For an isotropic and stationary gravitational wave
background, this covariance depends only on the angular separation between the
pulsars, and may be written as
\begin{equation}
    C_{ab} = \Gamma_{ab}(\zeta_{ab})\, \sigma_{_{\mathrm{GW}}}^{2},
\end{equation}
where $\sigma_{_{\mathrm{GW}}}^{2}$ encodes the overall amplitude of the
background and $\Gamma_{ab}(\zeta_{ab})$ is the geometrical overlap reduction function which depends only on the angular separation between the pulsars when the background is isotropic. In general relativity, only the two transverse--traceless tensor modes are 
present. In this special case, $\Gamma_{ab}$ reduces to a single, characteristic 
quadrupolar correlation function,
\be
\Gamma_{ab}^{\mathrm{HD}} = 
\frac{1}{2}+
\frac{3}{2} \left[ \frac{1 - \cos\zeta_{ab}}{2} \right]
\ln\!\left( \frac{1 - \cos\zeta_{ab}}{2} \right)
 - \frac{1}{4} \left[ \frac{1 - \cos\zeta_{ab}}{2} \right]
 + \frac{1}{2}\, \delta_{ab},
\ee
known as the Hellings--Downs (HD) curve~\citep{HellingsDowns:1983}. Notice that the term Hellings--Downs 
correlation therefore refers specifically to the ORF produced by an isotropic 
background of tensor gravitational waves in general relativity, and does not 
generally apply when additional polarizations or modified propagation effects 
appear in modified-gravity theories.

\begin{figure}
\centering
\includegraphics[width=\linewidth]{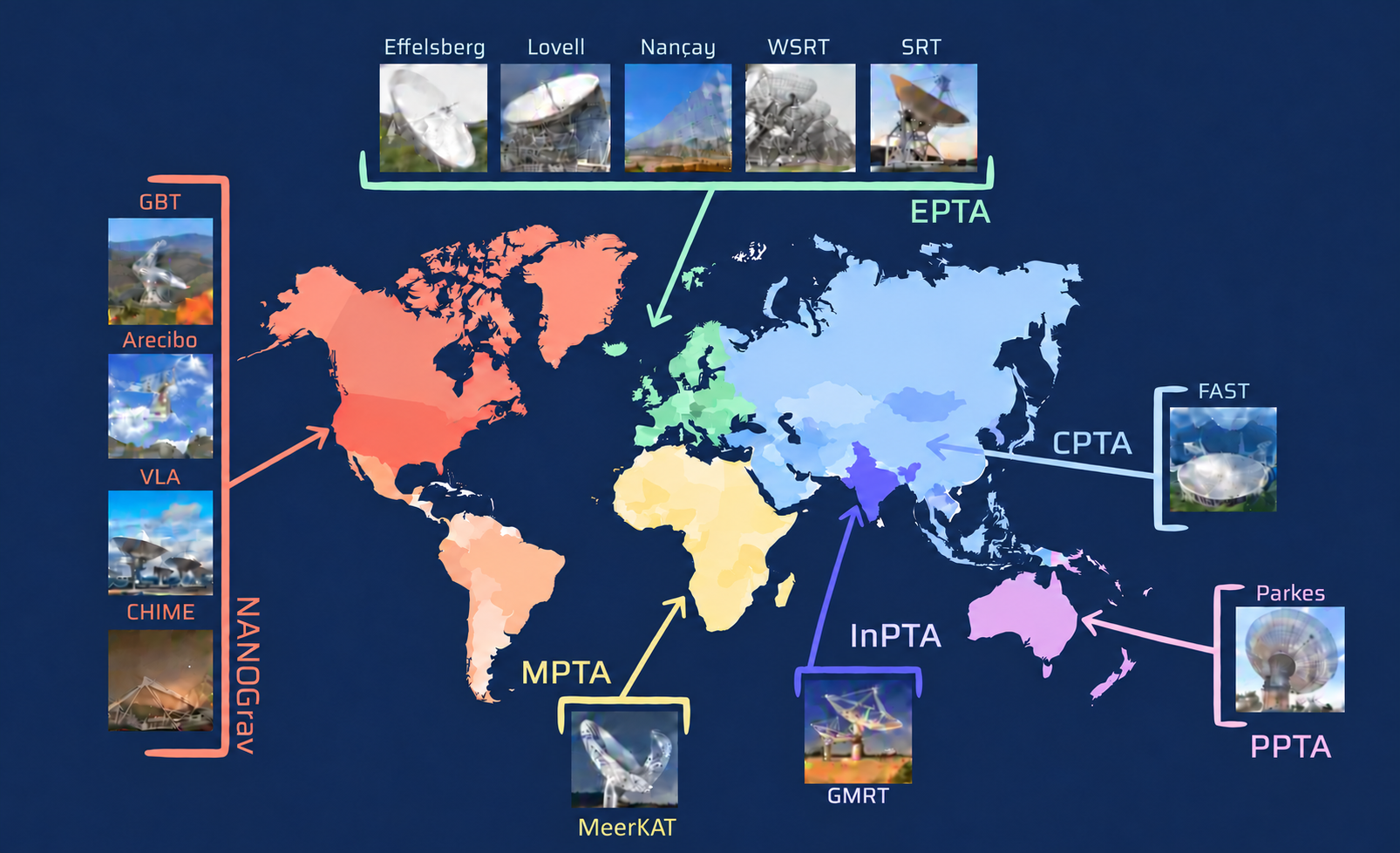}
\caption{The International Pulsar Timing Array (IPTA), a worldwide network of radio telescopes that times an ensemble of millisecond pulsars to search for nanohertz gravitational waves. Image credit: Adapted from H. Thankful Cromartie’s presentation at the 2023 SARA Eastern Conference.}
\label{fig:IPTA}
\end{figure}

Recent PTA measurements provide compelling evidence for a nanohertz gravitational wave background  (see left panel of \cref{fig:NanoGrav}). The NANOGrav 15-year data set \citep{NANOGrav:2023gor} identifies a common-spectrum process exhibiting the expected Hellings–Downs correlations~\citep{NANOGrav:2023gor}. Independent analyses by the European Pulsar Timing Array (EPTA) together with the Indian Pulsar Timing Array (InPTA) \citep{epta2023,Antoniadis:2023rjw,Rana:2025ano}, the Parkes Pulsar Timing Array (PPTA) \citep{Reardon:2023gzh,Zic:2023gta}, and the Chinese Pulsar Timing Array (CPTA) \citep{cpta2023} report consistent signals. Together, these regional collaborations form the International Pulsar Timing Array (IPTA), and collectively these results mark the emergence of gravitational-wave astronomy in the nanohertz regime (see \cref{fig:IPTA}), consistent with an origin in the cosmic population of slowly inspiraling supermassive black-hole binaries, although other astrophysical or cosmological contributions remain possible. The characteristic strain sensitivity curves of current PTA collaborations are shown in the right panel of \cref{fig:NanoGrav}. Future observations with the Square Kilometre Array (SKA) are expected to significantly improve timing precision and greatly enhance PTA sensitivity~\citep{ska_weltman2020,ska_janssen2015}.  Astrometric observations provide an alternative to pulsar timing arrays for probing the stochastic GWB and its angular properties~\citep{Caliskan:2023cqm,Garcia-Bellido:2021zgu}. Together with LISA and ground-based interferometers, PTAs anchor a coordinated multi-band gravitational wave network spanning nearly nine orders of magnitude in frequency, enabling a unified and comprehensive view of compact-object populations and their evolution across cosmic time.

%%%%%%%%%%%%%%%%%%%%%%%%%%%%%%%%%%%3-Laser

\subsection{Laser interferometry: Earth- and space-based GW observatories}\label{sec:Laser}

\begin{figure}[h]
\centering
\includegraphics[width=\textwidth]{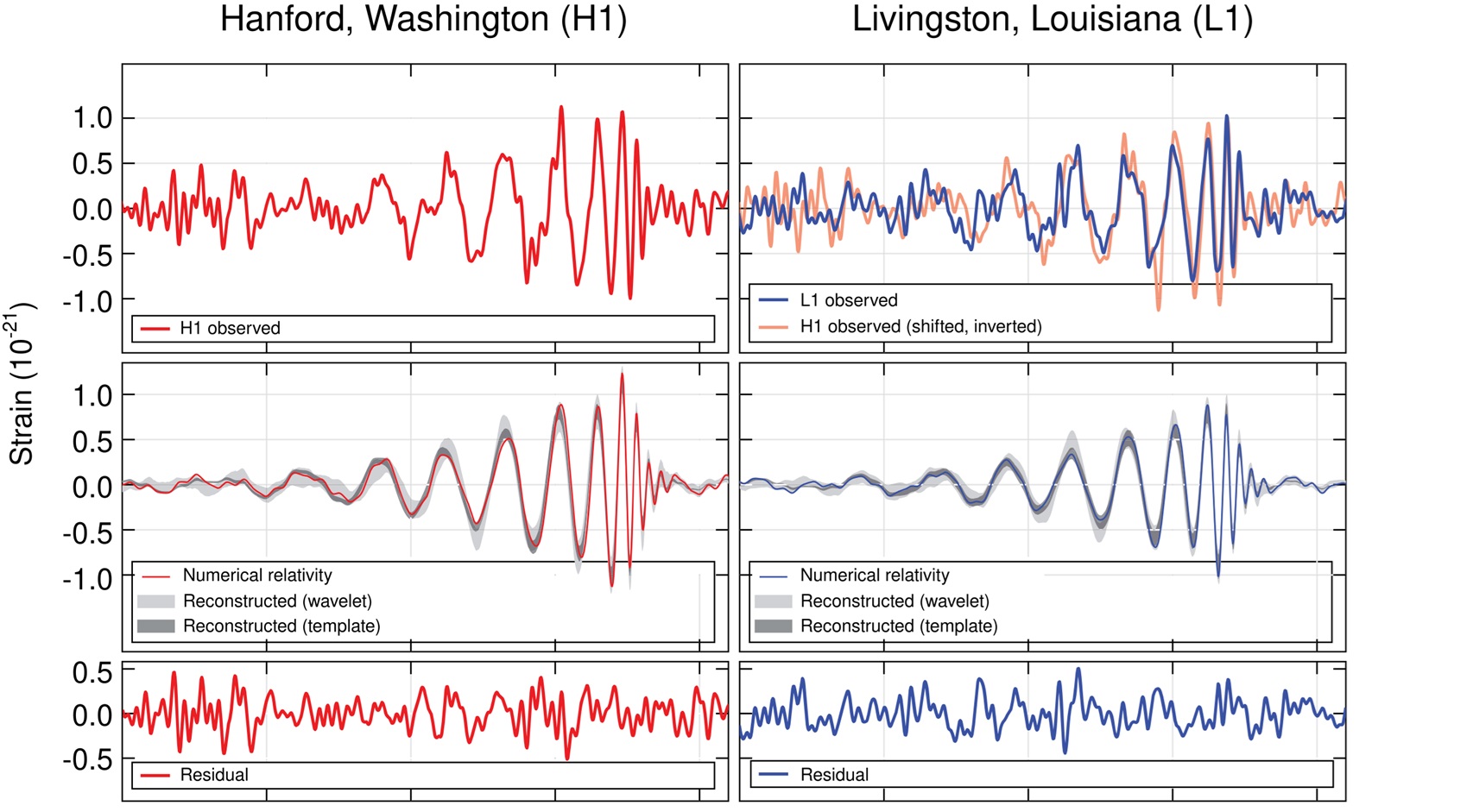}
\caption{Time-domain strain data for the first directly detected gravitational wave event, GW150914, as measured by the LIGO Hanford (H1) and Livingston (L1) detectors. The top panels show the observed detector strains, with the Livingston trace time-shifted and inverted to account for the light-travel delay and detector orientation. The middle panels compare the data with numerical-relativity waveforms and reconstructed signals obtained via wavelet and template methods, demonstrating the consistency of the inspiral–merger–ringdown signal with a binary black-hole coalescence. The bottom panels show the residuals after subtraction of the best-fit waveform.
Image credit: LIGO Scientific Collaboration and Virgo Collaboration \citep{Abbott:2016blz}.}  \label{fig:LIGO-detection}
\end{figure}

The idea of detecting gravitational waves using laser interferometry dates back to the early 1960s, ultimately culminating in the construction of the LIGO and Virgo observatories, whose landmark detections opened an entirely new observational window onto the Universe \citep{Abbott:2016blz} (see \cref{fig:LIGO-detection}). These achievements established laser interferometers as the premier tools for probing gravitational waves in the audio-band and demonstrated, for the first time, the direct measurability of spacetime dynamics. To set the stage for the methods reviewed in this section, let us begin with a brief overview of Michelson interferometers and the physical principles that govern their response to passing gravitational waves. This foundational framework will allow us to understand how modern detectors extract minute strain signals from noise and how their design has evolved into the sophisticated global network operating today.  

\bigskip

\begin{figure}[ht]
\centering
\includegraphics[width=0.8\textwidth]{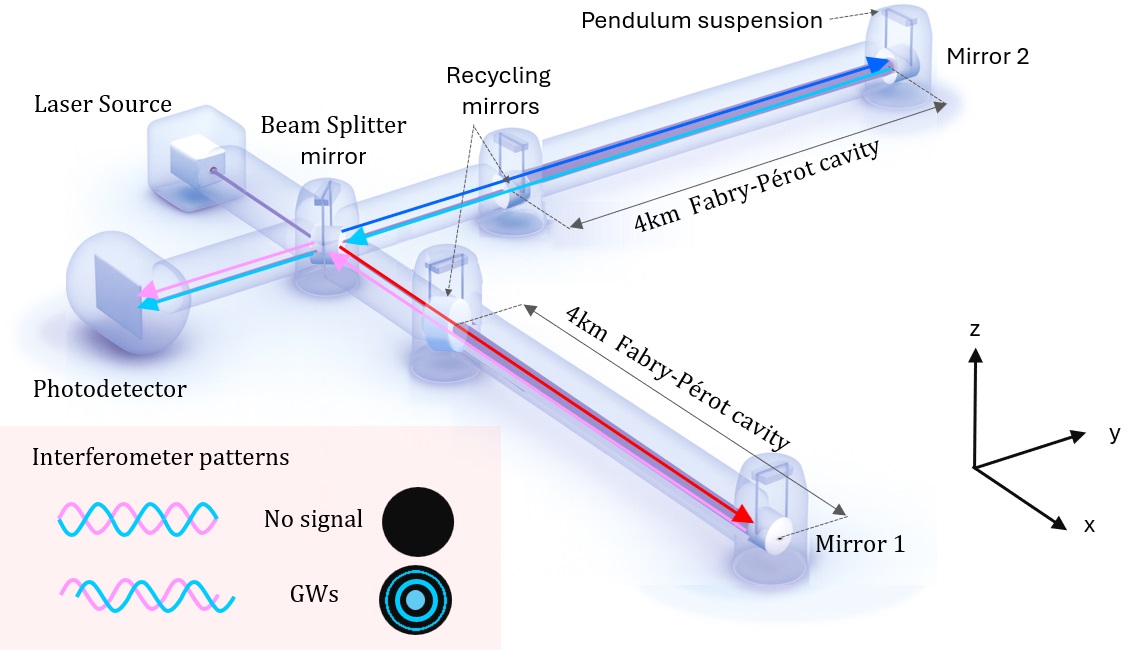}
\caption{The LIGO detector operates as a Michelson interferometer: passing gravitational waves alternately stretch one arm and compress the other, inducing a phase shift in the light at the beam splitter that produces an interference pattern at the output.} \label{fig:LIGO}
\end{figure}

\begin{tcolorbox}[colback=gray!10, colframe=gray!10, boxrule=0pt,
                  enhanced, breakable, halign=justify]
\subsubsection*{Michelson Interferometer} Consider a laser emitting monochromatic light in direction $\hat{x}$ with an electric field
\begin{align}
\vec{E}_{\text{in} ,1}(t,\boldsymbol{x}) =  \begin{pmatrix} E_{\text{in} ,1} \\ 0 \end{pmatrix} \quad \text{with} \quad  E_{\text{in} ,1} = E_0  e^{-i(\omega_L t - \boldsymbol{k}_L \cdot \boldsymbol{x})}, 
\end{align}
where $\omega_L$ is the laser frequency, $\boldsymbol{k}_L$ its wavevector, and $E_0$ the amplitude. The laser beam is directed onto a beam splitter, which divides it into two components of equal probability amplitude: one propagating along the $\hat{x}$-axis and the other along the orthogonal $\hat{y}$-axis. A lossless 50:50 beam splitter (half-reflecting, half-transmitting mirror) can be described by a $2 \times 2$ unitary matrix, the beam-splitter transfer matrix, as
\begin{equation}
\boldsymbol{U}= 
\frac{1}{\sqrt{2}}
\begin{pmatrix}
1 & ~i \\
i & ~1
\end{pmatrix}
.
\end{equation}
The factor of $i$ corresponds to a relative $\pi/2$ phase shift for the reflected beams. 
This convention ensures unitarity, i.e.  total intensity is conserved, and the beams interfere correctly. It thus splits the initial laser electric field into $\boldsymbol{E}_{\text{spl}, 1}$ as
\begin{align}
\boldsymbol{E}_{\text{spl}, 1} &=  \boldsymbol{U} \boldsymbol{E}_{\text{in}, 1}.
\end{align}
Next, each beam travels down its respective arm, reflects off a mirror at the far end, and returns to the beam splitter. We place the origin of the coordinate system at the beam splitter. The mirror terminating arm 1 is located at $(L_x, 0)$, while the mirror at the end of arm 2 is positioned at $(0, L_y)$. At this point we have
\begin{align}
E_{\text{ref},1} = -\frac{E_0}{\sqrt{2}} e^{-i(\omega_L t - 2k_L L_x)} , \quad 
E_{\text{ref},2}  = -i \frac{E_0}{\sqrt{2}}  e^{-i(\omega_L t - 2k_L L_y)} .
\end{align}
The beam splitter then recombines and re-splits the two beams. In the recombination stage, the returning beams impinge on opposite faces of the beam splitter compared to the forward pass.  
This geometric reversal flips the sign of the reflection coefficient for one arm, represented compactly by the diagonal surface matrix 
\be
\boldsymbol{S} = \boldsymbol{\sigma}_3 = \begin{pmatrix}1 & ~0 \\
0 & -1 \end{pmatrix}.
\ee
Hence, the total transformation for the returning fields is given by $\boldsymbol{U S}$, i.e.
\begin{align}
\boldsymbol{E}_{\text{out}} = \boldsymbol{U S} \, \boldsymbol{E}_{\text{ref}} = - \frac{E_0}{2} e^{-i\omega_L t} \,  \begin{pmatrix}
 e^{2i k_L L_y} + e^{2i k_L L_x} \\
i ( e^{2i k_L L_x} - e^{2i k_L L_y})
\end{pmatrix}.
\end{align}
Consequently, when the arm lengths are equal ($L_x = L_y$), one output port remains dark while the other is bright. We put the photodetector at the dark port. The total power recorded at the photodetector is proportional to  
\begin{align}
    |E_{\text{det}}|^2 = E_0^2 \, \sin^2\!\big(k_L (L_x - L_y)\big).
\end{align}  
Thus, any variation in the relative arm lengths induces a corresponding modulation in the detected power (see \cref{fig:LIGO}).

\end{tcolorbox}

\subsubsection*{Interaction of interferometer with gravitational waves}

We now turn to how this fundamental principle enables the detection of GWs.  
For clarity, let us consider a wave with only the plus polarization propagating along the $z$-axis, i.e. 
\begin{align}
ds^2 = -c^2 dt^2 + [1 + h_+(t,z)]dx^2 + [1 - h_+(t,z)]dy^2 + dz^2
,
\end{align}
where $h_+(t,z) = h_0 \cos(\omega_{_{gw}}t-k_{_{gw}}z)$. In the $z = 0$ plane of the interferometer, we therefore have
\bea
h_+(t) = h_0 \cos\omega_{_{gw}}t.
\eea
Photons propagate along null geodesics, $ds^2 = 0$. For light traveling in arm~1, the path is, to first order in $h_0$,  
\begin{align}
    dx = \pm c\, dt \, (1 - \tfrac{1}{2} h_+(t) ),
\end{align}  
where the plus sign corresponds to propagation from the beam splitter to the mirror, and the minus sign to the return trip.

In the TT gauge description, the coordinates of the mirrors and of the beam-splitter are not affected by the passage of the wave and the physical effect of the GW is manifested in the fact that it affects the propagation of light between these fixed points. 
 Consider a photon that
leaves the beam-splitter at a time $t_{\text{spl},1}$. It reaches the mirror, at the fixed
coordinate $x = L_x$, at a time $t_{\text{ref},1}$ obtained
\bea
L_x =  c \int^{t_{\text{ref},1}}_{t_{\text{spl},1}} dt [1 - \frac12 h_+(t)] =  c (t_{\text{ref},1}-t_{\text{spl},1}) - \frac{c}{2} \int_{t_{\text{spl},1}}^{t_{\text{ref},1}} h_{+}(t) dt.
\eea
Then the photon is reflected and reaches again the beam-splitter at a time $t_{\text{out},1}$ obtained integrating  with the minus sign
\bea
L_x =   c (t_{\text{out},1}-t_{\text{ref},1}) - \frac{c}{2} \int_{t_{\text{ref},1}}^{t_{\text{out},1}} h_{+}(t) dt.
\eea
As a result, the arrival time $t_{\text{out},1}$ after a round trip in the arm-1 is 
\bea
t_{\text{out},1} - t_{\text{spl},1} = \frac12 \int^{t_{\text{out},1}}_{t_{\text{spl},1}} h_{+}(t) dt + \frac{2L_x}{c}.
\eea
Given that the GW amplitude is small, we can solve perturbatively for $t_{\text{out},1}$ as
\bea
t_{\text{out},1} - t_{\text{spl},1} \simeq  \frac12 \int^{t_{\text{spl},1}+\frac{2L_x}{c}}_{t_{\text{spl},1}} h_{+}(t) dt + \frac{2L_x}{c}, 
\eea
which gives  
\bea
t_{\text{out},1} - t_{\text{spl},1} \simeq   \frac{L_x}{c} \frac{\sin(\omega_{_{gw}}L_x/c)}{\omega_{_{gw}}L_x/c} h_0 \cos\!\left[\omega_{_{gw}}(t_{\text{spl},1}+L_x/c)\right] + \frac{2L_x}{c}.
\eea
Since LIGO is fundamentally a time-delay interferometer, it is necessary to determine the corresponding travel time for the second arm. Doing the same analysis for arm-2, we find
\bea
t_{\text{out},2} - t_{\text{spl},2} \simeq -  \frac{L_y}{c} \frac{\sin(\omega_{_{gw}}L_y/c)}{\omega_{_{gw}}L_y/c} h_0 \cos\!\left[\omega_{_{gw}}(t_{\text{spl},2}+L_y/c)\right] + \frac{2L_y}{c}.
\eea
Now, we set 
\begin{align}
    L=L_x=L_y, \quad t=t_{\text{out},1}=t_{\text{out},2},
\end{align}
and find the total phase difference induced by GWs
in the Michelson interferometer as
\begin{align}
    \Delta\phi = \frac{2\omega_{L} L}{c} \frac{\sin(\omega_{_{gw}}L/c)}{\omega_{_{gw}}L/c} h(t-L/c).
\end{align}
Therefore, the total power observed at the photodetector,  $P \sim \lvert E_{\text{det}}\rvert^2$, is modulated
by the GW signal. 

\begin{figure}[ht]
\centering
\includegraphics[width=\textwidth]{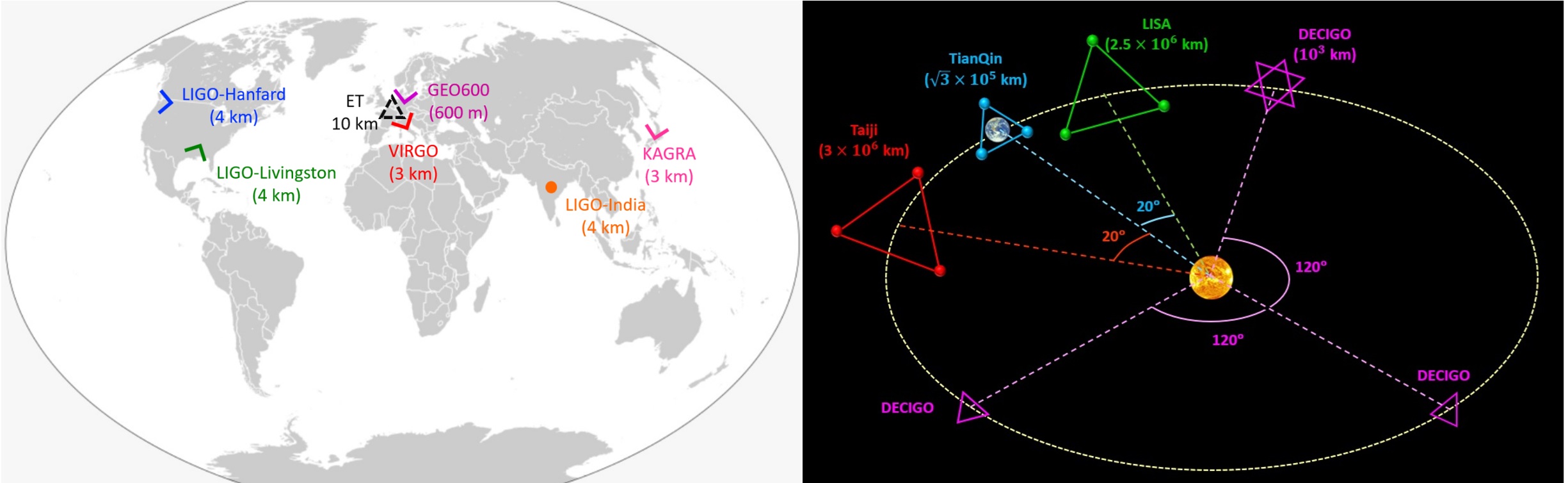}
\caption{Network of laser interferometer gravitational wave detectors on Earth (left) and in space (right). The terrestrial detectors are L-shaped Michelson interferometers with kilometre-scale arms, including LIGO, which began its first observing run in 2015, Virgo, which joined in 2017~\citep{Abbott:2016blz}, and KAGRA, operational since 2020~\citep{KAGRA:2018plz}, together establishing the first global ground-based gravitational wave network. Forthcoming facilities, LIGO–India, expected to begin observations in the early 2030s~\citep{Saleem:2021iwi}, and the Einstein Telescope, planned for the mid-2030s to 2040s~\citep{ET:2025xjr}, will further extend this network’s reach. In contrast, proposed space-based interferometers will form triangular constellations with arm lengths a million times larger, of order $10^6$ km, dramatically enhancing sensitivity to low-frequency gravitational waves. The LISA mission is scheduled for launch in 2035 \citep{Colpi:2024lisa}, while several complementary proposals, Taiji~\citep{Hu:2017mde, Ruan:2020smc}, TianQin~\citep{TianQin:2020hid}, and DECIGO~\citep{Kawamura:2020pcg}, aim to explore similar frequency bands with distinct orbital configurations and designs. Image credit: CERN Courier Feature (Aug 2023) \citep{MaleknejadRompineve2023}. }  \label{fig:laser-}
\end{figure}

For a simple Michelson interferometer, finite-arm effects are encoded in the transfer function
$\frac{\sin(\omega_{\mathrm{gw}}L/c)}{\omega_{\mathrm{gw}}L/c}$. For $L=4~\mathrm{km}$, this response is nearly flat throughout the main LIGO band and first vanishes at
$f=c/(2L)\simeq 3.75\times10^4~\mathrm{Hz}$.
LIGO therefore operates predominantly in the long-wavelength regime $(\omega_{\mathrm{gw}}L/c\ll1)$, 
where the interferometer response is nearly flat, and achieves high sensitivity through 
Fabry--Pérot arm cavities, which increase the photon storage time by a factor of $\mathcal{O}(10^{2})$ (see \cref{fig:laser-}).  
The effective optical path length is thus enhanced to $\sim 10^{3}\,\mathrm{km}$ while keeping the physical arms at $4~\mathrm{km}$, 
allowing the detector to probe frequencies across the $10$--$10^{3}~\mathrm{Hz}$ observational band \citep{Saulson:1994,Abbott:2016blz}.

\begin{figure}[h]
\centering
%%\includegraphics[width=\textwidth]{Laser-sensitivity.jpg}
%% FS: use original vector fig from arXiv: 
\includegraphics[width=\textwidth]{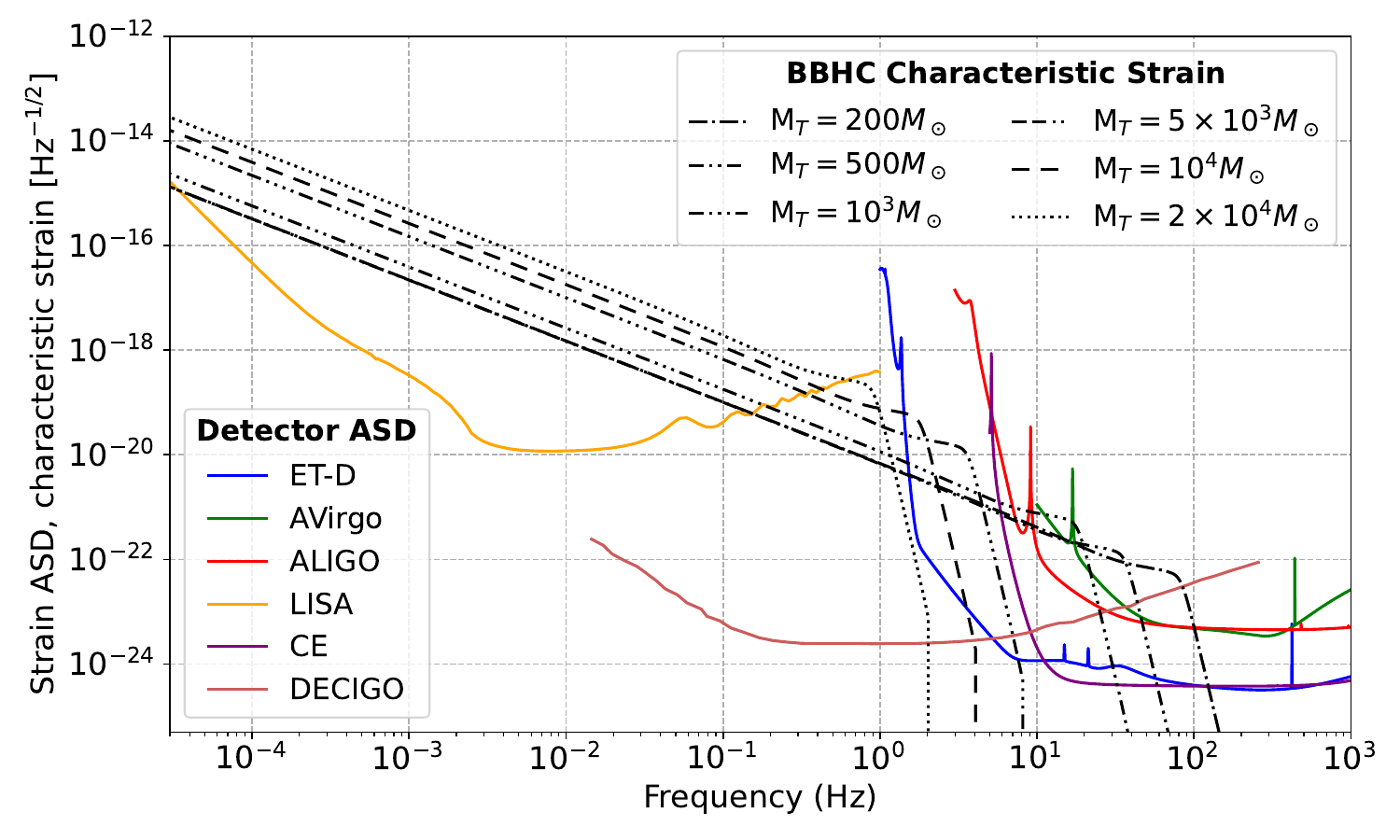}
\caption{Sensitivity curves of current and future laser-interferometer 
gravitational wave detectors (solid colored lines), including ET-D, 
Advanced Virgo, Advanced LIGO, LISA, Cosmic Explorer (CE), and DECIGO.
The dashed black curves correspond to the characteristic strain amplitude 
and associated frequencies of intermediate-mass black-hole binaries with 
total masses between $2\times 10^2$ to  $2\times10^{4}\,M_\odot$, 
assuming a luminosity distance of $d_L = 1\,\mathrm{Gpc}$. These trajectories 
illustrate how binaries of different masses sweep through the sensitivity 
bands of detectors across multiple frequency decades.
Image credit: The Science of the Einstein Telescope \citep{ET:2025xjr}.}
\label{fig:laser-sens}
\end{figure}

Advanced LIGO can detect compact binaries with total masses up to $\mathcal{O}(10^2)\,M_{\odot}$; however, more massive systems, such as supermassive black hole binaries, radiate at lower frequencies, beyond the reach of ground-based detectors.  
To extend sensitivity, next-generation observatories will employ advanced technologies to form a global network of terrestrial interferometers, probing gravitational waves across the ground-accessible frequency band ($\sim 1~\mathrm{Hz}$–$10~\mathrm{kHz}$).  
At still lower frequencies, they will be complemented by space-based missions, completing a multi-band gravitational wave detection network spanning from $10^{-4}~\mathrm{Hz}$ to the kilohertz regime. \cref{fig:laser-} illustrates the global network of ground-based laser-interferometer detectors, while their corresponding strain sensitivities are shown in \cref{fig:laser-sens}. Atom interferometry offers a complementary route to established laser-interferometric techniques for GW detection \citep{Graham2013_PRL, Dimopoulos2008_PRD, Hogan2011_GRG}. On Earth, several long-baseline cold-atom experiments are under construction or commissioning \citep{MIGA, ZAIGA, MAGIS}, while large-scale facilities such as ELGAR \citep{ELGAR} and AION \citep{AION, AION:2025nio} are being developed. In space, ambitious mission concepts such as AEDGE \citep{AEDGE} aim to extend atom-interferometric sensitivity toward lower GW frequencies. Together, these terrestrial and orbital instruments target the mid-frequency GW band, approximately $10^{-2}$ to 1 Hz.

%%%%%%%%%%%%%%%%%%%%%%%%%%%%%%%%%%%%%4-HFGW

\subsection{Photonic qumodes: a cutting-edge quantum platform for GW detection}\label{sec:cavity}

Gravitational waves have opened an unprecedented observational window onto the universe, yet this window currently extends only up to frequencies of order 10~kHz. The detection of higher-frequency gravitational waves (HFGWs) remains an open frontier and an area of active research and development. Such HFGWs may originate from a wide range of cosmological and astrophysical processes, and could even be generated in laboratory settings through the conversion of high-intensity laser fields (see \citealt{Aggarwal:2025noe} and the references therein). Their observation would provide access to rich new information about the early universe, particle physics, and potentially the quantum nature of gravity itself. Detecting gravitational waves at frequencies above that regime, however, requires experimental strategies fundamentally different from those employed by large-scale laser interferometers. As the GW frequency increases, the corresponding wavelength becomes much shorter than any practical detector baseline, leading to a rapid degradation of the strain sensitivity of conventional interferometric techniques. This limitation has motivated the exploration of alternative detection concepts that rely not on large spatial separations, but on resonant enhancement, electromagnetic interactions, and quantum-coherent effects. 

Joseph Weber pioneered the first experimental searches for gravitational waves using resonant bar detectors operating in the kilohertz regime \citep{Weber:1960zz}, thereby initiating the exploration of gravitational waves at the highest frequencies accessible at the time and laying the conceptual foundations for modern high-frequency GW detection. The main experimental approaches currently pursued for HFGW detection include high-frequency extensions of laser interferometers \citep{2017PhRvD..95f3002C,Patra:2024eke}, resonant mass detectors \citep{Vinante:2006uk}, modern resonant mechanical sensors \citep{Tobar:2023ksi, Aggarwal:2020umq,Domcke:2024mfu}, electromagnetic oscillators such as microwave cavities \citep{Berlin:2021txa,3094368}, resonant lumped-element LC detectors such as ABRACADABRA and DMRadio \citep{Domcke:2022LCGW,Domcke:2024LCGW,Pappas:2026ABRAGW}, photon (re-)generation experiments \citep{OSQAR:2015qdv,Albrecht:2020ntd,CAST:2017uph,IAXO:2020wwp,IAXO:2019mpb,Beacham:2019nyx,Kharzeev:2025lyu}, and other electromagnetic conversion–based schemes. Comprehensive reviews, sensitivity estimates, and discussions of experimental implementations can be found in \cite{Aggarwal:2025noe} and the references therein. Notably, ABRACADABRA-10\,cm has recently carried out the first dedicated HFGW search using a modified axion detector \citep{Pappas:2026ABRAGW}. In the following, we focus on a brief overview of photon (re-)generation experiments and highlight recent advances in this direction.

\bigskip

\begin{tcolorbox}[colback=gray!10, colframe=gray!10, boxrule=0pt,
                  enhanced, breakable, halign=justify]

\subsubsection*{Graviton–Photon Transitions: The Gertsenshtein Effect}
{\centering
\includegraphics[width=0.4\textwidth]{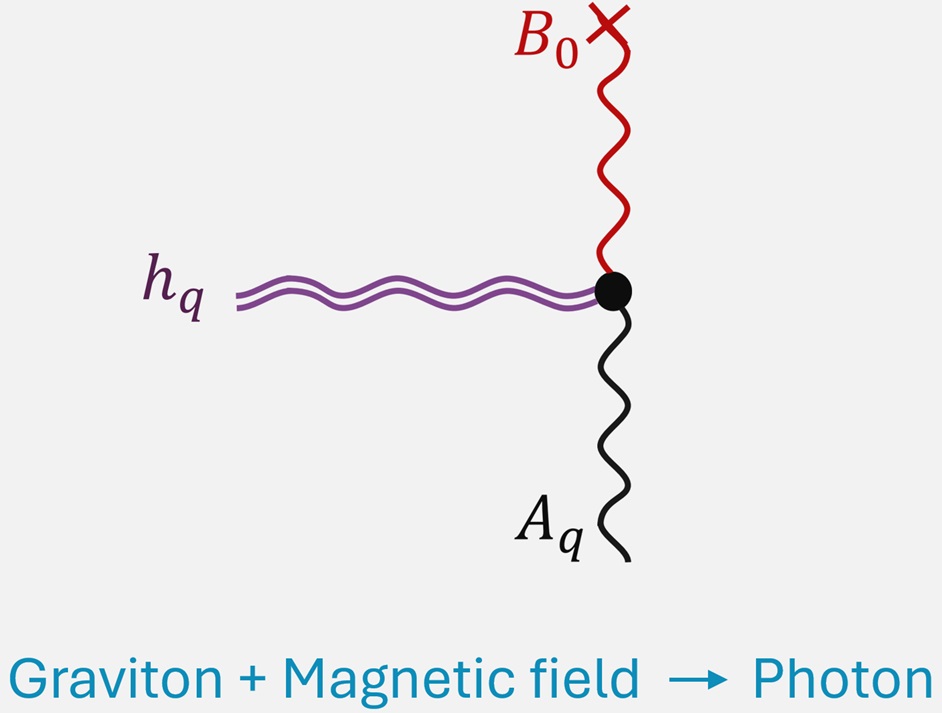}
\par}\vspace{0pt}

In a seminal paper, \cite{Gertsenshtein} pointed out that electromagnetic and gravitational waves can interconvert in the presence of a static magnetic field, i.e. the Gertsenshtein effect. This observation provides an early and concrete example of the coupling between electromagnetism and gravity at the level of field theory and forms the basis of many modern proposals for high-frequency gravitational wave detection. This effect can be understood from the Maxwell--Einstein action by considering electromagnetic fields propagating in a weakly curved spacetime. We write the spacetime metric as a small perturbation around flat space,
\be
g_{\mu\nu} = \eta_{\mu\nu} + \gamma_{\mu\nu}, \qquad |\gamma_{\mu\nu}| \ll 1 ,
\nonumber
\ee
where $\gamma_{\mu\nu}$ represents the gravitational perturbation. To leading order in $\gamma_{\mu\nu}$, the interaction Lagrangian density between the gravitational field and any field in nature is
\be
\mathcal{L}_\mathrm{int} = \frac12 \gamma_{\mu\nu} T^{\mu\nu},
\ee
where $T^{\mu\nu}$ is the energy--momentum tensor of a generic field.
For the electromagnetic field, this interaction Hamiltonian density can be written explicitly as
\begin{equation}\label{eq:L-int}
\mathcal{H}_\mathrm{int}
= -\frac{1}{2}\,\eta^{\lambda\sigma}
\left(\gamma^{\mu\nu} - \frac{1}{2}\,\gamma\,\eta^{\mu\nu}\right)
F_{\mu\lambda} F_{\nu\sigma},
\end{equation}
where \(F_{\mu\nu}\) is the electromagnetic field-strength tensor and \(\gamma \equiv \gamma^{\mu}{}_{\mu}\) is the trace of the metric perturbation.  Note here we adopt natural units and set \(c=\hbar=\epsilon_0=1\). The gravitational interaction \cref{eq:L-int} reveals how spacetime perturbations directly influence electromagnetic dynamics, opening a channel for graviton–photon conversion in a background field.
The transition amplitude from an initial state $\lvert i \rangle$ to a different final state $\lvert f \rangle$ is given by the S-matrix element
\begin{align}
 \mathcal{A}_{fi} = \langle f \lvert T \exp\!\left[-i \int d^4x \, \mathcal{H}_{\text{int}} \right] \rvert i \rangle,
 \label{eq:S-matrix}
\end{align}
where $T$ is the time-ordering operator. 

\end{tcolorbox}

Photon regeneration is a leading approach to probe HFGWs, which is based on this mechanism \citep{Ejlli:2019bqj}. In such experiments, GWs traverse a photon-shielded cavity permeated by a strong magnetic field, allowing for resonant graviton-to-photon conversion. Consider a cylindrical electromagnetic cavity of length $L$ and cross-sectional area $A$. We assume that the cavity walls are made of a perfect conductor. A uniform, static magnetic field is turned on inside the cavity, oriented along the $\hat{z}$ direction, i.e.
\be
\mathbf{B}(t,\mathbf{x}) = B_0\,\hat{z}.
\ee
At this stage, it is necessary to fix the gauge. The electromagnetic fields measured in the laboratory are most naturally described in the Detector Proper (DP) frame, i.e., the local frame adapted to the detector. By contrast, theoretical descriptions of gravitational waves often employ the TT gauge, in which freely falling test masses remain at fixed coordinate positions while the effect of the gravitational wave is entirely encoded in oscillations of the metric. In this work, we adopt the TT gauge, defined by
$h_{00} = h_{0i} = \partial_i h_{ij} = h^i{}_i=0$, where $i,j = 1,2,3$. We further assume that the background magnetic field $\mathbf{B}_0$ is static in the TT frame. Although in general, the transformation between the DP and TT frames in the presence of gravitational waves requires careful treatment, this approximation is well justified in the high-frequency regime $\omega L \gg v_s$~\citep{Ratzinger:2024spd}. Here $v_s$ denotes the sound velocity in the detector material, which is typically small, $v_s \lesssim 10^{-4}$.

\begin{figure}[H]
\centering
\includegraphics[width=0.6\textwidth]{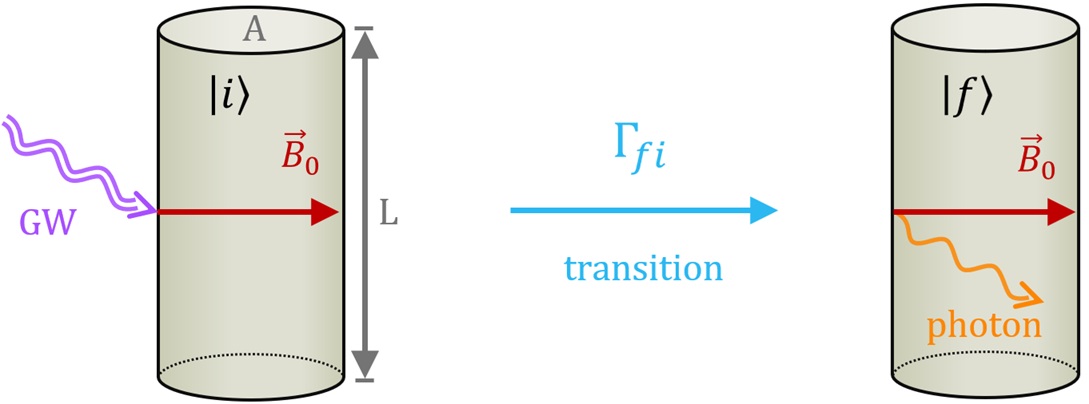} 
\caption{Illustration of a resonant cavity prepared in the electromagnetic state $\ket{i}$, which transitions to $\ket{f}$ as a passing gravitational wave induces graviton–photon conversion in a strong magnetic field.}\label{fig:cavity}
\end{figure} 

We begin by considering the interaction of the cavity with a monochromatic gravitational wave of frequency $\omega$, propagating in the direction $\hat{\bq}$ (with $\bq = \omega \hat{\bq}$) as  
\begin{align}
h_{ij}(t,\bx) = \sum_{\sigma}  \, e^{\sigma}_{ij}(\hat{\bq}) \,
 h_{\omega} e^{iqx} + h.c.,
\end{align}
the interaction Hamiltonian density can be written as
\begin{align}
\hat{\mathcal{H}}_\text{int} = -B_0 \sin\theta_{\!\bq}  \sum_{\sigma}  \delta \hat{B}_{i}(x) \, e^{\sigma}_{i}(\hat{\bq})  \,   h_{\omega}  e^{iqx},
\label{eq:H-int-}
\end{align}
where $\delta \hat{B}_{i}(x)$ denotes the magnetic-field operator associated with fluctuations of the electromagnetic field, and  
\be
e^{\pm}_{j}(\hat{\bq}) = \frac{1}{\sqrt{2}}(\cos\theta_{\!\bq} \cos\phi_{\!\bq} \mp i \sin \phi_{\!\bq}, \cos\theta_{\!\bq} \sin\phi_{\!\bq} \pm i \cos\phi_{\!\bq}, -\sin\theta_{\!\bq}),
\ee
in Cartesian coordinates. A cosmological stochastic background of unpolarized gravitational waves can be modeled as a superposition of plane waves with different frequencies and propagation directions as
\begin{align}
h_{ij}(x) = \sum_{\sigma=\pm} \int d\ln \omega \, e^{\sigma}_{ij}(\hat{\bq}) \, \int \frac{d^2\hat{\bq}}{2\pi} \,
 h_{\omega} e^{iqx} + h.c.,
 \label{eq:h-BG--}
\end{align}
 where $\bq= \omega \hat{\bq}$, $h_{\omega}$ is the mode function, and $\langle h_{ij}(t,\bx)h_{ij}(t,\bx) \rangle =2\int d\ln\omega \lvert h_\omega \rvert^2$. In this stochastic setting, the interaction Hamiltonian is obtained by superposing the monochromatic contributions \cref{eq:H-int-} over all frequencies and directions, with the cavity response effectively integrating over the accessible bandwidth.

Let us focus on electromagnetic transitions occurring inside the cavity in which a single photon is created, thereby increasing the occupation number of the mode at frequency $\omega$ by one. For normalized initial and final states $\ket{i}$ and $\ket{f}$, the transition rate for the process $\ket{i}\to\ket{f}$, evaluated to first order in perturbation theory with respect to the GW amplitude $h_\omega$, is given by
\begin{equation}
\Gamma_{fi} = \lim_{T\rightarrow \infty} \, \frac{1}{T} \, \Big\vert \int d^4x \, \bra{f}\mathcal{H}_\text{int}(x) \ket{i} \Big\vert^2.
\label{eq:Gammafi-}
\end{equation}
To proceed further and evaluate this expression explicitly, it is now necessary to specify the form of the initial state $\ket{i}$.  Existing and proposed experiments relevant to graviton--photon conversion include optical light-shining-through-wall experiments such as OSQAR \citep{OSQAR:2015qdv} and ALPS \citep{Albrecht:2020ntd}, X-ray helioscopes such as CAST \citep{CAST:2017uph}, BabyIAXO and IAXO \citep{IAXO:2020wwp,IAXO:2019mpb}, and the proposed JURA setup \citep{Beacham:2019nyx}. For the signal mode relevant to the conversion process, these conventional setups correspond effectively to an initially unoccupied state, $\ket{i}=\ket{0}$.
Recently, a qualitatively new approach has been put forward under the name QuGrav \citep{Kharzeev:2025lyu}. This proposal exploits highly occupied $n$-photon states (qumodes) inside the cavity, allowing the graviton-induced transition rate to be parametrically enhanced by Bose--Einstein statistics.
By populating the cavity with a large number of photons, the conversion process benefits from stimulated emission, leading to a substantial enhancement relative to vacuum-based schemes.  These experiments detect the photon count rate $\Gamma$ within an energy band $\Delta f=(f_f-f_i)$, with a single photon detection efficiency $\varepsilon$, and over a cross-sectional area $A$. The background noise is quantified in terms of the dark count rate in that frequency bin, i.e., $ \Gamma_D(f)$. It is the rate at which the detector falsely registers a photon even when no real photon is present.  In \cref{tab:experiments} we summarize these experiments and proposals.
 
\begin{table}[h]
\caption{Experimental parameters of electromagnetic conversion setups relevant to HFGW searches. All experiments assume a vacuum cavity state $\ket{0}$ except \textit{QuGrav}, which considers a general Fock state $\ket{n}$. Here $\Gamma_D(f)$ is the dark count rate in a frequency bin of width $\Delta f$, and $\varepsilon$ is the single-photon detection efficiency. Italicized entries indicate proposals.}
\label{tab:experiments}
\centering
%%\small
\begin{tabular}{llccccccc}
\toprule
 & \boldmath$f$ [Hz] & \boldmath$B_0$ [T] & \boldmath$L$ [m] & \boldmath$A$ [m$^2$] & \boldmath$\Gamma\!_{{\!D}}$ [mHz] & \boldmath$\varepsilon$ & Cavity state & Refs. \\
\midrule
OSQAR II       & $10^{15}$   & 9    & 14.3 & $5 \times 10^{-4}$  & $1.14$ & 0.9 &
\multirow{5}{*}{vacuum $\ket{0}$} & (1)\\ %%\citet{OSQAR:2015qdv} \\
ALPS II        & $10^{15}$   & 5.3  & 106  & $10^{-3}$           & $10^{-3}$ & 0.75 &  & (2)\\ %%\citet{Albrecht:2020ntd} \\
\textit{JURA}  & $10^{15}$   & 13   & 960  & $8 \times 10^{-3}$  & $10^{-6}$ & 1 &  & (3)\\ %%\citet{Beacham:2019nyx} \\
CAST           & $10^{18}$   & 9    & 9.26 & $2.9 \times 10^{-3}$ & $0.15$ & 0.7 &  & (4)\\ %%\citet{CAST:2017uph} \\
IAXO           & $10^{18}$   & 2.5  & 20   & 3.08                & $0.1$ & 1 &  & (5)\\ %%\citet{IAXO:2020wwp, IAXO:2019mpb} \\
\midrule
\textit{QuGrav} & $10^{10}$   & 10   & 1    & 1                   & $10^{-3}$ & 1 & Qumode $\ket{n}$ & (6)\\ %%\citet{Kharzeev:2025lyu} \\
\bottomrule
\end{tabular}
Refs.: (1) \citet{OSQAR:2015qdv}, (2) \citet{Albrecht:2020ntd}, (3) \citet{Beacham:2019nyx}, (4) \citet{CAST:2017uph}, (5) \citet{IAXO:2020wwp, IAXO:2019mpb}, (6) \citet{Kharzeev:2025lyu}
\end{table}

 %Some of these instruments operate well below the standard quantum limit. Ongoing research and development efforts aim to design next-generation instruments with enhanced power and sensitivity. That includes using stronger magnetic fields, bigger and higher quality cavities, and so on. For a recent review that covers these developments comprehensively, see \citet{Aggarwal:2025noe} and references therein (more and more ..... including \citet{Bringmann:2023gba,Domcke:2024eti}).

\bigskip

\begin{tcolorbox}[colback=gray!10, colframe=gray!10, boxrule=0pt,
                  enhanced, breakable, halign=justify]

\subsubsection*{From Qumodes to QuGrav}
 {{\centering
\includegraphics[width=0.5\textwidth]{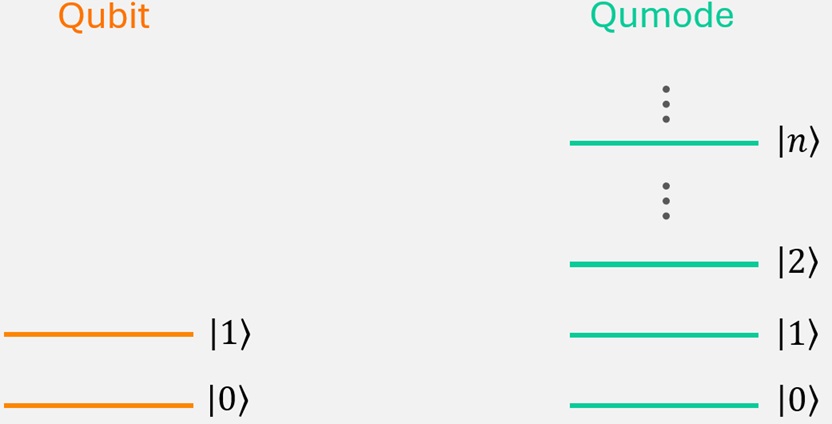}
\par}\vspace{0pt}}

In the context of quantum information science, qubits constitute the canonical building blocks for storing and processing quantum information, encoding logical states in arbitrary superpositions of the basis states $\ket{0}$ and $\ket{1}$. While this abstraction underlies much of quantum computation, many advanced applications benefit from moving beyond strictly two-level systems. Within this broader framework, quantum harmonic oscillators, commonly referred to as qumodes, provide a natural extension of the qubit paradigm. By replacing or supplementing qubits with bosonic modes, qumodes offer a countably infinite-dimensional Hilbert space spanned by Fock states $\ket{n}$, with $n=0,1,2,\ldots$ \citep{Blais:2020wjs,Stavenger:2022wzz,Araz:2024kkg}. This expanded state space enables richer encoding strategies and greater flexibility in the manipulation of quantum information, positioning qumodes as a powerful platform for continuous-variable quantum computation and cavity-based quantum technologies.

Recent technological breakthroughs have significantly expanded our ability to manipulate and measure complex quantum states of light. In particular, experimental platforms have begun to realize true 
$n$-photon qumodes with high fidelity and control \citep{qu-exp1,qu-exp2}, enabling exploration of rich, high-dimensional Hilbert spaces beyond the qubit paradigm. A key milestone in this direction is the demonstration of photon-number-resolving measurements capable of distinguishing photon counts up to $n=100$ \citep{Cheng:2022wgy}, far exceeding the few-photon regime of traditional detectors.
\vskip 0.2cm 

Building on these developments, Kharzeev, Shalamberidze, and the present author highlighted the transformative potential of this major advance in quantum measurement capability for the detection of HFGWs \citep{Kharzeev:2025lyu}. In particular, they proposed initializing the cavity in an $n$-photon Fock state rather than the vacuum, demonstrating that for large occupation numbers, $n\sim 10^2$, the transition rate can be enhanced by up to two orders of magnitude, corresponding to about one order of magnitude improvement in strain sensitivity. More precisely, given that $\lvert \bra{n+1} b^{\dag} \ket{n}\rvert^2 \propto (n+1)$, we have $\Gamma_{fi}\propto (n+1)$. This ambitious proposal was termed QuGrav, reflecting its quantum-enhanced approach to gravitational wave detection.

\end{tcolorbox}

Following the QuGrav proposal, we consider an $n$-photon Fock state as the initial state of the cavity in order to explore genuine Bose enhancements associated with finite photon occupation. The formalism, however, remains completely general: by setting $n=0$, one straightforwardly recovers the standard results corresponding to vacuum-initialized cavities. As an initial configuration, we consider an $n$-photon state at a single resonant frequency $\omega$,
\begin{align}
\ket{i} =   \ket{Q_{n}} = \ket{n(\omega)}_{\!\gamma},
   \label{eq:Qu-1}
\end{align}
where $\ket{n(\omega)}_{\!\gamma} = (b^{\dagger}_{\omega})^{n}\ket{0}_{\!\gamma}/\sqrt{n!}$. Now, using the above initial state in \cref{eq:Gammafi-}, and considering a monochromatic gravitational wave with the same frequency in \cref{eq:H-int-}, we find the transition rate to $\ket{f}=\ket{Q_{n+1}}$ as
\begin{align}
\Gamma_{fi}  =   2 (n+1)\,   Q  V   B^2_0   h^2_{\omega_g}\, \sin^2\theta_q,
\label{eq:rate-}
\end{align}
where $Q$ is the quality factor of the cavity. Remarkably, due to Bose–Einstein statistics, once the frequency of the GW matches the frequency of the n-photon mode inside the cavity, the probability of graviton-to-photon conversion is enhanced by a factor of $n+1$. This process is the EM analogue to the stimulated absorption of gravitons in the massive quantum acoustic resonators, with the process being most effective at frequencies that match the resonator's natural frequency \citep{Tobar:2023ksi}.

To complement this minimal setup and to enhance sensitivity to the more challenging gravitational wave background, we now introduce a more elaborate qumode configuration that illustrates the theoretical reach and long-term potential of this approach for broadband detection. Specifically, we consider a multimode qumode state in which each resonant cavity mode is prepared in an $n$-photon Fock state, with mode frequencies given by $\omega_l = \pi l / L$, namely
\begin{align}
   \overline{\ket{Q_n}} = \bigotimes_{l=1}^{\nint{2\Delta f L/c}} \ket{n(\omega_l)}_{\!\gamma}.
   \label{eq:Qu-2}
\end{align}
Here $\ket{n(\omega_l)}_{\!\gamma} = (b^{\dagger}_{\omega_l})^{n}\ket{0}_{\!\gamma}/\sqrt{n!}$, and $\nint{2\Delta f L/c}$ denotes the number of resonant modes supported by the cavity within the finite frequency bandwidth $\Delta f$ (with $f=\omega/2\pi$).

Considering the stochastic gravitational wave background described in \cref{eq:h-BG--}, and assuming for simplicity that the spectrum is approximately flat across this frequency interval, the total transition rate is obtained by summing the contributions from all resonant modes, as
\begin{align}
\Gamma_{\text{tot}}
= 2\, (n+1)\, \Delta f\, L\, V\, B_0^{\,2}\, h_{\omega}^{\,2}\, \sin^{2}\theta_{\!\bq}.
\label{eq:Gamma-bb}
\end{align}
Here $\Delta f$ characterizes the effective detection bandwidth of the cavity. While a bright coherent drive also enhances the generation rate by the factor 
$n+1$, it simultaneously introduces amplitude and phase noise. For a quantitative comparison, see \citet{Kharzeev:2025lyu}. 

For completeness, let us also consider a single graviton propagating in direction $\hat{\bq}$ with frequency $\omega_g$, i.e. $\ket{1_{\omega_g}}_{\!g}$, as 
\begin{align}
h_{ij}(t,\bx) =  \sum_{\sigma=\pm} \, e^{\sigma}_{ij}(\hat{\bq}) \, h_{\omega_g}^{\sigma} \, \hat{a}_{\omega_g, \sigma} e^{-i\omega_g t+ i\bq.\bx} + h.c.,
\end{align}
where  $\hat{a}^{\dag}_{\omega,\sigma}$ and $\hat{a}_{\omega,\sigma}$ are the graviton creation and annihilation operators, $[\hat{a}_{\omega,\sigma},\hat{a}^{\dag}_{\omega',\sigma'}]=\delta_{\sigma\sigma'}\delta(\omega-\omega')$.
In this case, we set the initial and final states as
 \begin{align}
     \ket{i} = \ket{Q_n}\ket{1_{\omega_g}}_{\!g}, \quad  \ket{f} = \ket{Q_{n+1}}\ket{0}_{\!g},
 \end{align}
 which is the conversion of a single graviton to a new photon inside the cavity. The transition rate in this case is equal to \cref{eq:rate-}.

\bigskip

\begin{tcolorbox}[colback=gray!10, colframe=gray!10, boxrule=0pt,
                  enhanced, breakable, halign=justify]
\subsubsection*{Single Gravitons}
 {{\centering
\includegraphics[width=0.5\textwidth]{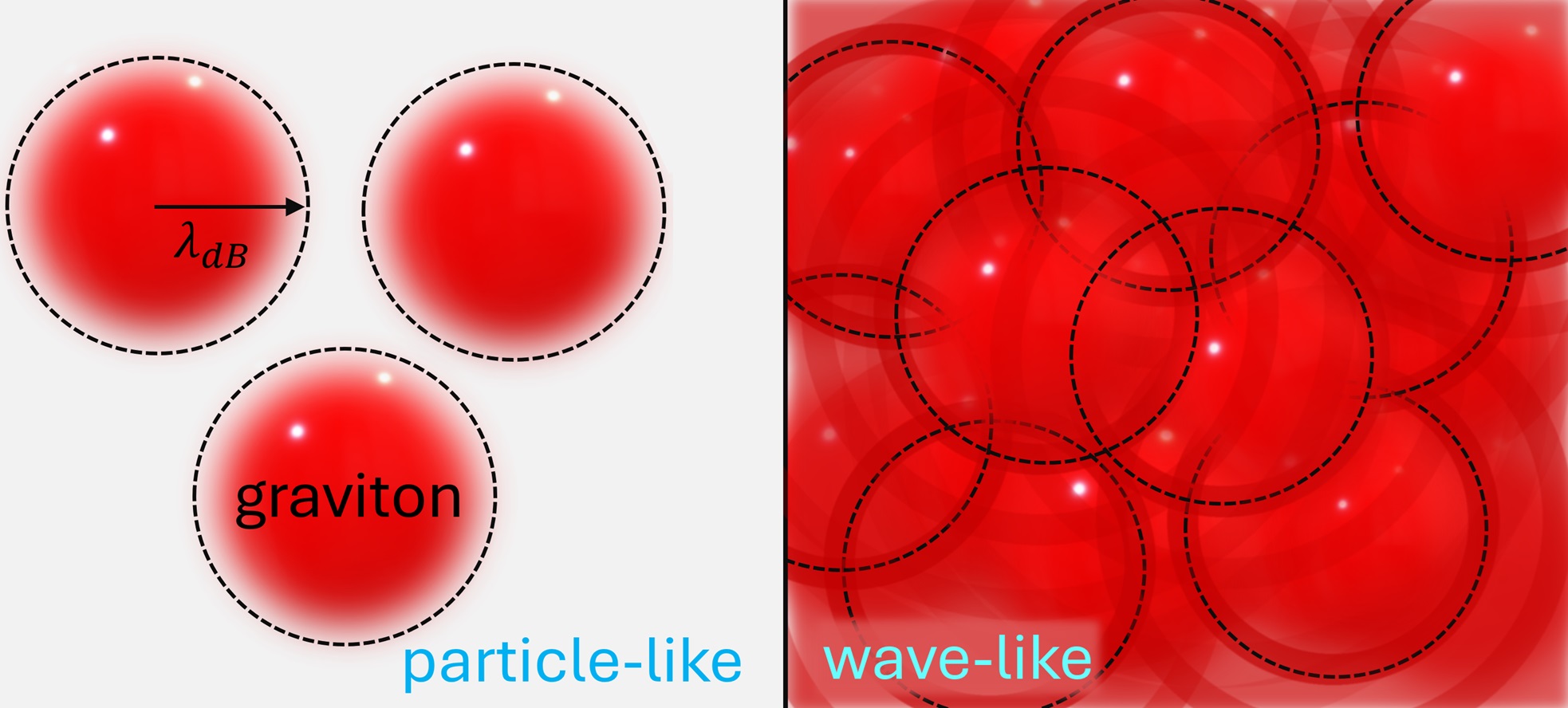}
\par}\vspace{0pt}}
 Modern detector architectures can detect single photons and even count them up to $n=100$ \citep{qu-exp1,qu-exp2}, but building a detector sensitive to single gravitons remains an extraordinary challenge. In fact, \cite{Dyson:2013hbl} and independently \cite{Rothman:2006fp} have argued that such detection is fundamentally impossible within certain experimental architectures, while remaining feasible in others. Recently, this question has been revisited in light of recent technological advances \citep{Carney:2023nzz,Tobar:2023ksi}.
 
  If gravity is quantized in the same manner as the other fundamental interactions, the number of gravitons contained within a single de Broglie volume may be estimated as \citep{Carney:2023nzz}
\begin{align}
N_{\text{grav}} = n_{\text{grav}} \lambda_{\text{dB}}^3 \simeq 2\times10^{35}
\left(\frac{h_{\omega}}{10^{-22}}\right)^2
\left(\frac{1\,\mathrm{kHz}}{f}\right)^2.
\end{align}
Thus a LIGO-scale signal with $h_{\omega}\sim10^{-22}$ at $f\sim1\,\mathrm{kHz}$ corresponds to $N_{\text{grav}}\sim2\times10^{35}$. Requiring less than a graviton per de Broglie volume, we arrive at 
\begin{align}
 h_{\omega} \leq 2\times 10^{-31}   \, \Big(\frac{f}{10^{3} \, \text{GHz}}\Big),
\end{align}
where below that limit, the gravitational field is more accurately described as a highly dilute graviton gas rather than a classical gravitational wave.

Naively, one might expect that observing a gravitational wave with an amplitude below this particle-like threshold would amount to the detection of a single graviton. However, establishing the genuinely quantum nature of gravitational radiation requires measurements that are far more demanding than the mere detection of individual gravitons. In QED, the photoelectric effect alone does not suffice to establish the photon as a quantum particle, since discrete detector clicks can also arise when quantum matter interacts with a classical electromagnetic field. More generally, observing individual quanta or scattering kinematics is not by itself sufficient to demonstrate that the incident field is quantized. An unambiguous distinction requires genuinely nonclassical observables, such as sub-Poissonian statistics or entanglement, which cannot be reproduced by a classical field ensemble \citep{Carney:2023nzz,Bao:2026quantumDM}. For gravitational radiation, this requires both the production of a genuinely nonclassical gravitational state and its detection with sufficiently high efficiency.

\end{tcolorbox}

We are now in a position to derive upper limits on the dimensionless
gravitational wave characteristic strain, $h(f)$, using the parameters
of current cavity experiments and proposed setups employing Qumode
technology. For each configuration, we estimate the sensitivity using the conservative requirement that the effective gravitationally induced transition rate satisfy
\be
\varepsilon\,\Gamma_{fi} > \Gamma_D.
\ee
More precisely, for a
gravitational wave signal to be detectable, the rate of real photon
production via gravitational transitions must exceed the detector’s
dark count rate, which accounts for false detection events. For resonant detectors with narrow bandwidth sensitivity, a monochromatic gravitational field, or a GW background with a sharply peaked spectrum, we find
 \begin{align}
      h_{\text{noise}}(f)  \approx \frac{1.4 \times 10^{-27}}{\sqrt{n+1}} \left(\frac{1 \, \text{m}}{L}\right) \left(\frac{10 \, \text{T}}{ B_0}\right) \sqrt{\left(\frac{1 \, \text{m}^2}{A }\right) \, \left(\frac{10^5}{\mathcal{F}}\right)}
    \sqrt{\frac{1}{\varepsilon}\left(\frac{\Gamma_D}{1 \, \mu\text{Hz}}\right) \left(\frac{1\, \text{GHz}}{f}\right)}.
\label{eq:SNR-nb}
\end{align}
where $\mathcal F$ is the cavity finesse.
 For the broadband case, using the theoretical benchmark Qumode state \cref{eq:Qu-2} with \cref{eq:Gamma-bb} for detectors, we have
\begin{align}
 h_{\text{noise}}(f)   \approx \frac{4.5 \times 10^{-25}}{\sqrt{(n+1)}}  \left(\frac{1 \text{m}}{L}\right) \left(\frac{10 \text{T}}{B_0 }\right) \sqrt{ \left(\frac{1 \, \text{GHz}}{\Delta f}\right) }   \sqrt{\frac{1}{\varepsilon}\left(\frac{\Gamma_D }{1 \, \mu\text{Hz}}\right)  \left(\frac{1 \, \text{m}^2}{A}\right)} .
  \label{eq:SNR-bb}
\end{align}  
The performance of a given detector operating in the Qumode
configuration can be directly contrasted with that of its standard
single-photon counterpart. In this case, the noise-equivalent
gravitational wave strains satisfy
\begin{align}
  \frac{h_{\text{noise}}^{\text{single}}}{h_{\text{noise}}^{\text{Qu}}}
  = (n+1)^{\frac{1}{2}},
\end{align}
which demonstrates that the Qumode implementation achieves a parametric
reduction in the effective strain noise as the photon occupation number
$n$ increases.

\begin{figure}[h!]
\centering
\includegraphics[width=\textwidth]{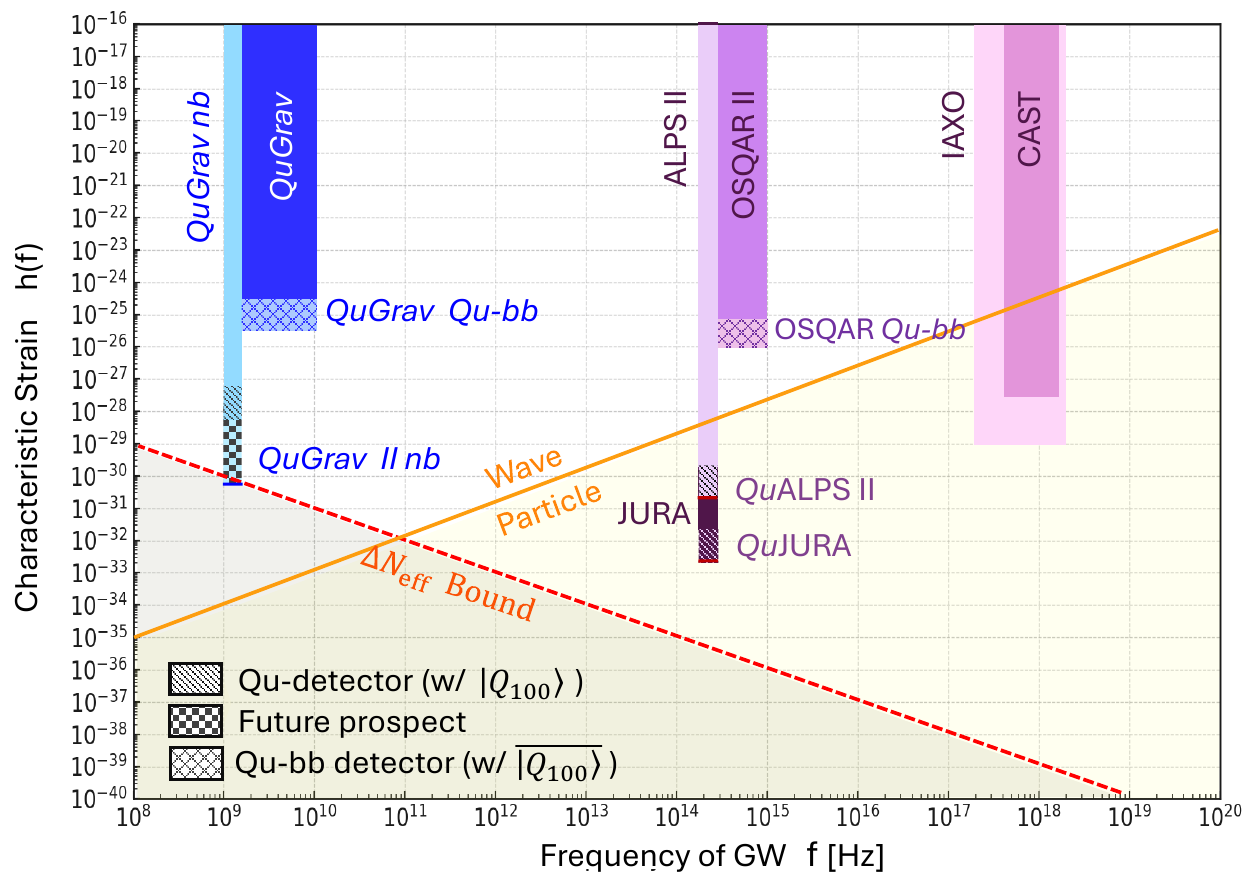} % Right plot
    \caption{Strain sensitivity of photon-regeneration--based HFGW detectors.
The region below the orange curve corresponds to the dilute-graviton regime, $n_\mathrm{grav}\lambda_\mathrm{dB}^3<1$, assuming the gravitational field is quantized. The dashed gray curve indicates the cosmological constraint
from $\Delta N_{\text{eff}}$. Hatched regions show the projected
sensitivity gains of existing experiments and proposed setups employing
a 100-photon Qumode ($\ket{Q_{100}}$, see \cref{eq:Qu-1}) within the
cavity. The checkered (black--blue) region represents additional
improvements achievable through near-term advances in cavity technology. Finally, the crossed regions illustrate the enhanced
broadband sensitivity attainable with the futuristic broadband
100-photon Qumode $\overline{\ket{Q_{100}}}$ introduced in
\cref{eq:Qu-2}. Image credit: \citet{Kharzeev:2025lyu}.}
   \label{fig:detector}
\end{figure}

In \cref{fig:detector}, we present the strain sensitivity of
photon-regeneration--based HFGW detectors, both in their conventional
configuration and with enhancements provided by qumode technology.
Remarkably, at optical frequencies, the implementation of a
$100$-photon qumode enables ALPS~II to probe well below the single-graviton
threshold, entering the dilute-graviton regime. At microwave
frequencies, the same approach has the potential to reach sensitivities
surpassing the Big Bang Nucleosynthesis (BBN) bound, opening a new window
on cosmologically motivated high-frequency gravitational wave signals.

%%%%%%%%%%%%%%%%%%%%%%%%%%%%%%%%%
% 
 \section{Physical mechanisms driving gravitational radiation}\label{sec:5}

As discussed earlier, gravitational waves arise as tensor perturbations of the 
spacetime metric, satisfying the linearized Einstein equation in the 
TT gauge. Their evolution is governed by 
\begin{equation}
    \Box \gamma_{ij} = -16\pi G\, \Pi^{\mathrm{TT}}_{ij},
\end{equation}
where $\Pi^{\mathrm{TT}}_{ij}$ denotes the transverse--traceless component of the 
anisotropic stress tensor. Here, for the $(-+++)$ convention, $\Box\equiv\nabla_\mu\nabla^\mu=-\partial_t^2-3H\partial_t+a^{-2}\nabla_{\mathbf{x}}^2$ in spatially flat FLRW, with $H\equiv\dot a/a$. \footnote{Beyond Einstein general relativity, a wide class of modified-gravity theories can alter the 
propagation of tensor modes as
$    \ddot{\gamma}_{ij}
    + \left(3H + \delta_{1}\right)\dot{\gamma}_{ij}
    + \left(c_{T}^{2}\frac{k^{2}}{a^{2}} + m_{T}^{2} + \delta_{2}\right)\gamma_{ij}
    = 16\pi G_{\mathrm{eff}}\, \Pi^{\mathrm{TT}}_{ij}$,
where $c_{T}$ is the tensor propagation speed, $m_{T}$ an effective graviton mass, 
and $\delta_{1}$, $\delta_{2}$ encode possible modifications to friction and 
dispersion arising in modified gravity
(e.g., \citealt{Clifton:2011jh, deRham:2014zqa, Lewandowski:2016yce, Capozziello:2017vdi, Creminelli:2017sry, Baker:2017hug, Ezquiaga:2017ekz}). These 
modified-gravity frameworks, however, lie beyond the scope of the present review, and we refer the interested reader to \citet{Frusciante:2019xia}.} In \cref{sec:1,sec:2}, we examined the 
propagation of gravitational waves in asymptotically flat spacetimes, in which 
local astrophysical systems provide the anisotropic stresses appearing on the 
right-hand side of the equation above. As shown in \cref{eq:gamma-Q}, the far-zone gravitational-wave amplitude is proportional, in the quadrupole approximation, to the second time derivative of the TT-projected mass quadrupole moment, establishing the familiar 
connection between gravitational wave generation and the non-spherical, 
accelerated motion of matter within such sources, schematically 
$\ddot{Q}^{\mathrm{TT}}_{ij}/r$. In contrast, in \cref{sec:3} we turned to cosmological 
backgrounds and focused on vacuum fluctuations for which the source 
term vanishes, $\Pi^{\mathrm{TT}}_{ij}=0$. The present section focuses on the various physical mechanisms capable of generating a nonzero 
$\Pi^{\mathrm{TT}}_{ij}$ and thus sourcing gravitational wave emission in both astrophysical 
and cosmological settings.

\subsection{Astrophysical sources}\label{sec:Astro}

\begin{figure}[h!]
\centering
\includegraphics[width=0.8\textwidth]{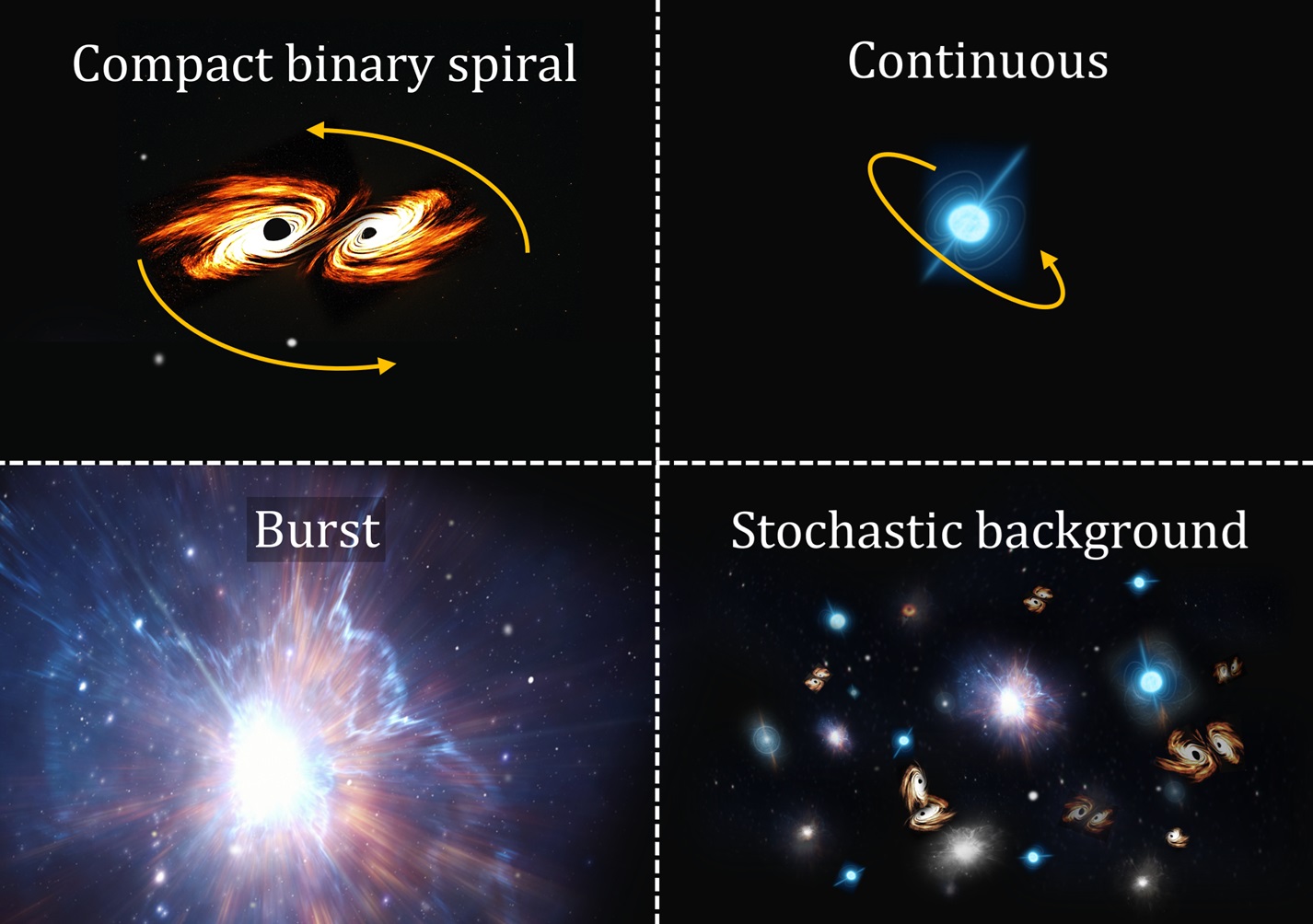} 
\caption{Illustration of the main astrophysical classes of gravitational wave sources, grouped by the duration of the signals they produce. Short-duration events (left) include late compact-binary inspirals and mergers and burst-like transients, while long-duration signals (right) arise from continuous-wave emitters and the stochastic gravitational wave background, together representing the broad range of astrophysical processes that generate gravitational waves.}\label{fig:astro-sources}
\end{figure}

A wide variety of astrophysical objects and dynamical phenomena can generate gravitational waves, each imprinting characteristic spectral features and signal morphologies. These sources span an enormous range of masses, environments, and timescales—from compact binaries to supernovae, isolated neutron stars, and stochastic backgrounds.  The main classes of gravitational wave sources are naturally grouped by the duration of the signals they produce. Short-duration events include late compact-binary inspirals and mergers, and burst-like transients, whereas long-duration signals arise from continuous-wave emitters and the stochastic gravitational wave background. Together, these categories capture the broad spectrum of astrophysical processes capable of radiating gravitational waves across cosmic time.  \cref{fig:astro-sources} illustrates the main categories of gravitational wave sources and their characteristic signal durations. In the following, we provide a brief summary of each.

\subsubsection{Compact binary coalescences} 

Compact binary coalescences (CBCs) are the most mature and well-understood astrophysical sources of gravitational waves. Binary neutron stars ($M\sim1$–$2 M_\odot$), stellar-mass black holes ($M\sim10$–$100 M_\odot$), and mixed neutron-star–black-hole systems lose orbital energy through quadrupole emission and ultimately merge in the strong-field regime. Black-hole binaries follow the characteristic inspiral--merger--ringdown sequence, while compact binaries populate the $\sim10$–$10^{3} \mathrm{Hz}$ band of ground-based interferometers. Neutron-star inspirals reach kilohertz frequencies near merger, while stellar-mass black-hole binaries radiate at tens to hundreds of hertz depending on total mass. The first LIGO–Virgo detections~\citep{Abbott:2016blz,Abbott:2017vtc} firmly established CBCs as the dominant sources (see \cref{fig:LIGO-detection}), enabling detailed population studies and stringent tests of general relativity~\citep{Barack:2018yly}.

At lower frequencies, supermassive black-hole (SMBH) binaries represent some of the most luminous predicted gravitational wave sources. Formed through hierarchical galaxy mergers~\citep{Begelman:1980,Volonteri:2010}, they radiate across the millihertz to nanohertz spectrum. For suitable masses and redshifts, their mergers fall within the $10^{-4}$--$10^{-1} \, \mathrm{Hz}$ band accessible to LISA~\citep{Amaro-Seoane:2012aqc}, providing a prime target for space-based detectors. LISA will also observe extreme mass-ratio inspirals (EMRIs), the gradual inspiral of a stellar-mass compact object into a $10^{5}$–$10^{7} \, M_\odot$ SMBH. EMRIs emit predominantly at millihertz frequencies and encode high-fidelity information about the spacetime geometry near massive black holes, offering precision tests of the Kerr hypothesis~\citep{Barack:2006pq,Babak:2017tow}. In addition, LISA will detect ultra-compact Galactic binaries, early inspirals of stellar-mass black-hole binaries, and potentially signals of primordial or exotic origin~\citep{LISA:2022yao}. The earliest inspiral stages of massive binaries produce nanohertz gravitational waves ($10^{-9}$--$10^{-7}\, \mathrm{Hz}$), which lie in the domain of pulsar timing arrays (PTAs)~\citep{BurkeSpolaor:2018}. These systems are expected to dominate the low-frequency gravitational wave sky and offer insight into galaxy growth, nuclear stellar dynamics, and black-hole coevolution~\citep{Mayer:2013}. As a window to fundamental physics, gravitational wave signals from compact binary coalescences may provide an ideal testing ground for potential deviations from classical horizon dynamics. In this context, possible modifications of near-horizon structure or horizonless ultracompact remnants can produce delayed post-merger GW echoes \citep{Cardoso:2016rao,Abedi:2016hgu,Abedi:2017isz,Abedi:2018npz}. Their properties are model-dependent and can be related to the assumed near-horizon structure and reflectivity \citep{Oshita:2019sat,Wang:2019rcf,Oshita:2023tlm,Saraswat:2019npa}. Dedicated searches have been performed, with no statistically significant evidence found in the GWTC-3 echo search \citep{Abedi:2021tti,LIGOScientific:2021sio}.

\subsubsection{Isolated neutron stars}

Isolated neutron stars furnish an additional class of promising sources through long-lived, nearly monochromatic continuous-wave emission \citep{Andersson:2000mf,Riles:2022wwz}. Such radiation, typically in the $\sim10^{2}$--$10^{3}\, \mathrm{Hz}$ range and set by the stellar rotation rate and deformation, remains undetected but is actively sought by all-sky searches and targeted pulsar analyses. Achieving astrophysically significant sensitivities would provide valuable constraints on neutron-star microphysics, crustal rigidity, and internal magnetic-field structure. Neutron stars can serve as natural astrophysical laboratories to search for new physics \citep{Kopp:2018jom,Nelson:2018xtr,Ellis:2017jgp,Baryakhtar2022,Hu:2023vsq,Grippa:2024ach}.

\subsubsection{Transient burst sources}

Highly dynamical and asymmetric stellar explosions also contribute to the gravitational wave landscape. Core-collapse supernovae produce short-duration, broadband bursts through turbulent convection, proto–neutron-star oscillations, rotation-driven instabilities, and standing accretion shock instabilities. Their gravitational wave spectra typically span $\sim10^{2}$–$10^{3} \, \mathrm{Hz}$, reflecting the characteristic dynamical timescales of proto–neutron-star cores. Although modeling remains challenging due to the nonlinear interplay of multidimensional hydrodynamics, neutrino transport, and magnetic fields, simulations indicate that next-generation detectors could observe signals from supernovae in the Milky Way or nearby galaxies~\citep{Kotake:2013,Ott:2009}. Additional transient phenomena, such as magnetar flares and gamma-ray-burst engines, may likewise produce gravitational radiation through violent magnetic or accretion-driven processes~\citep{Lasky:2015uia}.

\subsubsection{Astrophysical Contributions to the Stochastic GW Background}

Astrophysical sources, most notably compact-object mergers, give rise to the individually resolved GW events detected by current observatories. Taken collectively, these same populations also generate a stochastic gravitational wave background, produced by the superposition of numerous weak or distant events across cosmic history, each individually undetectable yet together forming a persistent, measurable signal. PTAs uniquely access the nanohertz band, where the background from the cosmic population of slowly inspiraling SMBH binaries is expected to dominate \citep{Sesana:2013}. Recent observations by NANOGrav, EPTA/InPTA, and the PPTA provide compelling evidence for a spatially correlated common-spectrum process consistent with a nanohertz gravitational-wave background~\citep{NANOGrav:2023gor,Antoniadis:2023rjw,Reardon:2023gzh} (see \cref{fig:NanoGrav}).

At higher frequencies, space-based laser interferometers such as LISA \citep{Colpi:2024lisa} will probe the millihertz band, where the astrophysical stochastic background is expected to be dominated at lower millihertz frequencies by unresolved Galactic compact binaries, especially white-dwarf binaries, with additional extragalactic contributions. At still higher frequencies, next-generation ground-based detectors such as the Einstein Telescope \citep{ET:2025xjr} will access the audio band with unprecedented sensitivity, enabling precise measurements of both individually resolved mergers and the astrophysical stochastic background generated by stellar-mass compact binaries across cosmological distances. Together, PTAs, LISA, and the Einstein Telescope provide complementary coverage of the gravitational-wave spectrum, enabling a truly multi-band characterization of the astrophysical gravitational-wave background.

\subsection{Cosmological sources}

A wide variety of processes in the early Universe can generate stochastic backgrounds of gravitational waves, offering a unique window into physics far beyond the reach of terrestrial experiments. Among the most extensively studied cosmological sources are inflation/reheating, first-order phase transitions, topological defects, primordial black holes, scalar-induced tensor modes, and magnetohydrodynamical phenomena in the radiation-era plasma. Each mechanism imprints characteristic spectral features that encode fundamental information about high-energy symmetries, inflationary dynamics, or hidden sectors. For comprehensive overviews, see, for example, the recent Einstein Telescope Science Books \citep{ET:2025xjr}, as well as the LISA Cosmology Reports \citep{LISACosmologyWorkingGroup:2022jok, Colpi:2024lisa}. In the following, we discuss the main classes of cosmological GW sources and the physics that underlies their distinctive signatures.

\subsubsection{Inflation and reheating}

 In addition to the standard vacuum tensor modes from slow-roll inflation discussed in \cref{sec:4}, signals associated with spectator fields, i.e., additional scalar or gauge fields present during inflation but not driving it, may source a stochastic gravitational wave background with distinctive spectral features.  For a recent comprehensive review of inflationary scenarios capable of enhancing the stochastic GWB to amplitudes observable by LISA, see \cite{LISACosmologyWorkingGroup:2024hsc}. Following inflation, similar dynamics can occur during reheating. During preheating, the coherent oscillations of the inflaton can transfer energy efficiently 
into scalar or gauge fields, sourcing gravitational waves through highly 
non-linear, anisotropic field configurations \citep{Dufaux2007,Figueroa:2017vfa,Adshead:2018doq,Lozanov:2019ylm}. 
The resulting GW backgrounds can be sufficiently large that the requirement of avoiding excess 
radiation energy density at BBN imposes important constraints on the 
viable parameter space of such models \citep{Clarke:2020bil,Pagano2016BBN,Caprini:2018mtu}.

Inflation also sources gravitational waves indirectly through scalar fluctuations. 
Large or resonantly enhanced curvature perturbations can generate second-order tensor 
modes already during inflation or shortly after horizon re-entry, producing scalar-induced 
contributions with potentially sharply peaked spectra. The underlying mechanism was studied 
in the seminal second-order calculations of \citet{Ananda2007,Baumann2007}, and later 
applied to scenarios featuring large small-scale scalar enhancements, such as those associated 
with primordial black-hole production \citep{SaitoYokoyama2009,KohriTerada2018}. Resonant or 
non-adiabatic dynamics during inflation can further amplify scalar fluctuations and source 
tensor modes directly, as shown in models of particle production and inflationary resonance 
e.g. \citet{Biagetti2013,Garcia-Bellido:2017aan}. Together, these mechanisms highlight inflation and 
reheating as fertile grounds for generating non-vacuum gravitational waves with characteristic 
signatures---sharp spectral features, potentially large non-Gaussianity, and potential chirality---not shared by the 
standard slow-roll vacuum tensor background. For comprehensive reviews of scalar-induced 
gravitational waves, see \cite{Domenech2021Review,YuanHuang2021Review}.

Related mechanisms also operate during radiation domination: tachyonic or parametric instabilities of scalar fields and coupled gauge sectors can generate peaked GW backgrounds, with frequencies set by the excited scales and production epoch rather than necessarily by inflation. Examples include axion--dark-photon production \citep{Machado:2018nqk}, axion fragmentation \citep{ChatrchyanJaeckel2021}, and kinetic-misalignment fragmentation \citep{EroncelEtAl2022}.

%%\vskip 0.2cm

\subsubsection*{Gauge fields in inflation and GWs} 

%%%%%%%%%%%%%%%%%%%%%%%%%%%%%%%    

Gauge fields are a foundational component of the SM and of nearly all well-motivated extensions of high-energy physics. Because cosmic inflation probes energy scales far beyond those accessible to terrestrial experiments, it is natural to view the early Universe as a particle collider capable of revealing new interactions—an idea formalized in the framework of the Cosmological Collider \citep{Chen:2009zp,Arkani-Hamed:2015bza}. This perspective motivates the question of whether gauge fields played a dynamical role during inflation and, if so, whether they left observable imprints in primordial correlators. Axion-like fields, protected by an approximate shift symmetry, provide compelling inflaton candidates and can couple to gauge sectors through the Chern--Simons interaction $\phi F_{\mu\nu}\tilde F^{\mu\nu}$. Originally introduced in the context of the Peccei–Quinn mechanism \citep{Peccei:1977hh,Peccei:1977ur,Weinberg:1977ma,Wilczek:1977pj} and later explored in cosmology \citep{Dine:1982ah,Abbott:1984fp} and inflation \citep{TurnerWidrow1988,Freese:1990rb,Adams:1992bn,Garretson1992}, this coupling naturally mediates interactions between inflaton dynamics and gauge degrees of freedom. In addition, axions may be realized as composite states originating from confining gauge dynamics \citep{Contino:2021ayn,Maleknejad:2022gyf}. Axions and axion-like particles play a central role in both cosmology and particle physics, and their theoretical foundations and detection prospects have been thoroughly reviewed in \cite{Marsh:2015xka,Graham:2015ouw,Irastorza:2018dyq,Ferreira:2020fam,Adams:2022pbo,Ringwald:2023yni,Ringwald:2024uds}. During inflation, the axion-gauge field interaction can lead to the exponential amplification of Abelian gauge-field fluctuations \citep{Anber:2006xt,Anber:2009ua}. Inflationary scenarios involving gauge fields therefore provide some of the earliest and most well-developed mechanisms for producing non-vacuum gravitational waves. These effects offer a rich arena for probing gauge sectors through primordial non-Gaussianity and gravitational-wave signatures \citep{Barnaby:2011qe,Komatsu:2022nvu}.

\begin{figure}[b]
\centering
\includegraphics[width=0.99\textwidth]{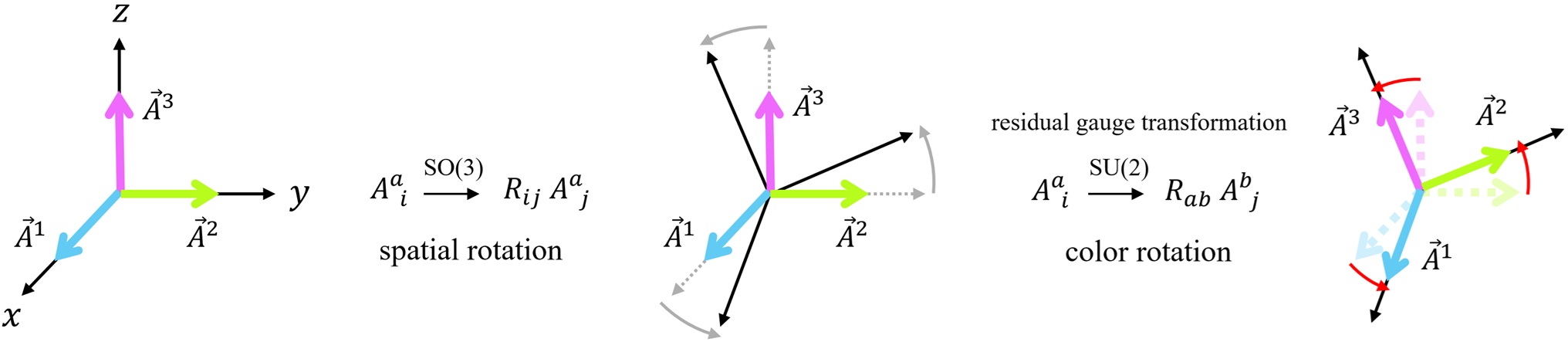} 
\caption{Coherent homogeneous $SU(2)$ background gauge field and the restoration of spatial isotropy. 
For the color--space-locked configuration $\bar A_i^a=a\psi\,\delta_i^a$, the gauge field transforms as a vector under spatial rotations. 
At the same time, the $SU(2)$ gauge symmetry contains residual transformations with the same algebraic structure, reflecting the Lie-algebra isomorphism $\mathfrak{su}(2) \cong \mathfrak{so}(3)$. 
Thus, any spatial rotation can be compensated by an appropriate $SU(2)$ gauge transformation, corresponding to the diagonal subgroup of $SO(3)_{\mathrm{rot}} \times SU(2)_{\mathrm{gauge}}$. 
Since physical observables are gauge invariant, they depend only on this diagonal action and therefore remain fully rotationally symmetric.
}\label{fig:color-}
\end{figure}

\begin{figure}[htbp]
 \centerline{\includegraphics[width=0.7\textwidth]{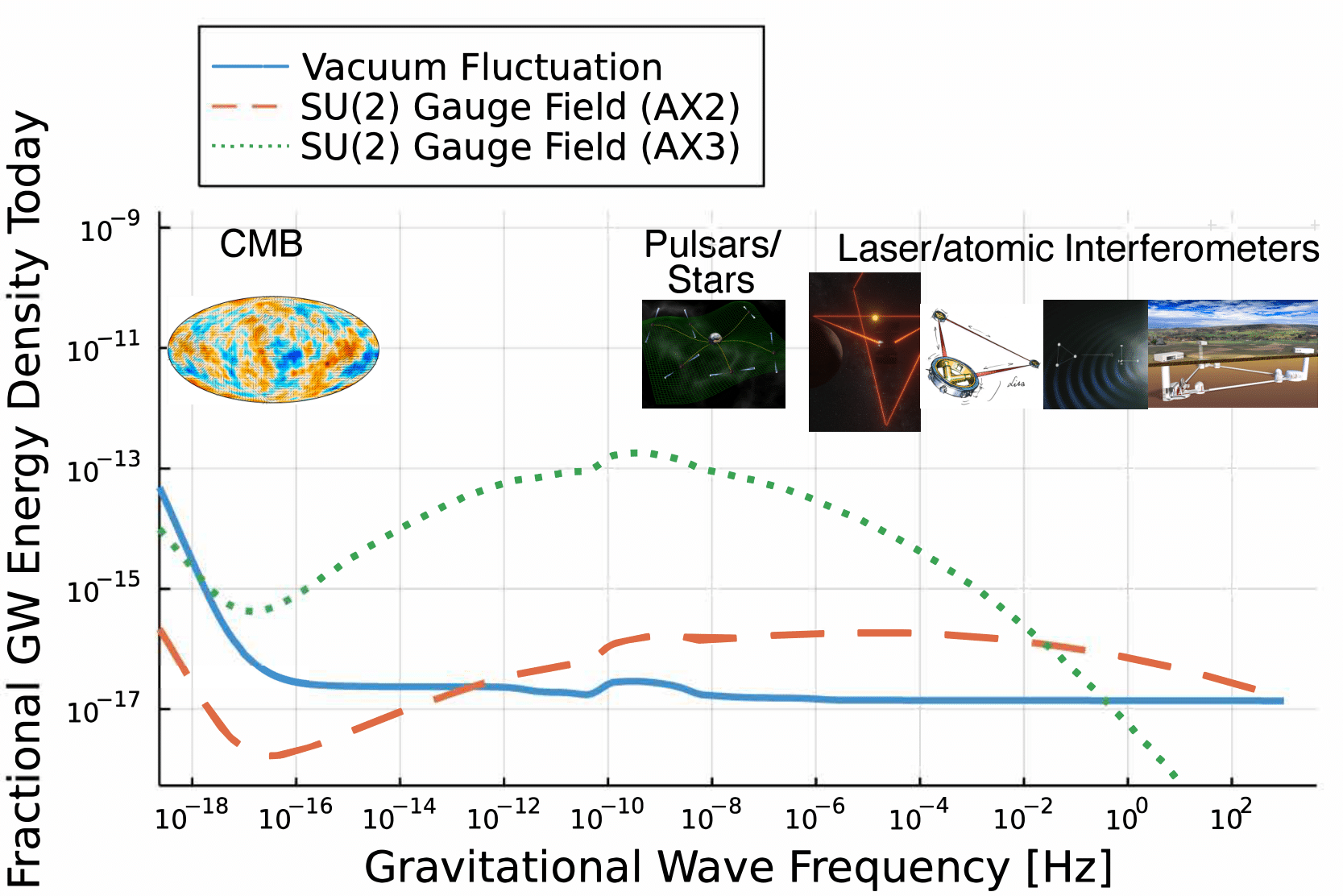}}
 \vspace{0.5cm}
 \centerline{\includegraphics[ width=0.7\textwidth]{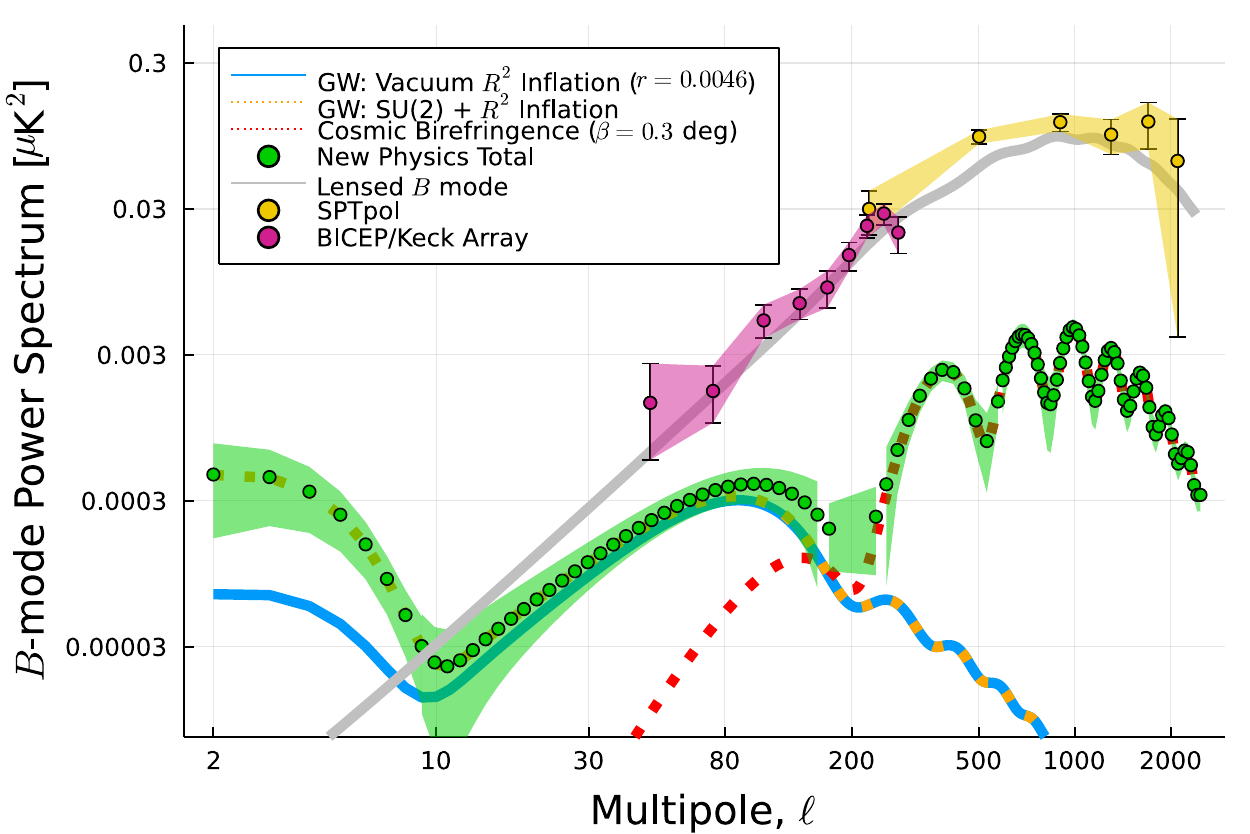}}
\caption{Characteristic imprints of SU(2)–axion inflation on the gravitational wave background and CMB B-mode polarization.
 Image credit: \citet{Komatsu:2022nvu}.}\label{fig:axion-SU2}
\end{figure}

\begin{figure}[h]
\centering
\includegraphics[width=0.5\textwidth]{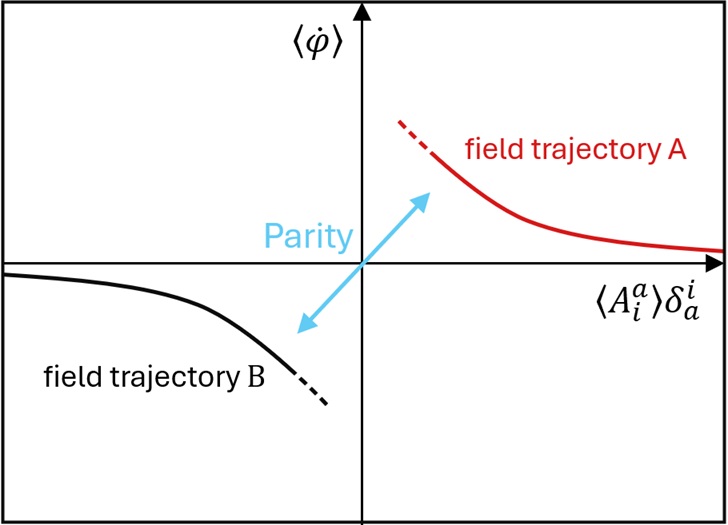} 
\caption{Schematic illustration of slow-roll field trajectories related by parity. In the axion–SU(2) gauge-field model, slow-roll solutions satisfy $\text{sign}(\dot\phi)=\text{sign}(A^a_i \delta^i_a)$, selecting one of the two parity-related branches.
}\label{fig:parity-axion}
\end{figure}

\cite{Maleknejad:2011sq,Maleknejad:2011jw} showed that non-Abelian gauge fields can contribute nontrivially to the dynamics of inflation while---unlike Abelian $U(1)$ background fields---preserving the observed cosmological isotropy within an $SU(2)$ subsector. In particular, the background gauge field can consistently acquire a homogeneous and isotropic vacuum expectation value of the form
\begin{equation}
\bar{A}_\mu(t) \equiv \bar{A}^a_\mu(t) T_a =
\begin{cases}
0, & \mu = 0, \\
a(t)\,\psi(t)\,\delta^a_i\, T_a, & \mu = i,
\end{cases}
\label{eq:SU2-VEV}
\end{equation}
where $\{T_a\}$ are the generators of the $SU(2)$ gauge algebra. This homogeneous and isotropic field configuration is possible due to the isomorphism between the $SU(2)$ gauge algebra and the spatial rotation algebra $SO(3)$, which allows gauge and spatial indices to be aligned (see \cref{fig:color-}). Perturbations of the gauge fields around this isotropic background contain an induced spin-2 mode that couples linearly to gravitational waves \citep{Maleknejad:2011sq,Maleknejad:2011jw}. In particular, the spatial gauge-field perturbations include a tensorial component of the form
\be
\delta A^a_i \supset t_{ij}\,\delta^{aj},
\ee
where $t_{ij}$ is symmetric, transverse, and traceless and transforms as a spin-2 degree of freedom. A rigorous mathematical proof of the spin-2 nature of this induced tensor mode can be found in Appendix B.1 of \citet{Maleknejad:2018nxz}. 
In the relevant parameter regime, this mode undergoes a brief phase of tachyonic instability near horizon crossing, leading to the formation of a chiral cloud of gauge-field tensor fluctuations localized around the cosmic horizon. Owing to its linear coupling to the metric tensor perturbations, this induced spin-2 mode acts as a chiral source for gravitational waves, corresponding to a nonvanishing transverse–traceless anisotropic stress, $\Pi^{\mathrm{TT}}_{ij} \neq 0$. As a result, this coupling enables the efficient production of chiral gravitational waves \citep{Adshead:2012kp,Maleknejad:2012fw,Dimastrogiovanni:2012ew,Maleknejad:2016qjz,Dimastrogiovanni:2016fuu,Adshead:2013nka,Adshead:2017hnc,Maleknejad:2018nxz}, accompanied by characteristic non-Gaussian tensor perturbations \citep{Agrawal:2017awz,Agrawal:2018mrg} (see \cref{fig:axion-SU2}). Consequently, these background solutions select a handedness (see \cref{fig:parity-axion}), with a pronounced asymmetry between left- and right-handed gravitational-wave modes imprinted at the level of primordial correlations.

%%%%%%%%%%%%%%%%%%%%%%%

%%%%%%%%%%%%%%%%%%%%%%%%

This framework admits several distinct realizations within inflationary model building, which have been explored extensively in the literature \citep{Maleknejad:2011sq,Adshead:2012kp,Maleknejad:2012dt,Maleknejad:2016qjz,Caldwell:2017chz,Dimastrogiovanni:2016fuu,Nieto:2016gnp,Adshead:2016omu,Adshead:2017hnc,McDonough:2018xzh,Holland:2020jdh,Watanabe:2020ctz,Maleknejad:2020yys,Iarygina:2021bxq,Dimastrogiovanni:2023juq}. In most cases, the isotropic configuration corresponds to an attractor solution \citep{Maleknejad:2011jr,Maleknejad:2013npa,Wolfson:2020fqz,Wolfson:2021fya}, except in scenarios involving Higgsed non-Abelian gauge fields with unequal masses for at least two gauge components \citep{Adshead:2018emn}. Variants of this setup have also been applied in the context of warm inflation \citep{Kamali:2019ppi,Berghaus:2019whh} and have found further applications in dark-energy model building \citep{Mehrabi:2015lfa,Heisenberg:2016wtr,Caldwell:2018feo,Guarnizo:2020pkj,Kaneta:2025kcn}. Beyond their distinctive tensor signatures, these models exhibit a rich phenomenology extending beyond standard scalar inflation, including the Schwinger effect during inflation \citep{Mirzagholi:2019jeb,Lozanov:2018kpk,Maleknejad:2019hdr,VicenteGarcia-Consuegra:2025lkh}, as well as mechanisms for baryogenesis and leptogenesis mediated by the chiral anomaly \citep{Maleknejad:2020yys,Maleknejad:2020pec,Elor:2022hpa}. For broader context and a systematic classification of Abelian and non-Abelian mechanisms, we refer the reader to the comprehensive review by \citet{Maleknejad:2012fw}, together with the recent perspectives presented in \citet{Komatsu:2022nvu} and \citet{LISACosmologyWorkingGroup:2024hsc}.

\begin{tcolorbox}[colback=gray!10, colframe=gray!10, boxrule=0pt,
                  enhanced, breakable, halign=justify]

\subsubsection*{Cosmic Chirality: Signatures of Broken Fundamental Symmetry in CMB, LSS, and GWs}

 Parity ($P$) is the spatial inversion $\bx \rightarrow -\bx$, and parity symmetry denotes invariance under this transformation. While parity is conserved by classical gravity and by electromagnetic and strong interactions, it is known to be maximally violated by weak interactions, establishing P violation as a fundamental feature of particle physics. This fact motivates the possibility that new physics beyond the SM may also violate parity. An open and profound question is therefore whether parity was ever an exact symmetry of the Universe at earlier stages of its evolution, or whether parity-violating interactions played a role throughout cosmic history. Cosmology provides a unique laboratory to address this question, offering observational access to physical processes across vastly different epochs. Parity violation in cosmology has consequently been explored across multiple epochs using CMB observations and LSS measurements, and more recently through proposed searches for stochastic GWBs.

Parity violation in the CMB can manifest as cosmic birefringence, a rotation of the linear polarization that induces nonvanishing $TB$ and $EB$ correlations. 
Measurements of uniform cosmic birefringence are limited by a degeneracy with absolute instrumental polarization-angle calibration. 
A recent breakthrough realization by \cite{MinamiKomatsu2020} showed how this degeneracy can be broken using correlations with Galactic foreground emission, leading to the first statistically significant hint of a nonzero cosmic birefringence signal in \textit{Planck} data. 
This result has revitalized efforts to probe fundamental parity-violating physics using the polarized sky.
 Subsequent studies using polarization data from \textit{WMAP}, \textit{Planck} (2018 and DR4), and the Atacama Cosmology Telescope have progressively tightened constraints on cosmic birefringence, with some reporting evidence for a nonzero uniform rotation angle while emphasizing the critical role of instrumental systematics \citep{MinamiKomatsu2020,DiegoPalazuelos2022,EskiltKomatsu2022,DiegoPalazuelosKomatsu2025}. Parity violation may also originate during inflation, where chiral gravitational waves sourced by parity-violating interactions generate intrinsically parity-odd signatures in CMB polarization, most prominently through nonzero $TB$ and $EB$ cross-correlations \citep{LueWangKamionkowski1999, Saito:2007kt, Sorbo:2011rz, Thorne:2017jft}. 
These correlations provide a probe of primordial parity violation, complementary to post-recombination birefringence. See \citet{Komatsu:2022nvu} for a clear and comprehensive recent review.

In addition, parity-violating physics during inflation or phase transitions can produce
circular polarization of gravitational waves, which can be probed by suitably configured
detector networks \citep{SetoTaruya2007,Cook:2011hg,Crowder:2005nr,
Thorne:2017jft,Chen:2024fto,Sato-Polito:2021efu}, as well as by colocated,
non-coplanar configurations such as SPIN \citep{Kharzeev:2026spin},
providing an independent and complementary probe of parity violation beyond
electromagnetic observations.
 
 At late times, parity violation has further been investigated in LSS through parity-odd components of higher-order galaxy correlation functions. Recent work has proposed galaxy intrinsic alignment as an additional probe~\citep{Kurita:2025hmp}. BOSS/SDSS analyses have reported differing conclusions about the significance of parity-odd signals \citep{Philcox2022,HouSlepianCahn2023,Krolewski2024}, while early results from DESI extend these tests \citep{Slepian2025}. Together, low-$\ell$ CMB $B$ modes, birefringence measurements, gravitational wave backgrounds, and LSS correlations form a complementary framework for testing parity-violating physics across cosmic history.

\end{tcolorbox}

\subsubsection{Cosmological phase transitions and hydrodynamical sources}

\begin{figure}[h!]
\centering
\includegraphics[width=\textwidth]{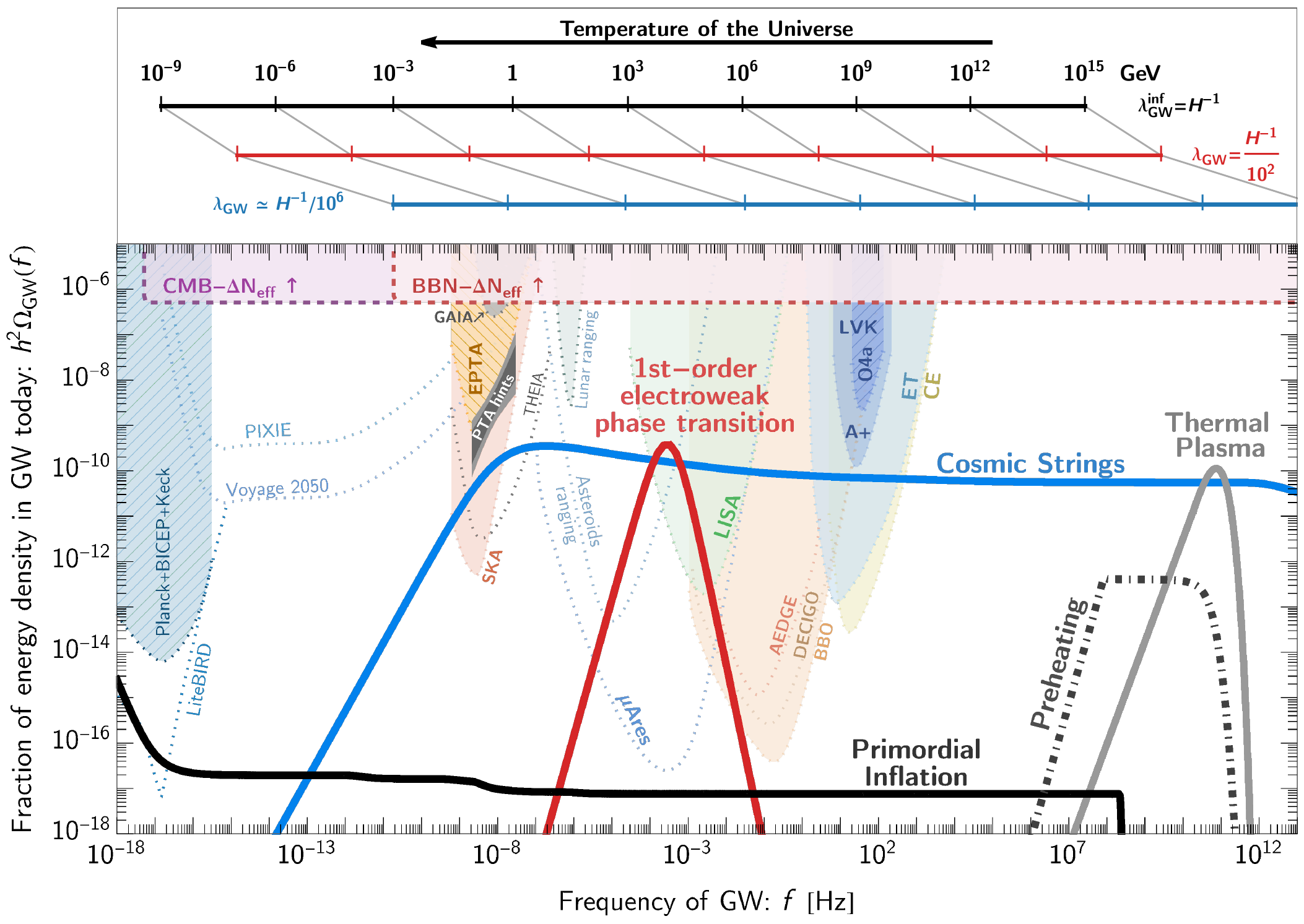} 
\caption{Sensitivity curves of current and future gravitational wave observatories across a broad frequency range, shown together with representative cosmological GW sources. The thick black curve denotes the nearly scale-invariant (vacuum) GW spectrum from inflation, while the blue and red curves represent GW backgrounds from cosmic strings and a first-order electroweak phase transition, respectively. The GAIA and THEIA curves are based on \citet{Caliskan:2023cqm} assuming SNR = 10 with observation times of 10 and 20 years, respectively. The colored lines at the top of the figure indicate the present-day GW frequencies corresponding to the production time for different source sizes $\lambda_\mathrm{GW}$ (colors match those of the spectra). For illustration, the cosmic string spectrum is shown for $G\mu = 10^{-11}$ ($\mu$ is the energy per unit length of a string and $G$ is Newton’s constant), and the first-order phase transition signal for $T_* = 150\,\mathrm{GeV}$, $\beta/H_* = 20$, and $\alpha = 0.5$.  Image credit: Peera Simakachorn  and \citet{Simakachorn:2022yjy}.}\label{fig:cosmo-sources-II}
\end{figure}

First-order cosmological phase transitions can act as powerful factories of gravitational waves: as bubbles of the true vacuum nucleate, expand, and collide, they generate tensor perturbations through the violent dynamics of bubble walls~\citep{Witten1984,Hogan1986}. For many non-runaway transitions in a thermal plasma, most of the released vacuum energy is transferred to the surrounding plasma, and acoustic waves can dominate GW production, with MHD turbulence providing an additional contribution~\citep{Kamionkowski1994,Apreda2001,Caprini:2007xq}. The resulting spectra are typically broad and peaked, with characteristic frequencies set by the bubble size and transition temperature, making electroweak-scale transitions a prime target for space-based interferometers such as LISA \citep{LISACosmologyWorkingGroup:2022jok,Caprini:2019egz,Servant2022EWbaryogenesis} (see \cref{fig:cosmo-sources-II}). Importantly, within the Standard Model, the electroweak phase transition is a smooth crossover rather than first order, and therefore does not generate significant GW production through this mechanism.\footnote{Anomaly-induced scalar gravitational radiation has also been studied assuming a first-order QCD transition \citep{Dai:2019ksi}. This is distinct from the tensor modes considered here; in the Standard Model at small baryon chemical potential, the QCD transition is a crossover \citep{Aoki:2006weu}.} Therefore, detectable signals require extensions of the Standard Model capable of strengthening the transition. Many extensions involving additional scalar degrees of freedom, hidden sectors, or supersymmetric frameworks predict strongly first-order transitions capable of producing observable gravitational waves. Consequently, cosmological phase transitions offer a unique observational window into BSM physics at energy scales ranging from the electroweak regime up to grand-unified theories \citep{Schwaller:2015tja,Breitbach:2018ddu,Croon:2018erz,Nakai:2020oit,Pasechnik:2024xkv}.

Primordial magnetic fields provide an additional MHD source of gravitational waves: their anisotropic stresses, together with the MHD turbulence and Alfvén waves they induce in the radiation-era plasma, generate a stochastic background whose amplitude and spectral shape depend on the field strength, coherence scale, and helicity \citep{CapriniDurrer2006,Caprini2009b}. Such magnetic fields may themselves originate from early-Universe phase transitions or other high-energy processes, making them a natural complement to hydrodynamical GW sources. Modern quantitative treatments combine bubble dynamics, plasma interactions, and numerical simulations to model these signals more reliably \citep{Hindmarsh2015,Hindmarsh2017}. Comprehensive recent reviews of phase-transition and MHD-sourced gravitational waves, and their implications for particle physics and cosmology, can be found in \citet{Caprini2016Review,Caprini:2018mtu,Mazumdar:2018dfl}.

\subsubsection{Topological defects}

Topological defects arise when a system with gauge (or global) symmetry undergoes spontaneous symmetry breaking during a phase transition. Let a theory have symmetry group $G$, acting on scalar fields $\Phi$, with a symmetry-breaking potential $V(\Phi)$. When the vacuum expectation value $\langle \Phi \rangle$ selects a ground state, the symmetry is broken to a subgroup $H \subset G$. The degenerate vacua form the vacuum manifold $\mathcal{M} = G/H $,
which is a coset space whose global topology determines the allowed defect structures. The topological defects correspond to topologically non-trivial mappings $\Phi:\; S^{n} \;\longrightarrow\; \mathcal{M}$,
where $S^{n}$ is an $n$-sphere enclosing the defect in physical space (for textures, $S^3$ corresponds to compactified physical space). If this map is non-contractible, the configuration cannot be continuously deformed to the vacuum. The relevant homotopy group $\pi_n(\mathcal{M})$ therefore classifies the defect. As shown by \cite{Kibble1976}, causality in an expanding universe implies that regions separated by more than the horizon scale at a phase transition choose vacua independently. Mismatches between these choices generically produce defects whenever the topology of $\mathcal{M}$ is nontrivial \citep{VilenkinShellard1994}.

\begin{table}[h]
\caption{Cosmologically relevant topological defects from gauge or global symmetry breaking. Defects are classified by the homotopy groups of the vacuum manifold $\mathcal{M}=G/H$, a coset space determined by the symmetry-breaking pattern. }
\centering
\begin{tabular}{llllll}
\toprule
Defect type & Vacuum topology  & Homotopy  & Cosmology  \\
\midrule
Domain walls &
Disconnected vacua &
$\pi_0(\mathcal{M})$ &
Broken discrete symmetry \\
%%\hline
Strings  &
Non-contractible loops  &
$\pi_1(\mathcal{M})$ &
Cosmic strings \\
%%\hline
Monopoles &
Non-contractible spheres $S^2$ &
$\pi_2(\mathcal{M})$ &
GUT magnetic monopoles \\
%%\hline
Textures &
Higher-order winding $S^3$  &
$\pi_3(\mathcal{M})$ &
Global textures  \\
\bottomrule
\end{tabular}
\end{table}

%--------------------------------------------

Topological defects can naturally arise during symmetry-breaking phase transitions in the early Universe, taking the form of domain walls, monopoles, cosmic strings, or textures depending on the topology of the vacuum manifold. Among these, cosmic strings are by far the most robust and observationally relevant gravitational wave sources. They may be produced in field-theoretic transitions or in a variety of beyond-the-Standard-Model scenarios. Early work by \cite{Vilenkin:1981bx} and \cite*{Brandenberger:1986xn}, together with later reviews of defect cosmology~\citep{Br94}, places cosmic strings within the wider context of early-Universe physics.  Once formed, cosmic strings rapidly organize into a long-lived scaling network that continuously produces string loops~\citep{Kibble1976,VilenkinShellard1994}. For thin local strings in the Nambu--Goto approximation, these loops oscillate relativistically under their tension and lose energy predominantly through GW emission, as established in the foundational works of \cite{Vilenkin:1981bx} and \cite{AS92}. 

GW production is enhanced at relativistic features on the loop—cusps, where a point on an ideal Nambu--Goto string momentarily reaches the speed of light, and kinks, discontinuities in the tangent vector typically generated by intercommutation. These features generate strong bursts whose superposition over cosmic history leads to a stochastic background with an approximately flat radiation-era plateau \citep{Siemens:2006yp} (see \cref{fig:cosmo-sources-II}), with characteristic spectral breaks encoding the expansion history and loop-size distribution~\citep{DV00,Blanco-Pillado:2013qja,Blanco-Pillado:2017rnf}. Extensions such as superconducting or current-carrying cosmic strings, originally proposed by \cite{Witten:1984eb}, further enrich the GW phenomenology by introducing current-induced energy-loss channels and modified loop dynamics, with detailed modern treatments presented in \cite{AuclairBlasi2023}. Several recent comprehensive reviews specifically address the gravitational wave phenomenology of cosmic-string networks \citep{Auclair:2019wcv,CSReviewNonStandard2019,Gouttenoire:2021jhk,Dimitriou:2025bvq,Schmitz:2024gds}. Other topological defects can also source GWs. Global textures generate smooth, nearly scale-invariant tensor spectra \citep{Turok1989,Durrer1999}, while monopole--string composite systems \citep{Preskill:1992bf,DunskyEtAl2022}  radiate during \citep{DunskyEtAl2022} or collapse \citep{VachaspatiVilenkin1987,BuchmullerDomckeSchmitz2021}. Domain walls, when present, can source GWs during their dynamical evolution and annihilation~\citep{Saikawa2017DomainWalls}; they can also drive powerful defect-induced phase transitions and corresponding GW signals~\citep{Blasi:2022woz}. Although typically more model-dependent than cosmic strings, such defects provide complementary probes of high-energy symmetry breaking.

\begin{figure}[ht]
\centering
\includegraphics[width=0.9\textwidth]{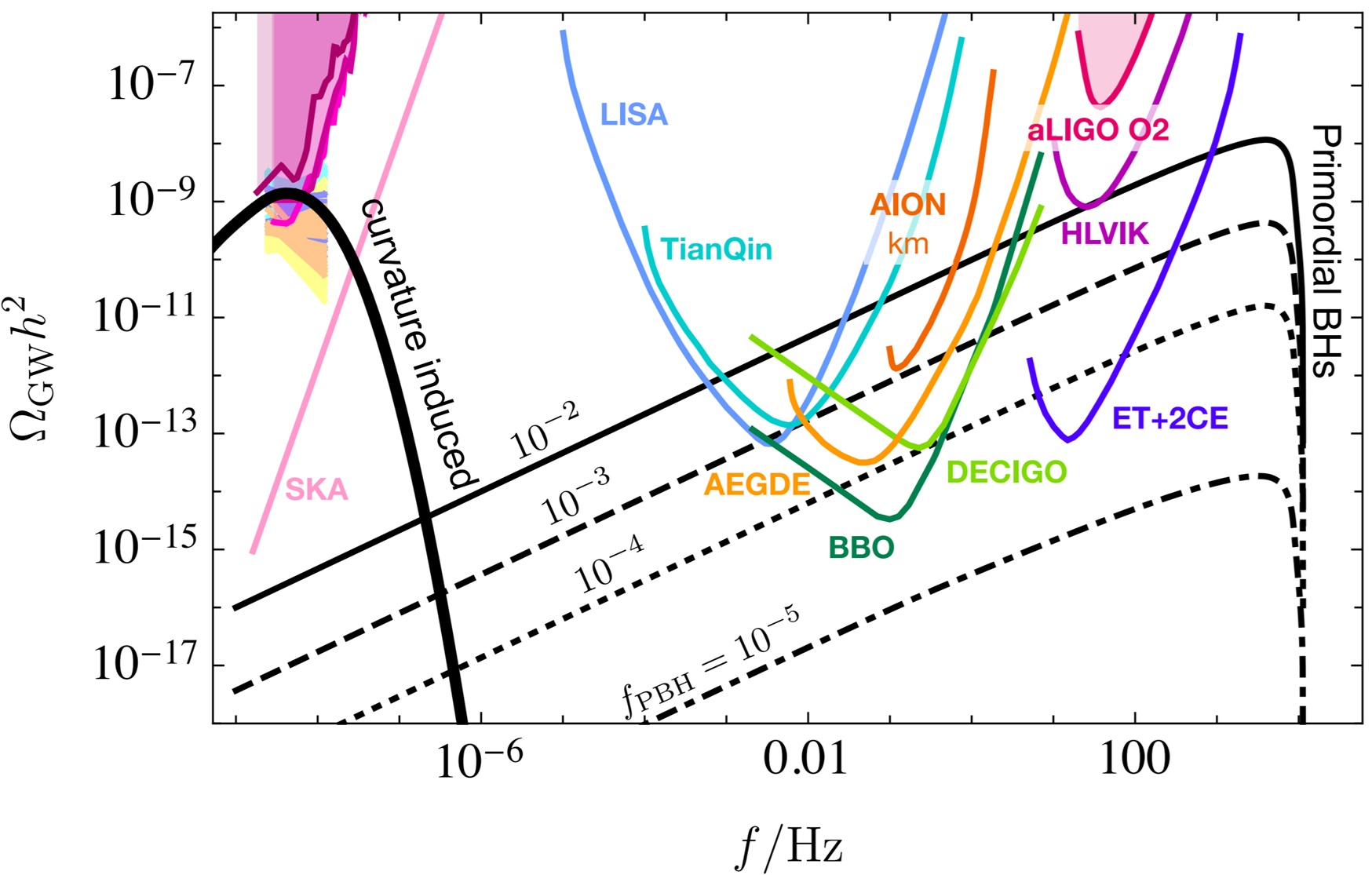} 
\caption{
Comparison of detector sensitivities with GW spectra associated with primordial black hole (PBH) scenarios and scalar-induced tensor modes. The thick black curve on the left shows the second-order curvature-induced GWs associated with PBH formation, while the thick black curve on the right shows the stochastic GW background from PBH binaries. Dashed black curves correspond to predicted backgrounds for different PBH abundance fractions. Image credit: Adapted from \citet{Kohri:2020qqd}.}\label{fig:cosmo-sources-I}
\end{figure}

\subsubsection{Primordial black holes and scalar-induced gravitational waves}

Primordial black holes (PBHs) are black holes formed in the early Universe, typically from the collapse of large overdensities shortly after horizon re-entry, rather than from stellar evolution~\citep{Hawking1971}. Their formation is most naturally described in terms of enhanced primordial curvature perturbations, whose associated overdensities can collapse when sufficiently large. The associated GW signatures arise through two widely studied channels. First, PBHs can form binary systems whose inspirals and mergers generate a population of resolvable GW events and a stochastic background, potentially contributing to the binary black-hole signals observed by ground-based interferometers~\citep{Bird2016}. Second, in this formation scenario, the large scalar perturbations associated with PBH formation source tensor perturbations at second order, leading to scalar-induced GWs with potentially peaked spectra that can fall in the PTA, LISA or ground-based frequency bands~\citep{SaitoYokoyama2009}. This connects PBH dark-matter scenarios directly with GW observations. Evaporating PBHs can also generate a gravitational-wave background through Hawking emission of gravitons \citep{Dong:2015yjs}. Scalar-induced gravitational waves (SIGWs) arise at second order in cosmological perturbation theory: large primordial curvature fluctuations, though not sourcing tensor modes at linear order, inevitably generate gravitational waves as they re-enter the horizon in the radiation era. This mechanism was studied in influential second-order calculations \citep{Ananda2007,Baumann2007} and later applied to scenarios with enhanced small-scale perturbations, such as those leading to primordial black-hole formation~\citep{SaitoYokoyama2009}. For the gravitational wave features in relation to the sensitivities of current probes, see \cref{fig:cosmo-sources-I}. The resulting GW spectra can be sharply peaked when the scalar enhancement is localized in scale, with frequencies determined by the horizon scale at re-entry, making SIGWs a powerful probe of inflationary physics on scales far smaller than those constrained by the CMB. For narrow scalar peaks producing asteroid-mass PBHs that account for all dark matter, the associated SIGW signal remains detectable by LISA-like observatories, even allowing local-type primordial non-Gaussianity \citep{CaiPiSasaki2019}.  For comprehensive discussions of PBHs and their GW signatures, including end-to-end treatments of inflationary production, binary formation, and induced backgrounds, see recent reviews~\citep{Bagui2025Review,Domenech2021Review,YuanHuang2021Review}.

%\section{Quantum Fields in GW Backgrounds: Leptogenesis and Dark Matter}

%\textcolor{red}{I am still working on this section, so it is not final yet.}

%%%%%%%%%%%%%%%%%%%%%%%%%%%%%%%%%
% Conclusions
 \section{Summary}\label{sec:summary}

In this review, we present a pedagogical study of gravitational waves, integrating theory, phenomenology, and observation. Gravitational waves offer a powerful probe of high-energy BSM physics and a unique window into the unexplored frontiers of gravity and the early Universe. Unlike photons, which could only free-stream after the formation of the cosmic microwave background ($T\sim 1$ eV), and neutrinos, which decoupled at $T\sim 1$ MeV, gravitational waves propagate essentially freely across cosmic history. As a result, they carry valuable and unique information about the Universe at energy scales far beyond those accessible by terrestrial particle colliders (see \cref{fig:cosmic-history}).

Spanning a gravitational wave rainbow with frequencies ranging from $10^{-17}$--$10^{4}$ Hz, corresponding to wavelengths from cosmological down to kilometer scales, these signals enable a truly multi-scale exploration of fundamental physics, providing insights impossible to obtain from any other messenger. We develop a coherent framework linking the mathematical foundations of gravitational radiation with the signals accessible to current and future detectors, aiming to bridge the gap between theoretical predictions and experimental measurements. Since gravitational waves are both generated and propagate through the Universe’s gravitational background, their description depends on the relevant spacetime geometry. On scales sufficiently small compared with the Hubble scale, the expansion of the Universe can often be neglected, and asymptotically flat spacetimes provide an accurate approximation to gravitational dynamics. On larger, cosmological scales, expansion becomes significant, and the FLRW geometry must be used. The nature of the signal introduces a complementary distinction: individual astrophysical sources often produce coherent, deterministic signals, whereas unresolved astrophysical populations and cosmological processes can generate stochastic backgrounds, requiring a statistical treatment (see \cref{fig:hierarchy}).

The review begins with the linearized theory of general relativity, deriving gravitational waves as solutions to the Einstein equations and analyzing their physical properties. Next, we provide a more geometrical and invariant characterization of gravitational waves by examining their nature through the Weyl tensor, including its principal null directions and Petrov classification. Within the Bondi–Sachs framework, this approach clarifies the radiative degrees of freedom of spacetime and the associated fluxes of energy and angular momentum. By combining algebraic and geometrical perspectives, this framework establishes a solid theoretical foundation for interpreting observational data and connecting theory with experiment.

We then explore cosmological gravitational waves, tracing their evolution across different cosmic epochs and their generation during inflation, phase transitions, and other early-Universe processes. Building on this foundation, we survey the principal detection strategies across a broad range of frequencies, including cosmic microwave background measurements, pulsar timing arrays, ground- and space-based laser interferometers, and resonant cavity experiments. We also highlight several ongoing developments and proposed concepts that could substantially enhance both the sensitivity and frequency reach of gravitational-wave observations. We then examine in more detail the physical mechanisms responsible for generating gravitational radiation, clearly distinguishing between astrophysical sources and cosmological origins.

The study of gravitational waves is a vibrant and rapidly advancing frontier, marked by open theoretical, numerical, and observational challenges and by progress on multiple fronts. By integrating theory, phenomenology, and observation, this review emphasizes the extraordinary power of gravitational waves to probe the Universe across a broad range of scales, granting access to energy regimes and physical processes far beyond the reach of conventional experiments. We hope this work will guide and inspire both newcomers and experienced researchers as gravitational-wave physics continues to open a transformative window onto the cosmos.

\section*{Acknowledgments}
\addcontentsline{toc}{section}{Acknowledgments}

First, I would like to express my gratitude to \textit{The Astronomy and Astrophysics Review} for the invitation to write this article. 
I owe particular thanks to Eiichiro Komatsu, whose encouragement and intellectual guidance not only sparked my enduring interest in the subject, but also greatly deepened my understanding of it. I am also grateful to Robert Brandenberger, Sadra Jazayeri, Eiichiro Komatsu, and Geraldine Servant for their valuable feedback and thoughtful suggestions, which significantly improved the clarity of this work. I thank Peera Simakachorn for helpful discussion and for providing the plot shown in \cref{fig:cosmo-sources-II}. I thank Niayesh Afshordi, Sabino Matarrese, Viatcheslav Mukhanov, Nicholas Rodd, Pedro Schwaller, and David Wands for their detailed comments on the first arXiv version and for suggesting additional references. I would further like to thank Dionysios Anninos, Daniel Baumann, Dmitri Kharzeev, Joachim Kopp, Nicholas Rodd, Andrew Strominger, and Alexander Zhiboedov, whose insights and discussions over the years have profoundly shaped my perspective on many aspects of this broad and evolving field. I am grateful to the Max Planck Institutes for Physics and for Astrophysics in Garching, and to Johannes Henn and Eiichiro Komatsu, for their kind hospitality during the initial stage of this work, as well as to DESY and Geraldine Servant for their warm hospitality during its final stage; both institutes provided an excellent environment for the completion of this manuscript. This work was supported by the Royal Society through a University Research Fellowship (Grant No. URF\textbackslash{}R1\textbackslash{}231782) and by the Deutsche Forschungsgemeinschaft under Germany’s Excellence Strategy (EXC 2121 Quantum Universe – 390833306).

\paragraph{Figures} Unless otherwise stated, all figures were prepared by the author. Any figures reproduced or adapted from other sources are credited in their captions.

%%\clearpage
%%%%%%%%%%%%%%%%%%%%%%%%%%%%%%%%%
% Bibliography

\bibliographystyle{sn-basic}
\bibliography{bib}

%%%%%%%%%%%%%%%%%%%%%%%%%%%%%%%%%

\end{document}